# High-precision $\alpha_s$ measurements from LHC to FCC-ee

Workshop Proceedings, CERN, Geneva, 12–13 October 2015

*Editors*

David d'Enterria (CERN), Peter Z. Skands (Monash)

*Authors*

S. Alekhin (U. Hamburg), A. Banfi (U. Sussex), S. Bethke (MPI, München), J. Blümlein (DESY), K.G. Chetyrkin (KIT, Karlsruhe), D. d'Enterria (CERN), G. Dissertori (ETH Zurich), X. Garcia i Tormo (Bern), A. H. Hoang (U. Wien), M. Klasen (U. Münster), T. Klijnsma (ETH Zurich), S. Kluth (T.U. München), J.-L. Kneur (U. Montpellier 2), B.A. Kniehl (U. Hamburg), D. W. Kolodrubetz (MIT), J. Kühn (KIT, Karlsruhe), P. Mackenzie (Fermilab), B. Malaescu (LPNHE, Paris), V. Mateu (U. Wien/U. A. Madrid), L. Mihaila (KIT, Karlsruhe), S. Moch (U. Hamburg), K. Mönig (DESY), R. Pérez-Ramos (Paris), A. Pich (U. València), J. Pires (U. Milano/MPP Munich), K. Rabbertz (KIT, Karlsruhe), G. P. Salam (CERN), F. Sannino (CP3-Origins, Odense), J. Soto i Riera (U. Barcelona), M. Srebre (U. Ljubljana), I. W. Stewart (MIT)

**Abstract**

This document provides a writeup of all contributions to the workshop on *"High precision measurements of $\alpha_s$: From LHC to FCC-ee"* held at CERN, Oct. 12–13, 2015. The workshop explored in depth the latest developments on the determination of the QCD coupling $\alpha_s$ from 15 methods where high precision measurements are (or will be) available. Those include low-energy observables: (i) lattice QCD, (ii) pion decay factor, (iii) quarkonia and (iv) $\tau$ decays, (v) soft parton-to-hadron fragmentation functions; as well as high-energy observables: (vi) global fits of parton distribution functions, (vii) hard parton-to-hadron fragmentation functions, (viii) jets in $e^{\pm}$p DIS and $\gamma$-p photoproduction, (ix) photon structure function in $\gamma$-$\gamma$, (x) event shapes and (xi) jet cross sections in $e^+e^-$ collisions, (xii) W boson and (xiii) Z boson decays, and (xiv) jets and (xv) top-quark cross sections in proton-(anti)proton collisions. The current status of the theoretical and experimental uncertainties associated to each extraction method, the improvements expected from LHC data in the coming years, and future perspectives achievable in $e^+e^-$ collisions at the Future Circular Collider (FCC-ee) with $\mathcal{O}\left(1-100\ \mathrm{ab}^{-1}\right)$ integrated luminosities yielding $10^{12}$ Z bosons and jets, and $10^8$ W bosons and $\tau$ leptons, are thoroughly reviewed. The current uncertainty of the (preliminary) 2015 strong coupling world-average value, $\alpha_s(m_Z^2) = 0.1177 \pm 0.0013$, is about 1%. Some participants believed this may be reduced by a factor of three in the near future by including novel high-precision observables, although this opinion was not universally shared. At the FCC-ee facility, a factor of ten reduction in the $\alpha_s$ uncertainty should be possible, mostly thanks to the huge Z and W data samples available.

This document is dedicated to the memory of Guido Altarelli.

Speakers

**A. Banfi (U. Sussex), S. Bethke (MPI, München), J. Blümlein (DESY), D. d'Enterria (CERN), X. Garcia i Tormo (Bern), A. Hoang (U. Wien), B.A. Kniehl (U. Hamburg), M. Klasen (U. Münster), S. Kluth (T.U. München), J.-L. Kneur (U. Montpellier 2), J. Kühn (KIT, Karlsruhe), P. Mackenzie (Fermilab), B. Malaescu (LPNHE, Paris) L. Mihaila (KIT), A. Mitov (Cambridge), K. Mönig (DESY), R.Pérez-Ramos (Paris), A. Pich (U. València), J. Pires (U. Milano, MPP Munich), K. Rabbertz (KIT, Karlsruhe), G. Salam (CERN), F. Sannino (CP3-Origins, Odense), P.Z. Skands (Monash), J. Soto i Riera (U. Barcelona), M. Srebre (U. Ljubljana)**

Additional Participants

**A. Ali (DESY), S. Amoroso (CERN), A. Blondel (U. Genève), M. González-Alonso (IPN, Lyon), C. Gracios (Puebla), K. Hamacher (Bergische Univ. Wuppertal), R. Hernández-Pinto (IFIC, València), P. Janot (CERN), M. Klute (MIT), I. Kolbe (U. Cape-Town), A. Larkoski (MIT/Harvard), J. Llorente-Merino (Univ. Autónoma Madrid), G. Luisoni (CERN), B. Meiring (U. Cape-Town), S. Menke (MPI, München), R. Morad (U. Cape-Town), A.N. Rasoanaivo (U. Cape-Town), P. Telles-Rebello (CBPF, Rio de Janeiro)**

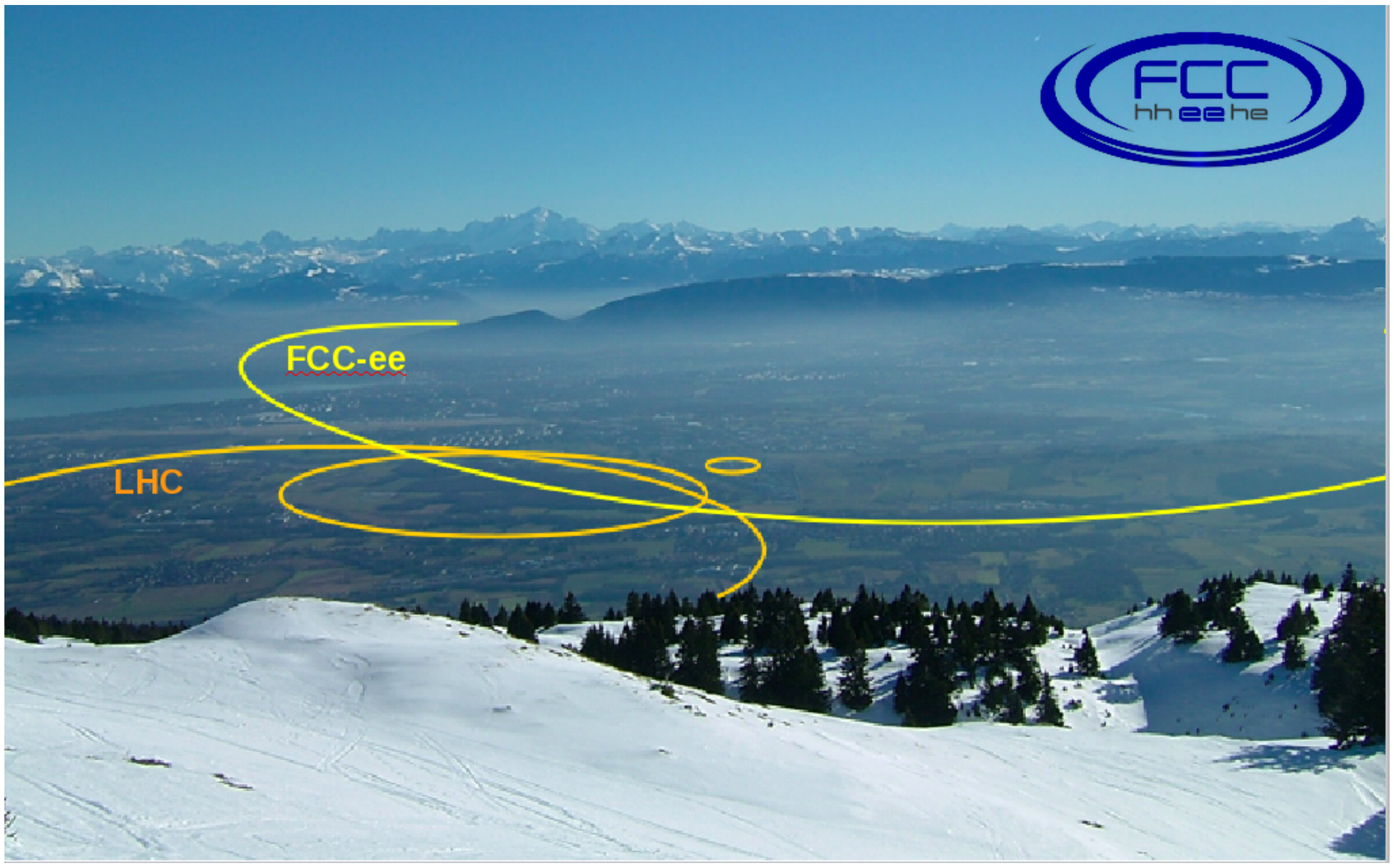

# 1 Introduction

The strong coupling $\alpha_s$ is one of the fundamental parameters of the Standard Model (SM), setting the scale of the strength of the strong interaction theoretically described by Quantum Chromodynamics (QCD). Its measured (2014) value amounts to $\alpha_s(m_Z^2) = 0.1185 \pm 0.0006$ at the reference Z pole mass scale. Given its current $\delta\alpha_s(m_Z^2)/\alpha_s(m_Z^2) \approx 0.5\%$ uncertainty—orders of magnitude larger than that of the gravitational ($\delta G/G \approx 10^{-5}$), Fermi ($\delta G_F/G_F \approx 10^{-8}$), and QED ($\delta\alpha/\alpha \approx 10^{-10}$) couplings—the strong coupling is the least precisely known of all fundamental constants in nature. Improving our knowledge of $\alpha_s$ is a prerequisite to reduce the theoretical uncertainties in the calculations of all high-precision perturbative QCD (pQCD) processes whose cross sections or decay rates depend on higher-order powers of $\alpha_s$, as is the case for virtually all those measured at the LHC. In the Higgs sector, in particular, the uncertainty on $\alpha_s$ is currently the second major contributor (after the bottom mass) to the parametric uncertainties of its dominant $H \to b\bar{b}$ partial decay. The same applies for the extraction of the charm Yukawa coupling via future $H \to c\bar{c}$ measurements.

The workshop *"High-precision $\alpha_s$ measurements from LHC to FCC-ee"* was held at CERN, October 12–13, 2015, as part of the FCC-ee *QCD and $\gamma$-$\gamma$ physics* working group activities in the context of the preparation of the FCC-ee Conceptual Design Report in 2016. The meeting brought together experts from several different fields to explore in depth the latest developments on the determination of the QCD coupling $\alpha_s$ from the key categories where high precision measurements are (or will be) available, and put its emphasis on the following issues:

- What is the current state-of-the-art of each one of the $\alpha_s$ determination methods, from the theoretical and experimental points of view?
- What is the current size of the theoretical (missing higher-order QCD and electroweak corrections, power corrections, hadronization corrections,...) and experimental uncertainties associated to each measurement?
- What is the expected $\alpha_s$ uncertainty in $\sim$10 years from now thanks to the ongoing (or expected) theoretical developments, plus $\mathcal{O}\left(1\,\mathrm{ab}^{-1}\right)$ collected p-p data at 14 TeV at the LHC?
- What are the improvements expected to be brought about by $e^+e^-$ collisions at the FCC-ee ($\sqrt{s} = 91$, 160, 240 and 350 GeV) with $\mathcal{O}\left(1 - 100\ \mathrm{ab}^{-1}\right)$ integrated luminosities yielding $10^{12}$ Z bosons and jets, and $10^8$ W bosons and $\tau$ leptons?
- What are the systematic errors that the FCC-ee detectors should target in order to match the expected statistical precision, or where that is not possible, what are the important theoretical targets that should be met or exceeded?

With those goals in mind, the workshop was organized along four broad sessions:

1. An introductory session, presenting the motivations of the workshop, the current status of the world average of the strong coupling, the impact of $\alpha_s$ on Higgs cross sections and branching ratios, and on new physics constraints.
2. A session dedicated to $\alpha_s$ determination at low energy including results from: lattice QCD, pion decay factor, $\tau$ decay, $Q\overline{Q}$ decays, and soft parton-to-hadron fragmentation functions.
3. A session dedicated to $\alpha_s$ determination at higher energy scales including: global fits of parton distribution functions, hard parton-to-hadron fragmentation functions, jets in deep-inelastic scattering and photoproduction in $e^{\pm}$-p collisions, $e^+e^-$ event shapes, $e^+e^-$ jets, hadronic Z and W decays, $\sigma(e^+e^- \to \mathrm{hadrons})$, and the SM electroweak fit,...

4. Recent experimental and theoretical results and plans for $\alpha_s$ measurements at the LHC via top-quark pair and jets cross sections.

One important goal of the workshop was to facilitate discussion between the different groups, and in particular to give speakers the opportunity to explain details that one would normally not be able to present at a conference, but which have an important impact on the analyses. There were about 50 physicists who took part in the workshop, and 24 talks were presented. Slides as well as background reference materials are available on the conference website

http://indico.cern.ch/e/alphas2015

The sessions and talks in the workshop program were organized as follows:

- Introduction
  - "Introduction and goals of the workshop", D. d'Enterria and P.Z. Skands
  - "World Summary of $\alpha_s$ (2015)", S. Bethke
  - "$\alpha_s$ and physics beyond the Standard Model", F. Sannino
  - "Impact of $\alpha_s$ on Higgs production and decay uncertainties", L. Mihaila
- Measurements of $\alpha_s$ at low energy scales:
  - "$\alpha_s$ from lattice QCD", P. Mackenzie
  - "$\alpha_s$ from the QCD static energy", X. Garcia i Tormo
  - "$\alpha_s$ from pion decay factor", J.-L. Kneur
  - "$\alpha_s$ from hadronic tau decays", A. Pich
  - "$\alpha_s$ from hadronic quarkonia decays", J. Soto i Riera
  - "$\alpha_s$ from soft parton-to-hadron fragmentation functions", R. Pérez-Ramos
- Measurements of $\alpha_s$ at high energy scales:
  - "$\alpha_s$ from global fits of parton distribution functions", J. Blümlein
  - "$\alpha_s$ from jets in DIS and photoproduction", M. Klasen
  - "$\alpha_s$ from scaling violations of hard parton-to-hadron fragmentation functions", B.A. Kniehl
  - "$\alpha_s$ from $e^+e^-$ event shapes", S. Kluth
  - "$\alpha_s$ from $e^+e^-$ C-parameter event shape", A. Hoang
  - "$\alpha_s$ from $e^+e^-$ jet cross sections", A. Banfi
  - "$\alpha_s$ from hadronic Z decays and from the full electroweak fit", K. Mönig
  - "$\alpha_s$ from hadronic W decays", M. Srebre
  - "$\alpha_s$ from $\sigma(e^+e^- \to \text{hadrons})$", J.H Kühn
- Measurements of $\alpha_s$ at the LHC and conclusions:
  - "$\alpha_s$ from top-pair cross sections at the LHC and beyond", A. Mitov
  - "$\alpha_s$ from top-pair cross sections at hadron colliders", G. Salam
  - "Future prospects of $\alpha_s$ from NNLO jets at the LHC and beyond", J. Pires
  - "$\alpha_s$ determinations from ATLAS (status and plans)", B. Malaescu
  - "$\alpha_s$ determinations from CMS (status and plans)", K. Rabbertz
  - "Worskhop summary and conclusions", D. d'Enterria

These proceedings represent a collection of extended abstracts and references for the presentations, plus a summary of the most important results and future prospects in the field. Contents of these proceedings will be incorporated into the FCC-ee Conceptual Design Report under preparation.

CERN, December 2015

David d'Enterria
Peter Skands

# 2 Proceedings Contributions

# World Summary of $\alpha_s$ (2015)

**Siegfried Bethke**[1], Günther Dissertori[2], and Gavin P. Salam[3,*]

[1]Max-Planck-Institut für Physik (Werner-Heisenberg-Institut), München, Germany
[2]Institute for Particle Physics, ETH Zurich, Switzerland
[3]CERN, PH-TH, CH-1211 Geneva 23, Switzerland

**Abstract:** This is a **preliminary** update of the measurements of $\alpha_s$ and the determination of the world average value of $\alpha_s(M_Z^2)$ presented in the 2013/2014 edition of the Review of Particle Properties [1]. A number of studies which became available since late 2013 provide new results for each of the (previously 5, now) 6 subclasses of measurements for which pre-average values of $\alpha_s(M_Z^2)$ are determined.

In the following, we list those new results which are used to determine the new average values of $\alpha_s$, i.e. which are based on at least complete NNLO perturbation theory and are published in peer-reviewed journals, as well as those which are used for demonstrating asymptotic freedom (although being based on NLO only)[†]:

- updated results from $\tau$-decays [2,3,4], based on a new release of the ALEPH data and on complete N$^3$LO perturbation theory,
- more results from unquenched lattice calculations, [5,6],
- more results from world data on structure functions, in NNLO QCD [7],
- results from $e^+e^-$ hadronic event shape (C-parameter) in soft collinear effective field theory (NNLO) [8],
- $\alpha_s$ determinations at LHC, from CMS data on the ratio of inclusive 3-jet to 2-jet cross sections [9], from inclusive jet production [10], from the 3-jet differential cross section [11], and from energy-correlations [12], all in NLO QCD, plus one determination in complete NNLO, from a measurement of the $t\bar{t}$ cross section at $\sqrt{s} = 7$ TeV [13];
- and finally, an update of $\alpha_s$ from a global fit to electro-weak precision data [14].

All measurements available in subclasses of $\tau$-decays, lattice results, structure functions, and $e^+e^-$-annihilation are summarized in Fig. 1. With the exception of lattice results, most results within their subclass are strongly correlated, however to an unknown degree, as they largely use the same data sets. The large scatter between many of these measurements, sometimes with only marginal or no agreement within the given errors, indicate the presence of additional systematic uncertainties from theory or caused by details of the analyses. In such cases, a pre-average value is determined with a symmetric overall uncertainty that encompasses the central values of all individual determinations ('*range averaging*'). For the subclass of lattice results, the average value determined in Ref. [5] is taken over. For the subclasses of hadron collider results and electroweak (ewk) precision fits, only one result each is available in full NNLO, so that no pre-averaging can be applied. Note, however, that more measurements of top-quark pair production at the LHC are meanwhile available, indicating that on average, a larger value of $\alpha_s(M_Z^2)$ is likely to emerge in the future [15]. The emerging subclass averages are plotted in Fig. 1, and summarized in Table 1.

*On leave from CNRS, UMR 7589, LPTHE, F-75005, Paris, France.

†Note that this does not fully account for all studies and results presented at this workshop, but rather reflects the restricted summary currently intended for the new edition of Ref. [1].

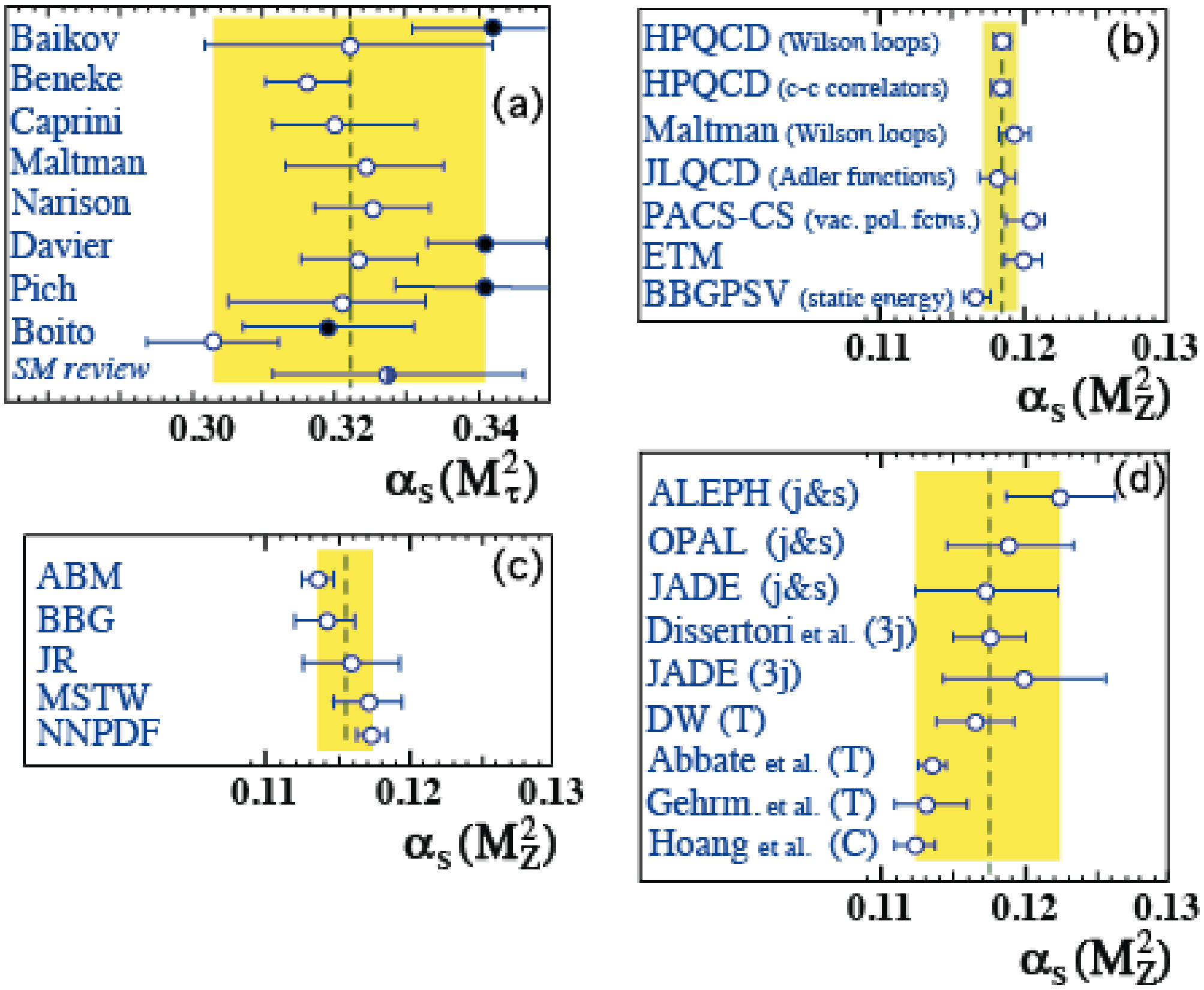

Figure 1: Summary of determinations of $\alpha_s$ from: (a) hadronic $\tau$-decays (full circles obtained using CIPT, open circles FOPT expansions, see text), (b) lattice calculations, (c) DIS structure functions, and (d) $e^+e^-$ annihilation. The shaded bands indicate the pre-average values explained in the text, to be included in the determination of the final world average of $\alpha_s$.

| Subclass | $\alpha_s(M_Z^2)$ |
|---|---|
| $\tau$-decays | $0.1187 \pm 0.0023$ |
| lattice QCD | $0.1184 \pm 0.0012$ |
| structure functions | $0.1154 \pm 0.0020$ |
| $e^+e^-$ jets & shapes | $0.1174 \pm 0.0051$ |
| hadron collider | $0.1151^{+0.0028}_{-0.0027}$ |
| ewk precision fits | $0.1196 \pm 0.0030$ |

Table 1: Pre-average values of subclasses of measurements of $\alpha_s(M_Z^2)$. The value from $\tau$-decays was converted from $\alpha_s(M_\tau^2) = 0.322 \pm 0.019$, using the QCD 4-loop $\beta$-function plus 3-loop matching at the charm- and bottom-quark pole masses.

Assuming that the resulting pre-averages are largely independent of each other, the final world average value is determined as the weighted average of the different input values. An initial uncertainty of the central value is calculated treating the uncertainties of all measurements as being uncorrelated and of Gaussian nature, and the overall $\chi^2$ to the central value is determined. If the initial value of $\chi^2$ is smaller than the number of degrees of freedom, an overall, a-priori unknown correlation coefficient is introduced and determined by requiring that the total $\chi^2$/d.o.f. equals unity. Applying this procedure to the values listed in Table 1 results in a **preliminary** new world average of

$$\alpha_s(M_Z^2) = 0.1177 \pm 0.0013 \; .$$

This value is in reasonable agreement with that from 2013/2014, which was $\alpha_s(M_Z^2) = 0.1185 \pm 0.0006$ [1], however at a somewhat decreased central value and with an overall uncertainty that has doubled. These changes are mainly due to the following reasons:

- the uncertainty of the lattice result, now taken from the estimate made by the FLAG group, is more conservative than that used in the previous review, leading to a larger final uncertainty of the new world average, and to a reduced fixing power towards the central average value;
- the decreased pre-average value of $\alpha_s(M_Z^2)$ from $\tau$-decays, due to the most recent re-evaluations and their unexplained, increased inconsistency with respect to each other;
- the relatively low value of $\alpha_s$ from the new sub-class of hadron collider results, which currently consists of only one measurement of the $t\bar{t}$ cross section at $\sqrt{s} = 7$ TeV, and which appears to be "lowish" if compared to further measurements at higher $\sqrt{s}$ [15].

Note that pending discussions about inclusion or exclusion of some of the most recent results, as well as refinements of the procedures applied may still change the final value of $\alpha_s(M_Z^2)$ and its assigned overall uncertainty. While there is still room for improved measurements and treatments of systematic uncertainties, the data and results, especially when including measurements which are available at NLO only, consistently demonstrate and prove asymptotic freedom and the running of $\alpha_s$, as predicted by QCD, up to energies beyond 1 TeV, see Fig. 2.

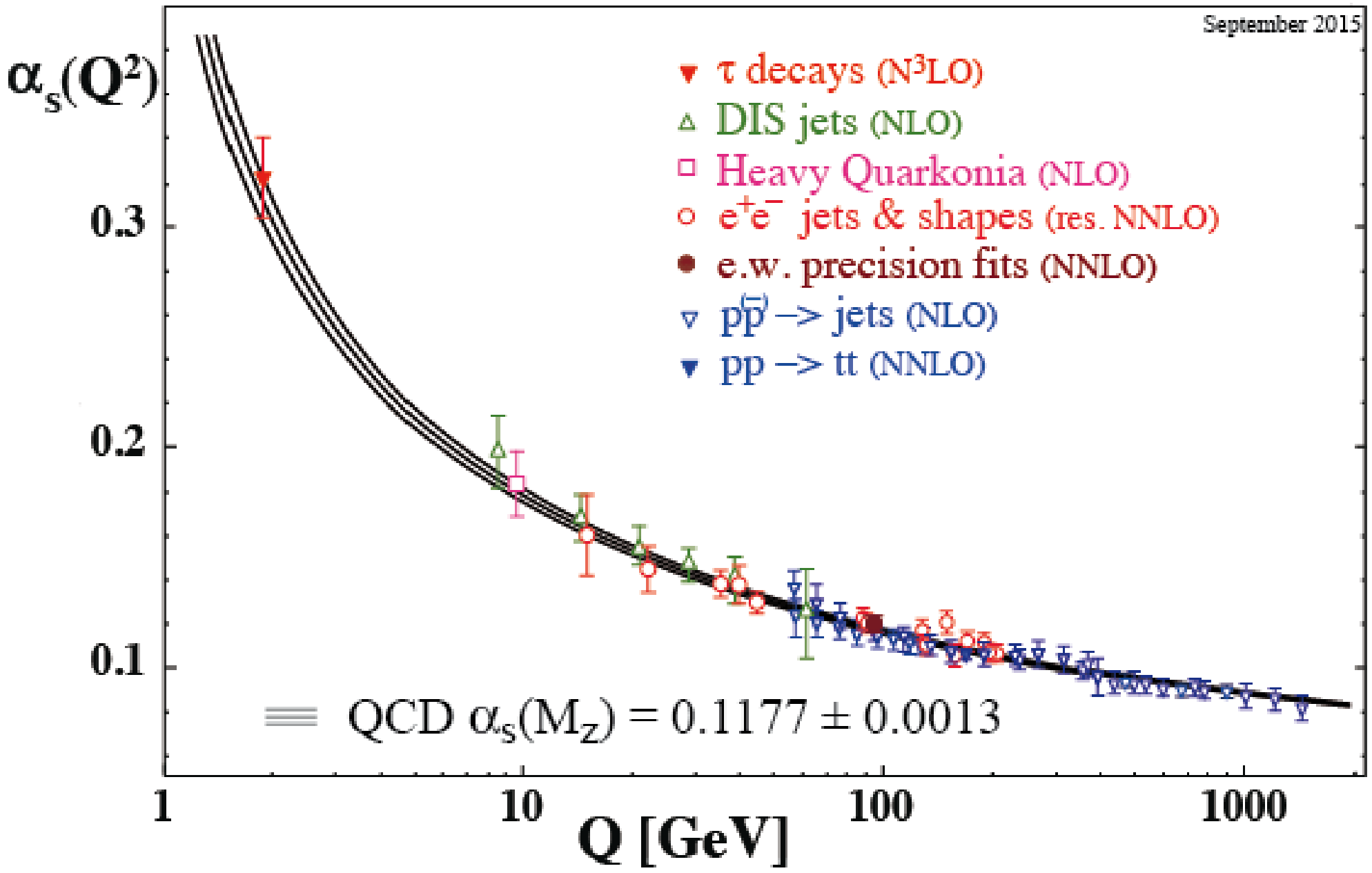

Figure 2: Summary of measurements of $\alpha_s$ as a function of the energy scale $Q$. The respective degree of QCD perturbation theory used in the extraction of $\alpha_s$ is indicated in brackets (NLO: next-to-leading order; NNLO: next-to-next-to leading order; res. NNLO: NNLO matched with resummed next-to-leading logs; N$^3$LO: next-to-NNLO).

# $\alpha_s$ at LHC: Challenging Asymptotic Freedom

Francesco Sannino

CP$^3$-Origins & Danish IAS, University of Southern Denmark

**Abstract:** Extensions of the standard model featuring new colored states are discussed together with the possibility that asymptotic freedom might be replaced by UV complete safe extensions of QCD at higher energies than explored so far. This motivates model-independent attempts to constraint new coloured states at the LHC, that are reviewed here.

## Introduction

Several extensions of the standard model feature new colored states that besides modifying the running of the QCD coupling could even lead to the loss of asymptotic freedom. Such a loss would potentially diminish the Wilsonian fundamental value of the theory. However, the recent discovery of complete asymptotically safe vector-like theories [1], i.e. featuring an interacting UV fixed point in all couplings, elevates these theories to a fundamental status and opens the door to alternative UV completions of (parts of) the standard model. If, for example, QCD rather than being asymptotically free becomes asymptotically safe there would be consequences on the early time evolution of the Universe (the QCD plasma would not be free). It is therefore important to test, both directly and indirectly, the strong coupling running at the highest possible energies. I will review here the attempts made in [2] to use pure QCD observables at the Large Hadron Collider (LHC) to place bounds on new colored states. Such bounds do not depend on the detailed properties of the new hypothetical states but on their effective number and mass. We will see that these direct constraints cannot exclude a potentially safe, rather than free, QCD asymptotic nature. A safe QCD scenario would imply that quarks and gluons are only approximately free at some intermediate energies, otherwise they are always in chains.

## The need to test QCD at higher energies

The standard model (SM) of particle interactions is an extremely successful theory at and below the Fermi scale. This scale is identified with the spontaneous breaking of the electroweak symmetry.

The mathematical structure of the SM contains a gauge sector associated to local invariance of the semi-simple group $SU(3) \times SU(2) \times U(1)$. As soon as an elementary scalar sector is introduced new accidental interactions emerge. These are the Yukawa interactions responsible for the flavour research program, and the Higgs scalar self interactions*. Accidental interactions are not associated to a gauge principle and their number and type is limited by global symmetries and the request of renormalisability. In four dimensions accidental symmetries are associated, at the tree level, to either relevant operators –from the infrared physics point of view– such as the Higgs mass term, or to marginal operators such as Yukawa interactions. Gauge sectors, on the other hand, lead only to marginal operators. These theories are known as Gauge-Yukawa theories and, especially after the discovery of the Higgs, it has become imperative to acquire a deeper understanding of their dynamics.

*The Higgs self-coupling is known when assuming the minimal SM Higgs realisation but it has not yet been directly measured.

One can further classify Gauge-Yukawa theories according to whether they admit UV complete (in all the couplings) fixed points. The presence of such a fixed point guarantees the fundamentality of the theory since, setting aside gravity, it means that the theory is valid at arbitrary short distances.

## Is QCD asymptotic free above the Fermi scale?

If the UV fixed point occurs for vanishing values of the couplings the interactions become asymptotically free in the UV [3,4]. The fixed point is approached logarithmically and therefore, at short distances, perturbation theory can be used. Asymptotic freedom is an UV phenomenon that still allows for several possibilities in the IR, depending on the specific underlying theory [5]. At low energies, for example, another interacting fixed point can occur. In this case the theory displays both long and short distance conformality. However the theory is interacting at short distances and the IR spectrum of the theory is continuous [6]. Another possibility that can occur in the IR is that a dynamical mass is generated leading to either confinement or chiral symmetry breaking, or both.

QCD does not possess an interacting IR fixed point because it generates dynamically a mass scale that can be, for example, read off any non-Goldstone hadronic state such as the nucleon or the vector meson $\rho$. Because, however, we have measured the strong coupling only up to sub-TeV energies [7] one cannot yet experimentally infer that QCD is asymptotically free. In fact it is intriguing to explore both theoretical and experimental extensions of the SM in which QCD looses asymptotic freedom at short distances. For example, asymptotic freedom can be lost by considering additional vector-like colored matter at or above the Fermi scale.

## Alternative safe QCD scenario

On the other hand, loosing asymptotic freedom would, at one loop level, unavoidably lead to the emergence of a Landau pole. A result that potentially diminishes the Wilsonian fundamental value of the theory. There is, however, another largely unexplored possibility. Namely that an UV interacting fixed point is re-instated either perturbatively or non-perturbatively. This equally interesting and safe UV completion of strong interactions would have far-reaching consequences when searching for UV complete extensions of the SM as well as cosmology. If experimentally true, in fact, it would radically change our view of fundamental interactions and profoundly affects our understanding of the early cosmological evolution of the universe (this is so since the primordial plasma would not be free at high temperatures).

That such an interacting UV fixed point can exists for nonsupersymmetric vector-like theory it has been recently established in [1,8]. Furthermore, no additional symmetry principles such as space-time supersymmetry [9] are required to ensure well-defined and predictive theories in the UV [10]. Instead, the fixed point arises dynamically through renormalisable interactions between non-Abelian gauge fields, fermions, and scalars, and in a regime where asymptotic freedom is absent. The potentially dangerous growth of the gauge coupling towards the UV is countered by Yukawa interactions, while the Yukawa and scalar couplings are tamed by the fluctuations of gauge and fermion fields. This has led to theories with "complete asymptotic safety", meaning interacting UV fixed points in all couplings [1]. This is quite distinct from the conventional setup of "complete asymptotic freedom" [11,12,13], where the UV dynamics of Yukawa and scalar interactions is brought under control by asymptotically free gauge fields; see [14,15] for recent studies.

It is also straightforward to engineer QCD-like IR behaviour in the theory investigated in [1], including confinement and chiral symmetry breaking. In practice one decouples, at sufficiently high energies, the unwanted fermions by adding mass terms or via spontaneous symmetry breaking in such a way that at lower energies the running of the gauge coupling mimics QCD. The use in [1] of the Veneziano limit of large number of colors and flavors was instrumental to prove the existence of the UV fixed point in all couplings of the theory within perturbation theory. Tantalising indications that such a fixed point exists nonperturbatively, and without the need of elementary scalars, appeared in [16], and they were further explored in [17,1]. Nonperturbative techniques are needed to establish the existence of such a fixed point when the number of colors and flavours is taken to be three and the number of UV light flavours is large but finite.

Interestingly the supersymmetric cousins of the theory investigated in [1] (technically super QCD with(out) a meson-like chiral superfield), once asymptotic freedom is lost, cannot be asymptotically safe [18,19]. The results were further generalised and tested versus a-maximisation [19].

One can envision several extensions of the SM that can lead to rapid changes in the running of the QCD coupling at and above the Fermi scale. Since I have shown that asymptotic freedom can be traded for asymptotic safety while leaving the fundamental properties of QCD untouched, it becomes crucial to test the high energy behaviour of strong interactions. In Figure 1 I present a cartoon version of how the QCD running coupling could look when going from the IR to the UV. Of course the final asymptotically safe value, to be reached below the Planck scale, does not have to be large.

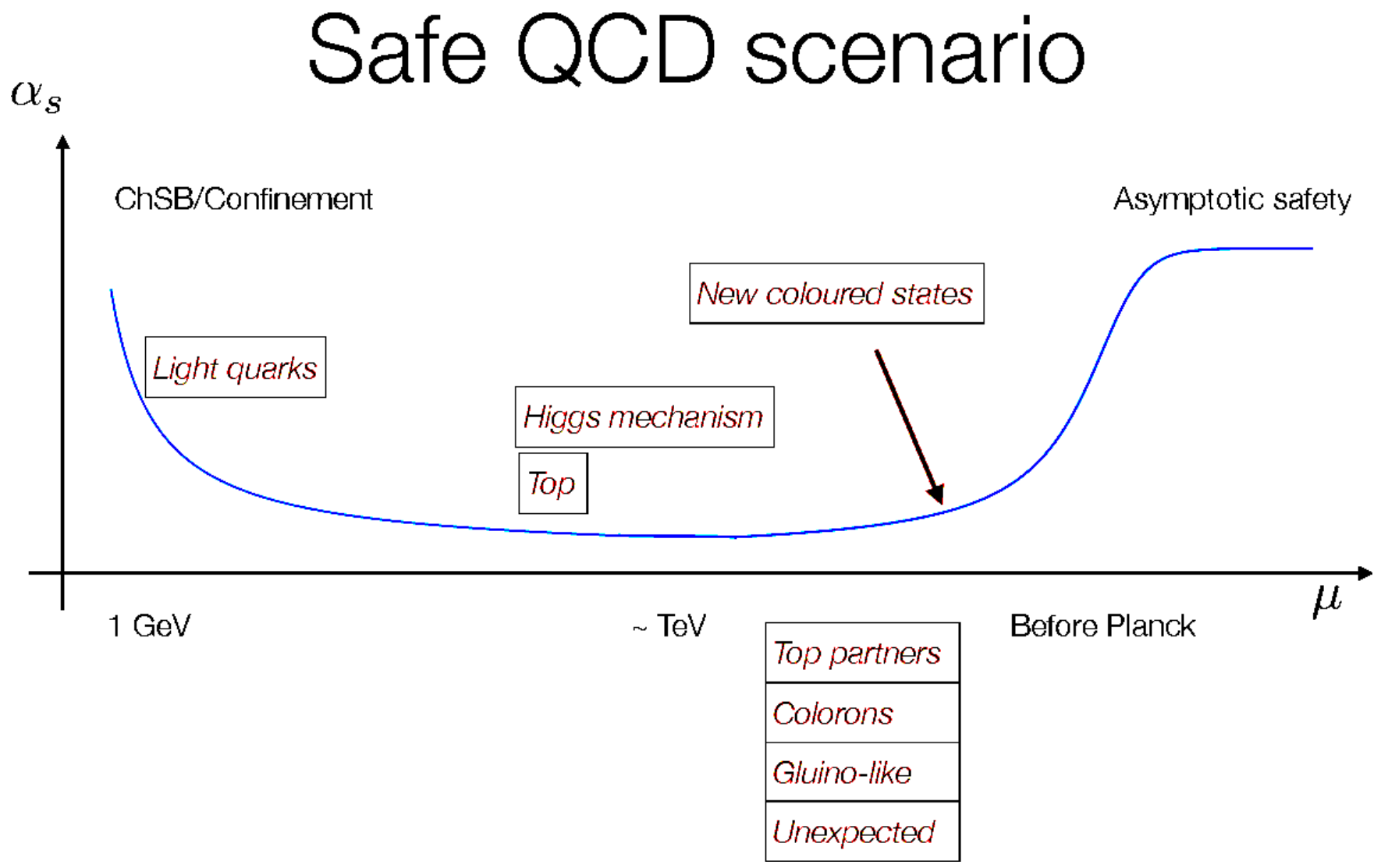

Figure 1: Asymptotically-safe scenario of QCD expressed in terms of the running of $\alpha_s$.

Although it would be extremely interesting to consider indirect constraints coming from cosmological and/or high energy astrophysical observations ranging from cosmic rays to compact objects one can investigate direct constraints coming from future (and current) LHC experiments. These, as we shall see, can help setting model-independent bounds on the effective number of new colored states around the Fermi scale.

## Constraining new colored matter at the LHC

The LHC experiments are, in fact, already probing the evolution of the strong coupling $\alpha_s$ up to the TeV scale. It was showed in [2] how the ratio of 3- to 2-jets cross sections is affected by the presence of new physics and argued that it can be used to place model-independent bounds on new particles carrying QCD color charge. It was also argued that such states need to be heavier than a few hundred GeVs. Stronger constraints can be derived model by model but, in this case, the results are not general.

In [20] the first determination of the strong coupling $\alpha_s(M_Z)$ from measurements of momentum scales beyond 0.6 TeV was presented. This determination was performed studying the behaviour of the ratio $R_{32}$ of the inclusive 3-jet cross section to the 2-jet cross section, defined in greater detail below. The result is in agreement with the world average value of $\alpha_s(M_Z)$.[†]

In [2] it has been argued that it is possible to constrain the presence of new colored states using such a measurement that probes quantum chromodynamics (QCD) at harder scales. It was also pointed out that such constraints should be taken with the grain of salt because of concerns regarding the validity of the interpretation given in the experimental analyses which warrants further studies.

Rather than dueling on the validity of the experimental analysis the focus in [2] has been on the large potential value of such an observable for placing bounds on new physics beyond the SM. In the process one gains insight related to the presence of new colored particles. For example, it was shown that their effect on the parton distribution functions is negligible. This is true, at least, when constructing ratios of cross-sections. Therefore in the absence of clear final states observables it was shown that the presence of new colored particles appears directly in the running of $\alpha_s$.

This approach provides complementary information with respect to typical direct limits, where several assumptions are made to specify production and decay of a given new colored particle. For instance if the new particles have the required quantum numbers, searches for di-jet resonances are particularly constraining [22], while there are models evading these bounds for which the results presented here may be relevant [23]. Furthermore the impact of $\alpha_s$ running only depends on the mass of the new states and on their color representation (and number). It is in this sense that an exclusion bound from such a measurement is, to a good approximation, model independent.

Efforts to constrain light colored states in the same spirit appeared, for example, in [24]. Here the effects of a gluino-like state on the global analysis of scattering hadron data were considered, while in [25] model-independent bounds on new colored particles were derived using event shape data from the LEP experiments. Finally, this type of approach generalises to other sectors of the SM, and the electroweak sector could for instance be constrained from measurements of Drell-Yan processes at higher energies [26].

Observables involving a low inclusive number of hard jets constitute ideal candidates to test QCD at the highest possible energy scales and therefore we focus on the ratio of 3- to 2-jets (differential) cross sections, $R_{32}$ [27,28,29,30,31,20]. Following CMS [20]:

$$R_{32}\left(\langle p_{T1,2}\rangle\right) \equiv \frac{\mathrm{d}\sigma^{n_j\geq 3}/\mathrm{d}\,\langle p_{T1,2}\rangle}{\mathrm{d}\sigma^{n_j\geq 2}/\mathrm{d}\,\langle p_{T1,2}\rangle}, \tag{1}$$

where $\langle p_{T1,2}\rangle$ is the average transverse momentum of the two leading jets in the event,

$$\langle p_{T1,2}\rangle \equiv \frac{p_{T1}+p_{T2}}{2}. \tag{2}$$

Other choices are possible regarding the kinematic variable [31].

In the original work [2] we considered the CMS analysis based on 5 fb$^{-1}$ of data collected at 7 TeV centre-of-mass energy [20]. Jets were defined requiring transverse momenta of at least 150 GeV and rapidities less than 2.5, using the anti-$k_T$ algorithm [32] with size parameter R = 0.7 and E-recombination scheme.

[†]Even more recent measurements of $\alpha_s$ at high energy scales have appeared after completion of this project, see [21] or other contributions to this report.

The computations for inclusive multijet cross sections include the next-to-leading order corrections in $\alpha_\mathrm{s}$ and $\alpha_W$ [33,34,35,36].[‡] NLO QCD corrections are implemented in NLOJet++ [38], that allows to evaluate the 3- and 2-jets cross sections at the parton-level within the Standard Model. The problem lies in the definition of the factorisation and the renormalisation scales identified with $\langle p_{T1,2}\rangle$ in the theoretical calculations presented by CMS. Since 3-jet events involve multiple scales, this simplified assignment may not represent the dynamics at play appropriately enough to allow a straightforward interpretation of the experimental data as a measurement of $\alpha_\mathrm{s}$ at $\langle p_{T1,2}\rangle$; the observable may be mainly sensitive to the value of the strong coupling at some fixed lower scale. Although the ideas presented here hinge on a resolution of this issue, finding the proper redefinition or reinterpretation of $R_{32}$ goes beyond the original scope of [2].

Hypothetical new colored particles can contribute to $R_{32}$ through a modification of the running of $\alpha_\mathrm{s}$ and of the PDFs, and as additional contributions to the partonic cross section at leading or next-to-leading order. It was argued in [2] that the most important of these effects is the change in $\alpha_\mathrm{s}$ and that the correspondence between $R_{32}$ and the strong coupling constant is reliable, even in the presence of new physics. The reader will find in [2] an in depth description of the validity and caveats of the approach. In general, new states may contribute at tree-level to the jet cross sections if their quantum numbers allow it. Although this would lead to important modifications in $R_{32}$ these contributions may partially cancel, the same way NLO corrections do as shown in the [2]. Notice that the colored states are stable, a new, heavy fermion in the final state could in principle be misidentified as a jet, since it would hadronise and end somewhere in the detector. There are however stringent constraints on the existence of such bound states [39].

Let's now consider the running of $\alpha_\mathrm{s}$when new colored fermions appear at high energies and therefore write the associated $\beta$ function [2]

$$\beta(\alpha_\mathrm{s}) \equiv \mu\frac{\partial\alpha_\mathrm{s}}{\partial\mu} = -\frac{\alpha_\mathrm{s}^2}{2\pi}\left(b_0 + \frac{\alpha_\mathrm{s}}{4\pi}\,b_1 + \ldots\right), \tag{3}$$

then the coefficients $b_0$ and $b_1$ in any mass-independent renormalisation scheme read

$$b_0 = 11 - \frac{2}{3}n_f - \frac{4}{3}n_X T_X, \tag{4}$$

$$b_1 = 102 - \frac{38}{3}n_f - 20\,n_X T_X\left(1 + \frac{C_X}{5}\right), \tag{5}$$

where $n_f$ is the number of quark flavours (i.e. $n_f = 6$ at scales $Q > m_t$), $n_X$ the number of new (Dirac) fermions, and $T_X$ and $C_X$ group theoretical factors depending in which representation of the color group the new fermions transform. Since the adjoint representation is real –like the gluino in the MSSM– a Majorana mass term can be written for a single Weyl fermion, and $n_X$ can take half-integer values. At leading order, the modification in the running of $\alpha_\mathrm{s}$ only depends on a single parameter $n_\mathrm{eff} \equiv 2n_X T_X$, counting the effective number of new fermions

$$n_\mathrm{eff} = n_{\mathbf{3}\oplus\overline{\mathbf{3}}} + 3\,n_{\mathbf{8}} + 5\,n_{\mathbf{6}\oplus\overline{\mathbf{6}}} + 15\,n_{\mathbf{10}\oplus\overline{\mathbf{10}}}, \tag{6}$$

where $n_{\mathbf{3}\oplus\overline{\mathbf{3}}}$, $n_{\mathbf{6}\oplus\overline{\mathbf{6}}}$ and $n_{\mathbf{10}\oplus\overline{\mathbf{10}}}$ are the number of new Dirac fermions in the triplet, sextet and decuplet representations respectively, and $n_{\mathbf{8}}$ the number of Weyl fermions in the adjoint representation. Asymptotic freedom is lost for $n_\mathrm{eff} > 10.5$. In view of our initial motivation we do not restrict ourselves to asymptotically free theories.

[‡]Recent progress, using new unitarity-based techniques, will allow for complete NNLO results in a near future [37].

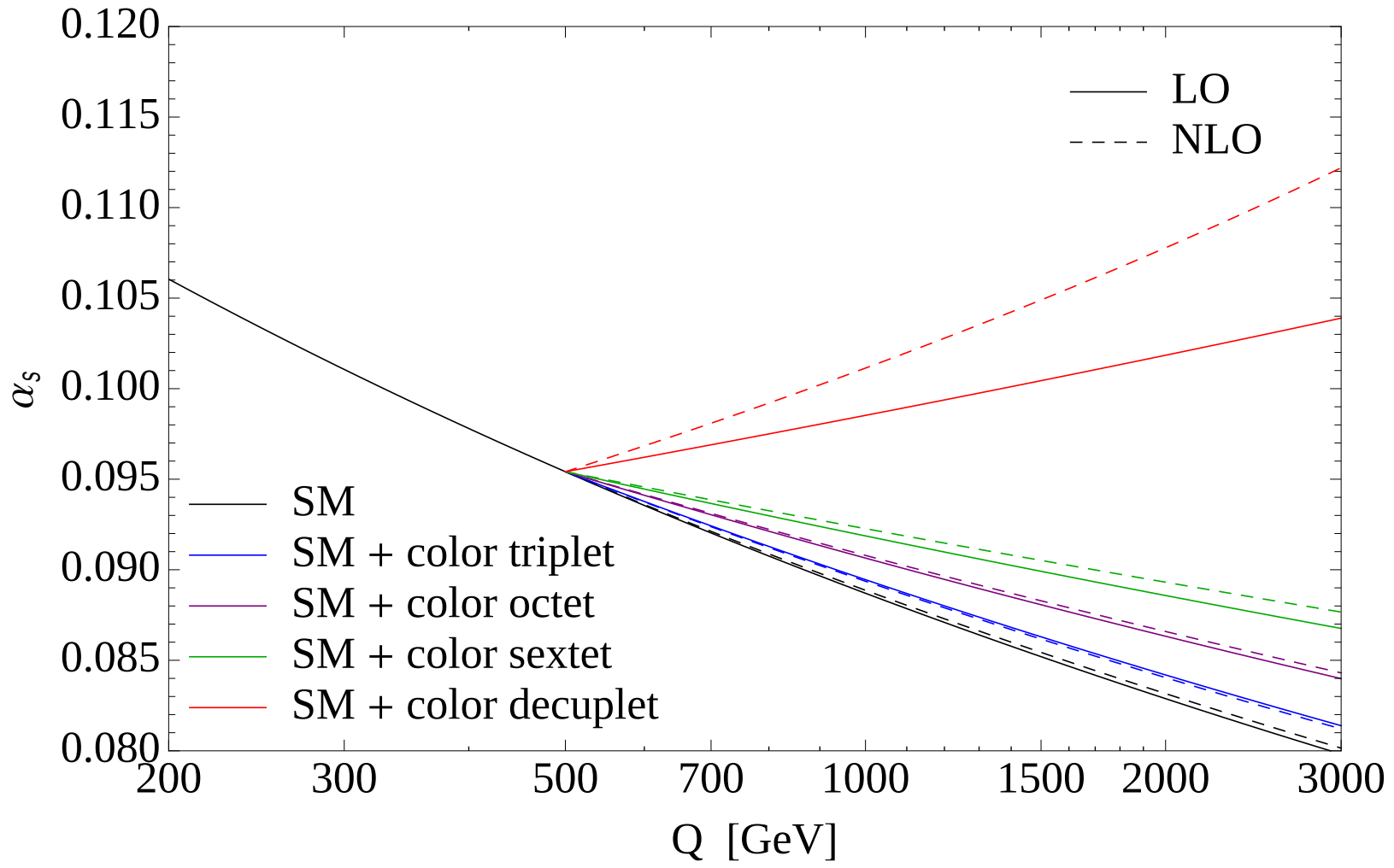

Figure 2: Example of the change in $\alpha_s$ induced by a new fermion of mass 500 GeV in various representations of the color gauge group. The running of $\alpha_s$ is performed at NLO, showing for comparison the running at LO from the mass of the new fermion.

| $Q$ [GeV] | $\alpha_s^{exp}(Q) \pm \sigma(Q)$ |
|---|---|
| 474 | $0.0936 \pm 0.0041$ |
| 664 | $0.0894 \pm 0.0031$ |
| 896 | $0.0889 \pm 0.0034$ |

Table 1: High-scale determinations of $\alpha_s$ from measurements of $R_{32}$ by CMS [20].

Furthermore, one Dirac fermion corresponds to four complex scalar degrees of freedom; scalar particles in the spectrum thus contribute to $n_{\text{eff}}$ four times less than corresponding Dirac fermions. For instance, the full content of the Minimal Supersymmetric Standard Model (1 adjoint Weyl fermion and 12 fundamental complex scalars) counts as $n_{\text{eff}} = 6$.

The running of $\alpha_s$ as given by the $\beta$ function above is only valid at energies larger than the mass of the new colored fermions –for simplicity, we assumed that they all have the same mass $m_X$ and that they are heavier than the top quark. One can now match $\alpha_s$ between the high-energy regime and the effective description without the new fermions at $m_X$.

The relative change in $\alpha_s$ induced by fermions in various representation can be assessed from Fig. 2. It is also useful to introduce the approximate expression [2]:

$$\frac{\alpha_s(Q)}{\alpha_s^{SM}(Q)} \approx 1 + \frac{n_{\text{eff}}}{3\pi}\alpha_s(m_X)\log\left(\frac{Q}{m_X}\right), \quad \text{for } Q \geq m_X, \tag{7}$$

where $\alpha_s^{SM}(Q)$ is the Standard Model value of the running coupling. The reader will find the detailed investigation of the impact of the new colored fermions on the PDFs in [2]. It is sufficient here to say that, at the precision level of the analysis, the modification of PDFs can thus be safely neglected for $R_{32}$. To illustrate the exclusion potential of high-scale measurements of $\alpha_s$ the bounds were presented for $n_{\text{eff}}$ depending on the scale of new physics $m_X$, and using CMS [20].

We assume the CMS estimates of $\alpha_s$ reproduced in Table 1) and further add to the analysis the world average measurement of the strong coupling $\alpha_s(M_Z) = 0.1185 \pm 0.0006$ [40]; since its uncertainty is much smaller than the ones of the other data points, we take as fixed input $\alpha_s(M_Z) = 0.1185$.

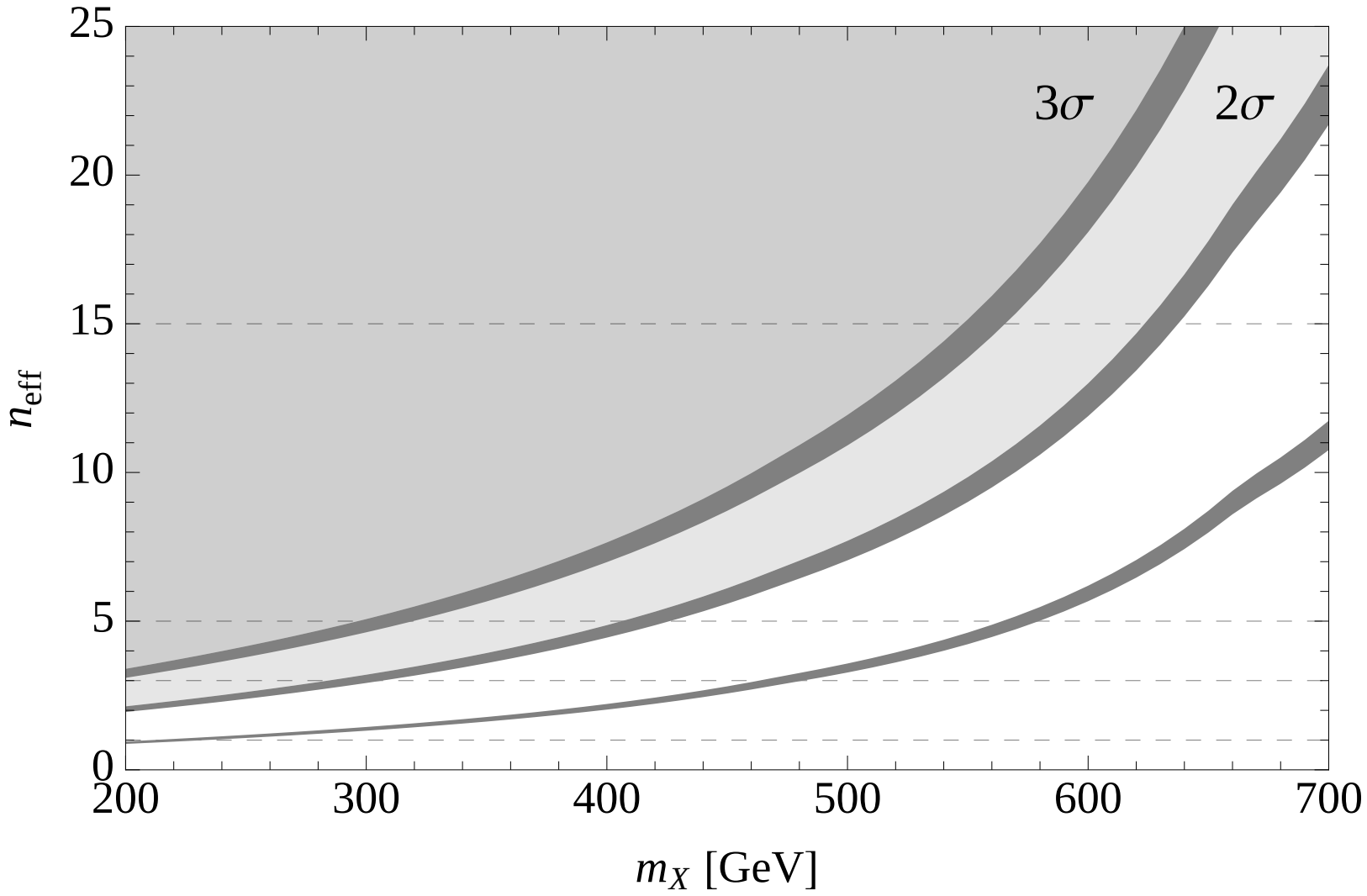

Figure 3: Upper bounds on $n_{\text{eff}}$ at $2\sigma$ and $3\sigma$ confidence levels (shaded regions), assuming the scale of new physics $m_X$ is known. They are delimited by grey bands whose width show the effect of varying the Casimir $C_X$. As further indication, the third band shows a $1\sigma$ limit. To guide the eye, the dashed horizontal lines indicate values of $n_{\text{eff}}$ corresponding to one fundamental, one adjoint, one two-index symmetric and one three-index symmetric fermion, see Eq. ((6)).

| color content | $n_{\text{eff}}$ | $m_X$ in GeV |
|---|---|---|
| Gluino | 3 | 280 |
| Dirac sextet | 5 | 410 |
| MSSM | 6 | 450 |
| Dirac decuplet | 15 | 620 |

Table 2: 95% CL mass exclusion bounds for various values of $n_{\text{eff}}$ according to a toy-analysis of the latest CMS measurement of $R_{32}$ [20].

The induced probability measure over the parameter-space to constrain is proportional to [2]

$$\exp\left[-\frac{1}{2}\sum_{Q}\left(\frac{\alpha_{\text{s}}^{exp}(Q)-\alpha_{\text{s}}^{th}(Q;n_{\text{eff}},m_X)}{\sigma(Q)}\right)^2\right]\times\text{priors}, \tag{8}$$

where $\alpha_{\text{s}}^{th}(Q;n_{\text{eff}},m_X)$ is the theoretical prediction for the value of the strong coupling at the scale $Q$, which is a function of $n_{\text{eff}}$ and $m_X$. The theoretical predictions for $\alpha_{\text{s}}$ are obtained by running up to $Q$ from the $Z$-mass at two-loop order, as described in Eq. ((3)), which is sufficient for our purpose. Beyond leading-order, $n_{\text{eff}}$ is not enough to parametrise the importance of new physics effects: the quadratic Casimir $C_X$ needs to be specified. In [2] its valued was varied between 4/3 and 6 –the values corresponding to fermions in the fundamental or decuplet representations, respectively– to demonstrate that it has little relevance. Assuming that the mass of the new states is known the upper bound on $n_{\text{eff}}$ is shown in Fig. 3. It is useful, following [2], to represent the exclusion potential of currently available experimental data as shown in Table 2.

### Conclusions

I first motivated the need to test QCD beyond the asymptotically free paradigm and then argued, based on the findings in [2], that pure QCD observables can help placing interesting bounds on new physics beyond the Standard Model. Such bounds are insensitive to the detailed properties of the new hypothetical states and mainly depend on their effective number $n_{\rm eff}$ and mass. Although these limits on colored particles might not be the most stringent for any specific model, are nevertheless unavoidable because of their model-independent nature. As an explicit example the ratio of 3- to 2-jets inclusive differential cross sections $R_{32}$ has been discussed along with the associated caveats [2]. With LHC running at almost double the centre-of-mass energy one expects that higher mass exclusion bounds will be available. The relative simplicity of the analysis suggested in [2] allows to swiftly extract limits on new physics as new data becomes available. It is clear from the above that direct constraints cannot exclude a potentially safe, rather than free, QCD asymptotic behavior. If the alternative safe QCD scenario were true quarks and gluons are never entirely free.

**Acknowledgments.** It is a pleasure for me to thank Celine Boehm, Richard Brower, Lance Dixon, Ken Intriligator, Klaus Rabbertz and Natascia Vignaroli for interesting discussions. I am indebted to Diego Becciolini, Marc Gillioz, Ken Intriligator, Daniel Litim, Matin Mojaza, Marco Nardecchia, Claudio Pica and Michael Spannowsky for collaborating on some of the results and ideas partially summarised here. I learnt a great deal from collaborating with them. I would like to note that the idea of a potential UV completion of QCD that is asymptotically safe rather than free is, to the best of my knowledge, original and it is not contained in references [1,2]. Last but not the least I thank David d'Enterria and Peter Z. Skands for having organised this topical and relevant workshop.

# Impact of $\alpha_s$ on Higgs production and decay uncertainties

Luminita Mihaila

Institut für Theoretische Physik, Universität Heidelberg, 69120 Heidelberg, Germany

**Abstract:** The impact of the parametric uncertainties induced by the strong coupling constant $\alpha_s$ on predictions of the Higgs boson production cross sections and decay rates are discussed.

## Introduction

After the discovery of the Higgs boson at the Large Hadron Collider (LHC), the central question about the particle discovered at 125.09 GeV [1] is whether this is "the Higgs boson" of the Standard Model (SM) or only one degree of freedom of a bigger theory. In general, additional degrees of freedom in the Higgs sector are expected to cause deviations in the Higgs couplings relative to SM predictions. Currently, the measured couplings are consistent with the SM expectations within uncertainties [2,3]. Therefore, it is expected that New Physics effects, if any, are comparable in size with the present theoretical and experimental uncertainties. In consequence, any conclusion derived from such analyses depends crucially on the magnitude of theoretical and parametric uncertainties entering the various calculations.

Furthermore, the measurement of Yukawa couplings to the top-, bottom- and charm-quarks with a precision at the few-percent level is one of the main tasks for the run II at the LHC. The necessity of precise knowledge of these couplings is obvious: i) it allows us to study the evolution of the SM potential up to the Planck scale and answer the question about the SM vacuum stability; ii) it is an essential ingredient for the determination of the Higgs self couplings; iii) the direct experimental measurement of the top-Yukawa coupling will allow us to distinguish New Physics effects in the gluon fusion channel; iv) deviations of Yukawa couplings w.r.t. the SM expectations can be used to constrain many New Physics models. This situation initiated a new era of precision analyses focused on the Higgs sector. During the last decade enormous theoretical and experimental efforts (for details see [4,5]) were devoted to the precise theoretical predictions and experimental measurements of the Higgs production cross sections and decay rates, the two main ingredients entering the determination of the Higgs couplings.

In this context, the goal of the present contribution is to study the impact of the parametric uncertainties induced by the strong coupling constant $\alpha_s$ on predictions of the Higgs boson production cross sections and decay rates. As it will be shown in the following, $\alpha_s$ uncertainty is for many analyses the limiting or one of the limiting factors as far as precision is concerned. For the shake of simplicity, the present study alludes only the SM as underlying theory. However, the main conclusions will remain valid also for other models that are compatible with the current experimental bounds and that do not predict unnatural enhancements of the QCD predictions.

## Higgs decay rates

Let us briefly review the current situation for the SM Higgs decay rates. The decay channels that are mostly affected by the $\alpha_s$ knowledge are the hadronic ones: $H \to b\bar{b}$, $H \to gg$, $H \to c\bar{c}$ and $H \to \gamma\gamma$. The other channels become sensitive to $\alpha_s$ only through mixed electroweak (EW)-QCD radiative corrections, that are however subdominant and lay well below the current theoretical uncertainties. Taking into account the newest calculations of the radiative corrections to $H \to b\bar{b}$ of order $\mathcal{O}(\alpha_s^4)$ [6], $\mathcal{O}(\alpha_s^5)$ [8] and $\mathcal{O}(\alpha\alpha_s)$ [9], the theoretical uncertainties reduce to below half a percent. This theoretical precision is sufficient w.r.t. the percent level experimental accuracy foreseen to be reached at a future lepton collider [5] and well below the $5-7\%$ expected at the LHC in run II. At this point, it is important to mention that the perturbative series converge very rapidly from 20% at next-to-leading order (NLO) in QCD to 4% at next-to-next-to-leading-order (NNLO) and further to 0.34% and $-0.1\%$ at N$^3$LO and N$^4$LO, respectively. The same observation holds also for the EW corrections, for which the numbers read $-1\%$ at NLO and $-0.3\%$ for mixed QCD-EW corrections. For the hadronic channel $h \to gg$ although the N$^3$LO are known [7], an overall theoretical uncertainty of about 3% is still present. The radiative corrections to the partial decay width are very large and amount to about 65% at NLO, 20% at NNLO and to 2% at N$^3$LO. The Higgs decay to two gluons is not (directly) accessible at the LHC, but the prospects for its measurement at an $e^+e^-$ Higgs factory indicate a precision of about 1% [5]. As it is also the case for the Higgs production in the gluon-fusion channel, the perturbative series for $H \to gg$ is not rapidly converging and the remnant scale dependence even at N$^3$LO is still significant, pointing to possible sizeable N$^4$LO contributions. Finally, for the last hadronic decay channel $H \to c\bar{c}$ is theoretically also well under control, similarly with the case of the $H \to b\bar{b}$ channel. The above discussion is summarized in Table 1, where the theoretical uncertainties due to QCD and EW radiative corrections for the hadronic Higgs decays are shown. This table is an update of a similar one present in Ref. [4].

| **Partial Width** | **QCD** | **Electroweak** | **Total** |
|---|---|---|---|
| $\mathrm{H} \to \mathrm{b\bar{b}/c\bar{c}}$ | $\sim 0.1\%$ | $\sim 0.3\%$ | $\sim 0.3\%$ |
| $\mathrm{H} \to \mathrm{gg}$ | $\sim 3\%$ | $\sim 1\%$ | $\sim 3\%$ |
| $\mathrm{H} \to \gamma\gamma$ | $< 1\%$ | $< 1\%$ | $\sim 1\%$ |

Table 1: Theoretical uncertainties for the Higgs hadronic and photonic decay rates, taking into account the most precise calculations currently available as described in text.

At this point of the discussion it is in order to study the magnitude of the parametric uncertainties induced by the strong coupling constant and the bottom- and charm-quark masses on the Higgs decay rates. For the numerical analysis the input values are $\alpha_s(M_Z) = 0.1184 \pm 0.002$, $m_c(m_c) = 1.279 \pm 0.013$ GeV and $m_b(m_b) = 4.163 \pm 0.016$ GeV. They are taken from Refs. [14] and [15], respectively. It is important to mention that a rather conservative choice for the uncertainty $\Delta\alpha_s$ was made, that render the present study compatible with the newest determination of $\alpha_s$ [12]. As can be read from Table 2, the uncertainty induced by the strong coupling constant is the second in size for $H \to b\bar{b}$ and the dominant one for $H \to c\bar{c}$ and $H \to gg$. They are in general a factor two to four larger than the prospects for the experimental uncertainties at a future lepton collider and compatible with the expected precision for the LHC.

## Higgs production cross sections

In the remaining part of this contribution the focus is on the impact of $\alpha_s$ on the uncertainty of the Higgs production cross section. As the Higgs production channel are collider dependent we start with a brief review on the current status at the LHC. The dominant Higgs production mode at the LHC, the gluon fusion channel, receives large radiative corrections that requires even the fourth term in the perturbative expansion. This production channel is probably the most studied one in the last decade, with the final conclusion summarized in the newest analysis [13] that translates into an amazing reached theoretical precision of $[-2.6\%; +0.32\%]$. Except for the associated $t\bar{t}H$ production that is still affected by theoretical uncertainties of about 10%, the other production modes are known at least at the percent level. The explicit numbers can be found in Table 3, that is an update of a similar one present in Ref. [4]. The total cross sections for the gluon fusion channel were obtained with the help of the computer code `SusHi` [10], that takes into account NNLO contributions. Similar results are delivered also by the other code designed for gluon fusion analysis, namely `HIGLU` [11]. The numerical results were obtained for the renormalization scale $\mu_r = M_H/2$, a convenient scale choice that renders the fixed order calculations very close to the resummed results and minimize the effect of higher order contributions [13].

| **Channel** | $M_{\rm H}$[GeV] | $\Delta\alpha_{\rm s}$ | $\Delta m_b$ | $\Delta m_c$ |
|---|---|---|---|---|
| $\rm H \to b\bar{b}$ | 126 | ± 0.4 % | ± 0.8% | |
| $\rm H \to c\bar{c}$ | 126 | ± 3.9 % | | ± 2.3 % |
| $\rm H \to gg$ | 126 | ± 4.1 % | | |

Table 2: Parametric uncertainties of the hadronic Higgs decay rates induced by $\alpha_{\rm s}$, $m_b$ and $m_c$.

For the study of the parametric uncertainties, two correlated sources have to be taken into account: the uncertainties induced by the parton density functions (PDFs) and the strong coupling constant. For the numbers given in the next-to-last column of Table 3, the recommendations of Ref. [16] concerning the PDFs variations between various groups were followed. For the numerical results in the last column of Table 3, $\alpha_{\rm s}$ was varied in the range $[0.117; 0.119]$. As can be read from the table, the uncertainty induced by the PDFs and $\alpha_{\rm s}$ is currently a factor three larger than the scale variation in gluon-fusion process. This is a completely new situation that requires combined efforts to improve PDFs predictions. The $\alpha_{\rm s}$ uncertainty alone is at present comparable in size with the theoretical uncertainties for all production channels at the LHC. As the theoretical predictions are currently known at least at the NLO accuracy, the present ratio between theoretical and $\alpha_{\rm s}$ induced uncertainties will probably remain valid for the next few years, until new results on the theory side will become available.

| **Process** | **Cross section(pb)** | **Scale**(%) | | **PDF** $+\alpha_{\rm s}$ | $\delta\alpha_{\rm s}$(%) |
|---|---|---|---|---|---|
| ggH | 49.87 | -2.61 | + 0.32 | -6.2 +7.4 | ± 3.7 |
| VBF | 4.15 | -0.4 | + 0.8 | ± 2.5 | ± 0.7 |
| WH | 1.474 | -0.6 | + 0.3 | ± 3.8 | ± 0.9 |
| ZH | 0.863 | -1.8 | + 2.7 | ± 3.7 | ± 0.9 |
| ttH | 0.611 | -9.3 | + 5.9 | ± 8.9 | ± 3.0 |

Table 3: Theoretical uncertainties estimated by scale variation and parametric uncertainties due to PDFs and $\alpha_{\rm s}$ for the Higgs production cross sections in pp collisions at $\sqrt{\rm s} = 14$ TeV at the LHC.

At lepton colliders, the impact of $\alpha_{\rm s}$ on Higgs production cross section is, with one exception,

subdominant. For the process $e^+e^- \to t\bar{t}H$, the QCD corrections become large when the $t\bar{t}$ pair is produced near threshold. The scale variation of the NLO cross section amount to about 5% near threshold. Comparable parametric uncertainties are induced by a variation of the strong coupling $\Delta\alpha_s = 0.002$. These numbers have to be compared with the experimental accuracy at the percent level expected to be reached at a future linear collider [5].

To conclude, the strong coupling constant plays an essential role in precision Higgs physics, especially in the determination of the Yukawa couplings. Of special importance is the impact of $\alpha_s$ on the decay rate $H \to b\bar{b}$, that is the dominant Higgs decay mode and enters all the branching ratios and Higgs coupling measurements. Currently, $\alpha_s$ and $m_b$ generate the dominant uncertainties of this decay rate.

# $\alpha_{\rm s}$ from lattice QCD

Paul Mackenzie

Fermilab, Batavia, IL, USA

**Abstract:** The most precise determinations of the strong coupling constant have been made using lattice QCD. They have been done using half a dozen different quantities to extract $\alpha_{\rm s}$, by several different groups using completely independent lattice methods. The results are robust in that if the determinations of the group claiming the highest precision are set aside, the averaged results of the remaining lattice determinations are consistent with the omitted results and almost as precise. Likewise, if the results of the non-lattice determinations are averaged, they agree with the lattice results to within uncertainties.

A lattice determination of $\alpha_{\rm s}$ consists of two components. The lattice spacing in a given simulation must be determined in physical units by calculating a dimensionful physical quantity such as a particle mass. Then, the coupling constant must be obtained by determining some short distance quantity nonperturbatively, and equating the result with the usual $\overline{\rm MS}$ perturbative expansion. For example, this could be done with the heavy quark potential at short distances, to take an example that is conceptually simple, though not the simplest numerically.

One reason that the lattice determinations are more precise is that experimental contributions to the uncertainty of a lattice determination of $\alpha_{\rm s}$ are usually negligible. Experimental uncertainty enters a lattice calculation through the determination of the lattice spacing in physical units. This can be done using any physical quantity that can be accurately determined both experimentally and with a lattice calculations, even, for example, the proton mass. The best such calculations can be done to sub-percent accuracy in both experiment and theory, leading to sub-percent uncertainty in $\Lambda_{\overline{\rm MS}}$ and a sub-0.1% contribution to the uncertainty of $\alpha_{\rm s}$. The bulk of the uncertainty in lattice determinations of $\alpha_{\rm s}$ usually arises in the determination of the coupling constant in the simulation, from discretization errors, finite volume errors, etc. However, these uncertainties are usually under better control than experimental uncertainties because the functional form of the asymptotic behavior can be understood on field theoretic grounds. For example, the approach to the zero-lattice spacing limit is expected on theoretical grounds to be a power series in powers of the lattice spacing (from the operator product expansion) and powers of the logarithm of the lattice spacing (that is, from the expansion in $\alpha_{\rm s}$):

$$Q = \sum_{m,n=0} c_{mn} a^m \alpha_{\rm s}^n. \tag{1}$$

The assumption that the calculation is in the needed short-distance domain can be checked by comparing the numerical approach to the short-distance limit with the expected functional form.

A second reason that lattice determinations can be done more precisely is that they are fully nonperturbative, so that they contain all orders of perturbation theory, plus the effects of any condensates. Perturbative errors arise only from the conversion to $\alpha_{\overline{\rm MS}}$. This means that the theoretical calculations contain information that can be used to bound the uncalculated higher order coefficients. To illustrate this, we consider one of the most precise determinations, HPQCD's method using current-current correlation functions to determine $\alpha_s$. These correlation functions have been used to obtain the most precise determinations of the masses of the $c$ and $b$ quarks. Their moments can be determined by fitting $e^+e^-$ annihilation data, and the lower moments have been calculated to third order in $\alpha_s$ by the Karlsruhe group [1]. These moments can be obtained even more precisely in nonperturbative lattice calculations, and fit to a power series like the one in Eq. (1) to obtain $\alpha_s(m_q)$. Unlike experimental determinations of the moments, which can be done only at the two physical heavy quark masses, the lattice determinations can be done at as many (fictitious) values of the heavy quark masses as desired. These fits can then be used to bound the size of the uncalculated higher order coefficients of the terms that survive in the continuum limit. This makes the size of the perturbative uncertainty smaller in such lattice calculations than in other determinations of $\alpha_s$ known to the same number of terms in perturbation theory. HPQCD has also obtained the strong coupling constant from small Wilson loops. These are closely related to the heavy-quark potential on the lattice. The results of these two determinations are $\alpha_{\overline{\rm MS}}(M_Z) = 0.1183(7)$ and $\alpha_{\overline{\rm MS}}(M_Z) = 0.1184(6)$, respectively, and are the most precise determinations of $\alpha_s$ so far. Both of these determinations used the approach just described to bound the uncalculated higher terms of perturbation theory.

There are several other lattice determinations that reach or exceed the precision of the best non-lattice determinations. These include determinations using the Adler function [3] (which also is derived from current-current correlation functions), the Schrödinger functional [4], the quark-gluon vertex [5], and the heavy-quark potential [6]. Ref. [7] reanalyzed a subset of the Wilson data generated in ref. [2] and obtained a consistent result with a somewhat lower estimate of the precision. The results have been quite consistent to within estimated uncertainties.

The state of the lattice determinations has been well summarized in the 2014 QCD review in the Review of Particle Properties [8] and I won't repeat that analysis here. Since the lattice results were independent and more or less consistent, they used the $\chi^2$ averaging method to combine them and obtained $\alpha_{\overline{\rm MS}}(M_Z) = 0.1185(5)$ after a small $\chi^2$ correction.

Another review of lattice results was done by the FLAG collaboration [9]. This review contains much more detail and useful information about the results just described. However, the uncertainty of the combined result was not based on the estimates of the results reviewed and was not justified with any scientific or statistical arguments. A combined uncertainty of 0.0012 was given, stating only that it is "conservative". This combined uncertainty was higher than the individual uncertainties of many of the results that went into it, and did not represent a consensus within FLAG. Therefore, the 2014 PDG combined result [8] in preferable.

Reference [10] estimated the prospects for improvement in the precision of $\alpha_s$ over the next ten years assuming various combinations of improvements in computing power and perturbation theory. The estimates employed the fitting program used in the correlation function method of determining $\alpha_s$ described above [2], but used with fake data with improved precision mimicking the improvements expected with improved computing power over the next ten years, and sometimes with a fourth order of perturbation theory. The current uncertainty in this determination is dominated by statistics and the unknown fourth and higher orders of perturbation theory. The coefficients of these terms have been bounded already by the current numerical data, and will be further improved with improved numerical data. This work estimated that with the improvements in computing power expected over the next ten years, the precision of this determination would be improved from ±0.0006 to ±0.0003. The calculation of a fourth order of perturbation theory would improve the precision of this determination to below ±0.0001. However, that will be a very large project taking more than ten years by one estimate [11]. It is not clear that it will be undertaken until there is more than one determination that can reach this accuracy so that the results can be compared.

# $\alpha_s$ from the QCD static energy

Xavier Garcia i Tormo

Albert Einstein Center for Fundamental Physics. Institut für Theoretische Physik,
Universität Bern, Sidlerstrasse 5, CH-3012 Bern, Switzerland

**Abstract:** The extraction of $\alpha_s$ from lattice calculations of the QCD static energy is reviewed. This contribution is mostly based on Refs. [1,2], to which we refer for further details.

The static energy, $E_0(r)$, in QCD, i.e. the energy between a static quark and a static antiquark separated a distance $r$, is used to determine the strong coupling $\alpha_s$. We want to focus on the short-distance perturbative region of $E_0(r)$, and obtain the value of $\alpha_s$ by comparing the theoretical perturbative expression for $E_0(r)$ with the outcome of lattice simulations. The general concept of performing such an extraction was suggested long ago, but it is only recent progress, both in the perturbative and lattice computations, that allowed for a practical realization of it.

## Perturbation theory

At tree level the perturbative expression for $E_0(r)$ is just the Coulomb potential, with an adequate color factor. Currently, corrections up to third order are completely known [3,4,5]. Apart from powers of $\alpha_s$, the expression for $E_0(r)$ at short distances also contains $\ln \alpha_s$ terms [6], the so-called ultrasoft logarithms, which first appear at three-loop order. They are completely known up to fourth-loop order [7,8,9], and have been resummed to sub-leading logarithmic accuracy [10,11]. All the perturbative results for the static energy at the currently known accuracy are collected in Ref. [12].

## Lattice data

The static energy with $n_f = 2+1$ light flavors has been computed on the lattice at relatively short distances by the HotQCD collaboration [13]. Note that one needs lattice data with at least three light flavors in order to be able to obtain $\alpha_s$ at, say, the $Z$-boson mass scale, $M_Z$, from the lattice-perturbation theory comparison; since we need to be able to perform the decoupling at the quark thresholds perturbatively. At present, the HotQCD data is the only $n_f \geq 3$ lattice data that is available for $E_0(r)$. We use it for this analysis.

## Analyses and results

The comparison of the perturbative results with the lattice data provides a determination of $\alpha_s$. In this kind of analyses, a crucial aspect is always to know whether or not the lattice data has really reached the purely perturbative regime, with enough precision to perform the $\alpha_s$ extraction. Since it is quite difficult to undoubtedly state this point, the procedure used to determine $\alpha_s$ is designed to test if this is indeed the case. It involves performing fits to the lattice data at different distance ranges, and selecting for the final analysis only the distances where a satisfactory perturbative behaviour is observed. We refer to Ref. [1] for the details. The main source of error in the final result comes from the perturbative uncertainty. This uncertainty is estimated by varying the scale in the perturbative expansion, and by adding generic higher-order terms. Several cross-checks are then performed to verify that the extraction is robust, and that all the sources of uncertainty have been properly estimated and taken into account.

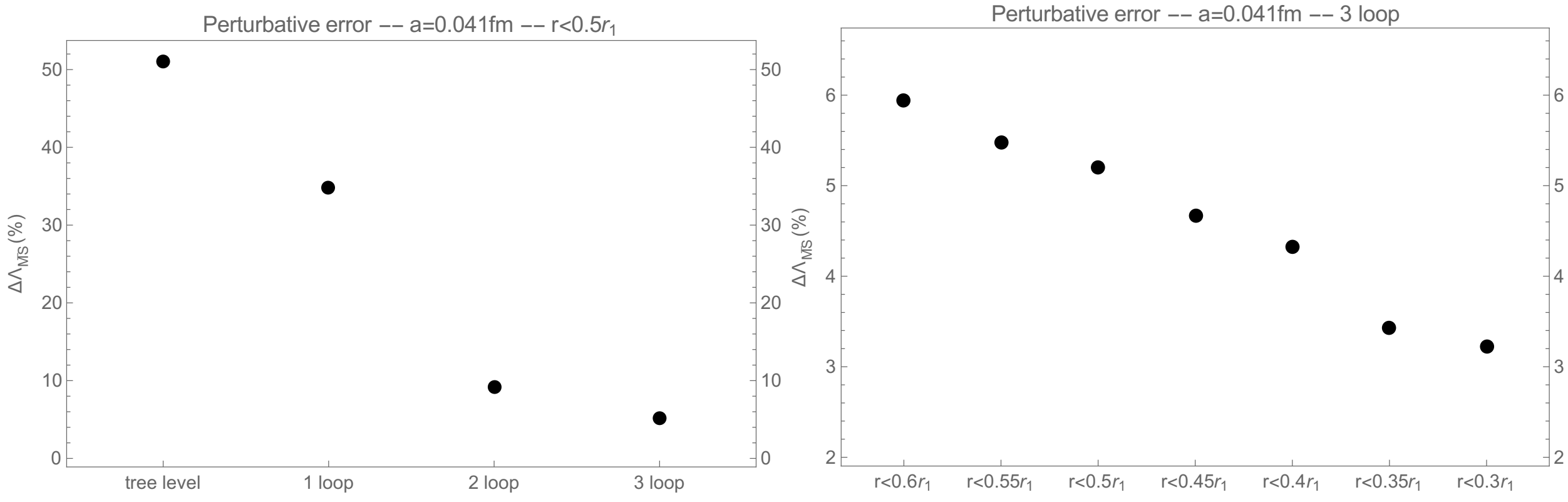

Figure 1: Perturbative error for $\Lambda_{\overline{\rm MS}}$, in percentage, as a function of the order in perturbation theory used in the theoretical expressions (left), and as a function of the fit distance range (right). See text for further explanations. The $r_1$ scale is $r_1 \simeq 0.311$ fm.

The outcome of the fits to the lattice is a value for the QCD parameter $\Lambda_{\overline{\rm MS}}$, which can then be translated to $\alpha_{\rm s}(M_Z)$ using perturbative running and decoupling. The result is

$$\Lambda_{\overline{\rm MS}} = 315^{+18}_{-12}\ \text{MeV}, \quad \alpha_{\rm s}(M_Z, n_f = 5) = 0.1166^{+0.0012}_{-0.0008}. \tag{1}$$

The above extraction is performed at three-loop accuracy, including resummation of logarithmic effects.

## Future

As mentioned before, the main source of uncertainty is the perturbative one. It seems unlikely that the fourth-order correction to the static energy will be available in the next few years, nevertheless this is unnecessary in order to reduce the current uncertainty. Although incorporating perturbation theory at fourth order would presumably reduce the errors in the $\alpha_{\rm s}$ extraction, this can also be achieved by incorporating more lattice data at shorter distances, since then the fits will involve regions where $\alpha_{\rm s}$ is smaller. We illustrate this in Fig. 1, where we show the perturbative errors of $\Lambda_{\overline{\rm MS}}$ for the finest lattice spacing we used, i.e. $a = 0.041$ fm. The left panel shows the relative error as a function of the perturbative order used, for the fit distance range $r < 0.5r_1$ which was the one used for our final result above, and makes manifest the importance of knowing higher orders in the perturbative expansion to obtain a precise extraction of the strong coupling. The right panel shows the relative error at three-loop order for several fit distance ranges. We can clearly see that the perturbative error reduces when the fits are restricted to shorter distances. Recall that these plots only show the perturbative error of the result, and not other errors which, with the current lattice data, are significant in some of these distance ranges, like fit parameter errors, discretization effects, ... In our current analysis, the fits with $r < 0.5r_1$ were used for the final result, since we wanted to use a distance range for which we could perform cross-checks excluding points at short distances that could suffer from larger discretization effects, have the statistical fit parameter errors under control, and be able to also use some of the available data at larger lattice spacings. The figure serves as guidance to what could be roughly expected for the size of the perturbative errors if future lattice data allows for reliable fits at these distances. Apart from that, it will also be interesting to perform the analysis with a different lattice data set, obtained with a different lattice action, in order to further reassure the robustness of the result.

**Acknowledgments** It is a pleasure to thank Alexei Bazavov, Nora Brambilla, Péter Petreczky, Joan Soto, and Antonio Vairo for collaboration on the work reported here.

# $\alpha_s$ from pion decay constant and renormalization group optimized perturbation

Jean-Loïc Kneur

Laboratoire Charles Coulomb (L2C), UMR 5221 CNRS-Univ. Montpellier, Montpellier, France

**Abstract:** A recent variant of optimized perturbation theory (OPT), making it consistent with renormalization group (RG) invariance, provides well-defined sequences of approximations of the ratio $F_\pi/\Lambda_{\overline{\rm MS}}$ of the pion decay constant and the basic QCD scale in the $\overline{\rm MS}$ scheme. From the precise experimental $F_\pi$ value it gives a new determination of $\Lambda_{\overline{\rm MS}}$, for $n_f = 2$ and $n_f = 3$ light quark flavors, thus of the QCD coupling constant $\overline{\alpha_s}(\mu)$ once combined with a standard perturbative RG evolution, reasonably reliable for $n_f = 3$. The empirical convergence properties of this RGOPT is exhibited from NLO to N$^3$LO order. We obtain $\overline{\alpha_s}(m_Z) = 0.1174^{+.0010}_{-.0005} \pm 0.001 \pm 0.0005_{evol}$, where the first errors are estimates of the intrinsic theoretical uncertainties of the method, and the second errors arise from present uncertainties in $F_\pi/F_0$ (mainly from lattice results), where $F_0$ is $F_\pi$ in the exact chiral SU(3) limit. We briefly discuss possible improvements in the future.

## Introduction

In spite of numerous determinations of $\overline{\alpha_s}(\mu)$ from confronting theoretical predictions to various data in a large range of $\mu$ scales [1], it is important to provide further independent analyses from other observables and theoretical methods, specially in the not so much explored infrared $\mu$ range. In the chiral symmetric (massless quarks) limit, the pion decay constant $F_\pi(m_q = 0) \equiv F_0$ is proportional to the basic QCD scale $\Lambda_{\overline{\rm MS}}$, in some chosen renormalization scale, here $\overline{\rm MS}$. $F_\pi(m_q = 0) \neq 0$ is the principal order parameter of spontaneous chiral symmetry breaking, and as such is an intrinsically nonperturbative quantity, indeed trivially vanishing in the massless limit of (ordinary) perturbation, which gives $F_\pi^2 \sim m_q^2 \sum_{n,p\geq 0} \alpha_s^n \ln^p(m_q/\mu)$. In the non exactly chiral real world, $F_\pi = 92.2 \pm 0.03 \pm 0.14$ MeV* is precisely measured from $\pi \to l\overline{\nu}(\gamma)$ decays) [1]. The sizable $F_\pi/F_0$ ratio, due to nonvanishing light quark masses, can be best extracted at present from lattice calculations with a reasonable (though not very precise) accuracy (see below). In a nutshell, the RGOPT [2,3] provides a well-defined sequence of non-trivial $F_0/\Lambda_{\overline{\rm MS}} \neq 0$ approximations, effectively resumming the perturbative expansion, up to N$^3$LO order at present. A crucial property of RGOPT is to naturally circumvent the vanishing chiral limit of ordinary perturbation, and in a way fully consistent with RG properties. Moreover, while one would naively expect the relevant renormalization scale to be presumably $\mu \sim \Lambda$, invalidating a standard perturbative expansion, as we will see the only relevant and uniquely determined renormalization scale implied by RGOPT is reasonably perturbative, stabilizing at NNLO and N$^3$LO at a scale roughly $\tilde{\mu} \simeq 1.2 - 1.3$ GeV, accidentally close to the charm quark mass.

## Optimized Perturbation (OPT)

The first key feature is to reorganize the standard QCD (massless) Lagrangian by "adding and subtracting" an *arbitrary* (quark) mass term, treating one mass piece as an interaction term. To organize this systematically at arbitrary perturbative orders, it is convenient to introduce an extra expansion parameter $0 \leq \delta \leq 1$, interpolating between $\mathcal{L}_{free}$ and $\mathcal{L}_{int}$, so that $m_q$ is traded for

*The normalization $f_\pi = \sqrt{2}F_\pi$ is often used in the rather recent literature an arbitrary trial parameter $m$ [4]. It is equivalent to taking a standard perturbative expansion in $g \equiv 4\pi\alpha_s$, with quark mass $m_q$ dependence, after renormalization, reexpanded in powers of $\delta$ (the so-called $\delta$-expansion) after substituting:

$$m_q \to m\,(1-\delta)^a, \quad g \to \delta\, g\ . \tag{1}$$

This is consistent with renormalizability [5,6] and gauge invariance [6]. Note the exponent $a$ in (1), reflecting a possibly more general interpolation form, but as we will see below $a$ is uniquely fixed from requiring [2,3] consistent RG-invariance properties. Applying Eq. (1) to a given perturbative (renormalized) expansion for a physical quantity $P(m,g)$, reexpanded in $\delta$ at order $k$, and taking *afterwards* the $\delta \to 1$ limit to recover the original *massless* theory, leaves a remnant $m$-dependence at any finite $\delta^k$-order. Note that while (1) drastically affects the mass dependence, it is a mild modification of the coupling dependence, which remains at least as much "perturbative" as in the original standard expansion as long as $0 \le \delta \le 1$. The arbitrary mass parameter $m$ is then conveniently fixed by seeking stationary points, from an optimization (OPT) prescription:

$$\frac{\partial}{\partial\, m} P^{(k)}(m,g,\delta=1)|_{m\equiv\tilde m} \equiv 0\ , \tag{2}$$

thus determining a nontrivial optimized mass $\tilde m(g) \neq 0$. It is legitimate to wonder if such a "cheap trick" with (1) and (2) always work, or when (and why) does it work? In simpler ($D=1$) models it may be seen as a particular case of the "order-dependent mapping" [7], and Eqs. (1),(2) have been proven [8] to converge exponentially fast for the $D=1$ $\Phi^4$ oscillator energy levels. For $D>1$ renormalizable models, the large $\delta$-orders behaviour is more involved and no rigorous convergence proof exists, but the procedure partially damps the factorially divergent (infrared renormalons) large perturbative order behaviour [9]. While (2) is not the only possible prescription to fix the arbitrary mass, it is the most convenient one when combined with RG properties, as it drastically simplifies the latter, as we shortly remind below. It also naturally gives $m$ of order $\Lambda \simeq \mu e^{-1/(\beta_0 g)}$, realizing in this way dimensional transmutation (in contrast with the original mass vanishing in the chiral limit). In various models with $O(N)$ or similar symmetries, the first $\delta$-order OPT is often equivalent to the large-$N$ approximation, and higher orders can give rather successful approximations beyond large-$N$.

## Renormalization Group consistency of OPT (RGOPT)

In most previous OPT applications however, the so-called linear $\delta$-expansion was used, assuming $a=1$ in Eq. (1) for simplicity. Moreover a well-known drawback of the OPT approach is that beyond lowest order Eq. (2) generally gives more and more solutions at increasing orders, and some being complex-valued. Thus without insight on the nonperturbative behaviour of the solutions, it can be difficult to select the right one, and non-real unphysical solutions are embarrassing. Our more recent construction [10,2,3] differs in two respects, which also drastically improve the convergence of the method. First, we combine OPT and RG properties, by requiring the ($\delta$-modified) expansion to satisfy, in addition to the mass optimization Eq. (2), a standard perturbative RG equation:

$$\mu\frac{d}{d\,\mu}\left(P^{(k)}(m,g,\delta=1)\right)=0, \tag{3}$$

where the RG operator is[†]

$$\mu\frac{d}{d\,\mu} = \mu\frac{\partial}{\partial\mu} + \beta(g)\frac{\partial}{\partial g} - \gamma_m(g)\, m\frac{\partial}{\partial m}\ . \tag{4}$$

[†]With normalization $\beta(g) \equiv dg/d\ln\mu = -2b_0 g^2 - 2b_1 g^3 + \cdots$, $\gamma_m(g) = \gamma_0 g + \gamma_1 g^2 + \cdots$ with $b_i$, $\gamma_i$ up to 4-loop [11].

Once combined with Eq. (2), the RG equation thus takes the reduced massless form:

$$\left[\mu\frac{\partial}{\partial\mu}+\beta(g)\frac{\partial}{\partial g}\right]P^{(k)}(m,g,\delta=1)=0\ , \tag{5}$$

which is also by definition the RG equation satisfied by $\Lambda_{\rm QCD}$. Now a crucial observation, overlooked in most previous OPT applications, is that after performing (1), perturbative RG invariance is generally lost, so that Eq. (5) gives a nontrivial additional constraint: but RG invariance can only be restored for a unique value of the exponent $a$, generally different from naive $a=1$ value, fully determined by the universal (scheme-independent) first order RG coefficients [2,3]:

$$a\equiv\gamma_0/(2b_0)\ . \tag{6}$$

Accordingly Eqs. (5) and (2) together completely fix *optimized* $m\equiv\tilde{m}$ and $g\equiv\tilde{g}$ values. (6) also guarantees that at arbitrary $\delta$ orders at least one of both the RG and OPT solutions $\tilde{g}(m)$ continuously matches the standard perturbative RG behaviour for $g\to 0$, i.e. Asymptotic Freedom (AF) for QCD:

$$\tilde{g}(\mu\gg\tilde{m})\sim\left(2b_0\ln\frac{\mu}{\tilde{m}}\right)^{-1}+\mathcal{O}\left(\left(\ln\frac{\mu}{\tilde{m}}\right)^{-2}\right), \tag{7}$$

moreover such solutions are often unique at a given $\delta$ order for both the RG and OPT equations. The same exponent (6) is also required for non-AF models [12], with appropriate RG coefficients (with always the property $\gamma_0/b_0>0$). Incidentally the link of $a$ with RG critical exponents had also been established before in the $D=3$ $\Phi^4$ model for the Bose-Einstein condensate critical temperature shift by two independent approaches [13,14]. Unfortunately, AF-compatibility and reality of solutions can be mutually incompatible beyond lowest orders, as is the case for $F_\pi$ in QCD. As we will see, a simple way out is to further exploit the RG freedom, considering a perturbative renormalization scheme change to possibly recover RGOPT solutions both AF-compatible and real [3].

## Application to $F_\pi/\Lambda_{\overline{\rm MS}}$ calculation

As a nonperturbative order parameter of chiral symmetry breaking ($SU(n_f)_L\times SU(n_f)_R\to SU(n_f)_{L+R}$, in practice for $n_f=2$ or $n_f=3$ massless quarks), $F_\pi$ can be conveniently defined from the $p^2\to 0$ axial current correlator [15]:

$$i\langle 0|TA^i_\mu(p)A^j_\nu(0)|0\rangle\equiv\delta^{ij}g_{\mu\nu}F^2+\mathcal{O}(p_\mu p_\nu) \tag{8}$$

where $A^i_\mu\equiv\overline{q}\gamma_\mu\gamma_5\frac{\tau_i}{2}\,q$, and $F$ is $F_\pi$ in the exact chiral limit, here for two light flavors. Note that for $p^2\to 0$ and in the chiral limit, $F$ is the only contribution to the right-hand-side of Eq. (8), not contaminated by higher mass resonances or other nonperturbative contributions. The advantage of this $F_\pi$ definition is that the left-hand side of (8) also has a well-defined standard perturbative QCD expansion in the $\overline{\rm MS}$ scheme, with non-zero quark masses $m\neq 0$, which can be extracted from already known 4-loop calculations [16]:

$$F^2_\pi=\frac{3m^2}{2\pi^2}\left[-L+\frac{\alpha_s}{4\pi}\left(8L^2+\frac{4}{3}L+\frac{1}{6}\right)+\left(\frac{\alpha_s}{4\pi}\right)^2[f_{30}L^3+f_{31}L^2+f_{32}L+f_{33}]+\mathcal{O}\left(\alpha_s^3\right)+\text{div.}\right] \tag{9}$$

where $L\equiv\ln\frac{m}{\mu}$ and the $f_{ij}$ coefficients are given in [3]. Before applying the RGOPT, a subtlety in dimensional regularization is that (9) requires an additive renormalization, after mass and coupling renormalization, giving additional finite contributions:

$$F^2_\pi(\text{RG-inv})\equiv F^2_\pi-S(m,\alpha_s);$$
$$S(m,\alpha_s)\equiv m^2(s_0/\alpha_s+s_1+s_2\alpha_s+\ldots) \tag{10}$$

To obtain a finite *and* RG-invariant expression, the perturbative coefficients $s_k$ in (10) are fixed from the coefficient of the $L$ term at order $k+1$ [3]. One finds $s_0 = 3/(16\pi^3(b_0-\gamma_0))$, etc.
We can now apply RGOPT to the (subtracted) RG-invariant perturbative series for $F_\pi$ in (10). Once given the elaborate three- and four -loop ordinary perturbative calculations for arbitrary mass values in [16], we stress that performing the RGOPT is a rather simple procedure: it amounts to Taylor-expand in $\delta$ after applying (1), then taking different (first) derivatives and solving the resulting (algebraic) equations for $L$ and/or $g = 4\pi\alpha_s$ according to Eqs. (2), (5). As a first simpler illustration, we truncate (9) at two-loop $\alpha_s$ order, and consider only the first RG order approximation, neglecting accordingly any non-logarithmic terms: then the RG and OPT Eqs. (5),(2) have a unique AF-compatible and real solution:

$$\tilde{L} \equiv \ln\frac{\tilde{m}}{\mu} = -\frac{1}{2b_0 g} = -\frac{\gamma_0}{2b_0};\ \tilde{\alpha_s} = \frac{1}{4\pi\gamma_0} = \frac{\pi}{2} \tag{11}$$

already giving a realistic value $F(\tilde{m},\tilde{\alpha_s}) = \left(\frac{5}{8\pi^2}\right)^{1/2}\tilde{m} \simeq 0.25\Lambda_{\overline{\rm MS}}$, where $\Lambda_{\overline{\rm MS}} = \mu\ e^{-1/(2b_0 g)}$ at this order. One might worry that the obtained optimal coupling $\tilde{\alpha_s}$ in (11) is quite strong, but at this stage it is a first order result using crude one-loop RG approximation. In fact considering at arbitrary $\alpha_s$ orders solely the first-order RG-dependence, (11) is always an exact solution, so this result is completelly stable, irrespectively of the apparently rather large optimized coupling. Moreover (11) is exactly scale-invariant: all coupling dependence is consistently absorbed into $\tilde{m} = \Lambda_{\overline{\rm MS}}$, such that the actual $\alpha_s$ value is irrelevant for the pure number $F/\Lambda_{\overline{\rm MS}}$ result. Considering next the two-loop pure RG dependence, a residual scale dependence reappears, i.e. not all the coupling dependence is absorbed in the appropriate (two-loop) expression of $\Lambda_{\overline{\rm MS}}$. But the optimal coupling is [3] $(4\pi\tilde{\alpha_s})^{-1} = \gamma_0 - (\gamma_1 - b_1)/(\gamma_0 - b_0)$ which gives $\tilde{\alpha_s} \simeq 0.335$ for $n_f = 2$, a drastic decrease to a much more perturbative value. Accordingly the renormalization scale $\mu$ is automatically fixed by optimization: $\ln\tilde{\mu}/\Lambda \simeq (2b_0\tilde{g})^{-1}$, to a reasonably perturbative value, rather than a naively expected much lower scale. Including higher order RG dependence and all non-logarithmic terms, the RG and OPT equations, polynomial in $(L, g)$, give at increasing $\delta$-orders many solutions. But selecting the AF-compatible solution is unambiguous, and it turns out to be unique. Also the residual scale dependence (not absorbed in $\Lambda_{\overline{\rm MS}}$) of those solutions is moderate (i.e. rather flat) around the optimized value. Unfortunately the intersection of AF-compatible RG and OPT branches is complex at $\delta^k, k \geq 1$ orders in the $\overline{\rm MS}$-scheme.

### Recovering real AF-compatible solutions

The non-reality of AF-compatible solutions is an artifact of solving exactly polynomial equations (2), (5) in $(g, L)$, and somewhat an accident for $F_\pi$ in the $\overline{\rm MS}$-scheme. Thus one expects to recover real solutions by performing a moderate (perturbative) deformation of the (AF-compatible) RGOPT solutions. A deformation consistent with RG properties is simply a standard perturbative renormalization scheme change (RSC). Since $m$ is basically an arbitrary parameter, we consider RSC affecting only $m$:

$$m \to m'(1 + B_1 g + B_2 g^2 + \cdots) \tag{12}$$

which also let the RG Eq. (5) and $\Lambda_{\overline{\rm MS}}$ unaffected. One may now move within the RSC parameter space with appropriate $B_k \neq 0$ in (12) until possibly reaching real solutions. To stay as nearly as possible to the reference $\overline{\rm MS}$-scheme, we require a *closest* contact solution between the RG and OPT curves (closest thus to $\overline{\rm MS}$ in RSC space), which can either be numerically searched, or is analytically given by collinearity of the tangent vectors of the two curves:

$$\frac{\partial\,{\rm RG}(g,L,B_k)}{\partial g}\frac{\partial\,{\rm OPT}(g,L,B_k)}{\partial L} - \frac{\partial\,{\rm RG}}{\partial L}\frac{\partial\,{\rm OPT}}{\partial g} \equiv 0 \tag{13}$$

to be solved together with the OPT and RG Eqs. (2),(5) now for three parameters $\tilde{L}, \tilde{g}, \tilde{B}_k$ at successive orders. As expected from general RSC properties the differences with respect to the original $\overline{\rm MS}$-scheme should decrease at higher perturbative order. Apart from recovering real solutions, RSC also provides us with well-defined uncertainty estimates: since there are non-unique RSC prescriptions (12), implying different results, we take those differences as intrinsic theoretical uncertainties of the method. The results for $n_f = 3$ are given in Table 1 (normalized with a 4-loop perturbative $\overline{\Lambda}_{4l}$ expression). Note that $\tilde{m}'/\overline{\Lambda}$ remains $\mathcal{O}(1)$ and $\tilde{\alpha}_{\rm s}$ stabilizes at 4-loop at reasonably perturbative values $\tilde{\alpha}_{\rm s} \simeq 0.37$, corresponding to an optimized remnant scale roughly of order the charm quark mass. For the similar $n_f = 2$ results [3], the theoretical RSC uncertainties are about twice larger (this can be traced to the larger RSC corrections needed to reach real solutions). Although the relative smallness of $\tilde{\alpha}_{\rm s}$ is relevant to expect a reasonable stability of the corresponding $F/\Lambda_{\overline{\rm MS}}(\tilde{\alpha}_{\rm s})$ expression, we stress that neither the optimal $\tilde{\alpha}_{\rm s}$ or $\tilde{m}$ have direct physical meaning: solely the $\Lambda_{\overline{\rm MS}}$ value extracted from the pure number $F/\Lambda_{\overline{\rm MS}}(\tilde{\alpha}_{\rm s}, \tilde{m})$ is used to determine the genuine $\alpha_{\rm s}(\mu)$.

Table 1: Main optimized results at successive orders for $n_f = 3$

| $\delta^k$ order | $\tilde{L}' \equiv \ln \tilde{m}'/\mu$ | $\tilde{\alpha}_{\rm s}$ | $F_0/\overline{\Lambda}_{4l}$ (RSC uncertainties) |
|---|---|---|---|
| $\delta$, RG-2l | $-0.523$ | 0.757 | $0.27 - 0.34$ |
| $\delta^2$, RG-3l | $-1.368$ | 0.507 | $0.236 - 0.255$ |
| $\delta^3$, RG-4l | $-1.760$ | 0.374 | $0.2409 - 0.2546$ |

**Explicit symmetry breaking and extracting $\alpha_{\rm s}(\mu)$**

We stress that the previous calculation is by construction for $F_0 \equiv F_\pi(m_q \to 0)$ in the exact chiral limit, so to obtain a final realistic result we have to account for the explicit (chiral symmetry breaking) effects from non-vanishing quark masses $m_u, m_d, m_s$, specially important in the SU(3) case. With $F$ and $F_0$ the $F_\pi$ values in the exact chiral SU(2) and SU(3) limits respectively, recently obtained lattice results are [17]:

$$\frac{F_\pi}{F} \sim 1.073 \pm 0.015; \qquad \frac{F_\pi}{F_0} \sim 1.172(3)(43) \tag{14}$$

where the $F$ value is robust, and $F_0$ was obtained in [18] using fits to NNLO Chiral Perturbation [19]. However one should keep in mind that somewhat different $F_0$ values are also advocated [17], pointing at a slower convergence of $n_f = 3$ Chiral Perturbation [20,19]. With the different theoretical uncertainties combined we thus obtain:

$$\overline{\Lambda}_{4l}^{n_f=2} \simeq 359^{+38}_{-26} \pm 5\,{\rm MeV}; \;\; \overline{\Lambda}_{4l}^{n_f=3} \simeq 317^{+14}_{-7} \pm 13\,{\rm MeV} \tag{15}$$

Those results compare reasonably well with different lattice determinations for $n_f = 2$ or $n_f = 3$ (see [3] for a detailed comparison with lattice results). Finally using $\overline{\Lambda}(n_f = 3)$ from (15) and a four-loop order standard perturbative evolution (including $m_c$, $m_b$ threshold effects [21]) we obtain [3]

$$\overline{\alpha_{\rm s}}(m_Z) = 0.1174^{+.0010}_{-.0005}|_{\rm th} \pm 0.0010\,_{\delta F_0} \pm 0.0005_{\rm evol} \tag{16}$$

where the first theoretical RGOPT error results from averaging three- and four-loop order results, and the second error comes from the above quoted lattice calculation uncertainties on $F_\pi/F_0$.

## Conclusions

A recent variant of OPT, consistent with RG properties, appears to drastically improve the convergence and stability properties of such a modified perturbative expansion. Calculating $F_\pi/\overline{\Lambda}$ at successive orders up to N$^3$LO, we have extracted a rather precise new independent determination of $\overline{\alpha_s} = 0.1174^{+.0010}_{-.0005}|_{\rm th} \pm 0.0010_{\,\delta F_0} \pm 0.0005_{\rm evol}$, also estimating the intrinsic theoretical uncertainties of the method.

We briefly mention some improvements that may be expected in the near or more distant future. First, the input value $F_\pi = 92.2 \pm 0.03 \pm 0.14$ MeV [1] (where the first error is experimental and the second is the theoretical accuracy of the $\pi \to l\overline{\nu}(\gamma)$ decay calculation) is unlikely to progress much, and the quoted errors are anyway well below the other theoretical errors in Table 1 and (14). The resulting theoretical RGOPT error in (16) may be slightly reduced, if instead of averaging three- and four-loop results, we less conservatively take only the (presumably more reliable) four-loop results. One could also in principle incorporate partly some of higher (five-loop) order (pure logarithmic) dependence from using RG recurrence properties (unless a full five-loop calculation would become available in the future). But the admittedly more fragile part of the calculation is reflected in the second error in (16), relying on a still rather challenging lattice calculation. Even if the latter should certainly progress in the near future, it will remain a priori difficult to reach the true chiral limit. A very foreseable alternative would be to replace the lattice $F_\pi/F_0$ calculation by incorporating explicit quark mass $m_{u,d,s}$ dependences directly within the RGOPT approach. While this is formally simple, in practice the analysis becomes much more involved, moreover not all the arbitrary $m_{u,d,s}$ mass dependence is known to three- and four-loop order at present. But if manageable, it could replace the $F_\pi/F_0 \sim 4-5\%$ uncertainty in (16) by a possibly improved one, from the more precise $\sim 2.7\%$ $m_s$ uncertainty [17].

# $\alpha_s$ from hadronic $\tau$ lepton decays

Antoni Pich

Departament de Física Teòrica, IFIC, Universitat de València – CSIC, València, Spain

**Abstract:** The decays of the $\tau$ lepton are ideally suited to test the strong interaction at low energies and extract the $\alpha_s$ coupling with high accuracy and precision [1]. Theoretical developments in the field are reviewed.

The inclusive character of the $\tau$ hadronic width renders possible an accurate calculation of the ratio [2,3,4,5]

$$R_\tau \equiv \frac{\Gamma[\tau^- \to \nu_\tau \,\text{hadrons}]}{\Gamma[\tau^- \to \nu_\tau e^- \overline{\nu}_e]} = R_{\tau,V} + R_{\tau,A} + R_{\tau,S}$$
$$= 12\pi \int_0^{m_\tau^2} \frac{ds}{m_\tau^2} \left(1 - \frac{s}{m_\tau^2}\right)^2 \left[\left(1 + 2\frac{s}{m_\tau^2}\right) \mathrm{Im}\Pi^{(1)}(s) + \mathrm{Im}\Pi^{(0)}(s)\right]. \tag{1}$$

The theoretical analysis involves the two-point correlators for the vector $V_{ij}^\mu = \overline{\psi}_j \gamma^\mu \psi_i$ and axial-vector $A_{ij}^\mu = \overline{\psi}_j \gamma^\mu \gamma_5 \psi_i$ colour-singlet quark currents ($i, j = u, d, s$; $\mathcal{J} = V, A$):

$$i \int d^4x \, e^{iqx} \langle 0 | T(\mathcal{J}_{ij}^\mu(x) \mathcal{J}_{ij}^\nu(0)^\dagger) | 0 \rangle = \left(-g^{\mu\nu} q^2 + q^\mu q^\nu\right) \Pi_{ij,\mathcal{J}}^{(1)}(q^2) + q^\mu q^\nu \, \Pi_{ij,\mathcal{J}}^{(0)}(q^2). \tag{2}$$

The $J = 0, 1$ spectral functions in Eq. (1) stand for the following combination of correlators:

$$\Pi^{(J)}(s) \equiv \sum_{q=d,s} |V_{uq}|^2 \left(\Pi_{uq,V}^{(J)}(s) + \Pi_{uq,A}^{(J)}(s)\right). \tag{3}$$

The $q = d$ terms correspond to $R_{\tau,V}$ and $R_{\tau,A}$, while $R_{\tau,S}$ contains the strange contributions. $\Pi_{ij,\mathcal{J}}^{(J)}(q^2)$ are analytic functions of $s$, except along the positive real $s$-axis where their imaginary parts have discontinuities. Weighted integrals of the spectral functions can then be written as contour integrals in the complex $s$-plane [4,5]:

$$\int_0^{s_0} ds \, w(s) \, \mathrm{Im}\Pi_{ij,\mathcal{J}}^{(J)}(s) = \frac{i}{2} \oint_{|s|=s_0} ds \, w(s) \, \Pi_{ij,\mathcal{J}}^{(J)}(s), \tag{4}$$

with $w(s)$ an arbitrary weight function without singularities in the region $|s| \leq s_0$. Several fortunate facts make $R_\tau$ particularly suitable for a precise theoretical analysis [4]:

i) The tau mass is large enough to safely use the the Operator Product Expansion (OPE), $\Pi^{(J)}(s) = \sum_{D=2n} C_D^{(J)}/(-s)^{D/2}$, to estimate the contour integral at $s_0 = m_\tau^2$.

ii) The OPE is only valid in the complex plane, away from the real axis where the physical hadrons sit. The contributions to the contour integral from the region near the real axis are heavily suppressed in $R_\tau$ by the presence in (1) of a double zero at $s = m_\tau^2$.

iii) For massless quarks, $s\,\Pi^{(0)}(s) = 0$ and only $\Pi^{(0+1)}(s)$ contributes to (1), with a weight function $w(x) = (1-x)^2(1+2x) = 1 - 3x^2 + 2x^3$ $[x \equiv s/m_\tau^2]$. Non-perturbative corrections originate then from operators with dimensions $D = 6$ and 8. The leading $D = 4$ contributions are strongly suppressed by a factor $\alpha_s^2$, which makes them negligible.

iv) While non-perturbative contributions to $R_{\tau,V}$ and $R_{\tau,A}$ are both suppressed by a factor $1/m_\tau^6$, the $D = 6$ contributions to the vector and axial-vector correlators have opposite signs leading to a partial cancelation in $R_{\tau,V+A}$.

The prediction for the Cabibbo-allowed combination $R_{\tau,V+A}$ can be written as [4]

$$R_{\tau,V+A} \;=\; N_C\, |V_{ud}|^2\, S_{\mathrm{EW}} \left\{1 + \delta_{\mathrm{P}} + \delta_{\mathrm{NP}}\right\}, \qquad (5)$$

where $N_C = 3$ and $S_{\mathrm{EW}} = 1.0201 \pm 0.0003$ contains the electroweak corrections [6].

The small non-perturbative contribution $\delta_{\mathrm{NP}}$ can be determined from the measured invariant-mass distribution of the final hadrons in $\tau$ decay, through the study of weighted integrals which are more sensitive to OPE corrections [5]. The predicted suppression of $\delta_{\mathrm{NP}}$ has been confirmed by ALEPH [7], CLEO [8] and OPAL [9]. The presently most complete and precise experimental analysis, performed with the ALEPH data, obtains [10]

$$\delta_{\mathrm{NP}} \;=\; -0.0064 \pm 0.0013\,. \qquad (6)$$

Combining the $\tau$ lifetime and $e/\mu$ branching fractions into a *universality-improved* electronic branching ratio, the *Heavy Flavor Averaging Group* values for the Cabibbo-allowed hadronic $\tau$ width imply $R_{\tau,V+A} = 3.4697 \pm 0.0080$ [11]. Using $|V_{ud}| = 0.97425 \pm 0.00022$ [12] and Eq. (6), the pure perturbative contribution to $R_\tau$ is determined to be [1]:

$$\delta_{\mathrm{P}} = 0.2009 \pm 0.0031\,. \qquad (7)$$

The theoretical prediction of $\delta_{\mathrm{P}}$ is very sensitive to $\alpha_s(m_\tau^2)$ [3,4], allowing for an accurate determination of the QCD coupling [1,10,12,13,14,15,16]. The result takes the form

$$\delta_{\mathrm{P}} \;=\; \sum_{n=1} K_n\, A^{(n)}(a_\tau) \;=\; \sum_{n=1} (K_n + g_n)\, a_\tau^n\,, \qquad (8)$$

where $K_0 = K_1 = 1$, $K_2 = 1.63982$, $K_3(\overline{\mathrm{MS}}) = 6.37101$ and $K_4(\overline{\mathrm{MS}}) = 49.07570$ [13] are the known perturbative expansion coefficients (for $n_f = 3$ flavours) of the logarithmic derivative of $\Pi^{(0+1)}(s)$, and $a_\tau \equiv \alpha_s(m_\tau^2)/\pi$. The contour integrals [5]

$$A^{(n)}(a_\tau) \;=\; \frac{1}{2\pi i}\oint_{|s|=m_\tau^2} \frac{ds}{s}\left(\frac{\alpha_s(-s)}{\pi}\right)^n \left(1 - 2\frac{s}{m_\tau^2} + 2\frac{s^3}{m_\tau^6} - \frac{s^4}{m_\tau^8}\right) \;=\; a_\tau^n + \mathcal{O}(a_\tau^{n+1}) \qquad (9)$$

can be numerically computed with high accuracy, using the exact solution (up to unknown $\beta_{n>4}$ contributions) for $\alpha_s(-s)$ given by the renormalization-group $\beta$-function equation. The resulting *contour-improved perturbation theory* (CIPT) series [5,17] has a very good perturbative convergence and is stable under changes of the renormalization scale.

A naive expansion of $A^{(n)}(a_\tau)$ in powers of $a_\tau$ (*fixed-order perturbation theory*, FOPT) [4] overestimates $\delta_{\mathrm{P}}$ by 12% at $a_\tau = 0.11$ [5]. The long running of $\alpha_s(-s)$ along the circle $|s| = m_\tau^2$ generates the $g_n$ coefficients in (8) which are larger than the $K_n$ contributions: $g_1 = 0$, $g_2 = 3.56$, $g_3 = 19.99$, $g_4 = 78.00$, $g_5 = 307.78$ [5]. FOPT suffers from a large renormalization-scale dependence. This bad perturbative behaviour makes compulsory the CIPT resummation of large logarithms $\log^n(-s/m_\tau^2)$ [5]. It has been argued that, once in the asymptotic regime (large $n$), the expected renormalonic behaviour of the $K_n$ coefficients could induce cancellations with the running $g_n$ corrections, which would be missed by CIPT. In that case, FOPT could approach faster the 'true' result provided by the Borel summation of the full renormalon series. Models of higher-order corrections with this behaviour have been advocated [14], but the results are however model dependent [18].

The main uncertainty in the $\tau$ determination of the strong coupling originates in the treatment of higher-order perturbative corrections. Using CIPT one gets from Eq. (7) $\alpha_s(m_\tau^2) = 0.341 \pm 0.013$, while FOPT would give $\alpha_s(m_\tau^2) = 0.319 \pm 0.014$ [1]. Combining the two results, but keeping conservatively the smallest error, we get [1]

$$\alpha_s^{(n_f=3)}(m_\tau^2) \;=\; 0.331 \pm 0.013\,. \tag{10}$$

A direct analysis of the ALEPH invariant-mass distribution [10] determines $\alpha_s(m_\tau^2)$ and the OPE corrections through a global fit of $R_{\tau,V+A}$ and four weighted integrals with $s_0 = m_\tau^2$ and weights $w(x) = (1+2x)(1-x)^3 x^l$ ($x = s/m_\tau^2$, $l = 0,1,2,3$). Using CIPT, this gives $\delta_{\rm NP}$ in Eq. (6) and $\alpha_s(m_\tau^2) = 0.341 \pm 0.008$, fully consistent with our CIPT result.

When the OPE is used to perform the contour integration (4), one neglects the difference $\Delta_{ij,\mathcal{J}}^{(J)}(s) \equiv \Pi_{ij,\mathcal{J}}^{(J)}(s) - \Pi_{ij,\mathcal{J}}^{(J),\rm OPE}(s)$. The missing correction (duality violation) [19,20]

$$\frac{i}{2}\oint_{|s|=s_0} ds\; w(s)\, \Delta_{ij,\mathcal{J}}^{(J)}(s) \;=\; -\int_{s_0}^{\infty} ds\; w(s)\, \mathrm{Im}\, \Delta_{ij,\mathcal{J}}^{(J)}(s) \tag{11}$$

is negligible in $R_\tau$; however, it could be more relevant for other weighted integrals. In order to maximize this effect, Refs. [21] analyze the weight $w(x)=1$ and fit the $s_0$ dependence of the $V$ and $A$ integrated distributions, in the range $s_{\rm min} \equiv (1.25~\mathrm{GeV})^2 \le s_0 \le m_\tau^2$, parametrizing the spectral functions with an ansatz which is then used to estimate (11). One pays a big price because (i) $s_{\rm min} \sim M_{a_1}^2$ is too low to be reliable; (ii) one directly fits the spectral functions along the real axis where the OPE is not valid, and (iii) the separate $V$ and $A$ correlators have larger non-perturbative contributions than $V+A$. In addition, one has a too large number of free parameters to be fitted to a highly correlated data set. In spite of all these caveats, reasonable values of the strong coupling are obtained (CIPT): $\alpha_s(m_\tau^2) = 0.310 \pm 0.014$ (ALEPH), $0.322 \pm 0.026$ (OPAL) [21]. Although the quoted uncertainties are clearly underestimated, this suggests a much better behaviour of perturbative QCD at low values of $s$ than naively expected. This had been already noticed longtime ago in the pioneering analyses of the $s_0$ dependence performed in Refs. [7,22].

The $R_\tau$ determination of $\alpha_s$, given in Eq. (10), is significantly larger ($16\,\sigma$) than the result obtained from the $Z$ hadronic width, $\alpha_s^{(n_f=5)}(M_Z^2) = 0.1197 \pm 0.0028$ [12]. After evolution up to the scale $M_Z$ the strong coupling in (10) decreases to $\alpha_s^{(n_f=5)}(M_Z^2) = 0.1200 \pm 0.0015$, in excellent agreement with the direct measurement at the $Z$ peak. The comparison of these two determinations provides a beautiful test of the predicted QCD running; i.e. a very significant experimental verification of asymptotic freedom:

$$\alpha_s^{(n_f=5)}(M_Z^2)\Big|_\tau - \alpha_s^{(n_f=5)}(M_Z^2)\Big|_Z \;=\; 0.0003 \pm 0.0015_\tau \pm 0.0028_Z\,. \tag{12}$$

Improvements of the $\alpha_s(m_\tau^2)$ determination would require high-precision measurements of the spectral functions, which should be possible at Belle-II and at a future FCC-ee, and a better theoretical understanding of higher-order perturbative corrections.

# $\alpha_s$ from hadronic quarkonia decays

Joan Soto i Riera

Dept. d'Estructura i Constituents de la Matèria
i Institut de Ciències del Cosmos,
Universitat de Barcelona
Martí i Franquès 1, 08028 Barcelona, Catalonia

**Abstract:** The current status of the strong coupling extraction from hadronic bottomonium decays, and the future theoretical and experimental developments needed to obtain a value of $\alpha_s$ at NNLO accuracy, are reviewed.

## Introduction

Bottom and charm quark masses are large enough so that perturbative QCD is applicable at those scales. Hence bottomonium and charmonium decays to light hadrons are sensitive to $\alpha_s$. In particular certain ratios of hadronic over radiative decays are proportional to $\alpha_s$ with known factors [1,2], up to power and higher-order $\alpha_s$ corrections. Hence, on the theoretical side, the discussion on how precisely we will be able to extract $\alpha_s$ from those ratios is focused on how well we will be able to pin down power and higher-order $\alpha_s$ corrections. The form and size of power corrections are well under control thanks to the NRQCD factorization framework [3,4], and are given in terms of a number of long-distance matrix elements (LDME). The LDME can be estimated using lattice NRQCD [5] (see [6]), or, for the lowest lying states, using weak coupling pNRQCD [7,8] (see [9]). Each LDME is multiplied by a short distance matching coefficient for which higher-order $\alpha_s$ corrections can be systematically calculated.

We shall focus our discussion here on the ratio $R_\gamma \equiv \Gamma(\Upsilon(1S) \to \gamma X)/\Gamma(\Upsilon(1S) \to X)$, $X$ being light hadrons, for the following reasons: (i) most of the LDME that appear in the denominator also appear in the numerator, and hence errors tend to cancel out (this is not the case if one uses leptonic decays either in the numerator or in the denominator [10]), (ii) experimental data is usually more accurate for the ground sate than for excited states, (iii) bottom is heavier than charm and hence the $\alpha_s(m_Q)$ expansion, $Q = b, c$, is more reliable in bottomonium than in charmonium, and (iv) for the ground state of bottomonium the weak coupling estimates of the LDME are more reliable than for excited states. A discussion for the charmonium case can be found in [11].

## Status

The most accurate data from which the $\alpha_s$ extraction has been performed from heavy quarkonium decays is due to the CLEO collaboration, about ten years ago [12]. References to earlier experiments may be found there. $\alpha_s(m_{\Upsilon(\mathrm{nS})})$, $n = 1, 2, 3$, was extracted from a theoretical expression for $R_\gamma$ that included the leading corrections in $\alpha_s$ and the leading relativistic corrections. The following numbers were put forward,

$$\begin{aligned} \Upsilon(1\mathrm{S}): \quad \alpha_s(M_Z) &= 0.1114 \pm 0.0002 \pm 0.0029 \pm 0.0053 \\ \Upsilon(2\mathrm{S}): \quad \alpha_s(M_Z) &= 0.1026 \pm 0.0007 \pm 0.0041 \pm 0.0077 \\ \Upsilon(3\mathrm{S}): \quad \alpha_s(M_Z) &= 0.113 \pm 0.001 \pm 0.007 \pm 0.008 \end{aligned} \tag{1}$$

where the errors are statistical, systematic and model dependence. The model dependence arises because accurate data for the direct photon spectrum is not available for soft photons and hence an extrapolation to the soft-photon region is required. Notice that this is the dominant error source.

The same data was used in [13] were the following improvements were introduced: (i) the theoretical expression was corrected to match the complete NLO NRQCD factorization formula by adding the color octet operators, (ii) the model dependence in the extrapolation to the soft-photon region was reduced by sticking to a QCD calculation [14], and (iii) the LDME were estimated using both lattice NRQCD and weak coupling pNRQCD calculations. The analysis used $\Upsilon(1S)$ data only, since (ii) and (iii) were only available for that state. The following value was obtained,

$$\alpha_s(M_Z) = 0.119^{+0.006}_{-0.005}\,, \tag{2}$$

where the error is dominated by the experimental error in $R_\gamma$. This number was competitive with the available extractions at that time and was included in the PDG average, until the latter was restricted to NNLO calculations in $\alpha_s$ only.

## Prospects

In order to get a competitive extraction of $\alpha_s$ with the current criteria, one needs first a more precise experimental value of $R_\gamma$. In particular a more accurate measurement of the direct photon spectrum. This will probably require the machine to run at the $\Upsilon(1S)$ pole or to simulate it by radiative return techniques.

Assuming that this will be achieved, let us next spell out the missing ingredients on the theoretical side. We have,

$$R_\gamma \equiv \frac{\Gamma(\Upsilon(1S)\to\gamma\, X)}{\Gamma(\Upsilon(1S)\to X)} = \frac{36}{5}\frac{e_b^2\alpha}{\alpha_s}\frac{N}{D}, \tag{3}$$

$e_b = -1/3$, and N and D include power and higher-order $\alpha_s$ corrections. In the NRQCD factorization approach, the size of the power corrections is dictated by the velocity ($v$) scaling rules, $v^2 \sim 0.1$ for $\Upsilon(1S)$. Numerically $\alpha_s(m_b)/\pi \sim v^2$, which provides the guideline to organize power and $\alpha_s$ corrections at a given order. Let us write

$$\begin{aligned} N &= 1 + N_{\rm NLO} + N_{\rm NNLO} + \dots \\ D &= 1 + D_{\rm NLO} + D_{\rm NNLO} + \dots \end{aligned} \tag{4}$$

$N_{\rm NLO}$ and $D_{\rm NLO}$ are known, and were used in [13] for the current NLO extraction. Symbolically they read,

$$N_{\rm NLO}\,, D_{\rm NLO} = \mathcal{O}(\alpha_s) + \mathcal{O}(v^2) + \mathcal{O}\left(\frac{v^4}{\alpha_s}\right) \tag{5}$$

$\mathcal{O}(\alpha_s)$ stands for the leading perturbative correction to the matching coefficient of the leading-order operator, $\mathcal{O}(v^2)$ for the leading relativistic correction encoded in a color-singlet operator and $\mathcal{O}(v^4/\alpha_s)$ for the contribution due to color-octet operators (notice the $1/\alpha_s$ enhancement that compensates the extra power of $v^2$). The NNLO contribution has the following form

$$N_{\rm NNLO}\,, D_{\rm NNLO} = \mathcal{O}({\alpha_s}^2) + \mathcal{O}(\alpha_s v^2) + \mathcal{O}\left(\alpha_s\frac{v^4}{\alpha_s}\right) + \mathcal{O}(v^4) + \mathcal{O}\left(\frac{v^6}{\alpha_s}\right) \tag{6}$$

We spell out the status of each contribution below,

- $\mathcal{O}({\alpha_s}^2)$: unknown. This is the next-to-leading perturbative correction to the matching coefficient of the leading-order operator, it requires the evaluation of the imaginary part of 4-loop integrals with one scale. It is known numerically for QED [15,16,17].

- $\mathcal{O}(\alpha_s v^2)$: unknown. This is the leading perturbative correction to the matching coefficient of the leading relativistic correction, it requires the evaluation of the imaginary part of 3-loop integrals with one scale.

- $\mathcal{O}(\alpha_s \frac{v^4}{\alpha_s})$: known. These are the leading perturbative corrections to the matching coefficients of the leading color-octet operators, they may be found in [18] for $D_{\rm NNLO}$ and in [19] for $N_{\rm NNLO}$.

- $\mathcal{O}(v^4)$: known for $D_{\rm NNLO}$. These are next-to-leading relativistic corrections encoded in color-singlet operators, they may be found in [20].

- $\mathcal{O}(\frac{v^6}{\alpha_s})$: known for $D_{\rm NNLO}$. These are tree-level contributions of higher order color-octet (and mixed singlet-octet) operators, they may be found in [6,18,21].

The missing $N_{\rm NNLO}$ contributions in the two last points should be easy to obtain from the know results of $D_{\rm NNLO}$.

The LDME entering at NNLO can be estimated using, for instance, weak coupling pNRQCD (see [22] for reviews). However we would need some precision (about 10%) for the LDME entering at NLO. This is not a problem for the $\mathcal{O}(v^2)$ operator, as it is mainly sensitive to the soft scale ($m_b\alpha_s$), at which weak coupling calculations are still reliable, but it maybe a problem for the color-octet operators $\mathcal{O}(v^4/\alpha_s)$, which are sensitive to the ultrasoft scale ($m_b\alpha_s^2$). More precise lattice NRQCD values would be welcome for those, or, alternatively an accurate extraction from experimental data. In that respect, a more precise measurement of the photon spectrum, a key ingredient in $R_\gamma$ anyway, may be instrumental. It turns out that the fragmentation contributions to the photon spectrum, which dominate the low energy region [23] and are already noticeable in the central region of the spectrum, depend on the same $\mathcal{O}(v^4/\alpha_s)$ LDME above [19]. A fit to the photon spectrum in the central to low energy region may pin down precise enough values for these LDME*. The fit also involves the quark to photon and gluon to photon fragmentations functions, which can be extracted from other processes (the quark to photon fragmentation function was measured at LEP [24]).

## Conclusions

A competitive extraction of $\alpha_s$ from hadronic decays of heavy quarkonium appears to be feasible, though not totally straightforward, in the coming years. The key experimental ingredient is a more precise number for $R_\gamma$. On the theoretical side, one needs (i) a couple of challenging perturbative calculations, (ii) an estimate of a number ($\sim 10$) of LDME, and (iii) somewhat more precise ($\sim 10\%$) values for the LDME of a color-singlet and three color-octet operators.

**Acknowledgments.** I thank Xavier Garcia i Tormo for the critical reading of this contribution, and for his collaboration in a number of related papers. I also thank Nora Brambilla and Antonio Vairo for their collaboration in ref. [13]. Financial support from the projects FPA2013-46570, FPA2013-43425-P, and CSD2007-00042 (Spain), and 2014-SGR-104 (Catalonia) is acknowledged. ICCUB is partially funded by the grant MDM-2014-0369.

*The missing $\mathcal{O}(\alpha_s)$ correction to the color-singlet operator in the quark to photon fragmentation function would also be needed for a consistent fit [14]

# $\alpha_s$ from soft parton-to-hadron fragmentation functions

David d'Enterria[1] and **Redamy Pérez-Ramos**[1,2,3]

[1] CERN, PH Department, CH-1211 Geneva 23, Switzerland
[2] Sorbonne Universités, UPMC Univ Paris 06, UMR 7589, LPTHE, F-75005, Paris, France
[3] CNRS, UMR 7589, LPTHE, BP 126, 4 place Jussieu, F-75252 Paris Cedex 05, France

**Abstract:** The QCD coupling $\alpha_s$ is extracted at approximate next-to-next-to-leading-order (NNLO$^*$) accuracy from the energy evolution of the first two moments (multiplicity and mean) of the parton-to-hadron fragmentation functions at low fractional hadron momentum $z$. Comparisons of the experimental $e^+e^-$ and DIS $e^{\pm}$p jet data to our NNLO$^*$+NNLL predictions, allow us to obtain $\alpha_s(m_Z^2) = 0.1205 \pm 0.0010^{+0.0022}_{-0.0000}$, in excellent agreement with the current world average.

## Introduction

For massless quarks and fixed number of colours $N_c$, the only fundamental parameter of quantum chromodynamics (QCD), the theory of the strong interaction, is its coupling constant $\alpha_s$. Starting from a value of $\Lambda_{QCD} \approx 0.2$ GeV, where the perturbatively-defined coupling diverges, $\alpha_s$ decreases with increasing energy $Q$ following a $1/\ln(Q^2/\Lambda^2_{QCD})$ dependence. The current uncertainty on $\alpha_s$ evaluated at the Z mass, $\alpha_s(m_Z^2) = 0.1185 \pm 0.0006$, is $\pm 0.5\%$ [1], making of the strong coupling the least precisely known of all fundamental constants in nature. Improving our knowledge of $\alpha_s$ is a prerequisite to reduce the uncertainties in perturbative QCD calculations of all partonic cross sections at hadron colliders, notably in Higgs physics [2], and for precision fits of the Standard Model. It has also far-reaching implications including the stability of the electroweak vacuum [3] or the scale at which the interaction couplings unify.

Having at hand new independent approaches to determine $\alpha_s$ from the data, with experimental and theoretical uncertainties different from those of the other methods currently used, is crucial to reduce the overall uncertainty in the combined $\alpha_s$ world-average value [1]. In Refs. [4,5] we have presented a novel technique to extract $\alpha_s$ from the energy evolution of the moments of the parton-to-hadron fragmentation functions (FF) computed at approximate next-to-leading-order (NLO$^*$) accuracy including next-to-next-to-leading-log (NNLL) resummation corrections. This approach has been extended to include full NLO plus a set of NNLO corrections [6]. Our new NNLO$^*$+NNLL theoretical results for the energy dependence of the hadron multiplicity and mean value of the FF are compared to jet fragmentation measurements in $e^+e^-$ and deep-inelastic $e^{\pm}$p collisions, and a high-precision value of $\alpha_s$ is extracted.

## Energy evolution of the soft parton-to-hadron fragmentation functions

The distribution of hadrons in a jet is encoded in a fragmentation function, $D_{i \to h}(z, Q)$, which describes the probability that parton $i$ fragments into hadron $h$ carrying a fraction $z = p_{hadron}/p_{parton}$ of the parent parton momentum. Usually one writes the FF as a function of the log of the inverse of $z$, $\xi = \ln(1/z)$, emphasizing the region of soft momenta that dominates the jet hadronic fragments. Indeed, due to colour coherence and gluon radiation interference, not the softest partons but those with intermediate energies multiply most effectively in QCD cascades, leading to a final FF with a typical "hump-backed plateau" (HBP) shape as a function of $\xi$ (Fig. 1), which can be expressed in terms of a distorted Gaussian (DG):

$$D(\xi, Y, \lambda) = \mathcal{N}/(\sigma\sqrt{2\pi}) \cdot e^{\left[\frac{1}{8}k - \frac{1}{2}s\delta - \frac{1}{4}(2+k)\delta^2 + \frac{1}{6}s\delta^3 + \frac{1}{24}k\delta^4\right]}, \text{ with } \delta = (\xi - \overline{\xi})/\sigma, \tag{1}$$

where $\mathcal{N}$ is the average hadron multiplicity inside a jet, and $\overline{\xi}$, $\sigma$, $s$, and $k$ are respectively the mean peak position, dispersion, skewness, and kurtosis of the distribution.

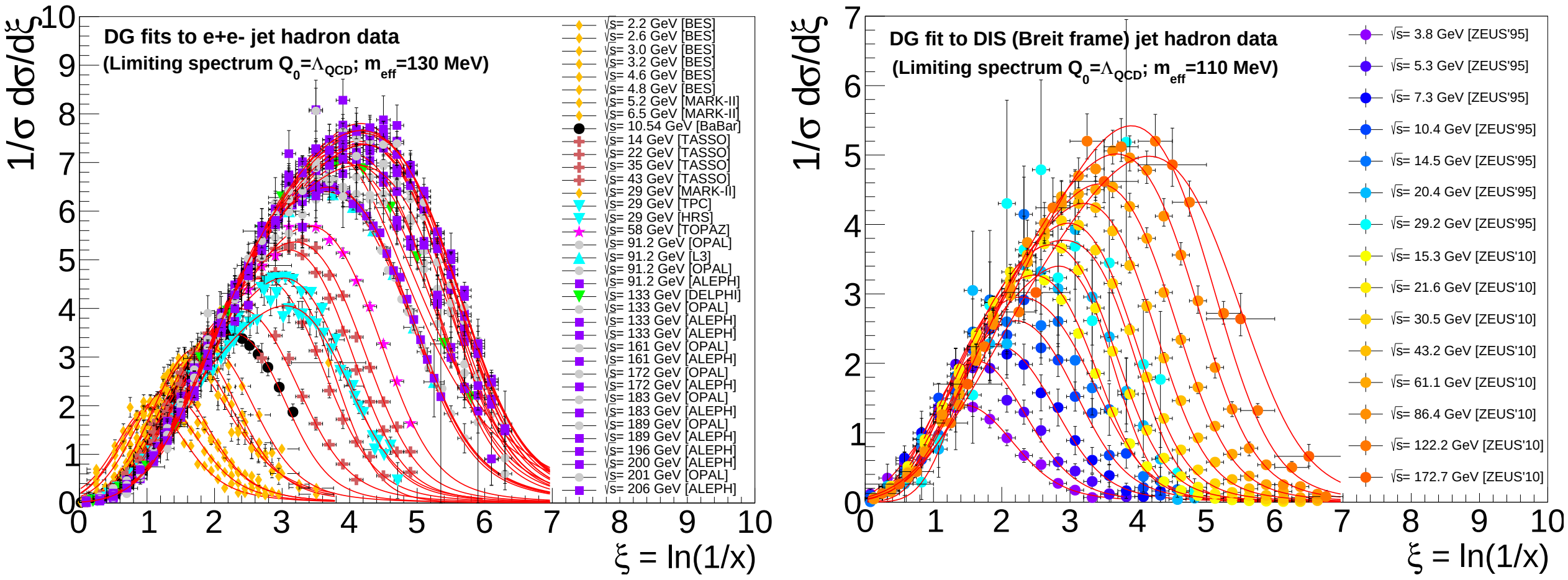

Figure 1: Charged-hadron spectra in jets as a function of $\xi = \ln(1/z)$ in $e^+e^-$ at $\sqrt{s} \approx$ 2–200 GeV (left), and $e^\pm, \nu$-p (Breit frame, scaled $\times 2$ for the full hemisphere) at $\sqrt{s} \approx$ 4–180 GeV (right), individually fitted to Eq. (1) with the hadron mass corrections ($m_{\text{eff}}$ = 130,110 MeV) quoted.

Starting with a parton at a given energy $Q$, its evolution to another energy scale $Q'$ is driven by a branching process of parton radiation and splitting, resulting in a jet "shower", which can be computed perturbatively using the DGLAP [7] equations at large $z \gtrsim 0.1$, and the Modified Leading Logarithmic Approximation (MLLA) [8], resumming soft and collinear singularities at small $z$. As for the Schrödinger equation in quantum mechanics, the system of equations for the FFs $D_{\text{i}\to\text{h}}(z, Q)$ can be written as an evolution Hamiltonian which mixes gluon and (anti)quark states expressed in terms of DGLAP splitting functions for the branchings $g \to gg$, $q(\overline{q}) \to gq(\overline{q})$ and $g \to q\overline{q}$, where $g$, $q$ and $\overline{q}$ label a gluon, a quark and an anti-quark respectively. Analytical solutions can only be obtained from the Mellin transform of the full-resummed regularized NNLL splitting functions [9]. The set of integro-differential equations for the FF evolution combining hard (DGLAP, MLLA, next-to-MLLA) and soft (DLA) radiation can be solved by expressing the Mellin-transformed hadron distribution in terms of the anomalous dimension $\gamma$: $D \simeq C(\alpha_\text{s}(t)) \exp\left[\int^t \gamma(\alpha_\text{s}(t'))dt\right]$ for $t = \ln Q$, leading to a perturbative expansion in half powers of $\alpha_\text{s}$: $\gamma \sim \mathcal{O}(\alpha_\text{s}^{1/2}) + \mathcal{O}(\alpha_\text{s}) + \mathcal{O}(\alpha_\text{s}^{3/2}) + \mathcal{O}(\alpha_\text{s}^{2}) + \mathcal{O}(\alpha_\text{s}^{5/2}) + \cdots$. The anomalous dimension $\gamma$ allows one to calculate the moments of the DG through:

$$\mathcal{N} = K_0, \quad \overline{\xi} = K_1, \; \sigma = \sqrt{K_2}, \; s = \frac{K_3}{\sigma^3}, \; k = \frac{K_4}{\sigma^4}; \text{ with } K_{n\geq 0}(Y, \lambda) = \int_0^Y dy \left(-\frac{\partial}{\partial\omega}\right)^n \gamma_\omega \Big|_{\omega=0}, \tag{2}$$

which are then inserted into Eq. (1). Corrections of $\gamma$ up to order $\alpha_\text{s}^{3/2}$ were computed in Refs. [4,5], followed by the full set of NLO $\mathcal{O}(\alpha_\text{s}^2)$ terms, including the two-loop splitting functions, in Ref. [6].

At NLO, the diagonalisation of the evolution Hamiltonian results in two eigenvalues $\gamma_{\pm\pm}$ in the $\mathcal{D}^{\pm}$ basis, where the relevant one for the calculation of the FF moments $\gamma_{++} \to \gamma_\omega^{\rm NLO+NNLL}$, reads:

$$
\begin{aligned}
\gamma_\omega^{\rm NLO+NNLL} &= \frac{1}{2}\omega(s-1) + \frac{\gamma_0^2}{4N_c}\left[-\frac{1}{2}a_1(1+s^{-1}) + \frac{\beta_0}{4}(1-s^{-2})\right] \\
&+ \frac{\gamma_0^4}{256N_c^2}(\omega s)^{-1}\Bigg[4a_1^2(1-s^{-2}) + 8a_1\beta_0(1-s^{-3}) + \beta_0^2(1-s^{-2})(3+5s^{-2}) \\
&- 64N_c\frac{\beta_1}{\beta_0}\ln 2(Y+\lambda)\Bigg] \\
&+ \frac{1}{4}\gamma_0^2\omega\left[a_2(2+s^{-1}+s) + a_3(s-1) - a_4(1-s^{-1}) - a_5(1-s^{-3}) - a_6\right],
\end{aligned}
\tag{3}
$$

where $\gamma_0^2 = \frac{4N_c\alpha_{\rm s}}{2\pi} = \frac{4N_c}{\beta_0(Y+\lambda)}$ is the LL anomalous dimension, $s = \sqrt{1+\frac{4\gamma_0^2}{\omega^2}}$, $\beta_i$ the QCD $\beta$-function coefficients, $a_{1,2}$ and hard constants obtained in [4], and $a_{3,4,5,6}$ are new constants obtained from the full-resummed NNLL splitting functions [9]. In addition, a fraction of the $\mathcal{O}(\alpha_{\rm s}^{5/2})$ terms from the NNLO $\alpha_{\rm s}$ running expression, have been now added [10]. Upon inverse-Mellin transformation, one obtains the energy evolution of the FF, and its associated moments, at NNLO$^*$+NNLL accuracy as a function of $\mathrm{Y} = \ln(E/\Lambda_{\rm QCD})$, for an initial parton energy $E$, down to a shower cut-off scale $\lambda = \ln(Q_0/\Lambda_{\rm QCD})$ for $N_f = 3, 4, 5$ quark flavors. The resulting formulae for the energy evolution of the moments depend on $\Lambda_{\rm QCD}$ as *single* free parameter. Relatively simple expressions are obtained in the limiting-spectrum case ($\lambda = 0$, i.e. evolving the FF down to $\mathrm{Q}_0 = \Lambda_{\rm QCD}$) motivated by the "local parton hadron duality" hypothesis for infrared-safe observables which states that the HBP distribution of partons in jets is simply renormalized in the hadronization process without changing its shape. Thus, by fitting the experimental hadron jet data at various energies to Eq. (1), one can determine $\alpha_{\rm s}$ from the corresponding energy-dependence of its FF moments.

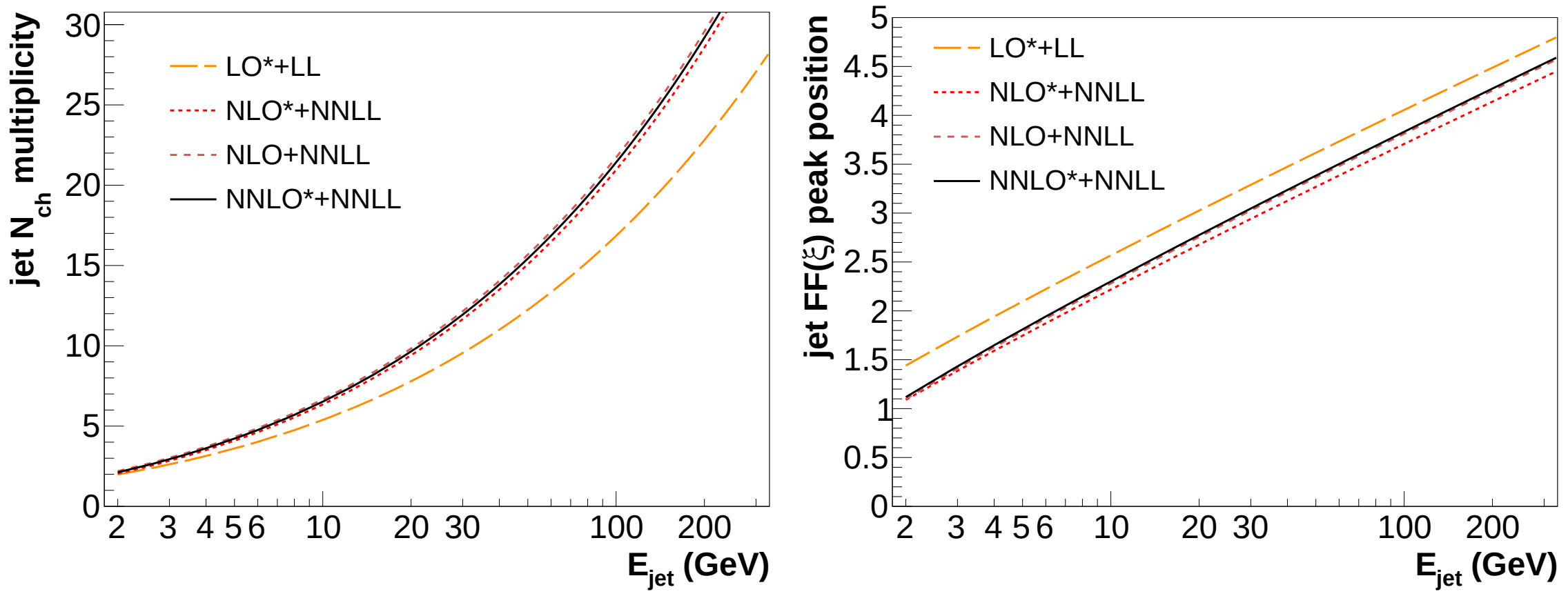

Figure 2: Comparison of theoretical predictions at increasing level of accuracy (LO$^*$ to NNLO$^*$) for the energy evolution of the jet charged-hadron multiplicity (left) and FF peak position (right).

Figure 2 shows the energy evolution of the zeroth (multiplicity) and first (peak position, closely connected to the mean of the distribution) moments of the FF, at four levels of accuracy (LO$^*$+LL, NLO$^*$+NLL, NLO+NNLL, and NNLO$^*$+NNLL). The hadron multiplicity and FF peak increase exponentially and logarithmically with energy, and the theoretical convergence of their evolutions are very robust as proven by the small changes introduced by incorporating higher-order terms.

## Data-theory comparison and $\alpha_s$ extraction

The first step of our procedure is to fit all existing jet FF data measured in $e^+e^-$ and $e^\pm,\nu$-p collisions at $\sqrt{s} \approx$ 2–200 GeV (Fig. 1) to Eq. (1), in order to obtain the corresponding FF moments at each jet energy. Finite hadron-mass effects in the DG fit are accounted for through a rescaling of the theoretical (massless) parton momenta with an effective mass $m_{\rm eff}$ as discussed in Refs. [4,5]. The overall normalization of the HBP spectrum ($\mathcal{K}_{\rm ch}$), which determines the average charged-hadron multiplicity of the jet, is an extra free parameter in the DG fit which, nonetheless, plays no role in the final $\Lambda_{\rm QCD}$ value given that its extraction *just* depends on the *evolution* of the multiplicity, and not on its absolute value at any given energy.

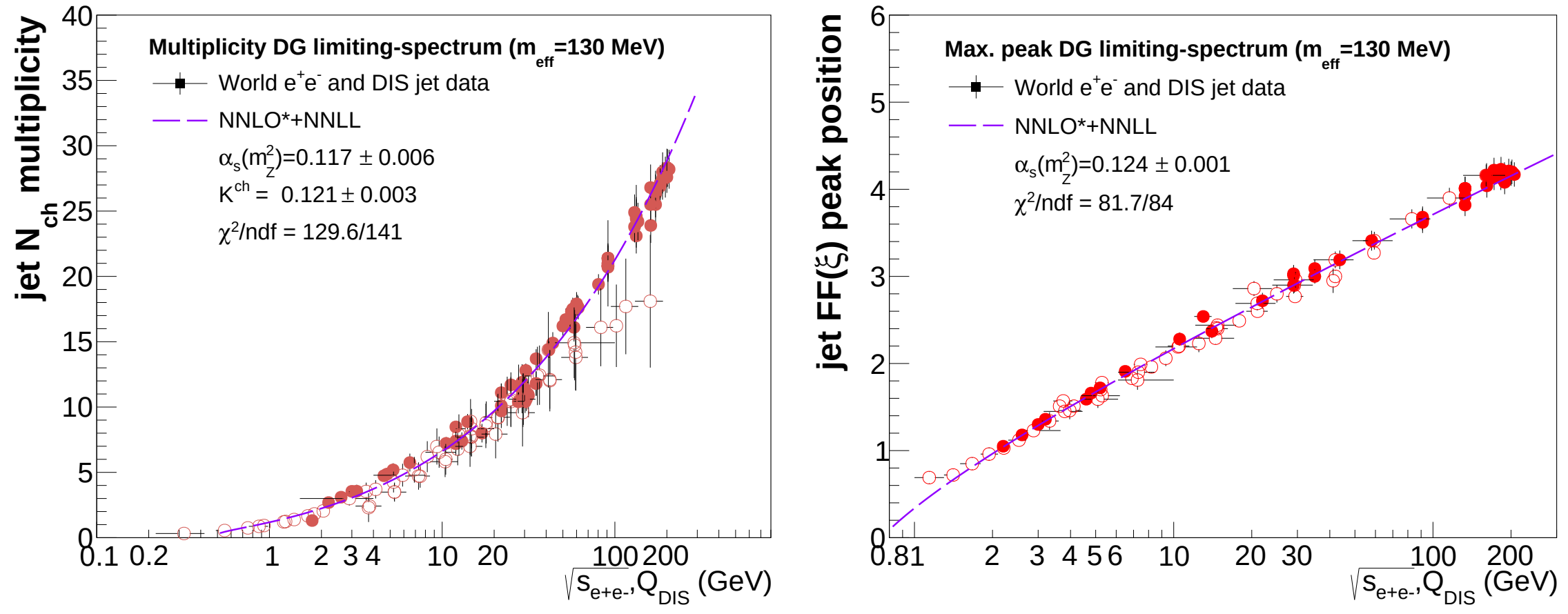

Figure 3: Energy evolution of the jet charged-hadron multiplicity (left) and FF peak position (right) in $e^+e^-$ and DIS jet data, fitted to the NNLO*+NNLL predictions. The obtained $\mathcal{K}_{\rm ch}$ normalization constant, the individual NNLO* $\alpha_s(m_Z^2)$ values, and $\chi^2$/ndf of the two fits, are quoted.

Once the FF moments have been obtained, we perform a combined fit of them as a function of the original parton energy (i.e. $\sqrt{s}/2$ in the case of $e^+e^-$, and the invariant 4-momentum transfer $Q_{\rm DIS}$ for DIS). The individual fits for the first two FF moments are shown in Fig. 3. The NNLO*+NNLL limiting-spectrum ($\lambda = 0$) predictions for $N_f = 5$ active quark flavours*, leaving $\Lambda_{\rm QCD}$ as a free parameter, reproduce very well the data. The most "robust" FF moment for the determination of $\Lambda_{\rm QCD}$ is the peak position $\xi_{\rm max}$ which proves quite insensitive to most of the uncertainties associated with the extraction method [5] as well as to higher-order corrections. The hadron multiplicities measured in DIS jets appear somewhat smaller (especially at high energy) than those measured in $e^+e^-$, due to limitations in the FF measurement only in half (current Breit) $e^\pm$p hemisphere and/or in the determination of the relevant $Q$ scale [5]. The value of $\alpha_s(m_Z^2)$ obtained from the combined multiplicity+peak fit yields $\alpha_s(m_Z^2) = 0.1205 \pm 0.0010$, where the error includes all uncertainties discussed in Ref. [5]. A conservative theoretical scale uncertainty of $^{+0.0022}_{-0.0000}$ (obtained in [5] at NLO accuracy only) is added. In Fig. 4 we compare our $\alpha_s(m_Z^2)$ value to all other NNLO results from the 2014 PDG compilation [1], plus that obtained from the $\pi$-decay factor [11], and the top-quark pair cross sections at the LHC [12]. The precision of our result ($^{+2\%}_{-1\%}$) is clearly competitive with the other measurements, with a totally different set of experimental and theoretical uncertainties. A simple weighted average of all these NNLO values yields: $\alpha_s(m_Z^2) = 0.1186 \pm 0.0004$, in perfect

*Few-% corrections are applied to deal with slightly different $N_f$ = 3,4 evolutions below charm,bottom thresholds.

agreement with the world-average, $\alpha_s(m_Z^2) = 0.1185 \pm 0.0006$, but with a 30% smaller uncertainty. Upcoming full-NNLO corrections of the energy evolution of the FF moments will allow the inclusion of our $\alpha_s$ result into the PDG world-average.

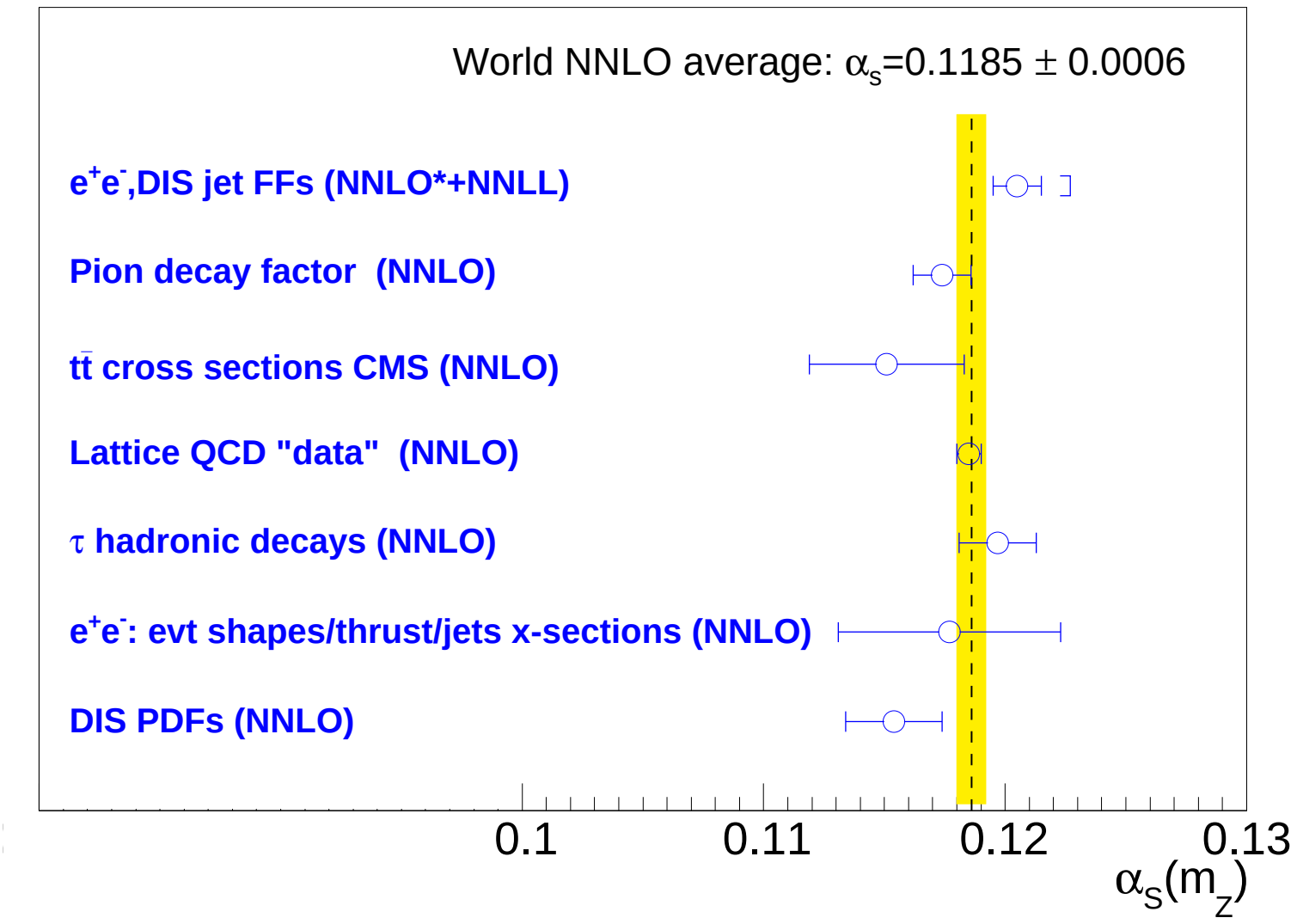

Figure 4: Summary of $\alpha_s$ determinations using different methods at NNLO$^{(*)}$ accuracy. The dashed line and shaded (yellow) band indicate the 2014 world-average and uncertainty (listed on top) [1].

# $\alpha_s$ from global fits of parton distribution functions

Sergey Alekhin[1], **Johannes Blümlein**[2], Sven Moch[1]

[1] II. Institut für Theoretische Physik, Universität Hamburg, Luruper Chaussee 149
D-22761 Hamburg, Germany
[2] Deutsches Elektronen-Synchrotron, DESY, Platanenallee 6, D-15738 Zeuthen, Germany

**Abstract:** The status of the determination of the strong coupling constant $\alpha_s(M_Z^2)$ from deep-inelastic scattering and related hard scattering data is reviewed.

## Introduction

Deep-inelastic scattering (DIS) is one of the cleanest ways to measure the strong coupling constant from the scaling violations of the structure functions, which form the hadronic tensor $W_{\mu\nu}$ in the differential scattering cross section [1]. In the unpolarized case for pure photon exchange one has

$$\begin{aligned} W_{\mu\nu}(q,P) &= \frac{1}{2x}\left(g_{\mu\nu} - \frac{q_\mu q_\nu}{q^2}\right) F_{\rm L}(x,Q^2) \\ &\quad + \frac{2x}{Q^2}\left(P_\mu P_\nu + \frac{q_\mu P_\nu + q_\nu P_\mu}{2x} - \frac{Q^2}{4x^2} g_{\mu\nu}\right) F_2(x,Q^2) \ . \end{aligned} \tag{1}$$

Here $q$ denotes the 4-momentum transfer from the lepton to the nucleon with $q^2 = -Q^2$, $P$ the nucleon momentum, and $x = Q^2/(2P.q)$ the Bjorken variable. The two structure functions $F_{\rm L,2}(x,Q^2)$ exhibit scaling violations due to the $Q^2$ dependence. They contain both massless and massive quark distributions (due to charm and bottom quarks).

A first comprehensive survey on the value of the strong coupling constant $\alpha_s(M_Z^2)$ as measured in different hard processes has been given by G. Altarelli [2]. At this time the value of $\alpha_s(M_Z^2)$ has been known in next-to-leading order (NLO) and the deep-inelastic value turned out to be low $\alpha_s^{\rm NLO}(M_Z^2) = 0.112 \pm 0.007$. This value resulted from data by BCDMS [3,4], a combined SLAC/BCDMS fit [5], $\rm EMC_H$ [6], $\rm CCFR_{Fe}$ [7,8,9], $\rm CHARM_{CaCO_3}$ [10] and NMC [11]. Except the value of NMC of $0.117 \pm 0.005$ all other experiments yielded values in the range of 0.108...0.114. An important test consisted in comparing the experimental results for the plot of $\partial F_2/\partial \ln(Q^2)$, cf. also [12], which failed to agree for CDHSW [13]. At this order in QCD the factorization and renormalization scale uncertainties are still large and amount to $\Delta\alpha_s(M_Z^2) = \pm 0.0050$, varying the scales $\mu_{F,R}^2 \in [Q^2/4, 4Q^2]$ [14]. In most of these analyses heavy flavor corrections have not been accounted for, as they were known to leading order only [15]. Rather they were fitted to the data using the massless NLO corrections for $N_F = 4$, using the $O(\alpha_s)$ Wilson coefficients [16] and the NLO anomalous dimensions [17], in early analyses. A few characteristic determinations of $\alpha_s(M_Z^2)$ at NLO are summarized in Table 1. Most of the values lay in the region between $\alpha_s^{\rm NLO}(M_Z^2) = 0.1150...0.1171$. In the polarized case, also values for $\alpha_s^{\rm NLO}(M_Z^2)$ were determined, which, however, have much larger errors, see [24,25] for details.

Because of the large scale variation uncertainties at NLO compared to the precision of the current world-data, we will discuss in the following only next-to-next-to leading order (NNLO) and next-to-next-to-next-to leading order (N$^3$LO) analyses, and refer to NLO analyses only briefly. Already the NLO analyses require a description of the heavy flavor corrections to $O(\alpha_s^2)$ [27]. The NNLO analyses rely on the massless $O(\alpha_s^2)$ Wilson coefficients [28] and the NNLO anomalous dimensions [29,30]. The heavy flavor corrections are required to $O(\alpha_s^3)$. Their calculation is underway for scales $Q^2/m_c^2 \gtrsim 10$ in case of the structure function $F_2(x,Q^2)$ [31]. A series of moments [32] and all logarithmic corrections have been calculated [33,34], as well as four of five contributing Wilson coefficients [35,34,36,37], which use in their representation the massless 3-loop Wilson coefficients [38]. The calculation of the last contributing Wilson coefficient is underway [39]. For an approximate NNLO representation see [40].

| | $\alpha_s(M_Z^2)$ | exp. | theory | Ref. |
|---|---|---|---|---|
| NLO | | | | |
| BCDMS | 0.1111 | $\pm0.0018$ | | [3,4] |
| H1 | 0.1150 | $\pm0.0017$ | $\pm0.0050$ | [18,19] |
| CTEQ6 | 0.1165 | $\pm0.0065$ | | [20] |
| A02 | 0.1171 | $\pm0.0015$ | $\pm0.0033$ | [21] |
| ZEUS | 0.1166 | $\pm0.0049$ | | [22] |
| MRST03 | 0.1165 | $\pm0.0020$ | $\pm0.0030$ | [23] |
| BB1 (pol) | 0.1135 | $\pm0.004$ | $^{+0.010}_{-0.009}$ | [24] |
| BB2 (pol) | 0.1132 | $^{+0.0043}_{-0.0051}$ | $^{+0.0056}_{-0.0097}$ | [25] |

Table 1: Some NLO results until $\sim$ 2003 on $\alpha_s(M_Z^2)$ from deep inelastic scattering, including also global PDF fits. The H1 value is subject to an additional error of $+0.0009/-0.0005$ and the ZEUS value of $\pm0.0018$ due to model dependence. We add results from deep-inelastic scattering off polarized targets; from [26].

In Section 2 we describe NNLO and N$^3$LO non-singlet analyses and turn to combined non-singlet and singlet analyses in Section 3. We always discuss the response to individual data sets in the different fits and investigate the perturbative stability going from NLO to NNLO (and to N$^3$LO). Section 4 contains the conclusions.

## $\alpha_s(M_Z^2)$ at NNLO from Non-Singlet Data

NNLO flavor non-singlet analyses have been carried out in Refs. [26,41,42,43]. They rely on the structure function differences $F_2^p(x,Q^2) - F_2^n(x,Q^2)$ and valence approximations typically in the region $x \gtrsim 0.35$. Here $F_2^n(x,Q^2)$ is measured from $F_2^d(x,Q^2)$ with account of the appropriate nuclear corrections. Also the target mass corrections [44] have to be applied. Figure 1 shows the size of sea-quark tail contributions, comparing to a consistent set of PDFs, i.e. to one whose scaling violations have a similar value of $\alpha_s(M_Z^2)$. An important issue is a necessary cut in $W^2 > 12.5$ GeV$^2$ and $Q^2 > 4$ GeV$^2$ to separate the higher twist contributions in the non-singlet case [41,46,43].

For the 4-loop non-singlet anomalous dimension the moments $N = 2,3,4$ are known [47,48,49,50]. In [41] it has been shown, that one may approximate the 4-loop anomalous dimension by the Pade-approximant

$$\gamma_n^{(3),\mathrm{approx}} = \frac{{\gamma_n^{(2)}}^2}{\gamma_n^{(1)}}, \qquad (2)$$

which agrees better than by 20% for the known moments for $N_F = 3$. Furthermore, we assigned a ±100% error to this quantity. In the fit this results into an uncertainty of $\delta\Lambda_{\rm QCD} = \pm 2$ MeV, far below the experimental uncertainty of $\Delta\Lambda_{\rm QCD} = \pm 26$ MeV. The 4-loop non-singlet anomalous dimension is therefore of lesser importance compared to the massless 3-loop Wilson coefficient. One has to consider also charm quark effects contributing with 2-loop order [51,52] which are below 0.35% in the valence region, which is similar for the asymptotic 3-loop corrections [36].

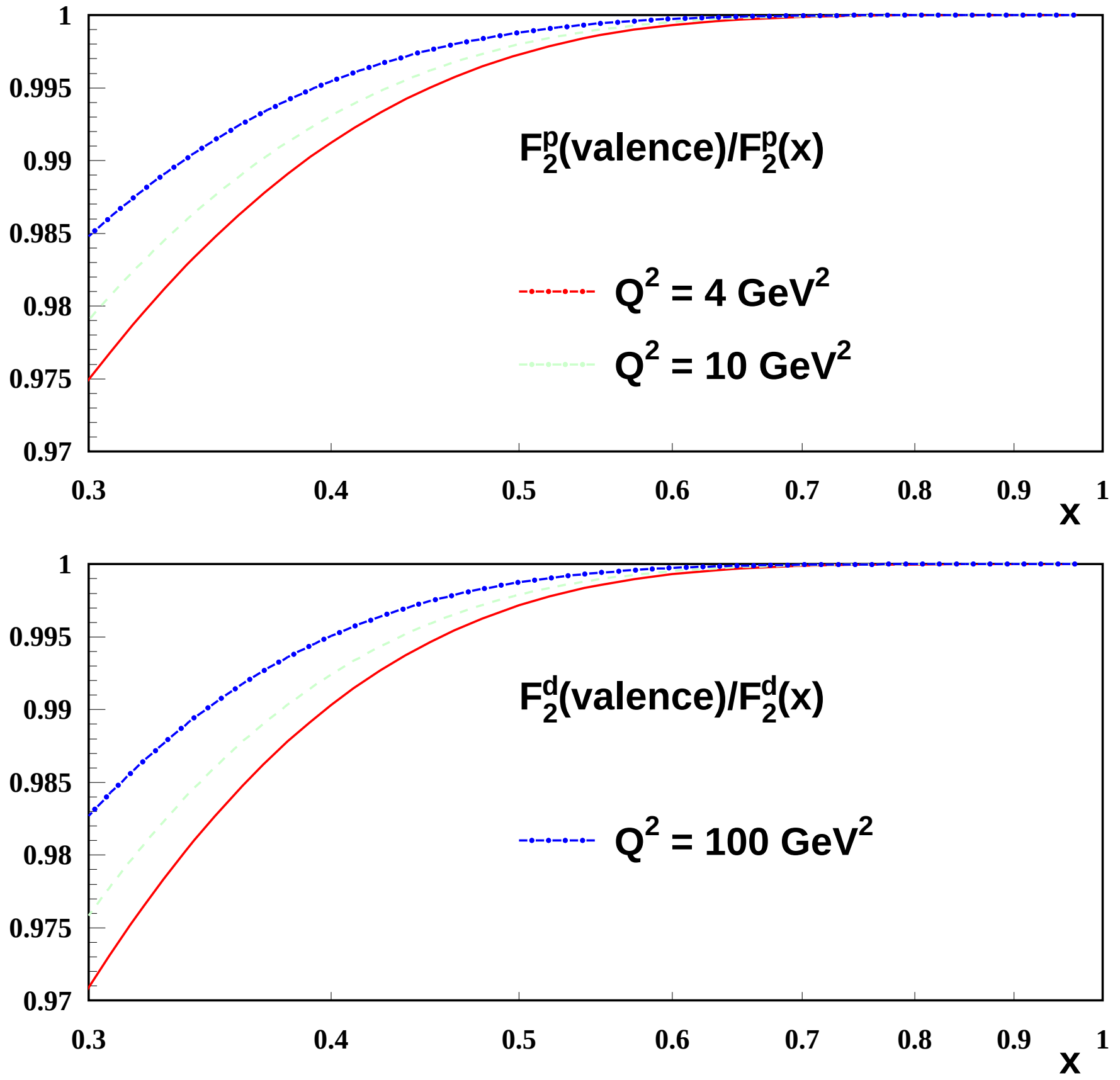

Figure 1: The ratio $F_2^{\rm p(d),val}/F_2^{\rm p(d)}$ for $Q^2 = 4, 10, 100$ GeV$^2$ using the ABKM09 parton distribution functions [45].

In the non-singlet analysis NNLO values of $\alpha_s^{\rm NNLO}(M_Z^2) = 0.1120...0.1134 \pm 0.0022$ are obtained and the corresponding NLO values are somewhat larger with $\alpha_s^{\rm NLO}(M_Z^2) = 0.1147 \pm 0.0021$, unlike the case for combined non-singlet and singlet analyses discussed in the next Section. Main results are summarized in Table 2.

The NLO results given by the experiments BCDMS and NMC are basically reproduced by the combined fit of the data from BCDMS, SLAC and NMC, as partial results. The steps from NLO to NNLO and N$^3$LO$^*$, the $^*$ standing for the yet not completely known 4-loop non-singlet anomalous dimension, are perturbatively stable. As NNLO value we finally quote

$$\alpha_s^{\rm NNLO}(M_Z^2) = 0.1132 \pm 0.0022 \qquad (3)$$

and for N$^3$LO$^*$

$$\alpha_s^{\rm N^3LO^*}(M_Z^2) = 0.1137 \pm 0.0022 \ . \qquad (4)$$

The difference of these values $\Delta\alpha_s(M_Z^2) = 0.0005$ may serve as an estimate of the size of the remaining theory uncertainty.

| Experiment | $\alpha_s(M_Z)$ | | | |
|---|---|---|---|---|
| | $\text{NLO}_{exp}$ | NLO | NNLO | $\text{N}^3\text{LO}^*$ |
| BCDMS | $0.1111 \pm 0.0018$ | $0.1138 \pm 0.0007$ | $0.1126 \pm 0.0007$ | $0.1128 \pm 0.0006$ |
| NMC | $0.117 \ {}^{+0.011}_{-0.016}$ | $0.1166 \pm 0.0039$ | $0.1153 \pm 0.0039$ | $0.1153 \pm 0.0035$ |
| SLAC | | $0.1147 \pm 0.0029$ | $0.1158 \pm 0.0033$ | $0.1152 \pm 0.0027$ |
| BBG | | $0.1148 \pm 0.0019$ | $0.1134 \pm 0.0020$ | $0.1141 \pm 0.0021$ |
| BB | | $0.1147 \pm 0.0021$ | $0.1132 \pm 0.0022$ | $0.1137 \pm 0.0022$ |

Table 2: Comparison of the values of $\alpha_s(M_Z)$ obtained by BCDMS [4] and NMC [53] at NLO with the results of the flavor non-singlet fits BBG [41] and BB [43] of the DIS flavor non-singlet world data, at NLO, NNLO, and $\text{N}^3\text{LO}^*$ with the response of the individual data sets, combined for the experiments BCDMS [3,4,54], NMC [11], and SLAC [55]; from [56].

## $\alpha_s(M_Z^2)$ at NNLO: General Analyses

Combined non-singlet and singlet analyses at NNLO have been performed by different groups starting with the year 2000. In the following we will first compare the results of more recent analyses [56,57,58,59], in particular also w.r.t. their response to individual data sets and give a summary of the present NNLO results at the end of this section.

| Experiment | $\alpha_s(M_Z)$ | | |
|---|---|---|---|
| | $\text{NLO}_{exp}$ | NLO | NNLO |
| BCDMS | $0.1111 \pm 0.0018$ | $0.1150 \pm 0.0012$ | $0.1084 \pm 0.0013$ |
| NMC | $0.117 \ {}^{+0.011}_{-0.016}$ | $0.1182 \pm 0.0007$ | $0.1152 \pm 0.0007$ |
| SLAC | | $0.1173 \pm 0.0003$ | $0.1138 \pm 0.0003$ |
| HERA comb. | | $0.1174 \pm 0.0003$ | $0.1126 \pm 0.0002$ |
| DY | | $0.108 \quad \pm 0.010$ | $0.101 \quad \pm 0.025$ |
| ABM11 [56] | | $0.1180 \pm 0.0012$ | $0.1134 \pm 0.0011$ |

Table 3: Comparison of the values of $\alpha_s(M_Z)$ obtained by BCDMS [4] and NMC [53] at NLO with the individual results of the fit in the present analysis at NLO and NNLO for the HERA data [60], the NMC data [11], the BCDMS data [61,4], the SLAC data [62,63,64,65,66,67], and the DY data [68,69]; from [56].

An NNLO analysis has been performed on the world deep-inelastic data on proton and deuteron targets and Drell-Yan (DY) data in Ref. [56]. A summary of results is given in Table 3. The NLO analysis results in a higher value of $\alpha_s(M_Z^2) = 0.1180$. Values of this size are also obtained in other analyses, cf. [57], while the value in [59] is even higher. A shift of $\Delta\alpha_s(M_Z^2) = -0.0045$ is observed from NLO to NNLO, which is larger than in case of the non-singlet analyses, cf. Section 2. This shift is thoroughly observed for all data sets. Here the Drell-Yan data have a rather low sensitivity to $\alpha_s$ and serve basically to constrain the different sea-quark distributions. In the NLO fit the partial BCDMS value moves up, but takes a low value again at NNLO. The NLO NMC value comes out consistently. We would like to mention that at the time of the BCDMS and NMC experiments no proper description of the longitudinal structure function has been used. It was not available yet to $O(\alpha_s^3)$ [38]. This information is, however, important and has been taken into account in the non-singlet analyses [41,43] and discussed in detail in Ref. [70] for the combined analyses.

| $\alpha_s(M_Z^2)$ | with $\sigma_{\rm NMC}$ | with $F_2^{\rm NMC}$ | difference |
|---|---|---|---|
| NLO | 0.1179(16) | 0.1195(17) | $+0.0026 \simeq 1\sigma$ |
| NNLO | 0.1135(14) | 0.1170(15) | $+0.0035 \simeq 2.3\sigma$ |
| NNLO + $F_{\rm L}$ $O(\alpha_s^3)$ | 0.1122(14) | 0.1171(14) | $+0.0050 \simeq 3.6\sigma$ |

Table 4: Fit results of the world deep-inelastic data at NLO and NNLO using the published values of either $F_2^{\rm NMC}$ or $\sigma^{\rm NMC}$ with a correct description of the longitudinal structure function; from [70].

In case of fitting $F_2^{\rm NMC}$ and not describing $F_{\rm L}(x,Q^2)$ at NNLO much larger values of $\alpha_s(M_Z^2)$ are obtained. Therefore we regard it as mandatory to fit the published differential scattering cross sections using $F_{\rm L}(x,Q^2)$ at $O(\alpha_s^3)$. Presently, the MMHT [71] analysis uses $F_{\rm L}(x,Q^2)$ only at NLO. One should note, however, that, the values of $F_{\rm L}(x,Q^2)$ at NNLO are significantly different in the small $x$ region.

| Experiment | $\alpha_s(M_Z)$ | | |
|---|---|---|---|
| | NLO$_{exp}$ | NLO | NNLO$^*$ |
| D0 1 jet | $0.1161 \ ^{+\,0.0041}_{-\,0.0048}$ | $0.1190 \pm 0.0011$ | $0.1149 \pm 0.0012$ |
| D0 2 jet | | $0.1174 \pm 0.0009$ | $0.1145 \pm 0.0009$ |
| CDF 1 jet (cone) | | $0.1181 \pm 0.0009$ | $0.1134 \pm 0.0009$ |
| CDF 1 jet ($k_\perp$) | | $0.1181 \pm 0.0010$ | $0.1143 \pm 0.0009$ |
| ABM11 [56] | | $0.1180 \pm 0.0012$ | $0.1134 \pm 0.0011$ |

Table 5: Comparison of the values of $\alpha_s(M_Z)$ obtained by D0 in [72] with the ones based on including individual data sets of Tevatron jet data [73,74,75,76] into the analysis at NLO. The NNLO$^*$ fit refers to the NNLO analysis of the DIS and DY data together with the NLO and soft gluon resummation corrections (next-to-leading logarithmic accuracy) for the 1 jet inclusive data, cf. [77,78]; from [56].

Another important issue concerns the description of higher twist contributions as discussed in the non-singlet analysis before. One has either to fit these terms within the combined analysis or to apply suitable cuts to remove part of the data being affected by these contributions. In Ref. [56] the cuts $W^2 > 12.5$ GeV$^2$ and $Q^2 > 2.5$ GeV$^2$, as used in [59], have been applied without fitting the higher twist terms. This resulted in a high value of $\alpha_s(M_Z^2) = 0.1191 \pm 0.0016$. Cutting more conservatively with $W^2 > 12.5$ GeV$^2$ and $Q^2 > 10$ GeV$^2$ led to $\alpha_s(M_Z^2) = 0.1134 \pm 0.0008$, a value well comparable to the one obtained in the ABM11 analysis [56]. We therefore recommend the latter cuts as mandatory. NNPDF [79] uses a cut of $Q^2 > 5$ GeV$^2$ removing part but not all of the higher twist contributions.

The ABM11 analysis considered also the effect of the Tevatron jet data on $\alpha_s(M_Z^2)$, fitting each of the individual data sets together with the deep-inelastic world-data, cf. Table 5. At NLO the jet data perfectly reproduce the DIS value. As the NNLO jet cross section is not yet known, we use a NNLO$^*$ description here, referring to threshold resummation [77,78]. We mention that the prescription deviates significantly at lower values of $p_\perp$ [80] and the approach did not account for the cone size dependence [81,80]. Yet the inclusion of the data sets from Tevatron did not change the DIS values of $\alpha_s(M_Z^2)$ at NNLO significantly. They resulted in an enhancement to values in the range 0.1134...0.1149, differing with data set and kind of jet algorithm used. Therefore JR and ABM did not include jet data in subsequent NNLO analyses, cf. [82,83], unlike MSTW [59], MMHT [71], NNPDF [58,79], and CTEQ [84]. The combined analysis of the world DIS data and the jet data from hadron colliders at NNLO will only be possible if the NNLO calculation of the jet cross section is completed, cf. [85].

| Experiment | $\alpha_s(M_Z)$ | | |
|---|---|---|---|
| | NLO$_{exp}$ | NLO | NNLO |
| BCDMS [61,4] | $0.1111 \pm 0.0018$ | $0.1204 \pm 0.0015$ | $0.1158 \pm 0.0015$ |
| NMC$_p$ [11] | | $0.1192 \pm 0.0018$ | $0.1150 \pm 0.0020$ |
| NMC$_{pd}$ [86] | $0.117 \begin{smallmatrix} +0.011 \\ -0.016 \end{smallmatrix}$ | | $0.1146 \pm 0.0107$ |
| SLAC [55] | | $> 0.124$ | $> 0.124$ |
| HERA I [60] | | $0.1223 \pm 0.0018$ | $0.1199 \pm 0.0019$ |
| ZEUS H2 [87,88] | | $0.1170 \pm 0.0027$ | $0.1231 \pm 0.0030$ |
| ZEUS F2C [89,90,91,92] | | $0.1144 \pm 0.0060$ | |
| NuTeV [93,94] | | $0.1252 \pm 0.0068$ | $0.1177 \pm 0.0039$ |
| E605 [68] | | $0.1168 \pm 0.0100$ | |
| E866 [95,96,69] | | $0.1135 \pm 0.0029$ | |
| CDF Wasy [97] | | $0.1181 \pm 0.0060$ | |
| CDF Zrap [98] | | $0.1150 \pm 0.0034$ | $0.1205 \pm 0.0081$ |
| D0 Zrap [99] | | $0.1227 \pm 0.0067$ | |
| CDF R2KT [73] | | $0.1228 \pm 0.0021$ | $0.1225 \pm 0.0021$ |
| D0 R2CON [75] | $0.1161 \begin{smallmatrix} +0.0041 \\ -0.0048 \end{smallmatrix}$ | $0.1141 \pm 0.0031$ | $0.1111 \pm 0.0029$ |
| NN21 [57,58] | | $0.1191 \pm 0.0006$ | $0.1173 \pm 0.0007$ |

Table 6: Comparison of the values of $\alpha_s(M_Z)$ obtained by BCDMS [4], NMC [53], and D0 [72] at NLO with the results of NN21 [57,58] for the fits to DIS and other hard scattering data at NLO and NNLO and the corresponding response of the different data sets analyzed; from [56].

Because of the significant difference between $\alpha_s^{\rm NLO}(M_Z^2)$ and $\alpha_s^{\rm NNLO}(M_Z^2)$, it has to be regarded as impossible to include data sets in a NNLO fit, which can be only described at NLO as jet data in $ep$ scattering and $p\overline{p}(p)$ scattering. Their inclusion usually will lead to a larger value of $\alpha_s^{\rm NNLO}(M_Z^2)$ if compared to present pure NLO analyses. Again JR and ABM refrain from including these data, while MRST/MMHT include both types of jet data and NNPDF and CTEQ include hadron collider jet data in a significant portion in their fit, see Tables 6,7 and Ref. [84].

In the more 'global' fits also deep-inelastic data off nuclei are used, cf. Tables 6,7 and [84]. This is problematic, since it is unknown whether QCD evolution proceeds the same way within large nuclei, cf. e.g. [126]. CTEQ also includes CDHSW data, which were regarded to be not unproblematic, cf. [2]. We mention that e.g. the NuTeV $\nu$Fe $F_2$ data request a high value of $\alpha_s(M_Z^2)$ [71].

The individual response in $\alpha_s(M_Z^2)$ for the data sets used in the fit of NNPDF and MRST are summarized in Tables 6,7. It is interesting to observe, that for MRST the partial fit results to the Tevatron jet data are obtained in the range $\alpha_s(M_Z^2) = 0.1133...0.1157$ at NNLO similar to the ABM11 results, while NNPDF finds values of $0.1111, 0.1205$ and $0.1225$.

In Table 8 we compare the pulls in $\alpha_s(M_Z^2)$ for the DIS and DY data sets for ABM11, BBG, NN21, and MSTW at NNLO. In case of the BCDMS data NN21 yields a higher value than obtained in the other analyses. In case of NMC the value obtained by MSTW is larger than in the other cases. For the SLAC data NNPDF obtains a large value, but MSTW obtains a lower and a larger value for *ep* and *ed* scattering, respectively. In case of the HERA data the ABM11 value is low, but those of MSTW and NNPDF are high and the DY values come out consistent between ABM11 and MSTW. Interestingly, the final values for NNPDF and MRST are very similar, but due to the above for very different reasons. They are higher than the values obtained by ABM11 and BBG by $\Delta\alpha_s(M_Z^2) = 0.0038$.

| Experiment | $\alpha_s(M_Z)$ | | |
|---|---|---|---|
| | NLO$_{exp}$ | NLO | NNLO |
| BCDMS $\mu p, F_2$ [61] | $0.1111 \pm 0.0018$ | – | $0.1085 \pm 0.0095$ |
| BCDMS $\mu d, F_2$ [4] | | $0.1135 \pm 0.0155$ | $0.1117 \pm 0.0093$ |
| NMC $\mu p, F_2$ [11] | $0.117\ ^{+\ 0.011}_{-\ 0.016}$ | $0.1275 \pm 0.0105$ | $0.1217 \pm 0.0077$ |
| NMC $\mu d, F_2$ [11] | | $0.1265 \pm 0.0115$ | $0.1215 \pm 0.0070$ |
| NMC $\mu n/\mu p$ [86] | | 0.1280 | 0.1160 |
| E665 $\mu p, F_2$ [100] | | 0.1203 | – |
| E665 $\mu d, F_2$ [100] | | – | – |
| SLAC $ep, F_2$ [62,101] | | $0.1180 \pm 0.0060$ | $0.1140 \pm 0.0060$ |
| SLAC $ed, F_2$ [62,101] | | $0.1270 \pm 0.0090$ | $0.1220 \pm 0.0060$ |
| NMC,BCDMS,SLAC, $F_L$ [61,11,55] | | $0.1285 \pm 0.0115$ | $0.1200 \pm 0.0060$ |
| E886/NuSea $pp$, DY [96] | | – | $0.1132 \pm 0.0088$ |
| E886/NuSea $pd/pp$, DY [69] | | $0.1173 \pm 0.107$ | $0.1140 \pm 0.0110$ |
| NuTeV $\nu N, F_2$ [102] | | $0.1207 \pm 0.0067$ | $0.1170 \pm 0.0060$ |
| CHORUS $\nu N, F_2$ [103] | | $0.1230 \pm 0.0110$ | $0.1150 \pm 0.0090$ |
| NuTeV $\nu N, xF_3$ [102] | | $0.1270 \pm 0.0090$ | $0.1225 \pm 0.0075$ |
| CHORUS $\nu N, xF_3$ [103] | | $0.1215 \pm 0.0105$ | $0.1185 \pm 0.0075$ |
| CCFR [93,94] | | 0.1190 | – |
| NuTeV $\nu N \to \mu\mu X$ [93,94] | | $0.1150 \pm 0.0170$ | – |
| H1 $ep$ 97-00, $\sigma_r^{\rm NC}$ [104,19,18,105] | | $0.1250 \pm 0.0070$ | $0.1205 \pm 0.0055$ |
| ZEUS $ep$ 95-00, $\sigma_r^{\rm NC}$ [106,107,108,109] | | $0.1235 \pm 0.0065$ | $0.1210 \pm 0.0060$ |
| H1 $ep$ 99-00, $\sigma_r^{\rm CC}$ [19] | | $0.1285 \pm 0.0225$ | $0.1270 \pm 0.0200$ |
| ZEUS $ep$ 99-00, $\sigma_r^{\rm CC}$ [110] | | $0.1125 \pm 0.0195$ | $0.1165 \pm 0.0095$ |
| H1/ZEUS $ep, F_2^{\rm charm}$ [89,90,111,112,113,114] | | – | $0.1165 \pm 0.0095$ |
| H1 $ep$ 99-00 incl. jets [115,116] | $0.1168\ ^{+\ 0.0049}_{-\ 0.0035}$ | $0.1127 \pm 0.0093$ | |
| ZEUS $ep$ 96-00 incl. jets [117,118,119] | $0.1208\ ^{+\ 0.0044}_{-\ 0.0040}$ | $0.1175 \pm 0.0055$ | |
| D0 II $p\bar{p}$ incl. jets [75] | $0.1161\ ^{+\ 0.0041}_{-\ 0.0048}$ | $0.1185 \pm 0.0055$ | $0.1133 \pm 0.0063$ |
| CDF II $p\bar{p}$ incl. jets [73] | | $0.1205 \pm 0.0045$ | $0.1165 \pm 0.0025$ |
| D0 II $W \to l\nu$ asym. [120] | | – | – |
| CDF II $W \to l\nu$ asym. [121] | | – | – |
| D0 II $Z$ rap. [99] | | $0.1125 \pm 0.0100$ | $0.1136 \pm 0.0084$ |
| CDF II $Z$ rap. [122] | | $0.1160 \pm 0.0070$ | $0.1157 \pm 0.0067$ |
| MSTW [59] | | $0.1202\ ^{+\ 0.0012}_{-\ 0.0015}$ | $0.1171 \pm 0.0014$ |

Table 7: Comparison of the values of $\alpha_s(M_Z)$ obtained by BCDMS [4], NMC [53], HERA-jet [115,117] (see also [123,124]), and D0 [72] at NLO with the results of the MSTW fits to DIS and other hard scattering data at NLO and NNLO and the corresponding response of the different data sets analysed, cf. Figs. 7a and 7b in [59]. Entries not given correspond to $\alpha_s(M_Z)$ central values below 0.110 or above 0.130; in case no errors are assigned these are larger than the bounds provided in form of the plots in [59,125]; from [56].

| Data Set | ABM11 | BBG | NN21 | MSTW |
|---|---|---|---|---|
| BCDMS | $0.1128 \pm 0.0020$<br>$(0.1084 \pm 0.0013)$ | $0.1126 \pm 0.0007$ | $0.1158 \pm 0.0015$ | $0.1101 \pm 0.0094$ |
| NMC | $0.1055 \pm 0.0026$<br>$(0.1152 \pm 0.0007)$ | $0.1153 \pm 0.0039$ | $0.1150 \pm 0.0020$ | $0.1216 \pm 0.0074$ |
| SLAC | $0.1184 \pm 0.0021$<br>$(0.1138 \pm 0.0003)$ | $0.1158 \pm 0.0034$ | $> 0.124$ | $\left\{\begin{array}{l} 0.1140 \pm 0.0060 \text{ ep} \\ 0.1220 \pm 0.0060 \text{ ed} \end{array}\right.$ |
| HERA | $0.1139 \pm 0.0014$<br>$(0.1126 \pm 0.0002)$ | | $\left\{\begin{array}{l} 0.1199 \pm 0.0019 \\ 0.1231 \pm 0.0030 \end{array}\right.$ | $0.1208 \pm 0.0058$ |
| DY | $(0.101 \pm 0.025)$ | – | – | $0.1136 \pm 0.0100$ |
| | $0.1134 \pm 0.0011$ | $0.1134 \pm 0.0020$ | $0.1173 \pm 0.0007$ | $0.1171 \pm 0.0014$ |

Table 8: Comparison of the pulls in $\alpha_s(M_Z)$ per data set between the ABM11 [56], BBG [41], NN21 [58] and MSTW [59] analyses at NNLO. The values in parenthesis of ABM11 correspond to the case where the shape parameters are not refitted which is also the case for BBG; from [56].

Since the world deep inelastic data contain a large contribution due to charm it is important to determine the charm quark mass $m_c$ correlated with $\alpha_s(M_Z^2)$ and the parameters of the parton distribution functions (PDFs), as well as the higher twist contributions in a NNLO analysis. Such analyses have been performed in [127] and [128] giving results in the $\overline{\text{MS}}$ scheme and the pole mass scheme, respectively. One may compare these values with those of the PDG [129] and precision determinations using $e^+e^-$ data [130,131]. The value obtained in [127] is in excellent agreement with the one given in [130,131] and the quarkonium $1S$ state [132]. In Ref. [128] the determination has been performed in the pole mass scheme using the general mass variable flavor scheme, obtaining a very low value of $m_c^{\text{Pole}} = 1.25$ GeV with $\alpha_s(M_Z^2) = 0.1167$. The value for $\chi^2$ turns out to be larger than that of Ref. [127], where the fixed flavor number scheme has been used. In other analyses by CTEQ [84] and NNPDF [79] pole mass values of 1.3 GeV and 1.275 GeV have been assumed as input, which are off the PDG-value.

| | | |
|---|---|---|
| $m_c^{\overline{\text{MS}}}$ | $1.24 \pm 0.03\ {}^{+0.03}_{-0.03}$ | ABDLM, DIS, FFNS, $\chi^2 = 61/52$ [127] |
| $m_c^{\overline{\text{MS}}}$ | $1.279 \pm 0.013$ | $e^+e^-$ [130] |
| $m_c^{\overline{\text{MS}}}$ | $1.288 \pm 0.020$ | $e^+e^-$ [131] |
| $m_c^{\overline{\text{MS}}}$ | $1.246 \pm 0.023$ | quarkonium $1S$ [132] |
| $m_c^{\overline{\text{MS}}}$ | $1.275 \pm 0.025$ | PDG [129] |
| $m_c^{\text{Pole}}$ | $1.67 \pm 0.07$ | PDG [129] |
| $m_c^{\text{Pole}}$ | 1.25 | GMVFNS; $\chi^2 = 72/52$ [128] |

Table 9: Comparison of the measurement of $m_c$ in DIS, $e^+e^-$ annihilation, and 1S quarkonium.

The use of the general mass variable flavor number scheme (GMVFNS) is advocated by several fitting groups [71,79,84], despite it has been shown in [133] that even in the kinematic range at HERA one may work in the fixed flavor number scheme. An ideal interpolation up to $O(\alpha_s^2)$ is possible in the BMSN-scheme [134], as has been shown in [45]. This is illustrated in Figure 2. Changing from $N_F \to N_F + 1$ at the scale $\mu^2 = m_{N_F+1}^2$ is usually impossible, since it is normally near to the production threshold, where the heavy quark is not ultrarelativistic.

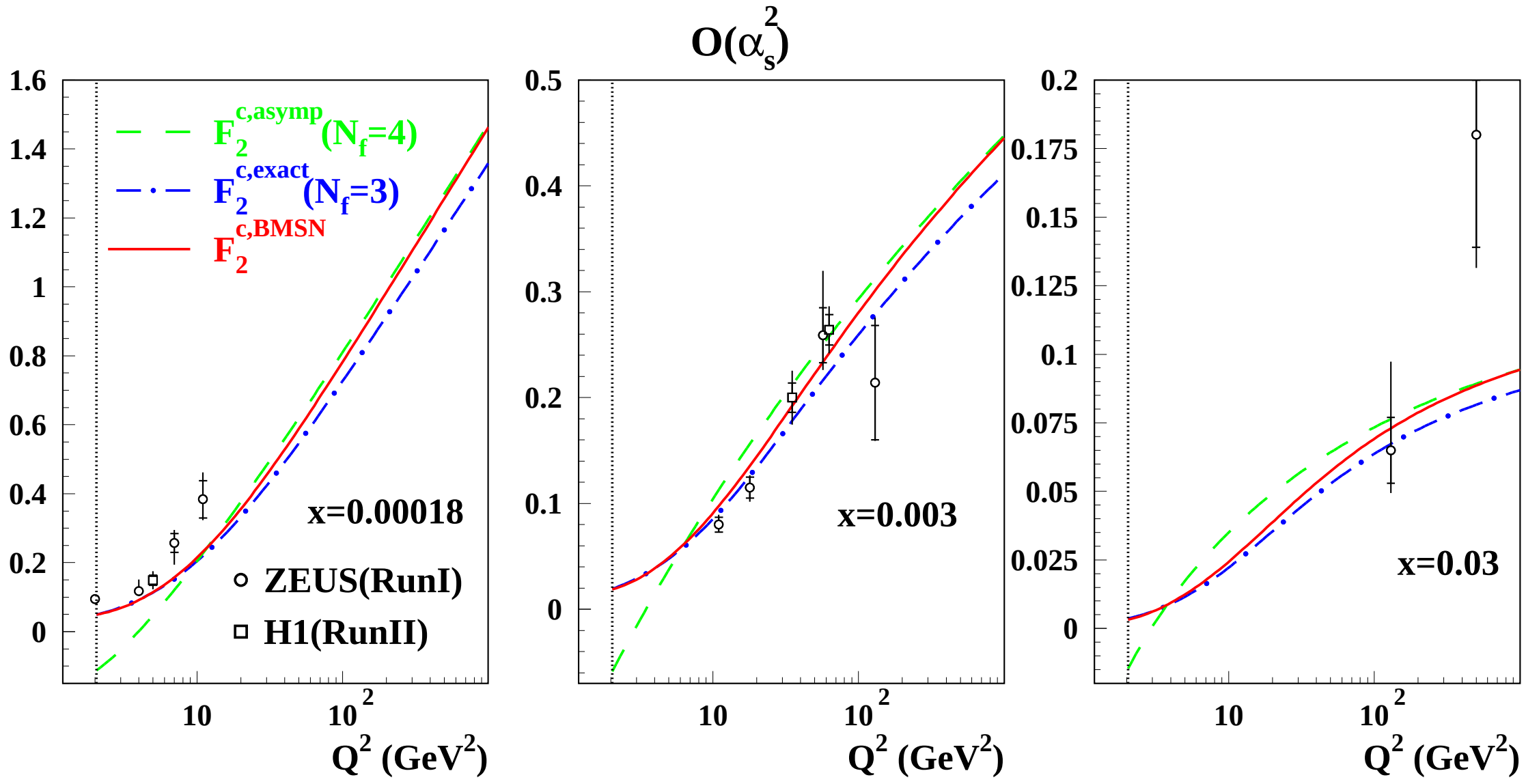

Figure 2: Comparison of $F_2^c$ in different schemes to H1- and ZEUS-data. Solid lines: GMVFN scheme in the BMSN prescription, dash-dotted lines: 3-flavor scheme, dashed lines: 4-flavor scheme. The vertical dotted line denotes the position of the charm-quark mass $m_c = 1.43$ GeV; from [45].

In some cases the transition of a massive quark to a massless quark at large values of $Q^2$ proceeds very slowly, cf. also [135]. This is illustrated in Figure 3 for the heavy quark contributions in case of the polarized Bjorken sum-rule to $O(\alpha_s^2)$, considering both the exact charm and the bottom quark contributions. Note that at low scales $\mu^2 = m_c^2$, the bottom contribution is even negative.

Let us now turn to the summary of the NNLO values for $\alpha_s(M_Z^2)$ having been determined since 2001, which we summarize in Table 10. An early determination is due J. Santiago and F. Yndurain [137], using moments of Bernstein polynomials. Here already an error $\delta\alpha_s(M_Z^2)/\alpha_s(M_Z^2) \simeq 1\%$ has been obtained analyzing the $F_2^{ep}$ data. Various analyses followed, mostly obtaining lower values of $\alpha_s(M_Z^2)$. We mention the MRST03 value with $\alpha_s(M_Z^2) = 0.1153 \pm 0.0020$, slightly lower than an earlier result of this group in [138]. Three values are higher with 0.1171...0.1174 by MSTW, NN21 and MMHT [59,58,71]. CTEQ finds a lower value of $\alpha_s(M_Z^2) = 0.1150$ quoting large errors of the size usually obtained by scale variation errors at NLO corresponding to its $\Delta\chi^2$ assignment. On the other hand, MMHT quote much smaller errors corresponding to $\Delta\chi^2 = 7.2$, which are correspondingly smaller than those of ABM performing the analysis with $\Delta\chi^2 = 1$. We would like to highlight the difference in the central values in the ABKM analysis [45] choosing either the FFNS or the BMSN description implying a theoretical uncertainty of $\delta\alpha_s(M_Z^2) = 0.0006$. This is one of the typical theory errors which cannot be undercut easily in size. It compares to the difference of the NNLO and NNLO* values in the non-singlet analyses [41] of $\delta\alpha_s(M_Z^2) = 0.0007$. CTEQ has performed an analysis leaving out the jet data [139] and obtains $\alpha_s(M_Z^2) = 0.1140$. Likewise, R. Thorne (MRST) [140] has fitted the DIS and DY data only and restricting higher twist at low $x$ leading to $\alpha_s^{\rm NLO}(M_Z^2) = 0.1179$ and $\alpha_s^{\rm NNLO}(M_Z^2) = 0.1136$, i.e. basically the same

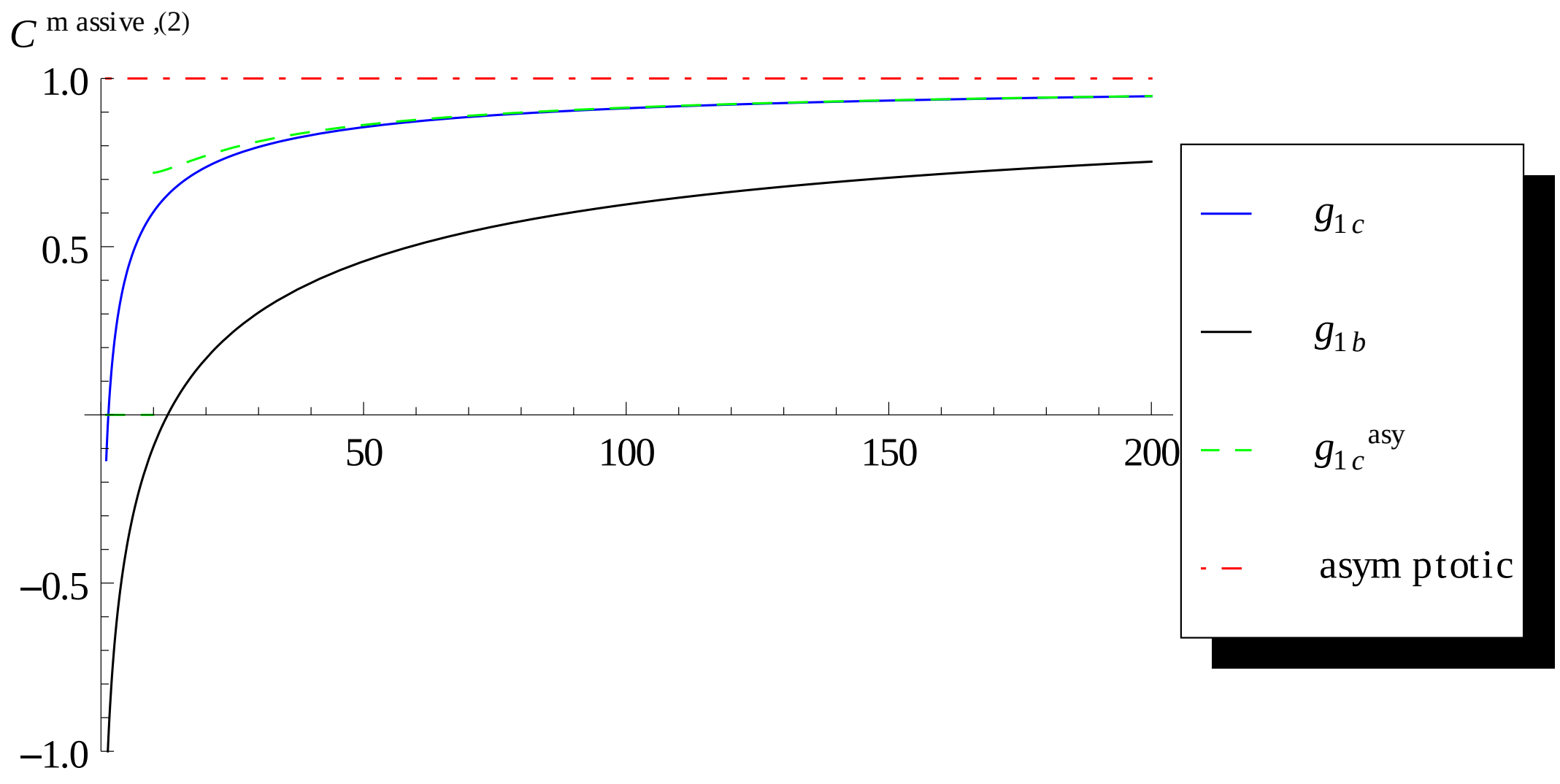

Figure 3: Coefficient of the massive contributions to the polarized Bjorken sum rule as a function of $\xi = Q^2/m^2$ for the $c$- and $b$-contribution (full lines). Dashed line: asymptotic formula for charm. Dash-dotted line: $N_F \rightarrow N_F + 1$ transition for charm; from [136].

values as having been given in Table 3.

Also in a series of other hard processes a lower value of $\alpha_s(M_Z^2)$ has been measured at NNLO, and even at NLO, see Table 11. At NNLO values in the range 0.1123...0.1151 were found for thrust and the $C$-parameter in $e^+e^-$ annihilation and the inclusive $t\bar{t}$ cross section by CMS. Furthermore there is a low lattice value of 0.1166, compared to other lattice determinations [153]. At NLO the lower values range as 0.1148...0.1160 and stem from jet measurements in $e^+e^-$ annihilation and $ep, pp$ and $p\bar{p}$-scattering. We clearly await the NNLO corrections for these processes for corresponding refined determinations of $\alpha_s(M_Z^2)$.

We finally would like to quote the estimated precision to measure $\alpha_s(M_Z^2)$ at LHeC [154]. For a proton beam energy of 7 TeV and an electron beam energy of 60 GeV at a luminosity of $\mathcal{L} = 10^{34}\mathrm{cm}^{-2}\mathrm{s}^{-1}$ using cuts of $Q^2 > 3.5~\mathrm{GeV}^2$ the relative precision of $\alpha_s(M_Z^2)$ would be 0.15%, while for $Q^2 > 10~\mathrm{GeV}^2$ 0.25% would be obtained [155], quoting full experimental uncertainties. This is up to an order of magnitude better than the present accuracy. Future precise determinations of $\alpha_s(M_Z^2)$ in DIS are also expected from EIC [156].

## Conclusions

At present NNLO analyses of the world deep-inelastic data can be supplemented only by the DY and $t\bar{t}$ data from hadron colliders as we still await the NNLO calculations for the $ep$ and hadronic jet cross sections to be finished. Non-singlet analyses can be carried out at N$^3$LO yielding $\alpha_s(M_Z^2) = 0.1137 \pm 0.0022$. These analyses are free of the gluon uncertainty. The corresponding value at NNLO is $\alpha_s(M_Z^2) = 0.1132 \pm 0.0022$. Severe cuts to prevent higher twist contributions are important. Correct NNLO analyses should be based on proton and deuteron data only because of large nuclear effects for heavy nuclei and leave out the jet data, since they would have to be dealt with at NLO only leading to an upward shift of $\alpha_s(M_Z^2)$. The analyses shall deliver $m_c$ together with $\alpha_s$, reproducing the measured values of $m_c$ determined in other high energy scattering processes. The BMSN-interpolation and the FFNS seem to be appropriate descriptions. A too early transition of charm to a massless quark in GMVFN-schemes would contradict the QCD description, since power correction terms are present in the important range at low $Q^2$. Assuming a fixed value of $\alpha_s(M_Z^2)$ within a PDF fit often leads to a biased determination of PDFs and other fitted parameters and the fit-result may be obtained off minimum since $\alpha_s$ is strongly correlated to various other fit parameters, notably to those of the gluon distribution. Various recent determinations yielded low values of $\alpha_s^{\rm NNLO}(M_Z^2)$.

| NNLO Analyses | | | |
|---|---|---|---|
| | | $\alpha_s(M_Z^2)$ | |
| SY | 2001 | $0.1166 \pm 0.0013$ | $F_2^{ep}$ [137] |
| SY | 2001 | $0.1153 \pm 0.0063$ | $xF_3^{\nu N}$ h. Nucl. [137] |
| A02 | 2002 | $0.1143 \pm 0.0020$ | [21] |
| MRST03 | 2003 | $0.1153 \pm 0.0020$ | [23] |
| BBG | 2004(06,12) | $0.1134\ ^{+0.0019}_{-0.0021}$ | valence analysis, NNLO [26,41,43] |
| GRS | 2006 | $0.112$ | valence analysis, NNLO [42] |
| AMP06 | 2006 | $0.1128 \pm 0.0015$ | [141] |
| JR | 2008 | $0.1128 \pm 0.0010$ | dynamical approach [142] |
| JR | 2008 | $0.1162 \pm 0.0006$ | including NLO-jets [142] |
| ABKM | 2009 | $0.1135 \pm 0.0014$ | HQ: FFNS $N_f = 3$ [45] |
| ABKM | 2009 | $0.1129 \pm 0.0014$ | HQ: BMSN [45] |
| MSTW | 2009 | $0.1171 \pm 0.0014$ | [59] |
| Thorne | 2013 | $0.1136$ | [DIS+DY+HT*] [140] |
| $\text{ABM11}_J$ | 2010 | $0.1134 - 0.1149 \pm 0.0012$ | Tevatron jets (NLO) incl. [143] |
| NN21 | 2011 | $0.1174 \pm 0.0006 \pm 0.0001$ | +h. Nucl. [57,58] |
| ABM11 | 2013 | $0.1133 \pm 0.0011$ | [56] |
| ABM11 | 2013 | $0.1132 \pm 0.0011$ | (without jets) [56] |
| CTEQ | 2013 | $0.1140$ | (without jets) [139] |
| JR | 2014 | $0.1136 \pm 0.0004$ | dyn. approach [82] |
| JR | 2014 | $0.1162 \pm 0.0006$ | standard fit [82] |
| CTEQ | 2015 | $0.1150\ ^{+0.0060}_{-0.0040}$ | $\Delta\chi^2 > 1$ +h. Nucl. [84] |
| MMHT | 2015 | $0.1172 \pm 0.0013$ | +h. Nucl. [71] |
| BBG | 2006 | $0.1141\ ^{+0.0020}_{-0.0022}$ | valence analysis, $\text{N}^3\text{LO}$ |

Table 10: $\alpha_s(M_Z^2)$ values at NNLO and $\text{N}^3\text{LO}$ from QCD analyses of the deep-inelastic world-data, partly including other hard scattering data.

| Other Lower $\alpha_s$ Values | | | |
|---|---|---|---|
| NNLO | | $\alpha_s(M_Z^2)$ | |
| Gehrmann et al. | 2009 | $0.1131\ {}^{+\ 0.0028}_{-\ 0.0022}$ | $e^+e^-$ thrust [144] |
| Abbate et al. | 2010 | $0.1140 \pm 0.0015$ | $e^+e^-$ thrust [145] |
| Hoang et al. | 2015 | $0.1123 \pm 0.0015$ | C-param. dist. [146] |
| Alioli et al. | 2015 | 0.1135 | DY matched parton showers [147] |
| Bazavov et al. | 2014 | $0.1166\ {}^{+\ 0.0012}_{-\ 0.0008}$ | lattice 2+1 fl. [148] |
| CMS | 2013 | $0.1151^{+\ 0.0028}_{-\ 0.0027}$ | $t\bar{t}$ [149] |
| NLO | | $\alpha_s(M_Z^2)$ | |
| Frederix et al. | 2010 | $0.1156^{+\ 0.0041}_{-\ 0.0034}$ | $e^+e^- \rightarrow 5$ jets [150] |
| H1 | 2009 | $0.1160^{+\ 0.0095}_{-\ 0.0080}$ | $ep$ jets [124] |
| D0 | 2010 | $0.1156^{+\ 0.0041}_{-\ 0.0034}$ | $p\bar{p} \rightarrow$ jets [72] |
| ATLAS | 2012 | $0.1151^{+\ 0.0093}_{-\ 0.0087}$ | jets [151] |
| CMS | 2013 | $0.1148 \pm 0.0052$ | 3/2 jet ratio [152] |

Table 11: Lower $\alpha_s(M_Z^2)$ values from other hard scattering processes in NNLO and NLO.

NLO analyses yield values of $\alpha_s(M_Z^2)$ which are systematically larger than those at NNLO. Therefore the results of these analyses cannot be averaged. Analyses in which part of the scattering cross sections are still described at NLO cannot be regarded as NNLO analyses. We have shown that the reasons for the larger value of $\alpha_s(M_Z^2)$ obtained by MRST/MMHT [59,71] are well understood, see also [140].

Many more QCD analyses at NNLO resulted in $\alpha_s(M_Z^2)$ lower than the present world average. This even applies to a number of NLO analyses. The next important step in the QCD analyses will be possible including the jet data from $ep$ scattering and hadron collider data at NNLO. The precise knowledge of $\alpha_s(M_Z^2)$ is of instrumental importance for the understanding of various hard scattering cross sections at the LHC, notably that of Higgs-boson production [157].

**Acknowledgment.** We would like to thank H. Böttcher, G. Falcioni, A. Hoang, M. Klein, R. Kogler, K. Rabbertz, E. Reya, R. Thorne and A. Vogt for communications and discussions. This work was supported in part by the European Commission through contract PITN-GA-2012-316704 (HIGGSTOOLS).

# $\alpha_s$ from jets in DIS and photoproduction

Michael Klasen

Institut für Theoretische Physik, Westfälische Wilhelms-Universität Münster,
Wilhelm-Klemm-Straße 9, D-48149 Münster, Germany

**Abstract:** The current status and future prospects of the determination of the strong coupling constant $\alpha_s$ from inclusive jet and multijet production in Deep-Inelastic Scattering (DIS) and photoproduction at the $ep$ collider HERA and at future colliders are reviewed. We also describe briefly the impact that inclusive DIS on photons, i.e. the measurement of $F_2^\gamma$ in particular at the $e^+e^-$ collider LEP, has had in the past and could again have in the future.

## Introduction

The HERA collaborations H1 and ZEUS have recently published their final analyses of jet production in DIS [1] and photoproduction [2], respectively, in which they also extracted values of $\alpha_s(M_Z)$ by comparing their data on multijets in DIS and inclusive jets in photoproduction with next-to-leading order (NLO) QCD calculations [3,4,5]. Since determinations in NLO QCD are currently not taken into account in the PDG world average of $\alpha_s(M_Z)$ [6], we have recently included next-to-next-to-leading order (NNLO) corrections [7,8] in our own NLO calculations of jet production in DIS [9,10], which have been shown to agree very well with those used by the H1 collaboration [4,10], and in photoproduction.

In the absence of full NNLO calculations, which are very difficult to obtain and where progress is rather slow, we rely on approximate NNLO (aNNLO) results extracted from threshold resummation in a unified approach to obtain the universal soft and virtual corrections [11]. They are known to be dominant, in particular close to threshold, when the ratio of final-state to initial-state invariant mass $z = (p_1 + p_2)^2/(p_a + p_b)^2$ approaches unity, and can in this region be resummed to all orders. Reexpansion of the resummed result then yields the aNNLO results, which still depend on the kinematics of the process (one-particle inclusive, 1PI, or pair invariant-mass, PIM) and factorization scheme (DIS or $\overline{\rm MS}$). Here we use the latter in both cases. In PIM kinematics, the higher-order cross sections are dominated close to threshold by the logarithms

$$D_l(z) \;=\; \left[\frac{\ln^l(1-z)}{1-z}\right]_+ , \qquad (1)$$

where $l = 0, 1$ in NLO and $l = 0, ..., 3$ in NNLO, and can be expressed in terms of universal master formulæ, neglecting non-universal, not logarithmically enhanced terms. At NLO, the partonic cross section

$$d\sigma_{ab} \;=\; d\sigma^B_{ab}\frac{\alpha_s(\mu)}{\pi}[c_3 D_1(z) + c_2 D_0(z) + c_1\delta(1-z)] \qquad (2)$$

$$+\; \frac{\alpha_s^{d_{\alpha_s}+1}(\mu)}{\pi}[A^c D_0(z) + T_1^c\delta(1-z)] \qquad (3)$$

is relatively compact, but already exhibits the typical dependence on the universal coefficients $c_i$, the decomposition into two parts for simple and complex color flow, and the $\alpha_s$-dependence of the latter, where $d_{\alpha_s} = 0, 1, 2, ...$ for an underlying Born cross section $d\sigma_{ab}^B$ of $\mathcal{O}(\alpha_s^{0,1,2,\cdots})$. The coefficients for DIS and direct photoproduction with simple color flow are identical and are given explicitly in Refs. [7,8], while those for the purely partonic processes pertinent to resolved photoproduction can be found in Ref. [11] together with the master formula at NNLO. In resolved photoproduction, care must be taken to disentangle the factorization scales in the photon and proton. We take into account the finite terms known from our explicit NLO calculations, but have to neglect of course the unknown additional finite terms at NNLO.

Since our resolved photoproduction calculations can also be applied to hadroproduction, we have checked explicitly that we can reproduce the numerical results in Fig. 2 of Ref. [12] for inclusive jet production as measured by D0 at the Tevatron up to transverse momenta $p_T$ of 500 GeV, where the cross section is moderately increased and the scale uncertainty is considerably reduced at aNNLO. Having gained confidence in our aNNLO results, we can apply them to DIS and photoproduction.

## Jets in DIS

The H1 data in DIS were taken from 2003 to 2007 at $\sqrt{S} = 319$ GeV with an integrated luminosity of 351 pb$^{-1}$, photon virtuality $Q^2 \in [150; 15000]$ GeV$^2$ and $y \in [0.2; 0.7]$, jet $p_T > 7$ GeV, rapidity $\eta \in [-1, 2.5]$ and distance parameter $R = 1$ [1]. In our numerical predictions, we set the squared renormalization and factorization scales to $\mu^2 = (Q^2 + p_T^2)/2$ and $\mu_p^2 = Q^2$, respectively, as in the H1 analysis, employ MSTW2008 NNLO proton parton density functions (PDFs) with $n_f = 5$ flavors [13] and vary $\alpha_s(M_Z)$ from 0.107 to 0.127. After hadronization corrections, taken from PYTHIA simulations performed by H1, we can reproduce their NLO simulations exactly and also approximately with the master formula given above. In DIS, the NNLO corrections reduce jet cross sections and their scale uncertainties at very large values of $Q^2$, which can be seen in Fig. 1.

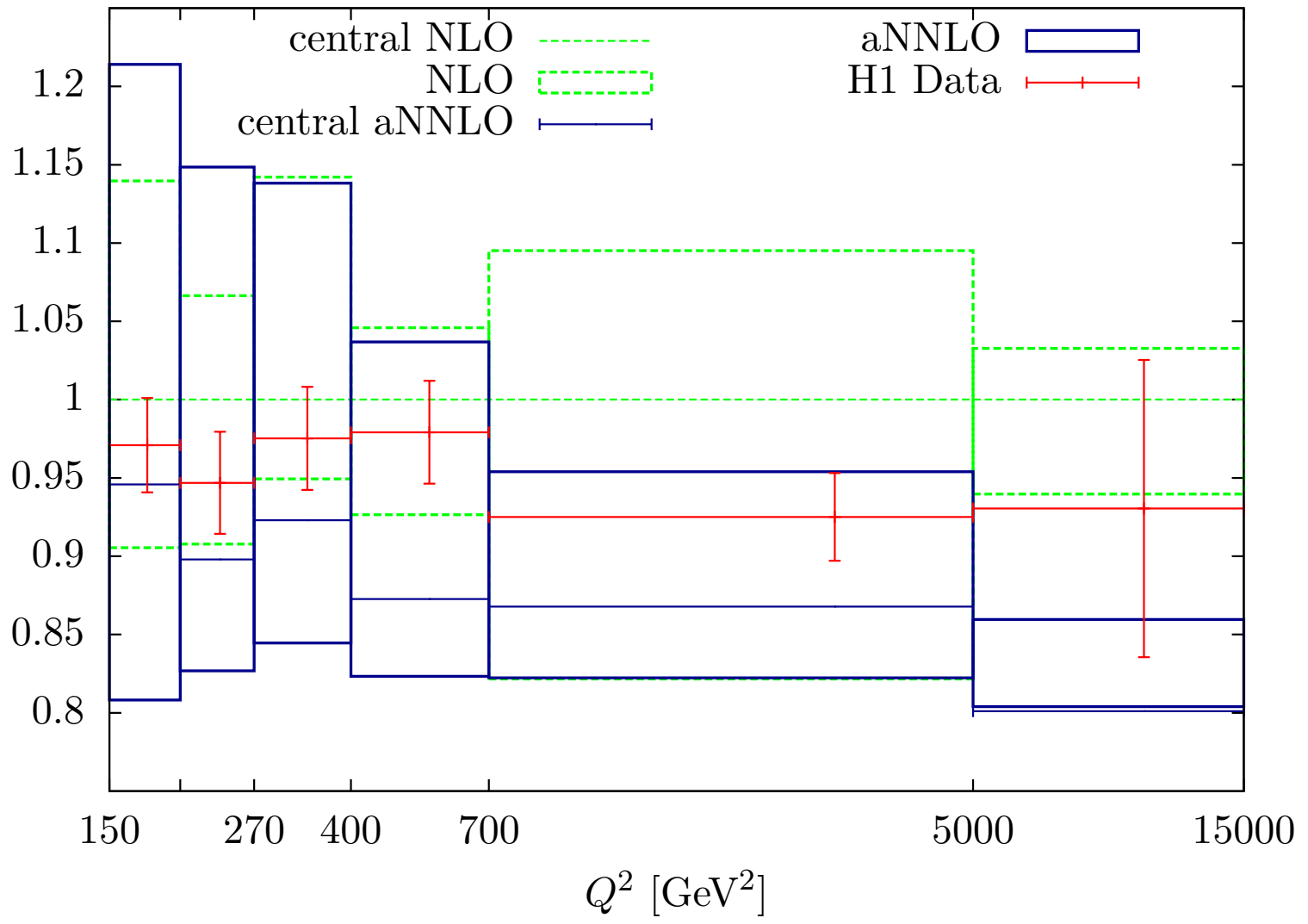

Figure 1: Single-differential inclusive jet cross sections as a function of photon virtuality $Q^2$ in NLO (green dashed lines) and aNNLO (blue full lines) with the corresponding scale uncertainty bands, obtained by varying $\mu_R$ and $\mu_F$ simultaneously by a factor of two up and down, normalized to the central NLO predictions and compared to the final H1 data (red points, color online).

While at NLO, the strong coupling is fitted to be $\alpha_s(M_Z) = 0.115 \pm 0.002(\text{exp.}) \pm 0.005(\text{th.})$, the determination at aNNLO yields a higher value of $\alpha_s(M_Z) = 0.122 \pm 0.002(\text{exp.}) \pm 0.013(\text{th.})$ to compensate for the reduced cross section. Unfortunately, in DIS the theoretical error is not reduced, which can be traced back to the fact that the jet transverse momentum is in the range of 7 to 50 GeV and thus relatively far from threshold.

## Jets in e-p photoproduction

Photoproduction data were taken by ZEUS from 2005 to 2007 with a somewhat smaller integrated luminosity of 300 pb$^{-1}$, at photon virtuality $Q^2 < 1$ GeV$^2$ and photon-proton energy $W \in [142; 293]$ GeV and with jets of larger $p_T > 17$ GeV, but identical rapidity range and distance parameter $R$. In the absence of a large $Q^2$, we identify all scales with $p_T$, use GRV-HO photon PDFs [14], transformed from the DIS$_\gamma$ to the $\overline{\text{MS}}$ scheme, and CT10 NNLO proton PDFs [15] with $n_f = 5$ and vary $\alpha_s(M_Z)$ from 0.112 to 0.124. After hadronization corrections, we obtain the transverse-momentum distributions in Fig. 2 (left), which clearly agree better with the ZEUS data at aNNLO than at NLO, where also the lowest data point can be accounted for within the scale uncertainty and where the latter is considerably reduced at large $p_T$. The rapidity distribution in Fig. 2 (right) differs even more from NLO to aNNLO and then reproduces the shape of the data much better, in particular in the notoriously difficult backward, but also in the forward region. In photoproduction, the strong coupling is fitted at NLO to be $\alpha_s(M_Z) = 0.121 \pm 0.002(\text{exp.})^{+0.005}_{-0.003}(\text{th.})$, while the determination at aNNLO yields a slightly smaller value of $\alpha_s(M_Z) = 0.120 \pm 0.002(\text{exp.})^{+0.003}_{-0.003}(\text{th.})$, which is not only closer to the world average, but also has a smaller theoretical error at aNNLO than at NLO. This is due to the fact that here higher $p_T$ values up to 71 GeV are probed.

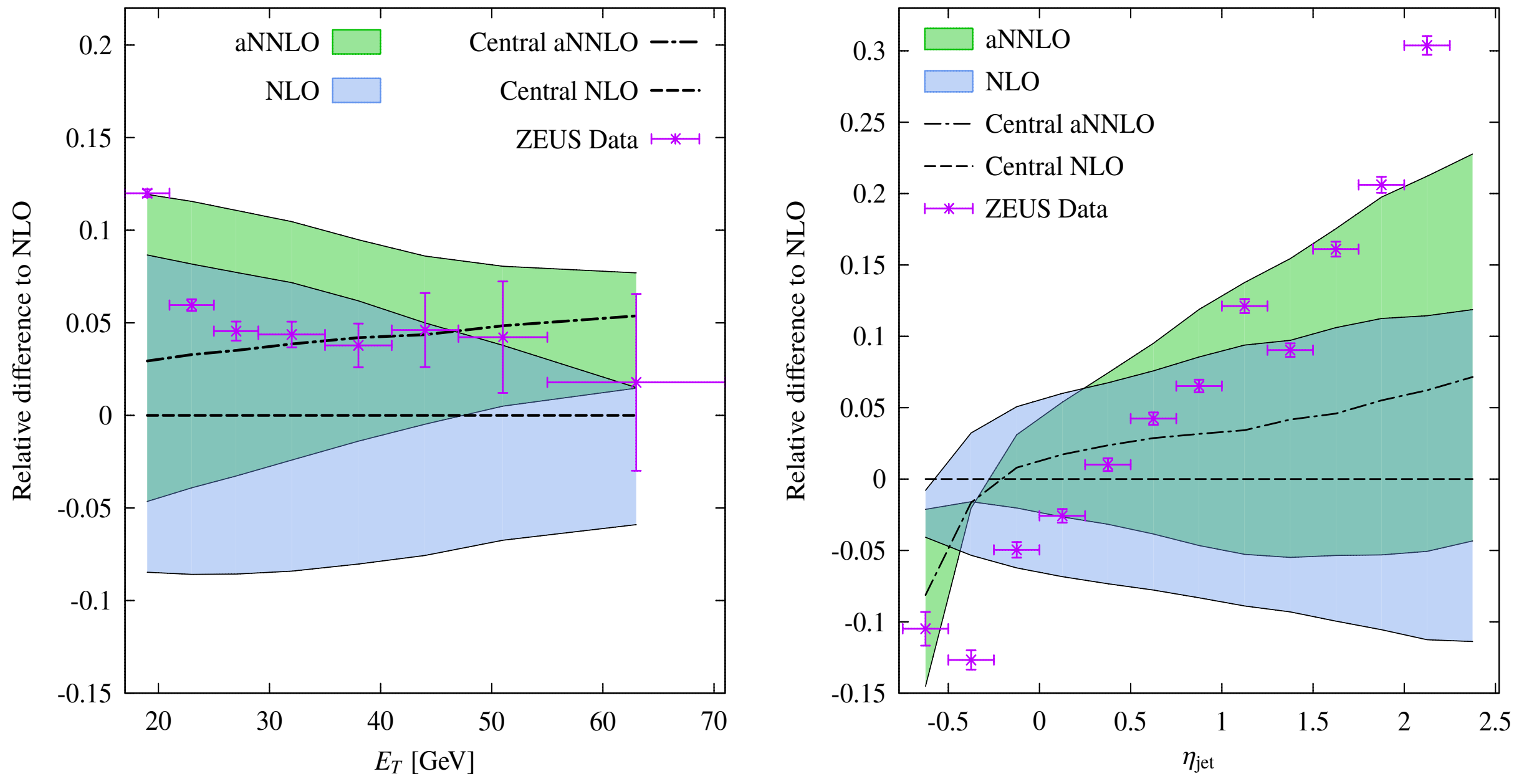

Figure 2: Relative difference of ZEUS photoproduction data, NLO and NNLO predictions with scales $\mu, \mu_\gamma, \mu_p = [0.5; 2] \times E_T$ to the central NLO prediction, as a function of jet transverse energy $E_T$ (left) and jet rapidity $\eta$ (right).

For more precise determinations of $\alpha_s$ in DIS and photoproduction at future colliders, several directions of research can be imagined. First, our approximate NNLO results will of course be more reliable at higher values of $Q^2$ and $p_{\rm T}$ that may become accessible in the future. We have demonstrated this explicitly by computing the $Q^2$-distribution for values beyond the reach of the H1 experiment [7]. Theoretically, the formalism that we have employed can be further improved by taking into account the finite mass of the produced jet [16]. This has been shown to extend the range of validity of the discussed approximate NNLO results down to lower values of $p_{\rm T}$ through a comparison to the only available full NNLO calculation, the one in the gluon-only channel at hadron colliders [17]. Experimentally, the continuation of the HERA program on jet production at a lepton-proton collider would allow for determinations of $\alpha_{\rm s}$ together with (simultaneous) fits of proton –and in the case of photoproduction also photon– PDFs.

Jet production in real or virtual photon collisions could be studied at an $e^+e^-$ collider. As we have shown in the past for LEP, also inclusive data on $F_2^\gamma$ can have a big impact on $\alpha_{\rm s}$, since the pointlike contribution dominates at high $Q^2$, making the fits independent of the photon PDFs [18]. In proton-proton collisions of very high energy, photoproduction (and maybe even DIS) will also become interesting due to an enhanced photon flux, which is even higher for nuclear collisions. Admixtures of diffractive exchanges must then be disentangled by kinematic cuts or normalizations to lepton-only final states.

**Acknowledgments.** The author thanks T. Biekötter, G. Kramer and M. Michael for their collaboration and the organizers of the CERN $\alpha_{\rm s}$ workshop for the kind invitation.

# $\boldsymbol{\alpha}_{\mathrm{s}}$ from scaling violations of hard parton-to-hadron fragmentation functions

Bernd A. Kniehl

II. Institut für Theoretische Physik, Universität Hamburg, Luruper Chaussee 149,
22761 Hamburg, Germany

**Abstract:** We assess the prospects of a high-precision determination of the strong-coupling constant $\alpha_{\mathrm{s}}$ from the scaling violations in fragmentation functions at the CERN Future Circular Collider operated in the $e^+e^-$ annihilation mode (FCC-ee).

## Introduction

The strong force acting between hadrons is one of the four fundamental forces of nature. It is now commonly believed that the strong interactions are correctly described by quantum chromodynamics (QCD), the SU(3) gauge field theory which contains colored quarks and gluons as elementary particles. The strong coupling constant $\alpha_{\mathrm{s}}^{(n_f)}(\mu^2) = g_s^2/(4\pi)$, where $g_s$ is the QCD gauge coupling, is a basic parameter of the standard model of elementary particle physics; its value $\alpha_{\mathrm{s}}^{(5)}(M_Z^2)$ at the $Z$-boson mass scale is listed among the constants of nature in the Review of Particle Physics [1]. Here, $\mu$ is the renormalization scale, and $n_f$ the number of active quark flavors, with mass $m_q \ll \mu$. In the modified minimal-subtraction ($\overline{\mathrm{MS}}$) scheme, the evolution of $\alpha_{\mathrm{s}}^{(n_f)}(\mu^2)$ is known through four loops [2] and its matching at the flavor thresholds through three [3] and even four loops [4].

There are a number of processes in which $\alpha_{\mathrm{s}}^{(5)}(M_Z^2)$ can be measured (see Ref. [1] and these workshop proceedings, for recent reviews). A reliable method to determine $\alpha_{\mathrm{s}}^{(5)}(M_Z^2)$ is through the extraction of the fragmentation functions (FFs) [5] in the $e^+e^-$ annihilation process

$$e^+e^- \to (\gamma, Z) \to h + X, \tag{1}$$

which describes the inclusive production of a single hadron $h$. Here, $h$ may either refer to a specific hadron species, such as $\pi^\pm$, $\pi^0$, $\eta$, $K^\pm$, $K_S^0$, $p/\overline{p}$, $\Lambda/\overline{\Lambda}$, or to the sum of all charged hadrons. In the parton model of QCD, the cross section of process (1), differential in the scaled hadron energy $x = 2E_h/\sqrt{s}$, may be evaluated, up to power corrections, as

$$\frac{d\sigma^h}{dx}(x) = \sum_a \int_x^1 \frac{dy}{y}\, \frac{d\sigma_a}{dy}(y,\mu^2,Q^2) D_a^h\left(\frac{x}{y}, Q^2\right), \tag{2}$$

where $a$ runs over the gluon and the quarks and antiquarks active at energy scale $\sqrt{Q^2}$. The partonic cross sections $d\sigma_a(x,\mu^2,Q^2)/dx$ pertinent to process (1) can entirely be calculated in perturbative QCD with no additional input, except for $\alpha_{\mathrm{s}}(\mu^2)$. They are known at next-to-leading order (NLO) [6] and even at next-to-next-to-leading order (NNLO) [7]. The subsequent transition of the partons into hadrons takes place at an energy scale of the order of 1 GeV and can, therefore, not be treated in perturbation theory. Instead, the hadronization of the partons is described by FFs $D_a^h(x,Q^2)$. Their values correspond to the probability that the parton $a$, produced at short distance of order $1/\sqrt{Q^2}$, fragments into the hadron $h$ carrying the fraction $x$ of the energy of $a$. For process (1), $\sqrt{Q^2}$ is typically of the order of the center-of-mass (CM) energy $\sqrt{s}$. Given their $x$ dependence at some scale $Q_0$, the evolution of the FFs with $Q^2$ may be computed perturbatively from the timelike Dokshitzer-Gribov-Lipatov-Altarelli-Parisi (DGLAP) equations [8],

$$\frac{dD_a^h(x,Q^2)}{d\ln Q^2} = \sum_b \int_x^1 \frac{dy}{y} P_{a\to b}^T(y,\alpha_s(Q^2)) D_b^h\left(\frac{x}{y},Q^2\right). \tag{3}$$

The timelike splitting functions $P_{a\to b}^T(x,\alpha_s(Q^2))$ are known at NLO [9] and NNLO [10].

This method to determine $\alpha_s^{(5)}(M_Z^2)$ is particularly clean in the sense that, unlike other methods, it is not plagued by uncertainties associated with hadronization corrections, jet algorithms, parton density functions (PDFs), *etc.* We recall that, similarly to the scaling violations in the PDFs, perturbative QCD only predicts the $Q^2$ dependence of the FFs. Therefore, measurements at different CM energies are needed in order to extract values of $\alpha_s^{(5)}(M_Z^2)$. Furthermore, since the $Q^2$ evolution mixes the quark and gluon FFs, it is essential to determine all FFs individually, which requires quark-flavor and gluon-jet tagging in the experimental data analysis.

This contribution is organized as follows. In Sec. 2, we discuss the $\alpha_s$ determination by fitting light-hadron FFs to experimental data of process (1) at NLO in the parton model of QCD. In Sec. 2, we present various ways how the standard NLO approach may be refined. In Sec. 2, we discuss the $\alpha_s$ determination by fitting the first Mellin moments of the FFs to experimental data of average hadron multiplicities in gluon and quark jets. Sec. 2, we assess the prospects of a high-precision determination of $\alpha_s$ at the FCC-ee.

## NLO fits of light-hadron FFs

NLO FFs for light hadrons were extracted via process (1) by several experimental [11,12] and theoretical [13,14,15,16,17] collaborations. Simultaneous determinations of $\alpha_s$ were performed in Refs. [11,14,15,16]. The latest one of the latter analyses [16] was focused on $\pi^\pm$, $K^\pm$, and $p/\overline{p}$ data. Collective charged-hadron data, including measurements of the gluon FF [12] and the longitudinal cross section, were excluded from the fit to avoid contaminations with charged particles other than the three lightest charged hadrons, but they were used for comparisons assuming the charged hadrons to be exhausted by $\pi^\pm$, $K^\pm$, and $p/\overline{p}$. The selected data were taken by TPC [18] at SLAC PEP, by ALEPH [19] and DELPHI [20] at CERN LEP, and by SLD [21] at SLAC SLC, and partially came as light-quark, charm, and bottom enriched samples. For the first time, the light-quark tagging probabilities obtained by OPAL [22] were included. The quality of the fit may be judged from Fig. 1. The fit result [16],

$$\alpha_s^{(5)}(M_Z^2) = 0.1176 \begin{array}{l} +0.0053 \\ -0.0067 \end{array} \begin{array}{l} +0.0007 \\ -0.0009 \end{array}, \tag{4}$$

is greatly dominated by the experimental error.

The fact that a high-precision determination of $\alpha_s$ from scaling violations in FFs is possible at all provides a stringent test of the DGLAP evolution. The universality of the FFs, as predicted by the factorization theorem of the QCD parton model, has been tested through extensive comparisons with experimental data of inclusive single hadron production in reactions other than process (1) [24,25].

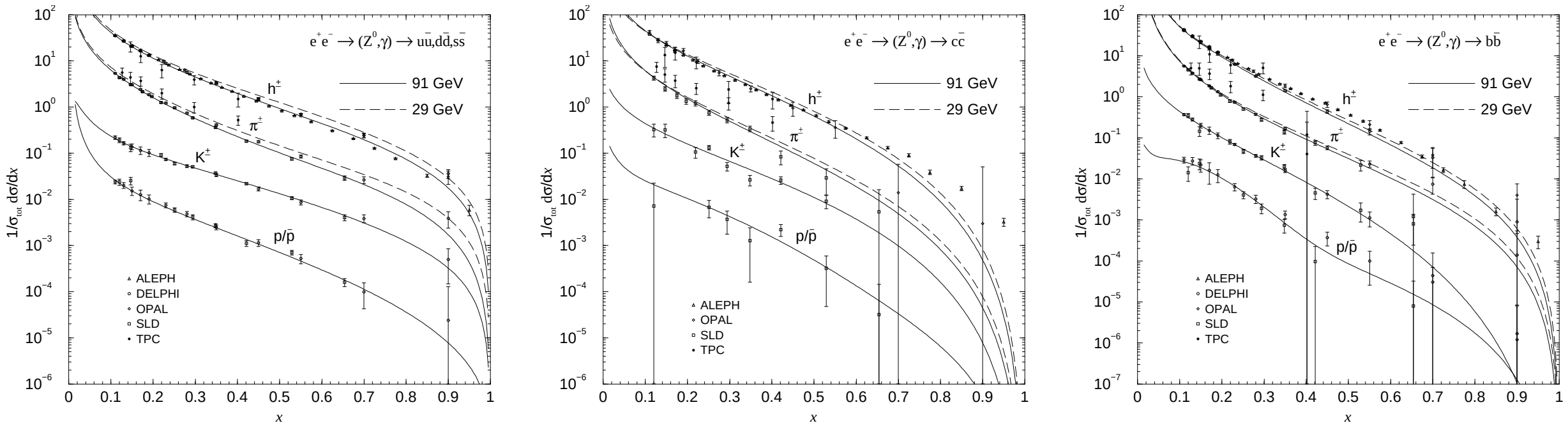

(a) (b) (c)

Figure 1: Comparisons of the cross sections of process (1) for $h = h^\pm, \pi^\pm, K^\pm, p/\overline{p}$ with (a) $a = u, d, s$, (b) $a = c$, and (c) $a = b$ evaluated using the fitted FFs with the experimental data sets included in the fit [18,19,20,21] and one by OPAL [23] without hadron identification [16]. The theoretical results for $h = h^\pm$ are taken to be the sums of the respective ones for $h = \pi^\pm, K^\pm, p/\overline{p}$. The theoretical results for $a = g$ are distributed among the respective ones for $a = u, d, s, c, b$ according to the flavor of the $q\overline{q}$ pair produced along with the gluon.

## Improvements

The fixed-order approach to FFs breaks down at small values of $x$ because of the appearance of soft-gluon logarithms (SGLs). They manifest themselves at $n$ loops in the timelike DGLAP splitting functions as terms which, in the limit $x \to 0$, behave as $(\alpha_s/2\pi)^n/x \ln^{2n-m-1} x$, where $m = 1, \ldots, 2n-1$ labels the SGL class. In Refs. [26,27], a general scheme for the evolution of FFs was introduced which resums both SGLs and mass singularities to any order in a consistent manner and requires no additional theoretical assumptions.

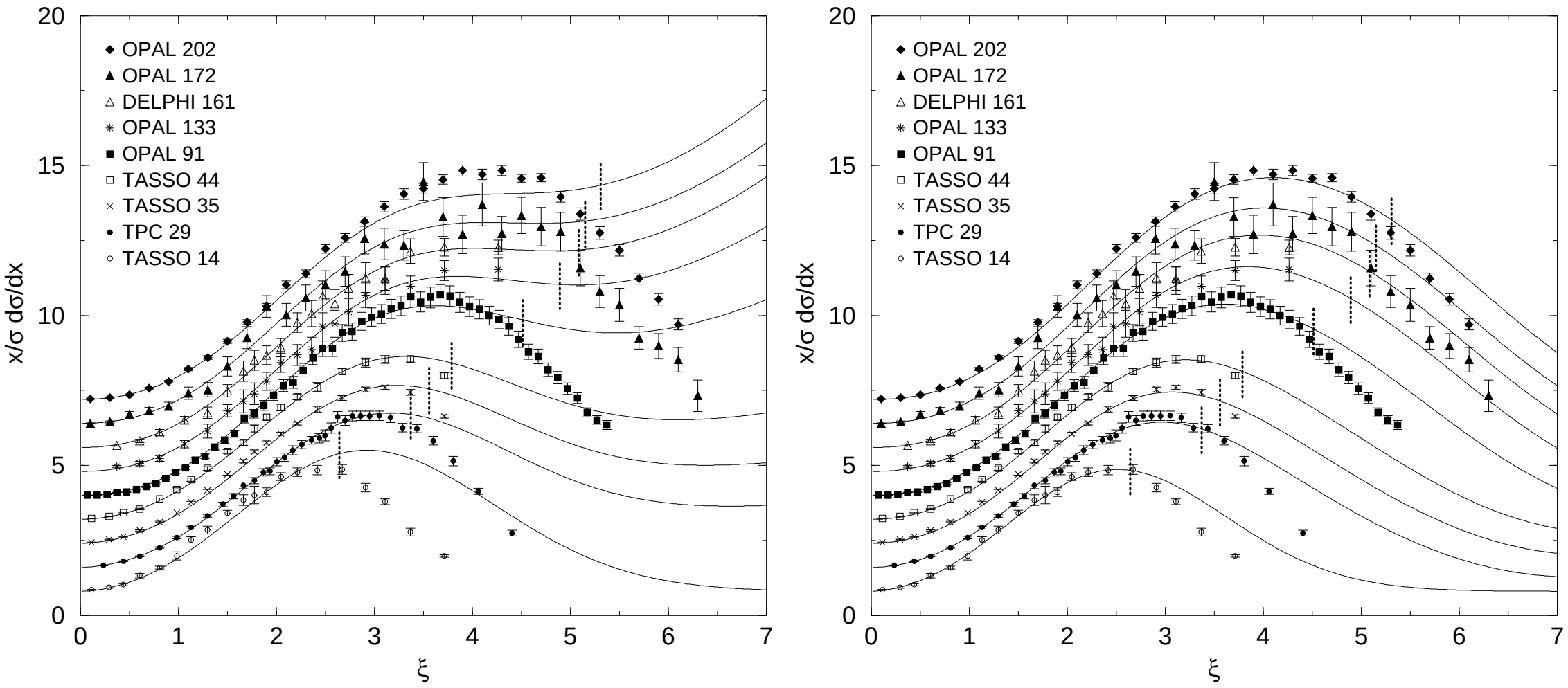

Figure 2: LO fits to small-$x$ data of process (1) without (left) and with (right) resummation of double logarithms [27].

The double and single logarithms, with $m = 1$ and $m = 2$, respectively, in the timelike DGLAP

splitting functions were resummed in Refs. [26,27,28]. The first approximation, with just double logarithms resummed and fixed-order contributions evaluated at LO, is already more complete than the modified leading logarithm approximation [29], which uses assumptions reaching beyond first principles of QCD. This dramatically improves the description of the *hump-backed plateau* as compared to the fixed-order treatment, as is illustrated in Fig. 2. The resummation of double logarithms in the coefficient functions of process (1) was explained in Ref. [30].

The fixed-order approach to FFs is spoiled by soft-gluon radiation also at large values of $x$. At $n$ loops in the timelike DGLAP splitting functions, the corresponding divergences take the form $(\alpha_s/2\pi)^n[\ln^{n-r}(1-x)/(1-x)]_+$, where $r = 0, \ldots, n$ labels the class of divergence. Starting from the NLO results [9], the resummation was performed through next-to-leading-logarithmic (NLL) accuracy, for $r = 0, 1$, in Ref. [31]. The corresponding NLL expressions for the coefficient functions of process (1) may be found in Ref. [32]. The NLL resummation enhances the cross section of process (1) at large $x$ values and reduces its theoretical uncertainty. In Ref. [17], this was found to improve the quality of the fit, as shown in Table 1 (left). In Ref. [31], the NLL effects were found to be comparable to the NNLO ones [10].

| Particle | Unresummed | Resummed | Fitted mass (MeV) | PDG mass (MeV) |
|---|---|---|---|---|
| $\pi^\pm$ | 519.0 | 518.7 | 154.6 | 139.6 |
| $K^\pm$ | 439.4 | 416.6 | 337.0 | 493.7 |
| $p/\overline{p}$ | 538.0 | 525.2 | 948.8 | 938.3 |
| $K_S^0$ | 318.7 | 317.2 | 343.0 | 497.6 |
| $\Lambda/\overline{\Lambda}$ | 325.7 | 273.1 | 1127.0 | 1115.7 |

Table 1: Left: Minimized $\chi^2$ values for the individual fits without charge-sign identification [17] without and with NLL resummation at large $x$ values. Right: Fit results for light-hadron masses obtained in Ref. [17].

Furthermore, the quality of the fits may be significantly improved by including hadron mass effects [17,25,27]. In fact, including the light-hadron masses among the fit parameters, one obtains values that are amazingly close to the true values [17], as may be seen from Table 1 (right).

Multiple photon radiation from the initial state may have an appreciable effect on the line-shape of the differential cross section $d\sigma/dx$ of process (1). This effect is more pronounced for $e^+e^-$ annihilation in the continuum than on resonances, such as the $Z$-boson one, because the photonic initial-state radiation shifts the CM energy of the hard $e^+e^-$ collision to lower values, where the cross section is typically somewhat larger in the former case, but significantly smaller in the latter one. QED initial-state radiation may be conveniently incorporated via radiator functions, which allows for the resummation of leading logarithms of the form $(\alpha/2\pi)^n \ln^n(s/m_e^2)$ to all orders $n = 1, 2, \ldots$ [33]. The appropriate formalism was presented in Ref. [34].

Due to charge-conjugation invariance, the cross section of process (1) is the same for the positively and negatively charged hadrons of a given species. The latter are, therefore, routinely combined in experimental analyses. By the same token, it is impossible to discriminate FFs for $h^+$ and $h^-$ hadrons by fitting to $e^+e^-$ data. This is also true for $p\overline{p}$ data. However, the situation is different for $ep$ and $pp$ data. In Ref. [17], the charge-sign asymmetries of single hadrons inclusively produced in $pp$ collisions as measured by BRAHMS, PHENIX, and STAR at BNL RHIC were used to constrain the valence-quark fragmentations. While this allows for a refined determination of $\alpha_s$, it introduces additional theoretical uncertainties, associated with the PDFs. Furthermore, the fitting procedure is more complicated because the PDFs themselves depend on $\alpha_s$.

## Average hadron multiplicities in gluon and quark jets

Global fits to average hadron multiplicities in gluon and quark jets of energy $\sqrt{Q^2}$ produced in $e^+e^-$ annihilation, $\langle n_h(Q^2)\rangle_a$ with $a = g, q$,* and their ratio $r(Q^2) = \langle n_h(Q^2)\rangle_g / \langle n_h(Q^2)\rangle_q$ also allow for high-precision determinations of $\alpha_s$ [35,36]. This requires the experimental ability to discriminate quark and gluon jets and the use of compatible jet algorithms in the experimental data analyses. By the definition of the FFs,

$$\langle n_h(Q^2)\rangle_a = \int_0^1 dx\, x^0 D_a^h(x, Q^2) = \tilde{D}_a^h(0, Q^2), \tag{5}$$

which is just the first Mellin moment of $D_a^h(x, Q^2)$. The respective expressions of the timelike DGLAP splitting functions, $\tilde{P}_a(0, \alpha_s(Q^2))$, are ill defined and require resummation. This was achieved with next-to-next-to-leading-logarithmic (NNLL) accuracy starting from the NNLO expressions [10] in Ref. [37]. The coupled system of the timelike DGLAP evolution equations in Mellin space is conveniently solved by diagonalization [35]. In this way, $\tilde{D}_a^h(0, Q^2) = \tilde{D}_a^{h+}(0, Q^2) + \tilde{D}_a^{h-}(0, Q^2)$ are decomposed into plus and minus components, which are numerically large and small, respectively. The ratios $r_\pm(Q^2) = \tilde{D}_g^{h\pm}(0, Q^2)/\tilde{D}_q^{h\pm}(0, Q^2)$ may be evaluated perturbatively in powers of $\sqrt{\alpha_s}$. At present, $r_+(Q^2)$ is known through $\mathcal{O}(\alpha_s^{3/2})$ [38] and $r_-(Q^2)$ through $\mathcal{O}(\alpha_s^{1/2})$ [36], i.e. through next-to-next-to-next-to-leading order (N$^3$LO) and NLO, respectively. Global fits of $\langle n_h(Q^2)\rangle_g$ and $\langle n_h(Q^2)\rangle_q$ to experimental data coming from CLASSE CESR with $\sqrt{s} = 10$ GeV, SLAC PEP with 29 GeV, DESY PETRA with 12–47 GeV, KEK TRISTAN with 50–61 GeV, SLAC SLC with 91 GeV, CERN LEP1 with 91 GeV, and CERN LEP2 with 130–209 GeV yielded the results listed in Table 2. The N$^3$LO$_{\text{approx}}$ + NLO + NNLL value of $\alpha_s^{(5)}(M_Z^2)$ therein corresponds to $\alpha_s^{(5)}(M_Z^2) = 0.1199 \pm 0.0026$ at the 68% CL. The quality of the fits may also be judged from Fig. 3.

| | N$^3$LO$_{\text{approx}}$ + NNLL | N$^3$LO$_{\text{approx}}$ + NLO + NNLL |
|---|---|---|
| $\langle n_h(Q_0^2)\rangle_g$ | $24.18 \pm 0.32$ | $24.22 \pm 0.33$ |
| $\langle n_h(Q_0^2)\rangle_q$ | $15.86 \pm 0.37$ | $15.88 \pm 0.35$ |
| $\alpha_s^{(5)}(M_Z^2)$ | $0.1242 \pm 0.0046$ | $0.1199 \pm 0.0044$ |
| $\chi^2_{\text{dof}}$ | 2.84 | 2.85 |

Table 2: Fit results for $\langle n_h(Q_0^2)\rangle_g$, $\langle n_h(Q_0^2)\rangle_q$, and $\alpha_s^{(5)}(M_Z^2)$ for $Q_0 = 50$ GeV obtained in the N$^3$LO$_{\text{approx}}$ + NNLL and N$^3$LO$_{\text{approx}}$ + NLO + NNLL approaches [36]. The errors refer to the 90% confidence level (CL).

We observe from Fig. 4 that the experimental uncertainty overwhelms the theoretical one in the case of $\langle n_h(Q^2)\rangle_q$, while they are comparable in the case of $\langle n_h(Q^2)\rangle_g$. Since the theoretical predictions are unlikely to advance beyond the NNLL accuracy in the foreseeable future, the potential of high-precision measurements of $\langle n_h(Q^2)\rangle_g$ and $\langle n_h(Q^2)\rangle_q$ to reduce the error in $\alpha_s^{(5)}(M_Z^2)$ will be limited.

## Conclusions

In this contribution, we explained how $\alpha_s$ may be determined from scaling violations in FFs by fitting the latter to experimental data of the inclusive production of single light hadrons in $e^+e^-$

*Here, $q$ denotes the quark singlet component.

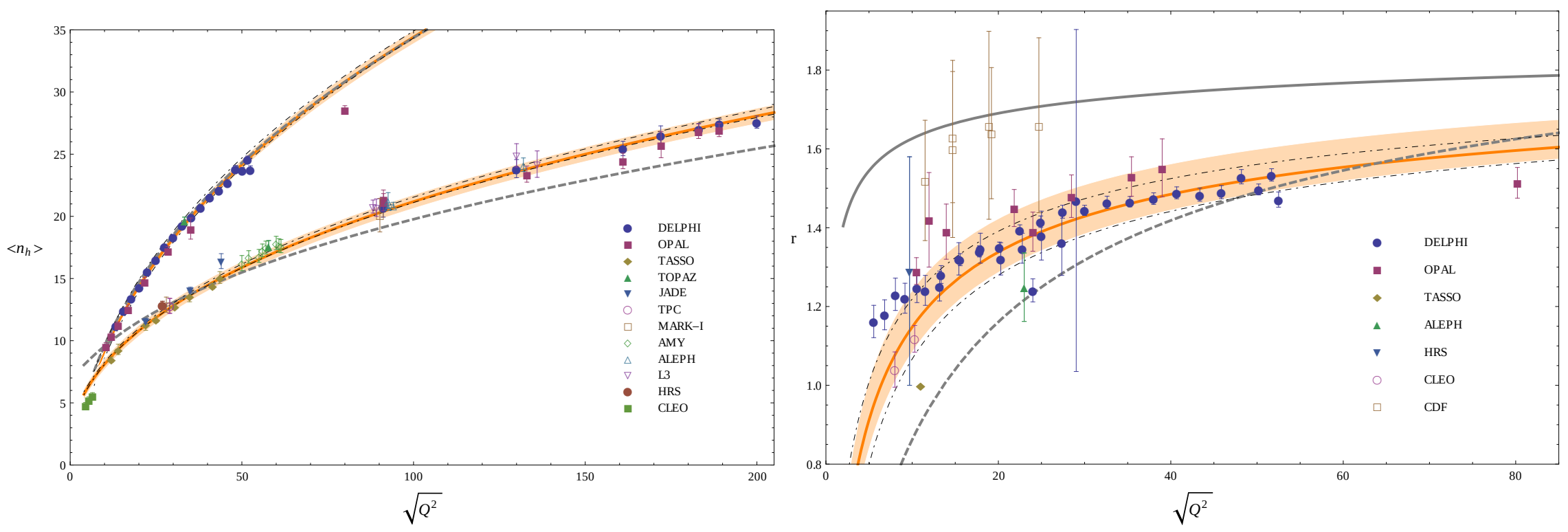

Figure 3: Comparisons of fit results for $\langle n_h(Q^2)\rangle_g$, $\langle n_h(Q^2)\rangle_q$ (left), and $r(Q^2)$ with experimental data (right) [36]. The experimental and theoretical uncertainties are indicated by the shaded/orange bands and the bands enclosed between the dot-dashed curves, respectively.

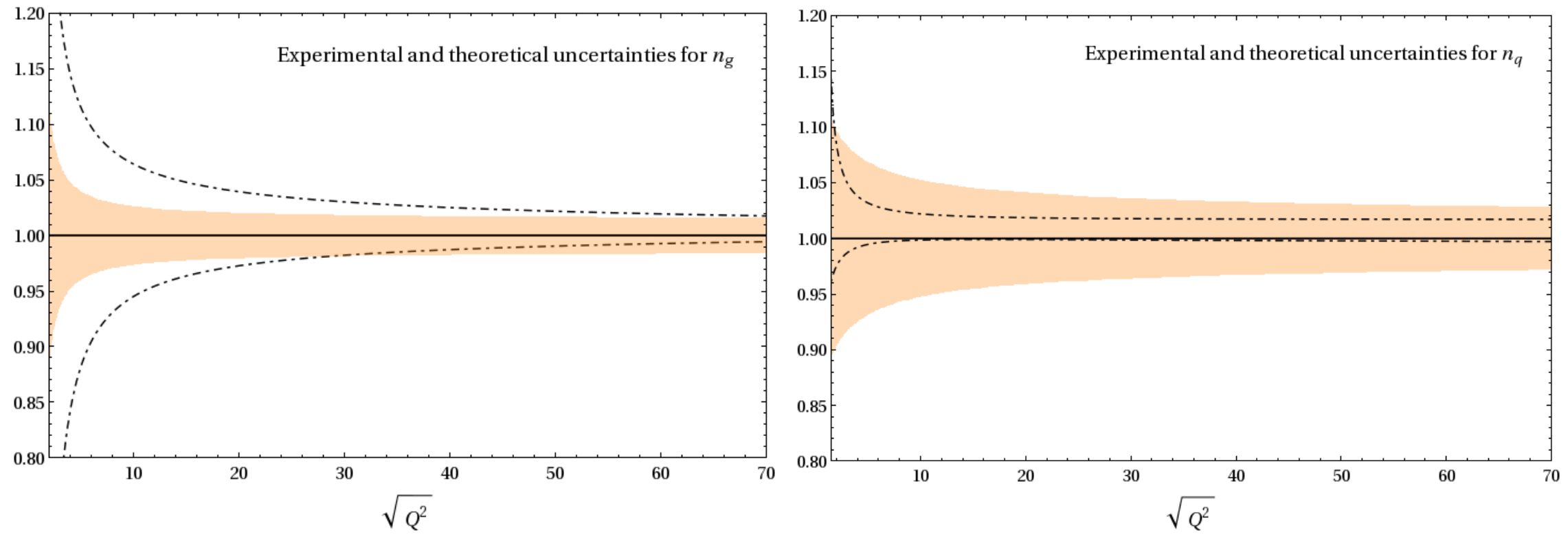

Figure 4: Experimental (shaded/orange bands) and theoretical (bands enclosed between dot-dashed curves) uncertainties in the $\mathrm{N^3LO_{approx} + NLO + NNLL}$ results for $\langle n_h(Q^2)\rangle_g$ (left), and $\langle n_h(Q^2)\rangle_q$ (right) normalized with respect to their default evaluations [36].

annihilation. We also discussed the analogous analysis based on the average hadron multiplicities in gluon and quark jets, which correspond to the first Mellin moments of the respective FFs. In both cases, we reviewed previous such $\alpha_s$ determinations in the literature and pointed out ways how they may be refined with regard to the theoretical description.

This allows us to usefully assess the prospects of a high-precision determination of $\alpha_s$ by fitting FFs or their first Mellin moments to experimental data to be collected at the FCC-ee. On the experimental side, it will be indispensable to measure the cross sections at widely separated values of $\sqrt{s}$ and with fine binnings in the $x$ variable, to identify the hadron species, to discriminate gluon and quark jets, and to tag the quark flavors in the latter case.

As for fits of the $x$ dependencies of the FFs, the experimental errors in $\alpha_s$ achieved so far are typically one order of magnitude larger than the theoretical ones, as may be seen from Eq. (4). By incorporating the theoretical improvements mentioned in Sec. 2, which was actually done in a determination of FFs for fixed value of $\alpha_s$ [17], and, to a lesser degree, by including the fixed-order corrections at the NNLO level, the theoretical error in $\alpha_s$ will be appreciably reduced, possibly by up to a factor of five or more. Moreover, the use of the generalized scheme of FF evolution [26,27] will allow one to also fully exploit the wealth of small-$x$ data, which must be excluded from fixed-order analyses because of the lack of SGL resummation. The enormous luminosity envisaged for the FCC-ee and the high quality of the detectors to be installed there will allow for the experimental error in $\alpha_s$ to be pushed well below the theoretical one after a few years of running at different CM energies. In conclusion, a combined error of order $\pm 0.0001$ in $\alpha_s$ appears to be reachable.

As for fits of the first Mellin moments of the gluon and quark FFs, the theoretical and experimental errors in $\alpha_s$ are presently comparable, as may be gleaned from Fig. 3. Since theoretical progress will be very hard to achieve in the foreseeable future, the theoretical error in $\alpha_s$ will soon limit the precision achievable via this method at the FCC-ee.

**Acknowledgments.** We thank David d'Enterria and Peter Skands for inviting us to this stimulating and perfectly organized workshop. We are indebted to A. Albino, J. Binnewies, P. Bolzoni, A. Kotikov, G. Kramer, and W. Ochs for collaboration on the work presented here. Work supported partly by the German Federal Ministry for Education & Research BMBF Grant No. 05H2015.

# $\alpha_s$ from $e^+e^-$ event shapes

Stefan Kluth

Max-Planck-Institut für Physik, Munich, Germany

**Abstract:** We give an overview of experimental procedures involved in measuring a distribution of an event shape observable in $e^+e^-$ collisions and estimates of corresponding uncertainties at high energies as expected to be available at a future FCC-ee facility. Then the procedures used to extract measurements of the strong coupling constant $\alpha_s$ are reviewed and possible uncertainties at future facilities are discussed.

## Measurement of event shape observable distributions

In the process of electron-positron annihilation to hadrons the measurement and analysis of event shape observables is one of the most fruitful ways to test the theory of strong interactions, quantum chromodynamics (QCD) [1]. As an example, the event shape observable *Thrust* $T$ [2,3] is defined by

$$T = \max_{\vec{n}} \frac{\sum_i \vec{p}_i \cdot \vec{n}}{\sum_i |\vec{p}_i|} \tag{1}$$

where $\vec{n}$ is the thrust axis (after maximisation), and the $\vec{p}_i$ are the momentum vectors of all particles in the event. In measurements and calculations usually the form $1-T$ is used such that the observable has small values for collimated two-jet events and progressively larger values with increasing jet multiplicity up to a maximal value $1-T = 0.5$ for a completely spherical event.

In $e^+e^-$ annihilation to hadrons at high energies above the the $Z^0$ peak and the W-pair production threshold the selection of genuine high energy events is the first step. One important background is so-called radiative return events where hard initial state photon radiation off the incoming beam particles reduces the effective centre-of-mass energy of the collision to be near the $Z^0$ peak. The algorithms developed by the LEP collaborations reconstruct the effective centre-of-mass energy of the hadronic event either from recoil against an energetic and isolated electromagnetic hit in the calorimeter or from the boosted kinematics of the event if it recoils against a hard photon collinear with the beam direction. The OPAL experiment, as an example, achieved a clean separation of background and signal [4] such that remaining background can be treated as part of the experimental corrections. At a future $e^+e^-$ annihilation experiment with much improved resolution of the electromagnetic calorimeter, the suppression of radiative events will thus not be a problem. However, the subsequent correction for residual initial state photon radiation effects will depend on the quality of the implementation in the Monte Carlo generators and also on avoiding double-counting if explicit electroweak corrections are applied later [5]. At LEP the uncertainty due to suppressing radiative return events and corresponding experimental corrections has never been dominant and thus with a well defined procedure and reliable modelling in the generators the same should be possible at a future $e^+e^-$ collider.

At energies above the W pair production threshold of $\sqrt{s} = 161$ GeV final states with both W bosons decaying hadronically to jets constitute the main background. The LEP collaborations developed sophisticated selection algorithms for their W boson physics programs which were also used to reject W pair events, see e.g. [6,7]. The residual W-pair dominated background in OPAL was between 2.1% and 6.2% for $161 < \sqrt{s} < 209$ GeV and was subtracted using simulated events. At large $\sqrt{s}$ and in the regions of the event shape distributions corresponding to multi-jet events the uncertainties from suppressing W-pair events were a significant part of the final experimental uncertainty on $\alpha_s$ of about 1-2% [7,8]. At a future $e^+e^-$ collider with energies above the LEP 2 limit of 209 GeV the identification of W-pair production in the hadronic W boson decay channel will improve due to the expected much better resolution for the measurement of jet energies. However, the propagation of W-pair background suppression uncertainties to a final result for $\alpha_s$ depends strongly on the specific analysis and the experiment and thus needs specific studies in the context of a concrete detector proposal. At a future $e^+e^-$ collider these uncertainties should be well below those achieved at LEP, i.e. below 1%.

Another important contribution to the experimental systematic uncertainties for the measurement of event shape distributions at LEP has been the model dependence of the experimental correction. This means that the results after application of the experimental corrections (particle or hadron level) depend on the Monte Carlo generator used to produce the events passed through the detector simulation. Briefly, the final states of simulated events before and after the detector simulation are used to calculate event shape observables and the corresponding corrections. The LEP experiments have applied bin-by-bin correction factors instead of more sophisticated unfolding procedures. At future $e^+e^-$ colliders a combination of advanced and well tuned Monte Carlo generators, a precise simulation of the detector response and adequate unfolding methods will be necessary to reach corresponding uncertainties well below those from the LEP collaborations.

## Massive b-quarks

At the transition between experimental and theoretical considerations is the issue of treatment of massive b-quarks in the determination of $\alpha_s$ values from event shape distributions. Compared to the available theory for $e^+e^-$ annihilation into massless quarks the theory for massive quarks is limited to NLO [9]. The measurements available so far are inclusive, i.e. including heavy quark events, with few exceptions. The re-analysed JADE data were corrected using Monte Carlo simulation for heavy quark events [10] while L3 has measured event shape distributions in flavour tagged samples at the $Z^0$ peak for b- and udsc quark events [11]. Therefore, in order to avoid biases, the theory includes NLO contributions with corresponding larger uncertainties scaled to the fraction of b-quark events, see e.g. [12], or the presence of b-quark events is corrected for as part of the hadronisation correction [7], with corresponding systematic uncertainties.

In the absence of theory progress, precision of $\alpha_s$ determinations will be limited to the NLO theory uncertainties scaled by the fraction of b-quark events, e.g. with a ca. 5% theory uncertainty in NLO at the $Z^0$ peak and a b-quark event fraction of 20% a limit of about 1% would result. At high energies of a future $e^+e^-$ facility the b-quark fraction is lower and the mass effects are suppressed leading to a negligible uncertainty. However, for a high precision $\alpha_s$ determination using possible very large samples of hadronic $Z^0$ decays at a future $e^+e^-$ facility this problem has to be taken into account.

For the re-analysed low-energy data from JADE flavour tagging was not an option. We note that according to the luminosity estimates for a future circular $e^+e^-$ collider rather short periods of data taking at $\sqrt{s} < m_{Z^0}$ would create equivalent samples, with precise flavour tagging and much better detector resolution [13]. The low energy data samples are important to constrain global analyses of event shape distributions at many energy points and could reduce experimental and theoretical uncertainties from global analyses significantly.

## Determination of $\alpha_s$ from event shapes

Once measurements of event shape distributions are established the determination of $\alpha_s$ proceeds by comparing the theory with corrections for non-perturbative effects (hadronisation corrections) to the data. The LEP experiments converged to a procedure where the hadronisation corrections are calculated using the Monte Carlo event generators, e.g. PYTHIA, HERWIG, ARIADNE or similar. The simulated event final states after the hard process and parton shower (parton-level) and after the hadronisation model (hadron or particle level) are used to calculate events shape observable values and thus to derive the corrections. The theory calculations are folded with the hadronisation corrections and then compared to the data.

A fundamental problem of this procedure is that the parton level predictions of the e.g. NNLO+NNLL theory and the e.g. LO+LL Monte Carlo generator are not consistent. OPAL has studied this in [8] by comparing the MC parton level with the theory (NNLO+NLL) parton level. The observation is that the inconsistency between the theory parton level and the MC generator parton levels is about as large as the inconsistency between the parton levels of the different MC generators. One thus concludes that the hadronisation uncertainty defined by using different MC generators to derive the corrections also covers uncertainties connected with the inconsistent parton levels between theory and MC generators.

The latest $\alpha_s$ analyses using this method and NNLO+NLL calculations had hadronisation uncertainties of about 1-3%, decreasing with centre-of-mass energy. Hadronisation effects are known to scale like $1/\sqrt{s}$ for event shape observables and thus with this method uncertainties could be expected to be about half this size at a future $e^+e^-$ facility running at $300 < \sqrt{s} < 500$ GeV.

In order to go beyond this uncertainty improvements of the procedure to determine $\alpha_s$ are needed. The MC hadronisation corrections have the advantage that they are universal and easily implemented. Alternatives are to provide a much closer integration of the perturbative and non-perturbative parts of the theory prediction. A. Hoang [14] reports on the procedure based on soft-collinear effective field theory (SCET) to yield a consistent prediction with NNLO calculations matched with NNLL resummation and a model for the non-perturbative effects with just $\alpha_s$ and an additional free non-perturbative parameter. Global fits of e.g. event shape data at low and high energies then quote much smaller non-perturbative uncertainties. The direct integration allows for a simultaneous fit of the perturbative and non-perturbative parts which treats the non-perturbative systematic uncertainty as a part of the fit error, leading to a substantial reduction. In addition, the perturbative part of the prediction is more complete, since it includes subleading logarithms and this implies that any remaining non-perturbative corrections are smaller compared to previous analyses leading to smaller uncertainties. The obvious disadvantage of this approach is that it is not universal and instead a new calculation is needed for each observable.

Another approach is to combine the most complete perturbative predictions with the hadronisation models available in the Monte Carlo generators. Recent progress in automation of NLO calcula-tions within MC generators has allowed to consistently match the fixed order calculation with the parton shower and to create merged samples of $e^+e^-$ annihilation to different parton multiplicities at the same time [15]. The result is a consistent description of event shape distributions for small and large jet multiplicities. No attempt to extract values of $\alpha_s$ using this method has been made yet. This procedure combines a consistent treatment of the parton level in the perturbative and non-perturbative part and allows for a simultaneous fit of the strong coupling $\alpha_s$ and the parameters of the hadronisation model.

A disadvantage of the procedure discussed above is that for the existing parton shower algorithms it is not known how exactly they correspond to the resummation of next-to-leading or subleading logarithms. This would cause an extra uncertainty in a possible $\alpha_s$ determination. The GENEVA algorithm [16] is a possible way to overcome this problem since for a given observable the parton shower is adapted to cover exactly up to NNLL effects. Together with automated fixed order calculations consistent predictions combined with a MC hadronisation model become a possibility. The ARES program [17] produces NNLL resummation for a large class of event shape observables such that it becomes a possibility to run the GENEVA algorithm for a set of event shape observables consistently. In the long term such an approach will be able to combine complete perturbative predictions with the well known and successful Monte Carlo hadronisation models in simultaneous fits. This approach is universal within the limits of the GENEVA and ARES algorithms.

Other theory aspects are the availability of NNLL resummation for many event shape observables. It would be interesting to study the effects of including these corrections in the $\alpha_s$ determination based on the ALEPH or OPAL data with the MC based hadronisation corrections, or combined with analytic non-perturbative corrections (power corrections), see e.g. [18].

### Summary

In order to reduce uncertainties on determinations of $\alpha_s$ from event shape observables in $e^+e^-$ annihilation to hadrons progress is needed on many fronts. The experimental uncertainties are expected to be reduced significantly at a future $e^+e^-$ facility compared to LEP results, due to improved resolution of the detectors, better experimental correction procedures and the availability of flavour tagging to identify heavy flavour (b-quark) events.

Theory uncertainties are expected to improve due to the availability of NNLO calculations together with resummation of next-to-next and subleading logarithms. Consistent combination with a model for non-perturbative effects will be needed in order to reduce uncertainties from this source. This can be achieved by e.g. SCET calculations for more observables, by combining analytic power corrections with improved perturbative calculations, and by leveraging automation of fixed order and resummation calculations and their combination in Monte Carlo generators. These would also provide consistent combination with a model for non-perturbative effects while still be universal.

**Acknowledgments.** The author would like to thank the organisers of the workshop for the invitation as well as for a lively workshop with interesting discussions.

# $\alpha_s$ from the $e^+e^-$ C-parameter Event Shape

**André H. Hoang**[1,2], Daniel W. Kolodrubetz[3], Vicent Mateu[1,4] and Iain W. Stewart[3]

[1]University of Vienna, Faculty of Physics, Boltzmanngasse 5, A-1090 Vienna, Austria
[2]E. Schrödinger Intl. Inst. Math. Phys., Univ. of Vienna, A-1090 Vienna, Austria
[3]Center for Theoretical Physics, MIT, Cambridge, MA 02139, USA
[4]Dept. Física Téorica and IFT-UAM/CSIC, Univ. Autónoma de Madrid, 28049 Madrid, Spain

**Abstract:** I report on our results for $\alpha_s(m_Z)$ from a global fit of $e^+e^-$ data for the C-parameter event shape at energies $Q$ between 35 and 207 GeV [1], and I also outline the theoretical input and the calculations that went into this analysis [2].

This work is part of a continuing project on strong coupling determinations from fits to $e^+e^-$ event shape data using factorization and log resummation within Soft-Collinear-Effective-Theory (SCET). There are two previous analyses along the same lines using data on the thrust distribution [3] and on moments of the thrust distribution [4]. Let me start by giving the definition of C-parameter:

$$C = \frac{3}{2}\,\frac{\sum_{i,j} |\vec{p}_i||\vec{p}_j| \sin^2\theta_{ij}}{\left(\sum_i |\vec{p}_i|\right)^2}\,, \tag{1}$$

where the double sum is over all final state hadronic particles and $\theta_{ij}$ is the angle between particles $i$ and $j$. The C-parameter is an infrared-safe and global jet distribution variable and quantifies how dijet-like an event is. For small back-to-back dijet-like event $C$ is small. For tree-level parton level predictions the distribution is a delta function at $C = 0$, so the observable distribution at $C > 0$ is a direct measure for gluon radiation governed by $\alpha_s$ as well as for hadronization effects.

The theoretical description on which our C-parameter analysis is based upon is a combination (i) of *partonic singular* contributions, which are described by a factorization theorem and contain the summation of large logarithmic terms, (ii) of *partonic non-singular* contributions, which contain fixed-order QCD corrections that are subleading in the dijet limit and not treated within factorization, and (iii) of a *non-perturbative power correction* that can be incorporated through a convolution with a soft-model function. In the tail of the C-distribution, outside the peak region, the theoretical description can be based on an OPE, where only $\alpha_s$ and the first moment $\Omega_1$ of the soft model function enter as relevant theoretical parameters in the fits. Within this setup our theoretical prediction is based on the following ingredients:

1. Matrix elements and hard Wilson coefficient entering the singular cross section as well as the fixed order nonsingular terms are included up to $\mathcal{O}(\alpha_s^3)$. The non-logarithmic $\mathcal{O}(\alpha_s^3)$ corrections to the jet and soft functions in the singular cross section are unknown and contribute to the perturbative error estimate. The $\mathcal{O}(\alpha_s^3)$ non-singular contributions are obtained using [5].

2. Resummation of large logarithmic terms in the singular contributions with N$^3$LL accuracy using SCET. The uncertainty due to the unknown 4-loop cusp anomalous dimension is an irrelevant fraction of our theoretical error.

3. Profile functions ($C$-dependent renormalization scales $\mu_J$, $\mu_S$, $R$, $\mu_{ns}$ entering the various partonic components) that govern the proper summation of large logarithms and account for the multijet boundary condition, which ensures that fixed-order results are used in the far tail endpoint multi-jet region.

4. Description of the nonperturbative and partonic soft contributions based on field theory matrix elements of Wilson lines with switching from the $\overline{\mathrm{MS}}$ to the R-gap infrared subtraction scheme for the partonic soft function [6,7]. The R-gap scheme allows rendering the perturbative calculation and $\Omega_1$ free of $\mathcal{O}(\Lambda_{\mathrm{QCD}})$ renormalon ambiguities and entails an RGE in the scale R that sums large logarithms arising in the perturbative renormalon subtractions.

5. Treatment of effects of finite hadron masses based on [8]. Hadron mass effects can in general spoil the universality of power corrections entering event shapes, but only have small effects in C-parameter and thrust. So universality can still be used as consistency check for results from C-parameter and thrust.

We fit simultaneously for $\alpha_s(m_Z)$ and $\Omega_1$ in the tail region of the C-parameter distributions. Our default 404 bin dataset contains the ranges $25\,\mathrm{GeV}/Q \leq C \leq 0.7$. We use a normalization for $\Omega_1$ such that equality to the corresponding measured value for thrust means that universality is satisfied. For a single fixed Q there is a strong degeneracy between $\alpha_s(m_Z)$ and $\Omega_1$. It is lifted by the global fit with different Q values. To estimate perturbative uncertainties we performed a 500 point random scan over 16 parameters, refitting for each point ("parameter scan"). The scan accounts for theory uncertainties from higher orders via varying scales, statistical theory errors, and unknown perturbative coefficients. As shown in Fig. 1 (left panel), we observe very good convergence when increasing the orders in perturbation theory, and achieve $\chi^2/\mathrm{dof} = 0.988$ when including all theoretical ingredients.

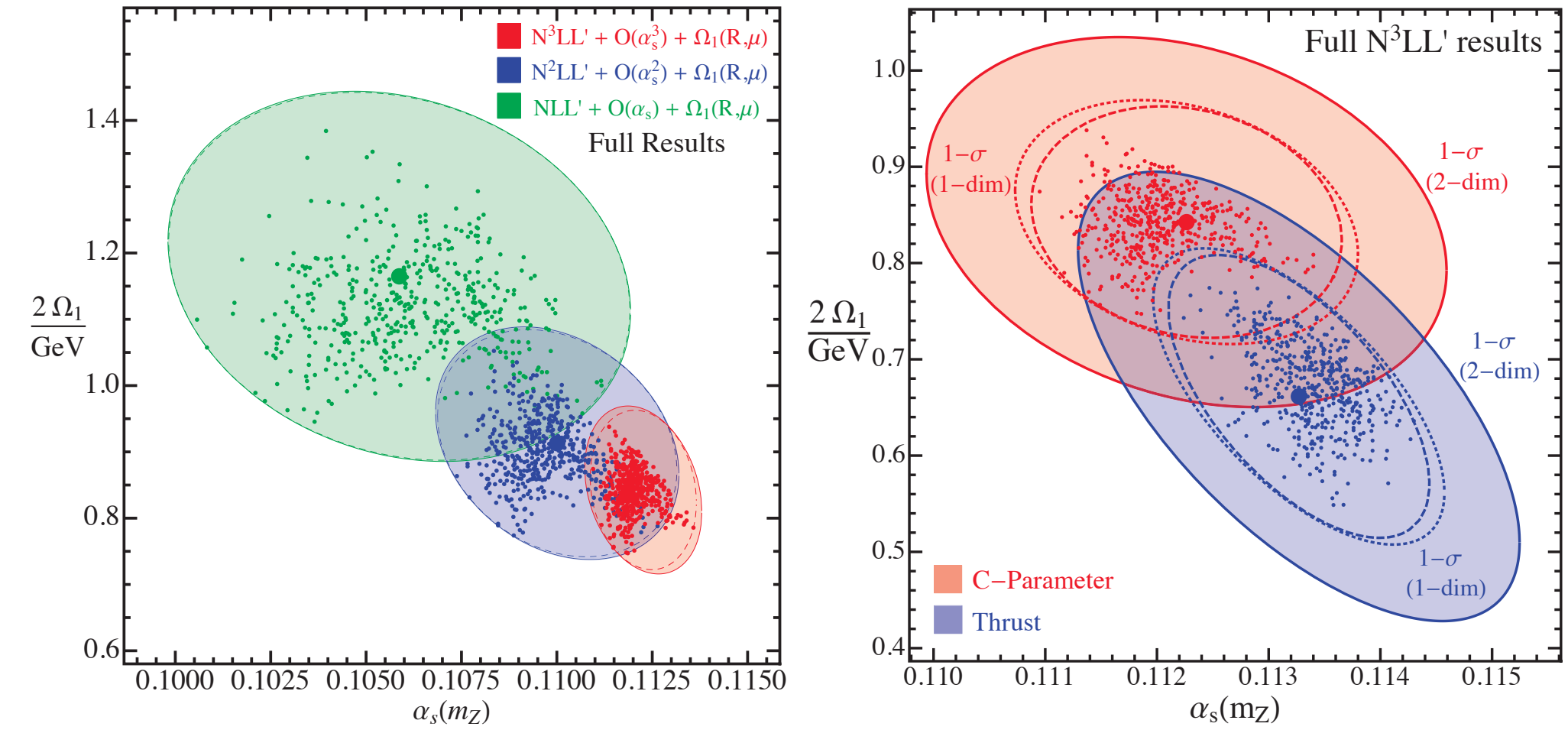

Figure 1: Left: Distribution of best-fit points in the $\alpha_s(m_Z)$-$\Omega_1$ plane resulting from the parameter scan at different orders for the C-parameter analysis. The dashed lines correspond to the theoretical (1-dim) 1-$\sigma$ uncertainties and are related to the ellipse fitting the enclosing contour of the scan points. The solid ellipses also include the (1-dim) 1-$\sigma$ experimental uncertainties. Right: Distribution of best fit points at highest order from the C-parameter and an updated thrust analysis. Outer solid lines correspond to (2-dim) 1-$\sigma$ combined theoretical and experimental uncertainties.

In Fig. 2 the evolution of the extracted fit results for $\alpha_s(m_Z)$ is shown when adding various components of the theoretical calculation. The most dramatic effect is the $-8\%$ decrease of $\alpha_s(m_Z)$ from around 0.122 to about 0.113 when adding the $\Omega_1$ nonperturbative fit parameter at stage 3. The effects from adding renormalon subtractions result in a small reduction of the perturbative uncertainty but are very small otherwise. The effects from including finite hadron masses is small as well. As shown in Fig. 1 (right panel), the fit results for $\Omega_1$ and $\alpha_s(m_Z)$ from the C-parameter analysis are fully consistent with the corresponding results from our tail fit thrust analysis. There is also full consistency to the thrust moment analysis carried out in [4] and the indicated uncertainties also cover the dependence on variations of the data sets used in the fits. The consistency of the results demonstrates that: (a) the theoretically expected universality of the non-perturbative power corrections for C-parameter and thrust is confirmed in the data analysis, and (b) the theoretical description as a whole, and of the non-perturbative effects in particular, are under control.

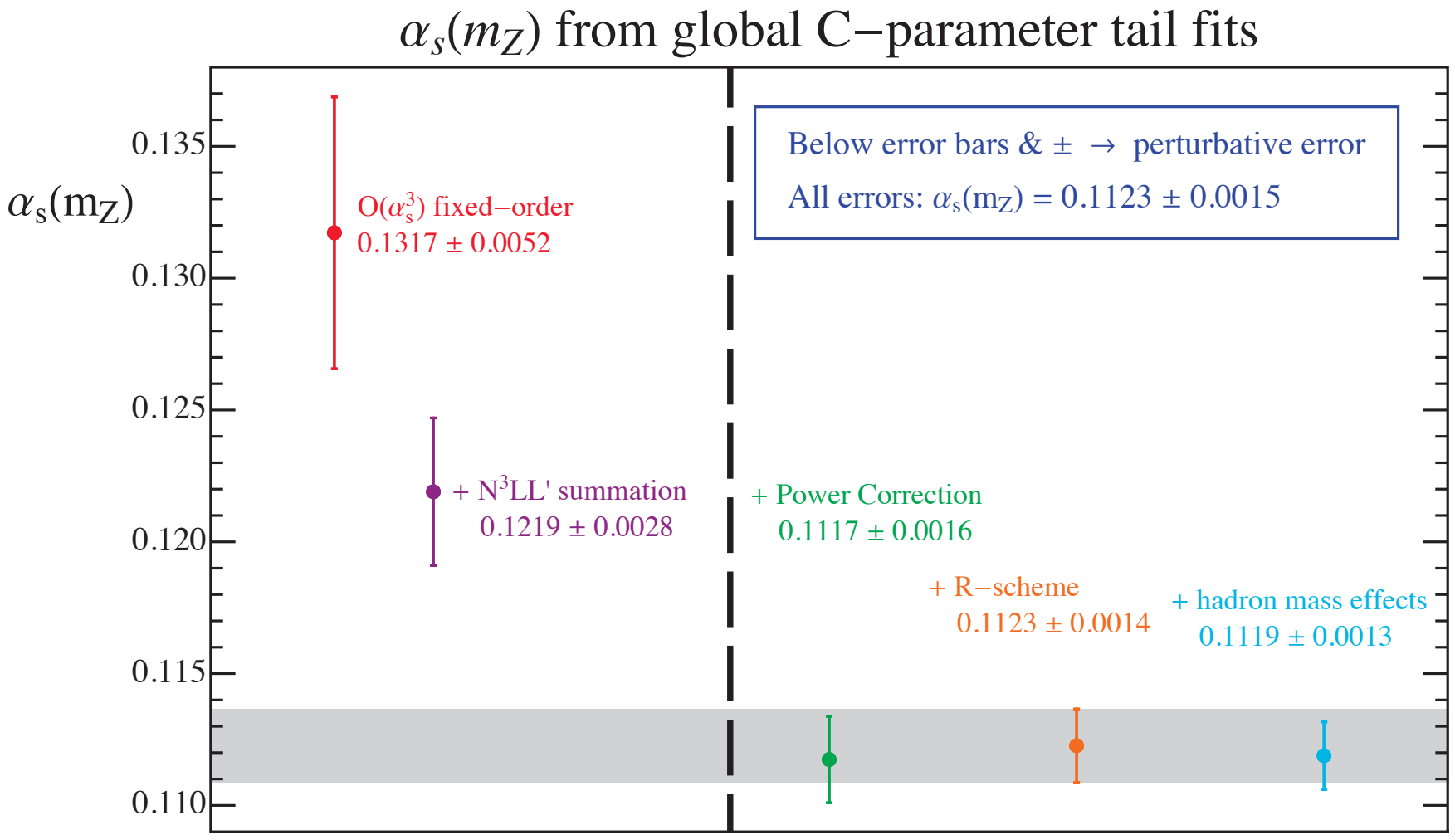

Figure 2: Evolution of the result for $\alpha_s(m_Z)$ adding different components of the theoretical calculation. The two left points do not contain any uncertainty from ignoring hadronization effects.

The final results from the C-parameter analysis in [1] read:

$$\alpha_s(m_Z) = \quad 0.1123 \pm 0.0002_{\rm exp} \pm 0.0007_{\rm hadr} \pm 0.0014_{\rm pert}\,, \tag{2}$$

$$\Omega_1(R_\Delta, \mu_\Delta) = \quad 0.421 \pm 0.007_{\rm exp} \pm 0.019_{\alpha_s} \pm 0.060_{\rm pert}\,{\rm GeV}\,, \tag{3}$$

where experimental (exp) and perturbative (pert) as well as the uncertainties coming from the simultaneous fits of $\alpha_s$ and $\Omega_1$ are indicated separately. The result for $\alpha_s(m_Z)$ belongs to the most precise individual determinations of the strong coupling entering the strong coupling world average. However, its central value is substantially lower. Apart from our analyses in [1,3,4] this feature has also been observed in the results of the event shape analyses [9,10,11], all of which have in common that $\alpha_s(m_Z)$ and the non-perturbative effects were obtained in simultaneous fits (often called the "method of analytic power-corrections"). The smallness of $\alpha_s(m_Z)$ in these analyses can be traced back to the following facts:

- There is a negative correlation of the dependence of the theoretical event shape description in the tail region on $\alpha_s$ and $\Omega_1$. So for a given Q, an increase in $\alpha_s$ can be compensated by a decrease in $\Omega_1$. Simultaneous fits to data from different Q values are necessary to lift this degeneracy.
- The outcome of the simultaneous fits to data from different Q values results in best fits having low $\alpha_s$ values and sizeable values for the non-perturbative parameter $\Omega_1$, but still fully compatible with the theoretical expectation $\Omega_1 \sim \Lambda_{\rm QCD}$.

High precision measurements of event shape distributions using the (existing) off-resonance data from B-factories might help to learn more about the systematics of the global fits, and would therefore be very welcome. High precision data from $Q \gg m_Z$ for a possible future $e^+e^-$ collider might have a similar impact.

The analytic power correction method is in sharp contrast with the method of estimating non-perturbative effects from the difference of hadron and parton level Monte Carlo (MC) event shape results. Event shape analyses based on this method all result in larger values for $\alpha_s(m_Z)$ in better agreement with the world average since the hadronization effects in MC descriptions of event shape distributions are small. However, the MC method to determine the size of non-perturbative effects has the property that it is based on leading order partonic (matrix element and parton shower) calculations, which raises doubts that it can be used in the context of high-order perturbative and resummed QCD calculations.

# $\alpha_s$ from $e^+e^-$ jet cross sections

Andrea Banfi

University of Sussex, Brighton, United Kingdom

**Abstract:** Jet rates, fractions of the total number of events having a definite number of jets, have been widely exploited for measurements of $\alpha_s$ at $e^+e^-$ colliders. Here we aim at giving an overview of the state-of-the art of theoretical predictions for jet rates, as well as of the most precise extractions of $\alpha_s$ from such observables available at the moment. We also present recent developments on the theoretical side, namely electroweak corrections, that are not exploited in current measurements, and discuss some prospects for future colliders.

At $e^+e^-$ colliders, jets are found by applying sequential clustering procedures. At the moment, we have data for the JADE [1], the Durham [2], both in its plain and angular-ordered (AO) [3] versions, and the Cambridge [3] algorithms. All of these procedures are characterised by a resolution parameter $y_{\rm cut}$. For a given $y_{\rm cut}$, the number of jets is the number of pseudo-particles left when the sequential procedure stops. This happens when a suitably defined distance between all pairs of pseudo-particles is larger than $y_{\rm cut}$. For instance, for the Durham algorithm, the distance between a pair of particles of momenta $p_i$ and $p_j$ is given by

$$y_{ij}^{(D)} = 2\frac{\min\{E_i^2, E_j^2\}}{Q^2}(1-\cos\theta_{ij})\,, \tag{1}$$

with $Q=\sqrt{s}$ the centre-of-mass energy of the $e^+e^-$ collision. The $n$-jet rate $R_n(y_{\rm cut})$ is defined as the fraction of events having $n$ jets.

An important difference between event-shape distributions and jet rates is the size of hadronisation corrections. These are believed to be smaller than the ones affecting event-shape distributions, although this is so far supported only by simulations with Monte Carlo event generators. Having data with higher centre-of-mass energy will make it possible to neglect hadronisation effects, thus enabling a more reliable determination of $\alpha_s$.

## Theoretical predictions

The observables that are used to extract $\alpha_s$ are either jet rates, as a function of $y_{\rm cut}$, or differential distributions in the jet-resolution parameters $y_n$. Let us now focus on jet rates, and be aware that similar considerations hold for distributions in the jet-resolution parameters. For $y_{\rm cut} \sim 1$, fixed-order QCD predictions give reliable estimates of these observables. With decreasing $y_{\rm cut}$, large logarithms of $y_{\rm cut}$ become important, and need to be resummed at all orders so as to have finite theoretical predictions. Resummed calculations involve combinations of Sudakov form factors, which vanish for $y_{\rm cut} \to 0$, consistent with the vanishing probability of observing coloured particles without accompanying QCD radiation.

On the fixed-order side, the two-jet rate $R_2$ starts from 1 at lowest order, and its deviation from one is known up to three orders in $\alpha_s$ (NNLO) [4,5]. The same fixed-order calculation is applied to $R_3$, which starts at $\alpha_s$, and is also known at NNLO accuracy (i.e. up to order $\alpha_s^3$). Both the four [6,7,8] and five [9] jet rates are known at NLO, i.e. up to order $\alpha_s^3$ and $\alpha_s^4$ respectively. NLO calculations exist also for the six- and seven- jet rates [10], but in the limit of a large number of colours.

Large logarithms $L = \ln(1/y_{\mathrm{cut}})$ have been resummed in the two-jet rate up to the so-called next-to-leading logarithmic (NLL) accuracy [11], in which all terms $\alpha_s^n L^n$ in the *logarithm* of $R_2$ are under control. All the other jet rates have been resummed at the lower $\mathrm{NLL}_\Sigma$ accuracy, which correctly reproduces all terms $\alpha_s^n L^{2n}$ and $\alpha_s^n L^{2n-1}$ in the corresponding perturbative expansions [2].

## Extractions of $\alpha_s$

Here we present current determinations of $\alpha_s$ using jet rates, and related differential distributions, from lower to higher jet multiplicity.

The most accurate determinations of $\alpha_s$ from the two-jet rate combine NNLO and NLL resummation. Actually what is considered is the differential distribution in the three-jet resolution $y_3$, which is just the derivative of the two-jet rate. An interesting point observed in Ref. [12] is that, for this observable, Monte Carlo predictions at parton level agree quite well with perturbative QCD predictions. This suggests that determining hadronisation corrections as the ratio of hadron-level versus parton-level Monte Carlo predictions gives a sensible estimate of these non-perturbative effects. A summary of the available determinations is given in Table 1.*

| Q [GeV] | $\alpha_s(m_Z)$ | $\Delta\alpha_s$(stat.) | $\Delta\alpha_s$(exp.) | $\Delta\alpha_s$(had.) | $\Delta\alpha_s$(theo.) |
|---|---|---|---|---|---|
| 14–44 [13] | 0.1199 | 0.0005 | 0.0013 | 0.0046 | 0.0023 |
| 91 [12] | 0.1186 | 0.0002 | 0.0011 | 0.0017 | 0.0029 |
| 91–209 [14] | 0.1192 | 0.0010 | 0.0026 | 0.0004 | 0.0021 |

Table 1: Summary of $\alpha_s$ determinations using the $y_3$ distribution. The result from Ref. [12] is shown for $\sqrt{s} = m_Z$ only.

We notice that, except for low JADE energy, the theoretical error, determined by varying all relevant scales in perturbative QCD predictions, is the dominant one. To reduce this uncertainty, it would be very useful to have NNLL resummation, as is available for all $e^+e^-$ event shapes through the ARES method [15]. From results for event shapes in Ref. [15], we see that uncertainties are halved from NLL to NNLL resummation. Also, preliminary results obtained with the ARES code show that increasing the energy from 91 GeV to 400 GeV reduces uncertainties again by 50%. Assuming these results hold for the two-jet rate as well, we expect to be able to extract $\alpha_s$ with an accuracy of 1% using data from a future high-energy $e^+e^-$ collider.

Another accurate determination of $\alpha_s$ comes from the three-jet rate [16], using ALEPH data only. The extraction is performed for a single bin of $R_3(y_{\mathrm{cut}})$, with $-4.0 < \ln(y_{\mathrm{cut}}) < -3.8$, the one which has the lowest error (see Table 2). Hadronisation corrections are negligible, giving an error that is less than one per-mille. The statistical error has the same size, and can be neglected. We notice that in this case, the theoretical error, around 1.3%, is less than the experimental error. Resummation, relevant for $\ln(y_{\mathrm{cut}}) \lesssim -4.5$, is only $\mathrm{NLL}_\Sigma$ accurate. A NLL prediction would be very useful to fit the three-jet rate in a wide range, as done for the two-jet rate. Another

*Ref. [12] contains fitted values of $\alpha_s(\sqrt{s})$ for all energies, without providing the corresponding value of $\alpha_s(m_Z)$. Here we do not attempt to extrapolate the provided errors to obtain $\alpha_s(m_Z)$ for each centre-of-mass energy, but rather report its determination for $\sqrt{s} = m_Z$ only.

| Q [GeV] | $\alpha_s(m_Z)$ | $\Delta\alpha_s$(stat.) | $\Delta\alpha_s$(exp.) | $\Delta\alpha_s$(had.) | $\Delta\alpha_s$(theo.) |
|---|---|---|---|---|---|
| 22–44 [17] | 0.1199 | 0.0010 | 0.0021 | 0.0054 | 0.0007 |
| 91 [16] | 0.1175 | – | 0.0020 | – | 0.0015 |

Table 2: Summary of $\alpha_s$ determinations using the three-jet rate.

determination [17], with a larger error, performed by the JADE collaboration, uses indeed NLL$_\Sigma$ resummation, which is important at JADE energies to correctly describe the shape of the three-jet rate. Hadronisation uncertainties here dominate over theoretical ones (see again Table 2).

Two determinations of $\alpha_s$ using the four-jet rate exist at the moment. They both use NLO predictions provided by the program DEBRECEN [6], an earlier version of NLOJET++ [18]. The analysis by DELPHI [19] does not include NLL$_\Sigma$ resummation. There, in order to be able to have a reasonable description of data with the NLO calculation only, one needs to fit also the central value of the renormalisation scale, which is not taken by default equal to $\sqrt{s}$. The fit returns quite low values, corresponding to $\mu_R = 0.015\, m_Z \simeq 1.4\,\mathrm{GeV}$ for the Durham and $\mu_R = 0.042\, m_Z \simeq 3.8\,\mathrm{GeV}$ for the Cambridge algorithm. Such low scales suggest that resummation is important for this observable. In fact, NLL$_\Sigma$ resummation has been added by OPAL [20] and JADE [21] collaborations. Notice that OPAL quotes an extra uncertainty associated with the mass of the $b$-quarks, based on conservative estimate on the contribution of $b$-quarks to the four-jet rate (see [20] for details).

| Q [GeV] | $\alpha_s(m_Z)$ | $\Delta\alpha_s$(stat.) | $\Delta\alpha_s$(exp.) | $\Delta\alpha_s$(had.) | $\Delta\alpha_s$(theo.) | $\Delta\alpha_s$(mass) |
|---|---|---|---|---|---|---|
| 91 (Durham) [19] | 0.1178 | – | 0.0012 | 0.0031 | 0.0014 | – |
| 91 (Cambridge) [19] | 0.1175 | – | 0.0010 | 0.0027 | 0.0007 | – |
| 91–197 [20] | 0.1182 | 0.0003 | 0.0015 | 0.0011 | 0.0012 | 0.0013 |
| 22–44 [21] | 0.1159 | 0.0004 | 0.0012 | 0.0024 | 0.0007 | – |

Table 3: Summary of $\alpha_s$ determinations using the four-jet rate.

There exists also a determination of $\alpha_s$ with five-jet events, using ALEPH data [9]. Results are summarised in Table 4. Hadronisation corrections, obtained with SHERPA [22] implementing the CKKW [23] multi-jet merging procedure, are very small. Neglecting them results in a tiny increase of the perturbative uncertainty, without changing the overall uncertainty in the $\alpha_s$ determination. The results are quite sensitive to the choice of the fit range. This effect is quantified by means of the extra uncertainty $\Delta\alpha_s$(range). The combination of the results in Table 4 [9] gives the competitive

| Q [GeV] | $\alpha_s(m_Z)$ | $\Delta\alpha_s$(stat.) | $\Delta\alpha_s$(exp.) | $\Delta\alpha_s$(had.) | $\Delta\alpha_s$(theo.) | $\Delta\alpha_s$(range) |
|---|---|---|---|---|---|---|
| 91 [9] | 0.1159 | $^{+0.0002}_{-0.0002}$ | $^{+0.0027}_{-0.0029}$ | $^{+0.0012}_{-0.0012}$ | $^{+0.0062}_{-0.0043}$ | $^{+0.0014}_{-0.0014}$ |
| 183–206 [9] | 0.1155 | $^{+0.0015}_{-0.0016}$ | $^{+0.0008}_{-0.0008}$ | – | $^{+0.0029}_{-0.0020}$ | $^{+0.0028}_{-0.0028}$ |

Table 4: Summary of $\alpha_s$ determinations using the five-jet rate and the differential distribution in the five-jet resolution parameter $y_5$.

determination

$$\alpha_s(m_Z) = 0.1156^{+0.0041}_{-0.0034}\,. \tag{2}$$

A summary of the presented results is shown in Fig. 1. The total error on each determination is obtained by adding in quadrature the individual errors. All results are compatible with the world average (yellow band in the figure) [24]. Errors are between 2 and 5%, and, except for JADE energies where hadronisation effects are more important, are dominated by theoretical uncertainties on perturbative QCD calculations.

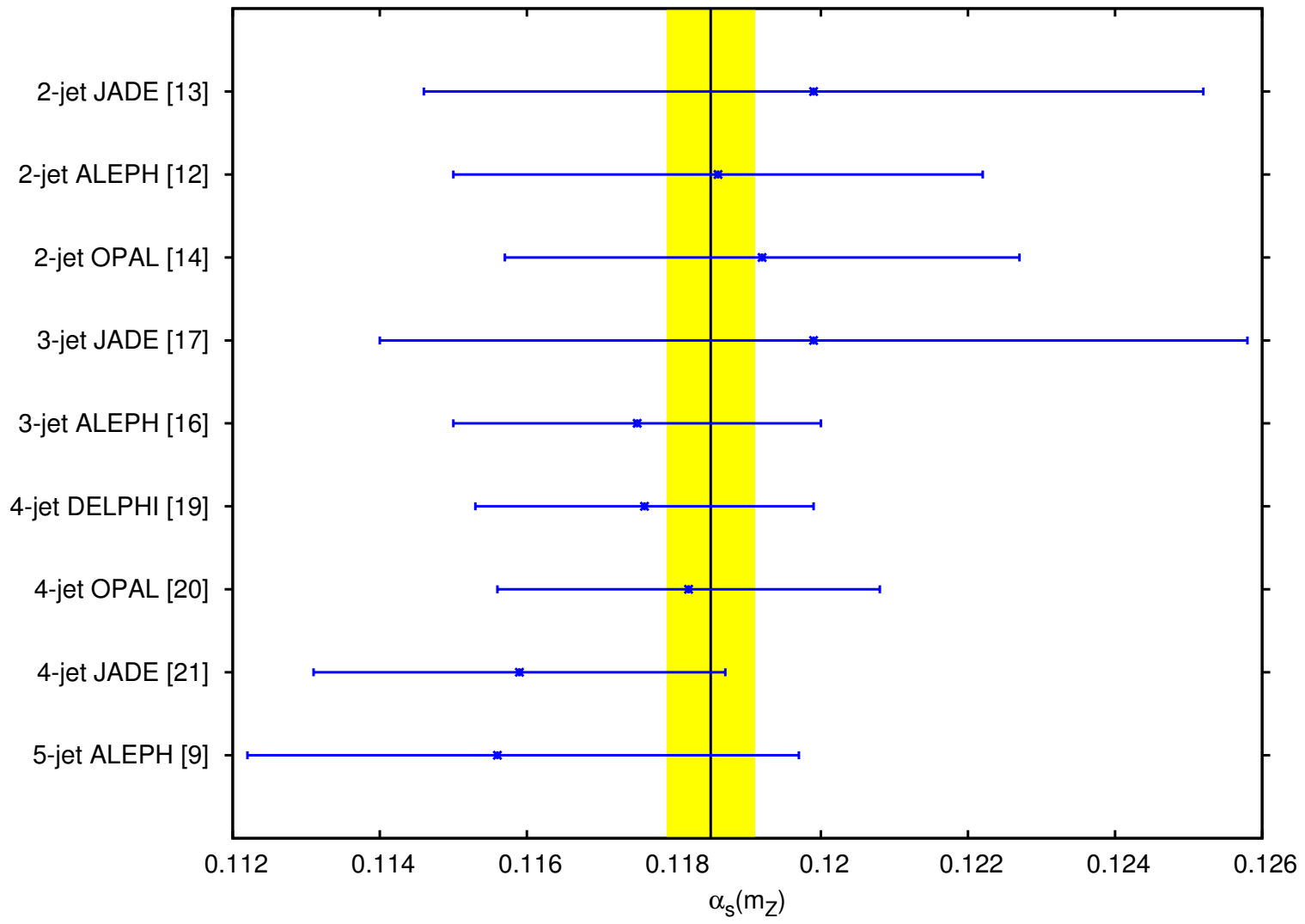

Figure 1: Summary of the $\alpha_s(m_Z)$ determinations presented here. The label for each extraction contains the number of jets, the source of the data, and the corresponding reference. The yellow band indicates the current $\alpha_s$ world average.

## Prospects

All the analyses described so far use QCD predictions only. Actually, at the present level of accuracy (NNLO), electroweak corrections start to matter, since $\alpha \sim \alpha_s^2$. In all current determinations of $\alpha_s$, photon radiation is corrected for using Monte Carlo simulations, and the contributions of di-boson production is subtracted off as background (see e.g. [25] for details). At the moment however a calculation of the three-jet rate including electroweak effects at order $\alpha$ does exist [26]. The corresponding theoretical analysis contains various important messages. The first is that, while the three-jet cross section depends crucially on the experimental cuts on the photons, when normalising by the total $e^+e^-$ cross section, as is appropriate to obtain the three-jet rate, corrections induced by photon radiation largely cancel, leaving an overall effect between 1 and 3% at moderate values of $y_{\text{cut}}$. At smaller values of $y_{\text{cut}}$, and at centre-of-mass energies above 172 GeV, corrections due to the radiation of a hard photon become largish, between 5 and 10%. Pure weak corrections are very small, below the per-mille level. Another interesting feature of QED corrections is that they give rise to a peak structure in the distribution in the three-jet resolution $y_3$ (visible at $\log_{10} y_3 \simeq -1$ in the left-hand plot of Fig. 2). The reason for this enhancement is that there might be photons that escape photon isolation cuts, and get clustered with a soft gluon, making up a photon jet that enters the jet-clustering procedure. These photons recoil against a quark-antiquark pair whose invariant mass is close to $m_Z$ (radiative return). From the analysis of Ref. [26], one obtains that the position of the peaky structure shifts to lower values of $y_3$ with increasing centre-of-mass energy. In fact, it becomes invisible at $\sqrt{s} = 500\,\mathrm{GeV}$ (see the right-hand plot of Fig. 2).

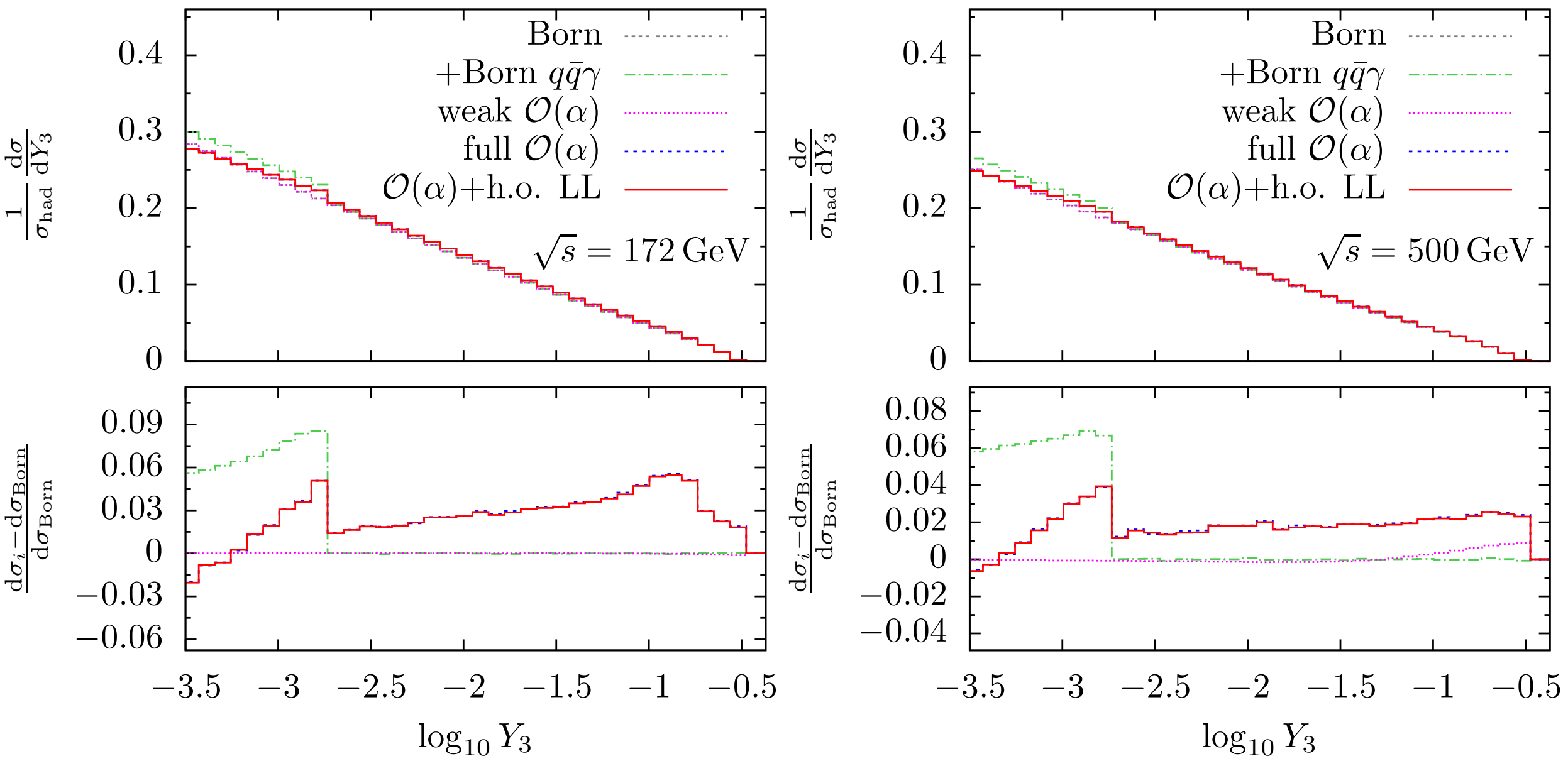

Figure 2: The differential distribution in the three-jet resolution for $\sqrt{s} = 172\,\mathrm{GeV}$ (left) and $\sqrt{s} = 500\,\mathrm{GeV}$ (right).

To conclude, increasing the centre-of-mass energy of $e^+e^-$ collisions brings various advantages to $\alpha_s$ determination using jet resolution parameters. First, it gets rid of hadronisation corrections, which are poorly known, and rely exclusively on Monte Carlo event generators. This fact, together with advances in QCD resummation techniques, has the potential to bring the theoretical error on $\alpha_s$, so far the dominant one, to around 1%, at the level of the experimental one. Furthermore, potentially large electroweak contributions due to radiative return get smaller when increasing the centre-of-mass energy. Since electroweak corrections are theoretically under control, we very much hope that these could be included in future determinations of $\alpha_s$.

**Acknowledgements.** I would like to thank the organisers of the workshop for the kind invitation to contribute to it. I am also grateful to P.F. Monni and G. Luisoni for help in the preparation of my presentation, and to G. Zanderighi and G. Salam for comments and suggestions on the talk and on the proceeding.

# $\alpha_s$ from hadronic Z decays and the full electroweak fit

Klaus Mönig

DESY, Zeuthen, Germany

**Abstract:** The strong coupling $\alpha_s$ is extracted from different experimental observables at the Z mass pole ($R^0_\ell$, $\sigma^{\rm had}_0$ and $\Gamma_Z$) using the most uptodate theoretical and experimental inputs. Prospects for future $e^+e^-$ colliders (ILC and FCC-ee) are discussed.

QCD corrections to the cross section $\sigma(e^+e^- \to \text{hadrons})$ are known since long. At lower energies usually the ratio $R = \frac{\sigma(e^+e^-\to\text{hadrons})}{\sigma(e^+e^-\to\mu^+\mu^-)}$ has been used to determine $\alpha_s$ [1]. Similar corrections arise at the Z-resonance. These corrections modify the partial width of the Z decaying to hadrons ($\Gamma_{\rm had}$) and through them, the total Z-width ($\Gamma_Z$). At centre of mass energies close to the Z-resonance, relevant observables for the $\alpha_s$ determination are: (i) the ratio of hadronic to leptonic Z-decays, $R^0_\ell = \frac{\Gamma_{\rm had}}{\Gamma_\ell}$, (ii) the hadronic pole cross section, $\sigma^{\rm had}_0 = \frac{12\pi}{m_Z}\frac{\Gamma_e\Gamma_{\rm had}}{\Gamma^2_Z}$, where the sensitivity is reduced because the QCD correction appears in the denominator and the numerator, and (iii) the total Z-width, $\Gamma_Z$, which is measured with complementary systematics. It is often noted that a very sensitive observable is the leptonic pole cross section, $\sigma^0_\ell = \frac{12\pi}{m_Z}\frac{\Gamma^2_\ell}{\Gamma^2_Z}$. In a global fit to the first three observables this is however already fully included and must not be taken in addition.

At the Born level, the partial width of the Z decaying into a fermion pair $f\overline{f}$ is proportional to the squared sum of the vector and axial-vector couplings, i.e. $\Gamma_f \propto (g^2_{V,f} + g^2_{A,f})$, where $g_{A,f}$ is simply given by the third component of the weak isospin, while $g_{V,f}$ is modified by the weak mixing $g_{V,f} = g_{A,f}(1 - 4|q_f|\sin^2\theta_W)$. Including higher orders, the couplings can be written as $g_{A,f} \to \sqrt{1+\Delta\rho_f}\, g_{A,f}$, $\sin^2\theta_W \to \sqrt{1+\Delta\kappa_f}\sin^2\theta_W = \sin^2\theta^f_{\rm eff}$, which means that unknown standard model (SM) and beyond-SM parameters modify the predictions. In general, the $\Delta\rho_f$ and $\Delta\kappa_f$ parameters are flavour independent apart from small constant terms and some possible contributions to the b-quark observables. After the discovery of the Higgs boson [2,3], the electroweak sector is completely defined and $\Delta\rho_f$ and $\Delta\kappa_f$ can be calculated. In an alternative approach, $\sin^2\theta^l_{\rm eff}$ can be measured from various asymmetries at LEP and SLD. In this case only $R^0_\ell$ and $\sigma^{\rm had}_0$ can be used for the $\alpha_s$ determination since $\Gamma_Z$ is affected by $\Delta\rho$ which cannot be measured independently.

All theory input is known by now to a precision better than the experimental uncertainties. The QCD corrections to the hadronic Z-width are known to fourth order [4]. The electroweak corrections to $\Gamma_f$ are known to 2nd order for the fermionic corrections plus some higher order terms [5], $\sin^2\theta^l_{\rm eff}$ is known to full 2-loop order with leading 3- and 4-loop corrections $\mathcal{O}(\alpha\alpha_s^2)$, $\mathcal{O}((\alpha m_t)^2\alpha_s)$, $\mathcal{O}((\alpha m_t)^3)$, $\mathcal{O}(\alpha m_t \alpha_s^3)$ [6], and $m_W$ is known to the same precision as $\sin^2\theta^l_{\rm eff}$ [7].

## $\alpha_s$ extraction with current data

The main experimental inputs are the data from the LEP energy scans between 1991 and 1995. The Z-lineshape parameters have been obtained from precise measurements of the hadronic and leptonic cross sections at energies close to the Z-mass and from extremely precise measurements of the beam energies [8]. The results, combined for the four LEP experiments, are: $m_Z = 91.1875 \pm$

0.0021 GeV, $\Gamma_Z = 2.4952 \pm 0.0023$ GeV, $\sigma_0^{\rm had} = 41.540 \pm 0.037$ nb and $R_\ell^0 = 20.767 \pm 0.025$. As additional inputs to fix $\Delta\rho$ and $\Delta\kappa$, the following data are used: $\sin^2\theta_{\rm eff}^l$ measurements from LEP/SLD, $\sin^2\theta_{\rm eff}^l = 0.23153 \pm 0.00016$ [8], the W-mass from LEP/Tevatron, $m_W = 80.385 \pm 0.015$ GeV [9], the top-mass from the Tevatron and the LHC, $m_t = 173.34 \pm 0.76$ GeV [10], and the hadronic vacuum polarisation in the running of $\alpha$, $\Delta\alpha_{\rm had}^{(5)}(m_Z) = 0.02757 \pm 0.00010$ [11]. For the Higgs mass a private average, $m_H = 125.14 \pm 0.24$ GeV, has been used where all results of the electroweak fit are completely insensitive to the details of the Higgs mass assumption.

The theoretical uncertainties for the electroweak calculations have been taken from the original publications [5,6,7]: $\Delta R_\ell^0 = 5 \cdot 10^{-3}$, $\Delta\sigma_0^{\rm had} = 6$ pb, $\Delta\Gamma_Z = 0.5$ MeV and $\sin^2\theta_{\rm eff}^l = 4.7 \cdot 10^{-5}$, which is in all cases significantly smaller than the present theoretical uncertainty. For the QCD corrections the full $\mathcal{O}(\alpha_s^4)$ term has been assumed as uncertainty, and for the top quark mass an additional error of 0.5 GeV has been added to account for the ambiguity in the mass definition.

The final extraction of $\alpha_s$, from all the ingredients mentioned above, has been carried out with Gfitter [12] which contains all known electroweak and strong corrections to the Z-lineshape and takes correctly into account all correlations between the experimental inputs. The global fit to all data yields

$$\begin{aligned} \alpha_s(m_Z^2) &= 0.1196 \pm 0.0028_{\rm exp} \pm 0.0006_{\rm QCD} \pm 0.0006_{\rm EW} \\ &= 0.1196 \pm 0.0030, \end{aligned}$$

where $\alpha_s$ is the only unconstrained parameter in the fit. The $\chi^2/ndf$ was 17.8/14 corresponding to a probability of 22% which shows that the global fit is a valid approach. As two extreme alternative approaches either all asymmetries and $m_W$ have been omitted from the fit (so that all electroweak inputs are taken from the calculations) or $m_W$, $m_t$ and $m_H$ have been omitted (so that only the measured $\sin^2\theta_{\rm eff}^l$ is used for $R_\ell^0$ and $\sigma_0^{\rm had}$). In these fits, $\alpha_s(m_Z^2) = 0.1202 \pm 0.0030$ and $\alpha_s(m_Z^2) = 0.1200 \pm 0.0031$ have been obtained which is completely consistent with almost identical errors.

## $\alpha_{\rm s}$ extraction with future data

Detailed studies exist for the precision achievable for the input parameters in an $\alpha_{\rm s}$ determination at the ILC. For the most important input variable, $R_\ell^0$, an error on $4 \cdot 10^{-3}$ can be expected, corresponding to an improvement of more than a factor six, including experimental uncertainties [13]. For $\sin^2\theta_{\rm eff}^l$ an improvement of a factor 15 is possible leading to $\Delta\sin^2\theta_{\rm eff}^l = 1.3 \cdot 10^{-5}$ [14]. The W-mass precision can be improved to 5 MeV and the top mass precision to 100 MeV using threshold scans [14]. Most importantly also the top mass error due to the mass ambiguity drops out in this case. In addition, it has also been assumed that all theory uncertainties can be reduced by a factor four and that $\Delta\alpha_{\rm had}^{(5)}(m_Z)$ will be known with a precision of $4 \cdot 10^{-5}$. With these assumptions an error of

$$\begin{aligned} \Delta\alpha_s &= 0.00065_{\rm exp} \oplus 0.00023_{\rm QCD} \oplus 0.00025_{\rm EW} \\ &= 0.00070 \end{aligned}$$

can be expected. In the more model independent case, where the measured $\sin^2\theta_{\rm eff}^l$ is used to predict $R_\ell^0$, the error increases to 0.00075. The expected $\alpha_s$ precision is graphically depicted in Fig. 1.

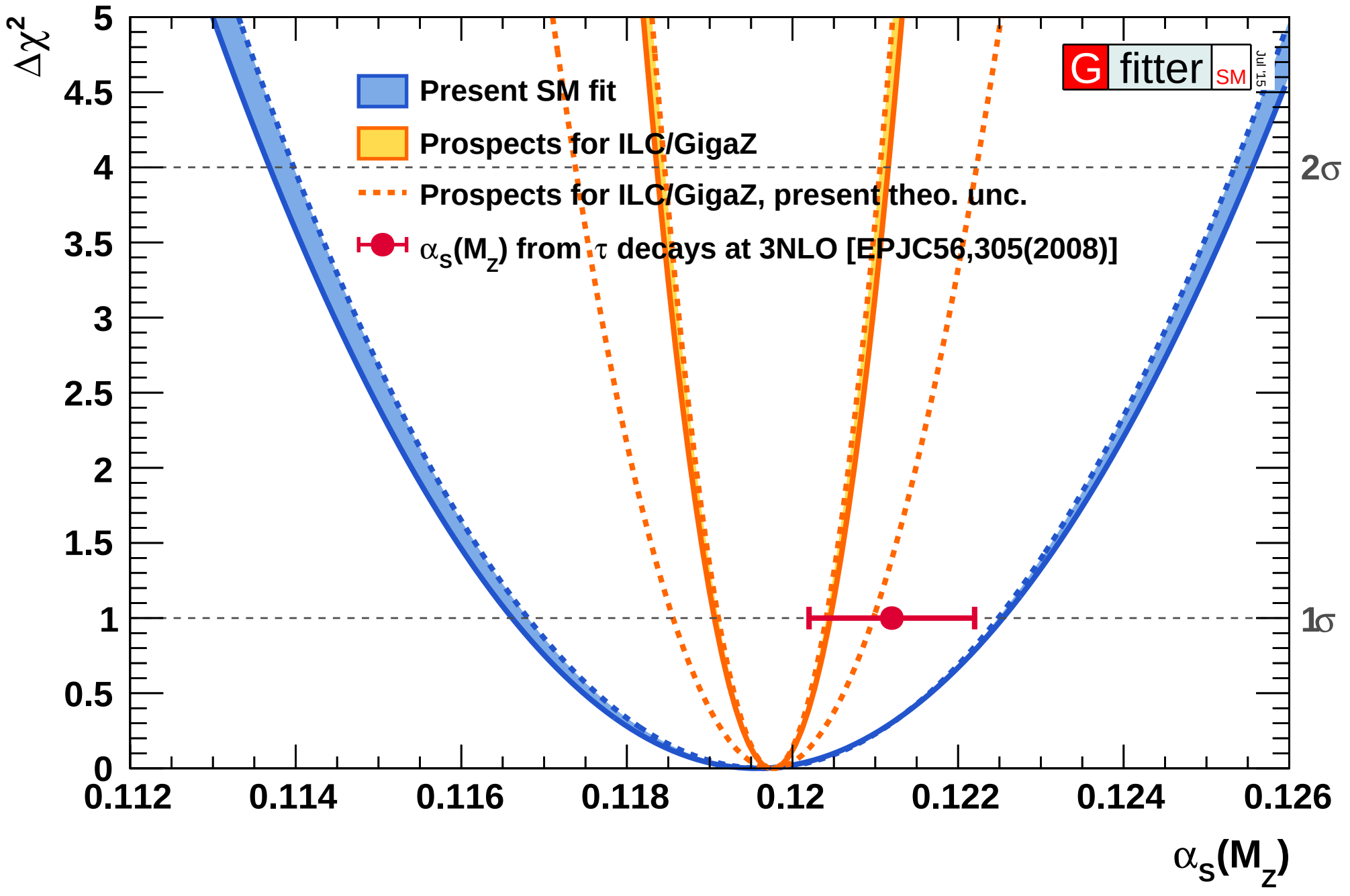

Figure 1: Precision of $\alpha_s$ from the electroweak fit at present and after the GigaZ phase of ILC.

At the FCC-ee, the following uncertainties are considered reasonable [15]:

$$\begin{aligned}\Delta m_Z &= 0.1\,\text{MeV}\,,\\ \Delta\Gamma_Z &= 0.05\,\text{MeV}\,,\\ \Delta R_\ell^0 &= 1\cdot 10^{-3}\,,\end{aligned}$$

where improvements in $m_Z$ and $\Gamma_Z$ become achievable thanks to the possibility to perform a threshold scan including energy calibration with resonant depolarisation. Using these values, the uncertainty on $\alpha_s$ reduces to $\Delta\alpha_s = 0.00030$ using the same theory uncertainties as above and $\Delta\alpha_s = 0.00016$ without theory uncertainties. The error is again mainly determined by the precision on $R_\ell^0$.

In summary $\alpha_s$ at the Z-pole is currently known from global fits to the electroweak data to be $\alpha_s(m_Z^2) = 0.1196 \pm 0.0030$ including theory uncertainties. At ILC this can be improved to $\Delta\alpha_s = 0.00070$, while at FCC-ee a precision of $\Delta\alpha_s = 0.00016$ is possible, neglecting theory uncertainties.

**Acknowledgements:** I'd like to thank Roman Kogler for providing the fits for the FCC-ee projections.

# $\alpha_s$ from hadronic W boson decays

David d'Enterria[1] and **Matej Srebre**[1,2]

[1]CERN, Physics Department, Geneva, Switzerland
[2]University of Ljubljana, Slovenia

**Abstract:** The strong coupling is extracted from hadronic W boson decays using the most recent theoretical and experimental inputs, obtaining $\alpha_s(m_Z^2) = 0.117 \pm 0.043_{exp} \pm 0.001_{th}$ (assuming CKM matrix unitarity), where a detailed estimation of the associated uncertainties is provided for the first time. Prospects for future $\alpha_s$ determinations from W data at the LHC, and in $e^+e^- \to W^+W^-$ collisions at FCC-ee are highlighted.

## Introduction

The hadronic decay widths of the electroweak bosons ($\Gamma_{Z,had}$, $\Gamma_{W,had}$) are high-precision theoretical and experimental observables from which an accurate extraction of $\alpha_s$ can be obtained. Whereas $\Gamma_{Z,had}$ provides, together with other Z-pole hadronic measurements, one of the most precise constraints on the current $\alpha_s$ world-average [1], no extraction of $\alpha_s$ from $\Gamma_{W,had}$ has been performed so far. The reasons for that are two-fold. First, while $\Gamma_{Z,had}$ has been experimentally measured with 0.1% uncertainties, $\Gamma_{W,had}$ has much larger experimental uncertainties of order $\sim$2%, or 0.4% in the case of the hadronic branching ratio $BR_{W,had} \equiv (\Gamma_{W,had}/\Gamma_{W,tot})$, and the sensitivity to $\alpha_s$ of the W and Z hadronic decays comes only through small higher-order (loop) corrections. Secondly, although $\Gamma_{Z,had}, \Gamma_{W,had}$ are both theoretically known up to $\mathcal{O}\left(\alpha_s^4\right)$, a complete expression of $\Gamma_{W,had}$ including all computed higher-order terms was lacking until recently. This situation has now been corrected with the work of [2] that obtained $\Gamma_{W,had}$ at N$^3$LO accuracy including so-far missing mixed QCD+electroweak $\mathcal{O}\left(\alpha_s\alpha\right)$ corrections, improving upon the previous calculations of one-loop $\mathcal{O}\left(\alpha_s\right)$ QCD and $\mathcal{O}\left(\alpha\right)$ electroweak terms [3,4,5], and two-loop $\mathcal{O}\left(\alpha_s^2\right)$, three-loop $\mathcal{O}\left(\alpha_s^3\right)$ [6], and four-loop $\mathcal{O}\left(\alpha_s^4\right)$ [7] QCD corrections. Despite the clear progress, the work of [2] still contains a range of approximations (e.g. one-loop $\alpha_s$ running between $m_W$ and $m_Z$, massless quarks, and CKM matrix set to unity for the calculation of higher-order corrections and renormalization constants), plus no real estimation of the associated theoretical uncertainties, which hinder its use to extract $\alpha_s$ from a comparison to the experimental data. We remove such approximations here and provide $\alpha_s$ values extracted from current and future W boson hadronic decay data [8].

## $\alpha_s$ from current $\Gamma_{W,had}$ and $BR_{W,had}$ data

The theoretical expression for the hadronic decay width, including QCD terms up to order $\mathcal{O}(\alpha_s^4)$, plus electroweak $\mathcal{O}(\alpha)$ and mixed $\mathcal{O}(\alpha\alpha_s)$ corrections, reads

$$\Gamma_{W,had} = \frac{\sqrt{2}}{4\pi} G_F m_W^3 \sum_{\text{quarks i,j}} |V_{i,j}|^2 \left[1 + \sum_{k=1}^{4} \left(\frac{\alpha_s}{\pi}\right)^k + \delta_{\text{electroweak}}(\alpha) + \delta_{\text{mixed}}(\alpha\alpha_s)\right]. \quad (1)$$

Compared to previous works, we implement finite quark masses in QCD corrections up to order $\mathcal{O}(\alpha_s)$, we use NNLO (instead of LO) $\alpha_s$ running, and for the numerical evaluation we use the 2015 PDG world average values [9] for $\alpha, G_{\rm F}, m_{\rm q}, m_\ell, m_{\rm W}, m_{\rm Z}, m_{\rm H}$, and CKM matrix elements $|V_{\rm i,j}|$ (summed over $q = u, c, d, s, b$ since the top-quark is not kinematically accessible). Our final theoretical result amounts to $\Gamma_{\rm W,had} = 1428.84 \pm 22.61_{\rm par} \pm 0.03_{\rm th}$ MeV (using the experimentally measured $|V_{\rm i,j}|$ values), and $\Gamma_{\rm W,had} = 1411.58 \pm 0.79_{\rm par} \pm 0.03_{\rm th}$ MeV (assuming CKM matrix unitarity), well in agreement with the current experimental value $\Gamma^{\rm exp}_{\rm W,had} = 1405 \pm 29$ MeV [9]. The (small) theoretical uncertainties quoted include the estimation of higher-order corrections, non-perturbative effects –suppressed by $\mathcal{O}\left(\Lambda^4_{\rm QCD}/m^4_{\rm W}\right)$ power corrections– and finite quark mass (beyond LO) [8]. The parametric uncertainties quoted have been obtained by individually varying each parameter in Eq. (1) within 1-$\sigma$ of its associated error and adding all the differences from the central $\Gamma_{\rm W,had}$ in quadrature. The dominant parametric uncertainties are those associated with the charm-strange quark mixing element $|V_{\rm cs}|$ (which propagates into $\pm 22$ MeV in $\Gamma_{\rm W,had}$) followed by $m_{\rm W}$ (which propagates into $\pm 0.7$ MeV). The functional dependence of $\Gamma_{\rm W,had}$ on $\alpha_{\rm s}$ alone is shown in Fig. 1 (left). Setting $\Gamma_{\rm W,had}$ in Eq. (1) to its measured experimental value $\Gamma^{\rm exp}_{\rm W,had}$ and fixing all other parameters (except $\alpha_{\rm s}$) to their PDG values, allows one to extract the strong coupling values listed in Table 1.

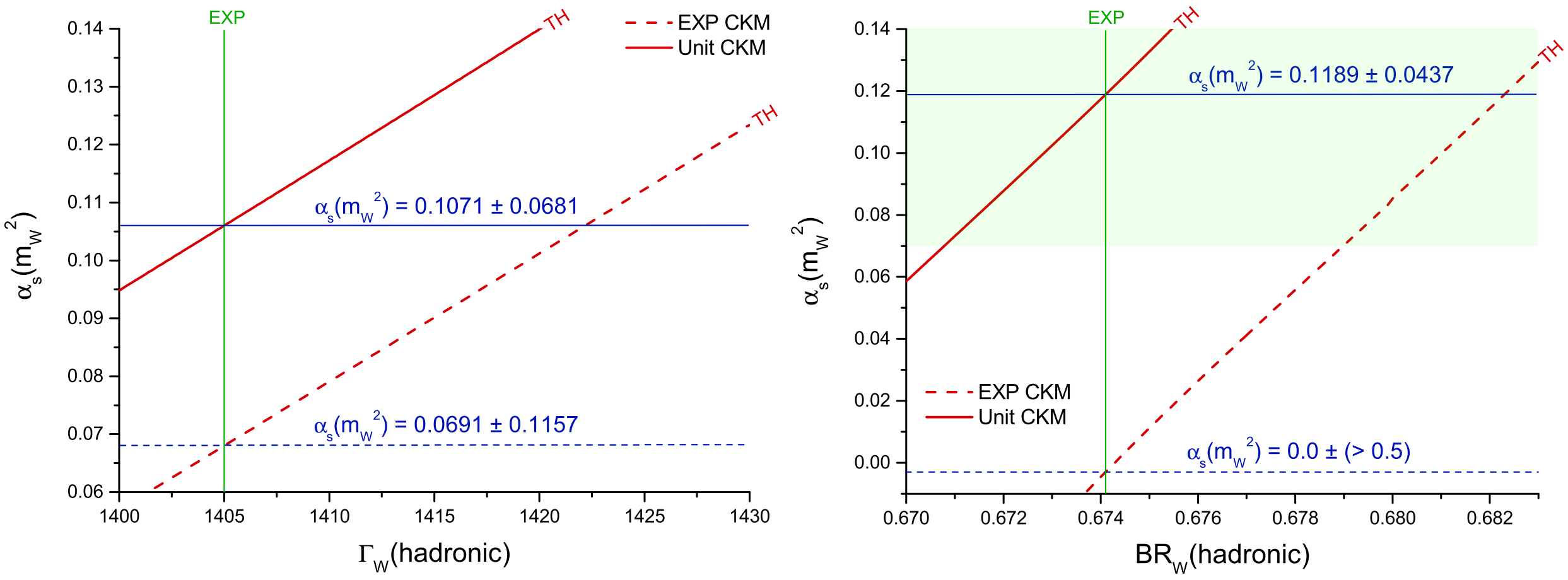

Figure 1: Functional dependences of $\alpha_{\rm s}$ on the W boson hadronic width $\Gamma_{\rm W,had}$ (left) and branching ratio $\rm BR_{W,had}$ (right), given by Eq. (1) imposing CKM unitarity (solid curves) or to their measured values (dashed curves). The vertical lines are the experimental $\Gamma_{\rm W,had}$ and $\rm BR_{W,had}$ values.

| $\Gamma_{\rm W,had}$ | $\alpha_{\rm s}(m^2_{\rm W})$ | $\alpha_{\rm s}(m^2_{\rm Z})$ |
|---|---|---|
| Experimental CKM | $0.069 \pm 0.065_{\rm exp} \pm 0.051_{\rm par}$ | $0.068 \pm 0.064_{\rm exp} \pm 0.051_{\rm par}$ |
| CKM unitarity | $0.107 \pm 0.066_{\rm exp} \pm 0.002_{\rm par}$ | $0.105 \pm 0.065_{\rm exp} \pm 0.002_{\rm par}$ |
| $\rm BR_{W,had}$ | $\alpha_{\rm s}(m^2_{\rm W})$ | $\alpha_{\rm s}(m^2_{\rm Z})$ |
| Experimental CKM | $0 \pm 0.038_{\rm exp} \pm 0.495_{\rm par}$ | $0 \pm 0.038_{\rm exp} \pm 0.454_{\rm par}$ |
| CKM unitarity | $0.119 \pm 0.043_{\rm exp} \pm 0.001_{\rm par}$ | $0.117 \pm 0.043_{\rm exp} \pm 0.001_{\rm par}$ |

Table 1: Values of $\alpha_{\rm s}$ (and associated experimental and parametric uncertainties) at the W and Z scales, extracted from the experimental values of $\Gamma_{\rm W,had}$ (top rows) and $\rm BR_{W,had}$ (bottom rows), setting the CKM matrix to the experimental values or imposing CKM unitarity.

As we can see from the top rows of Table 1, the current experimental uncertainty of $\Gamma_{\mathrm{W,had}}$ and the parametric uncertainties of Eq. (1), mostly from the $|V_{\mathrm{cs}}|$ element, propagate into a huge uncertainty on $\alpha_{\mathrm{s}}$. An alternative approach, with reduced experimental and theoretical uncertainties, is to extract $\alpha_{\mathrm{s}}$ through the hadronic W branching ratio, $\mathrm{BR}_{\mathrm{W,had}} \equiv (\Gamma_{\mathrm{W,had}}/\Gamma_{\mathrm{W,tot}})$, which is experimentally known to within $\pm0.4\%$ and for which the $m_{\mathrm{W}}$ parametric uncertainty cancels out in the ratio. We compute $\mathrm{BR}_{\mathrm{W,had}}$ from the ratio of Eq. (1) to the NNLO expression for $\Gamma_{\mathrm{W,tot}}$ from [10,11], obtaining $\mathrm{BR}_{\mathrm{W,had}} = 0.6824 \pm 0.0108_{\mathrm{par}}$ (experimental CKM matrix) and $\mathrm{BR}_{\mathrm{W,had}} = 0.6742 \pm 0.0001_{\mathrm{par}}$ (assuming CKM matrix unitarity), both results in very good accord with the experimental value $\mathrm{BR}^{\mathrm{exp}}_{\mathrm{W,had}} = 0.6741 \pm 0.0027$. The functional dependence of $\mathrm{BR}_{\mathrm{W,had}}$ on $\alpha_{\mathrm{s}}$ alone is shown in Fig. 1 (right). As done for $\Gamma_{\mathrm{W,had}}$, by comparing the experimental value of $\mathrm{BR}^{\mathrm{exp}}_{\mathrm{W,had}}$ to the theoretical predictions, we can extract the values of $\alpha_{\mathrm{s}}$ listed in the bottom rows of Table 1. For $V_{\mathrm{ij}}V_{\mathrm{jk}} = \delta_{\mathrm{ik}}$, the extracted $\alpha_{\mathrm{s}} = 0.117 \pm 0.043_{\mathrm{exp}} \pm 0.001_{\mathrm{par}}$ has now a relative uncertainty of 37%. Thus, even if the current experimental uncertainty of $\mathrm{BR}_{\mathrm{W,had}}$ is lower than that of $\Gamma_{\mathrm{W,had}}$, it still propagates into a large uncertainty on $\alpha_{\mathrm{s}}$.

## $\alpha_{\mathrm{s}}$ from future $\Gamma_{\mathrm{W,had}}$ and $\mathrm{BR}_{\mathrm{W,had}}$ data

Future measurements at the LHC and at FCC-ee will provide $\Gamma_{\mathrm{W,had}}$ and $\mathrm{BR}_{\mathrm{W,had}}$ with higher accuracy and precision. First, it is not unreasonable to reduce the $\Gamma_{\mathrm{W,had}}$ uncertainty to approximately $\pm12$ MeV via high-statistics LHC measurements of the large transverse mass spectra of W decays (sensitive to $\Gamma_{\mathrm{W,tot}}$ and thereby to $\Gamma_{\mathrm{W,had}} = \mathrm{BR}_{\mathrm{W,had}} \cdot \Gamma_{\mathrm{W,tot}}$). Combining this result with an improved determination of the $|V_{\mathrm{cs}}|$ matrix element commensurate with that of $|V_{ud}|$ (since the other diagonal CKM member, $|V_{tb}|$, is not kinematically relevant for W decays), would change the relative uncertainty of our extracted $\alpha_{\mathrm{s}}$ value from 37% to 23%. Additional LHC and Tevatron W leptonic data (combined with the total width) can further reduce it to 10% (first row of Table 2). At the FCC-ee, the W hadronic branching ratio would be measured in huge samples of $e^+e^- \to \mathrm{W}^+\mathrm{W}^-$ collisions yielding $5\times10^8$ W bosons (a thousand times more than the $5\times10^5$ collected at LEP). This would reduce the statistical uncertainty of $\mathrm{BR}_{\mathrm{W,had}}$ to around 0.005%. The $\mathrm{BR}_{\mathrm{W,had}}$ measurement at the FCC-ee would thus significantly improve the extraction of $\alpha_{\mathrm{s}}$ with propagated experimental uncertainties of order 0.3%, which could be even lowered to 0.15% combining, in addition, this result with other closely related W-decay observables as done for the Z boson [1].

| | $\delta\alpha_{\mathrm{s}}(m^2_{\mathrm{W}})$ | $\delta\alpha_{\mathrm{s}}(m^2_{\mathrm{Z}})$ |
|---|---|---|
| LHC ($\delta\Gamma_{\mathrm{W,had}} \approx 12$ MeV, improved $\mathrm{BR}_{\mathrm{W,lept}}$) | $\pm0.0120_{\mathrm{exp}} \pm 0.0004_{\mathrm{par}}$ | $\pm10\%$ |
| FCC-ee ($\delta\mathrm{BR}_{\mathrm{W,had}} \approx 0.005\%$) | $\pm0.0004_{\mathrm{exp}} \pm 0.0004_{\mathrm{par}}$ | $\pm0.3\%$ |

Table 2: Uncertainties on the $\alpha_{\mathrm{s}}$ values at the W and Z scales, extracted from future measurements of $\Gamma_{\mathrm{W,had}}$ at the LHC (top row) and $\mathrm{BR}_{\mathrm{W,had}}$ at FCC-ee (bottom row), for a value of the $|V_{\mathrm{cs}}|$ matrix element with experimental uncertainty similar to that of $|V_{ud}|$ today.

As a short side-project, we have also extracted a value of $|V_{\mathrm{cs}}|$ comparing the experimental $\Gamma_{\mathrm{W,had}}$ and $\mathrm{BR}_{\mathrm{W,had}}$ results to their theoretical predictions fixing $\alpha_{\mathrm{s}}$ to the current world average [8]. The resulting $|V_{\mathrm{cs}}|$ values are listed in Table 3 with the experimental and parametric uncertainties propagated as described before. As we can see from the values tabulated, the extracted $|V_{\mathrm{cs}}|$ has an uncertainty of $\pm0.6\%$, which is about 3 times smaller than the $\pm1.6\%$ of the current experimental measurement, $|V_{\mathrm{cs}}|^{\mathrm{exp}} = 0.986 \pm 0.016$ [9].

| Extraction through | $\|V_{cs}\|$ |
|---|---|
| $\Gamma_{W,had}$ | $0.969 \pm 0.021_{exp} \pm 0.002_{par}$ |
| $BR_{W,had}$ | $0.973 \pm 0.004_{exp} \pm 0.002_{par}$ |

Table 3: Values of the charm-strange CKM element obtained from the theoretical and experimental values of $\Gamma_{W,had}$ and $BR_{W,had}$, for a value of $\alpha_s$ fixed at its current world average.

To summarize, we have extracted $\alpha_s$ from the hadronic W decay width ($\Gamma_{W,had}$) and its hadronic branching ratio ($BR_{W,had}$) using the experimental values of both quantities compared to the theoretical predictions computed at $N^3LO$ and NNLO accuracies respectively. The current experimental and parametric uncertainties on both $\Gamma_{W,had}$ and $BR_{W,had}$ are too large today to extract a precise enough $\alpha_s$ value (the best result obtained is $\alpha_s(m_Z^2) = 0.117 \pm 0.043_{exp} \pm 0.001_{th}$, assuming CKM matrix unitarity). Measurements with reduced uncertainties at the LHC and, in particular, at the FCC-ee will allow one to extract $\alpha_s$ with uncertainties as low as 0.15%, providing a new independent value of the strong coupling to be weighted-averaged with all other methods discussed in this document.

**Acknowledgements.** We are grateful to J. F. Kamenik, D. Kara, D. Nomura, and T. Riemann for useful discussions.

# $\alpha_s$ from $\sigma(e^+e^- \to$ hadrons)

Konstantin G. Chetyrkin and **Johann H. Kühn***

Institut für Theoretische Teilchenphysik
Karlsruher Institut für Technologie (KIT), D-76128 Karlsruhe, Germany

**Abstract:** The extraction of $\alpha_\mathrm{s}$ from the electron-positron annihilation cross section into hadrons is reviewed, including the status and perspectives of calculations at and above the Z resonance, as well as in $e^+e^- \to$ Z + H($\to$ hadrons).

## Introduction

The cross section for electron-positron annihilation into hadrons provides one of the cleanest ways for measuring the strong coupling constant. In principle a wide range of energies exists which can be exploited for this purpose, ranging from the region below the charm threshold, say around 3 GeV, up to the highest energies available at electron-positron colliders. Apart from the regions very close to the charm-, bottom- and top-thresholds, the cross section can be evaluated by taking the limit of massless quarks. The leading term for QCD corrections is thus given by the extremely simple form $(1 + a_s)$ with $a_s \equiv \alpha_\mathrm{s}/\pi$. Higher order terms, however, exhibit a non-trivial dependence on the number of active flavours. Also the difference between vector- and axial vector current induced reactions starts (in the massless limit) at order $\alpha_\mathrm{s}^2$. Closely related to this quantity are QCD-corrections to $\tau$-lepton decays. Although in principle extremely clean and simple, the predictions are, in this case, at the limit of applicability of perturbative QCD. This fact is well visible from the difference between the predictions based on "Fixed Order" and "Contour Improved" (see [1] and references therein) and are discussed in more detail in [2].

The following discussion will, therefore, be limited to decays of the virtual photon into hadrons, i.e. the famous $R$-ratio, or the corresponding object at higher energies, in particular the decay rate of the $Z$-boson into hadrons, $\Gamma(Z \to \text{hadrons})$, and the closely related quantities for the $W$ or the Higgs-boson decay.

## The R-ratio below the bottom threshold

The $R$-ratio, given by $\sigma(e^+e^- \to \text{hadrons})/\sigma(e^+e^- \to \mu^+\mu^-)$, can be measured at vastly different energies. In the low energy region the annihilation proceeds through the virtual photon only. A combined fit [3] to CLEO results [4], based on calculations at four-loop accuracy, gives

$$\alpha_\mathrm{s}^{(4)}(9^2\,\mathrm{GeV}^2) = 0.160 \pm 0.024 \pm 0.024 \tag{1}$$

which, at the scale of $M_Z$, corresponds to

$$\alpha_\mathrm{s}^{(5)}(M_Z^2) = 0.110^{+0.014}_{-0.017}. \tag{2}$$

*Presented by Johann H. Kühn

This CLEO result, in turn, can be combined with earlier measurements by BESS [5], MD-1 [6] and CLEO [7], and leads to [3]

$$\alpha_s^{(4)}(9^2\,\mathrm{GeV}^2) = 0.182^{+0.022}_{-0.025} \tag{3}$$

corresponding to

$$\alpha_s^{(5)}(M_Z^2) = 0.119^{+0.009}_{-0.011}. \tag{4}$$

This result is in nice agreement with a more recent measurement of the BESS collaboration [8]

$$R_{uds} = 2.224 \pm 0.019 \pm 0.089 \tag{5}$$

taken at energy values below the charm threshold. Let us emphasize again that more precise measurements in this region, based on the enormous statistic of the BESS and BELLE experiment, would be highly welcome.

## Status and perspectives for $e^+e^- \to$ hadrons at the $Z$ resonance

### a) QCD corrections

At present the most precise determination of $R$ is based on the measurement of $Z$ decays, more precisely on the measurement of $\Gamma_{had}$ and $\Gamma_{had}/\Gamma_{lept}$, where corrections have been calculated to $\mathcal{O}(\alpha_s^4)$ [1,9,10,11]. Effects from non-vanishing charm and bottom quark masses will be mentioned below. Even in the massless limit there are three different types of contributions shown in Fig. 1:

i) The non-singlet vector and axial vector correlators $r_{NS}$ [1,9].

ii) The singlet correlator $r_S^V$, which starts at $\mathcal{O}(\alpha_s^3)$ and is present for the vector current only [10,11].

iii) A contribution to the axial correlator, $r_S^A$, resulting from the top-bottom doublet with $4m_t^2 \gg s \gg 4m_b^2$, which starts at order $\alpha_s^2$ [12,13,11].

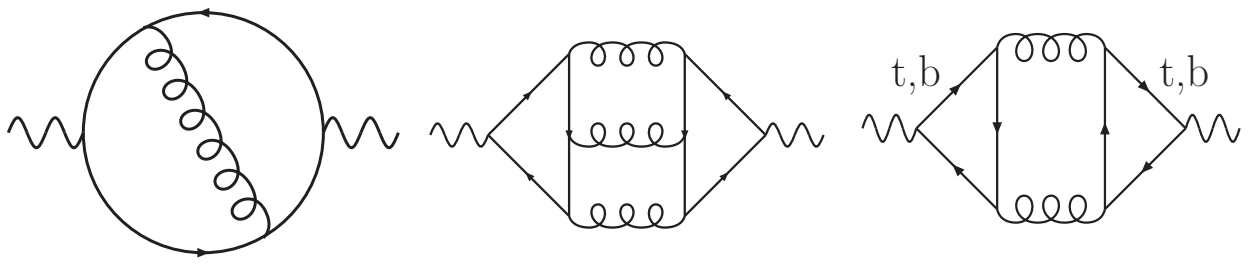

Figure 1: Non-singlet & singlet, vector & axial correlators.

The corresponding predictions are available to order $\alpha_s^4$. For an extremely conservative error estimate based on the variation of the scale parameter $\mu$ between $M_Z/3$ and $3M_Z$ this corresponds to the variation

$$\delta\Gamma_{NS} = 101\,\mathrm{keV}, \tag{6}$$

(7)

$$\delta\Gamma_S^V = 2.7\,\mathrm{keV}, \tag{8}$$

(9)

$$\delta\Gamma_S^A = 42\,\mathrm{keV} \tag{10}$$

and, correspondingly, to a variation of the strong coupling around $3 \cdot 10^{-4}$. This theory error is completely negligible compared to the present experimental uncertainty around 0.026. In fact, it is even comparable to the error of $\delta\Gamma_{had} \approx 100\,\mathrm{keV}$ expected for an analysis at the FCC-ee.

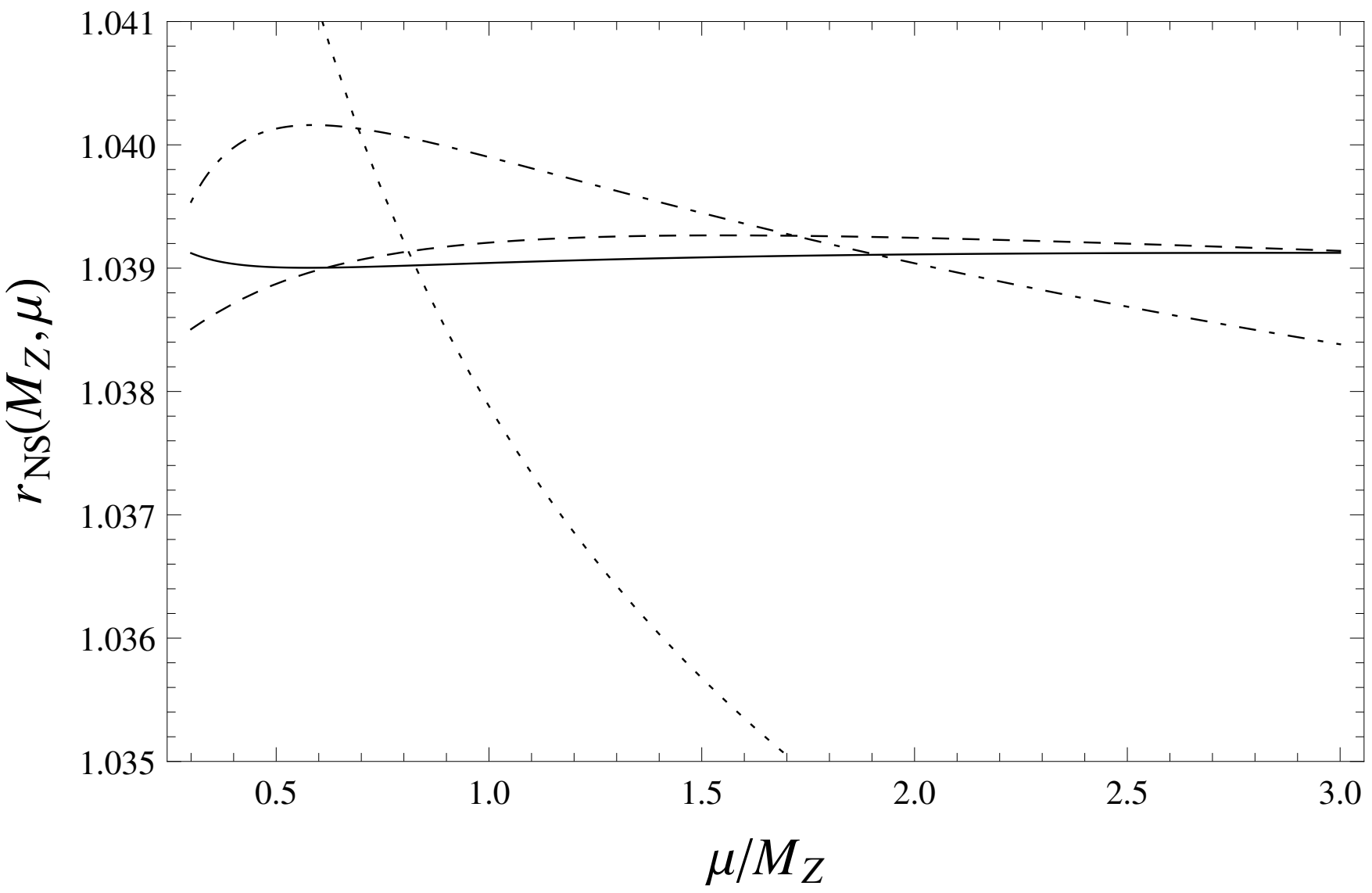

Figure 2: Predictions for $r_{NS}$ at increasing orders of $\alpha_s$ [10].

Various additional small corrections must be applied, which, however, have been calculated some time ago: quark mass corrections up to order $m_b^2\alpha_s^4$ and $m_b^4\alpha_s^3$ [14,15] and mixed QCD and electroweak corrections that will be discussed in the following.

**b) Mixed electroweak and QCD corrections**

For light quarks terms of order $\alpha\alpha_s$ have been evaluated some time ago [16]. The difference between the correct two-loop terms and the product of the corresponding two one-loop terms

$$\Delta\Gamma \equiv \Gamma(\text{two loop (EW} \star \text{QCD})) - \Gamma_{\text{Born}}\delta_{\text{EW}}^{\text{NLO}}\delta_{\text{QCD}}^{\text{NLO}} = -0.59(3) \qquad (11)$$

is again of high importance, if a precision of $\delta\Gamma \approx 0.1\mathrm{MeV}$ should be reached at some point in the future. The next, presently unknown three-loop term is smaller by another factor of order $\alpha_s/\pi$, times an unknown coefficient which should not exceed 5 in order to keep this term under control.

Similar comments apply to $\Gamma(Z \to b\bar{b})$. The present result, $\Gamma(Z \to b\bar{b})/\Gamma_Z \equiv R_b = 0.1512 \pm 0.0005$, obtained at LEP corresponds to an error of about 1.3 MeV. Pushing the relative error $\delta\Gamma(Z \to b\bar{b})/\Gamma_Z$ down to $2 \cdot 10^{-5}$, would imply a precision of 0.05 MeV for the partial width $\Gamma(Z \to b\bar{b})$. This should be compared to the present result of order $\alpha\alpha_s$, which is given by

$$\Gamma(Z \to b\bar{b}) - \Gamma(Z \to d\bar{d}) = (-5.69 - 0.79)_{\mathcal{O}(\alpha)} + (+0.50 + 0.06)_{\mathcal{O}(\alpha\alpha_s)}\,\mathrm{MeV} \qquad (12)$$

separated into $m_t^2$ enhanced terms [17] and the rest [18].

The size of these terms has even motivated the evaluation of the $m_t^2$-enhanced corrections [19] of order $G_F m_t^2 \alpha_s^2$ with the result $\delta\Gamma_b \approx 0.1\,\mathrm{MeV}$ for the non-singlet contributions, a result drastically below the current precision and comparable in size to the planned precision of the FCC-ee.

Let us mention that many corrections seem to be significantly smaller if expressed in terms of the $\overline{\rm MS}$-mass [20,21,22]

$$\overline{m}_t(\overline{m}_t) = m_{pole}(1 - 1.33\,a_s - 6.46\,a_s^2 - 60.27\,a_s^3 - 704.28\,a_s^4). \tag{13}$$

Note that the $\overline{\rm MS}$-mass may well be directly accessible at an $e^+e^-$ collider with high precision, e.g. through a measurement of the potential subtracted mass with a precision of 20 – 30 MeV.

## $M_W$ from $G_F$, $M_Z$, $\alpha$ and the rest

The precision of the present $M_W$-measurement amounts to 23 MeV  For the FCC-ee option a precision of 0.5 to 1 MeV is foreseen. In the Standard Model $M_W$ can be considered a derived quantity that can be calculated from $G_F$, $M_Z$, $M_t$, $\Delta\alpha$ and $m_H$. Let us, for the moment, focus on the top mass dependence, which is obtained from the following chain of equations

$$\begin{aligned}
M_W^2 &= f(G_F, M_Z, m_t, \Delta\alpha, \ldots) && (14)\\
&= \frac{M_Z^2}{2(1-\delta\rho)}\left(1+\sqrt{1-\frac{4\pi\alpha(1-\delta\rho)}{\sqrt{2}G_F M_Z^2}\left(\frac{1}{1-\Delta\alpha}+\ldots\right)}\right), && (15)\\
\delta M_W &\approx M_W \frac{1}{2}\frac{\cos^2\theta_w}{\cos^2\theta_w - \sin^2\theta_w}\delta\rho \approx 5.7\times 10^4\,\delta\rho[\text{MeV}], && (16)\\
\delta\rho_t &= 3X_t\left(1 - 2.8599\left(\frac{\alpha_s}{\pi}\right) - 14.594\left(\frac{\alpha_s}{\pi}\right)^2 - 93.1\left(\frac{\alpha_s}{\pi}\right)^3\right). && (17)
\end{aligned}$$

In this case the three-loop term amounts to 9.5 MeV, the four-loop term to 2.1 MeV [24,25]. In three loop approximation also purely weak ($\sim X_t^3$) and mixed (QCD*electroweak) of order $\alpha_s X_t^2$ three-loop terms are available, corresponding to $\delta M_W = 0.2$ MeV and 2.5 MeV respectively. These are shown in Fig. 3 together with the terms of order $\alpha_s X_t^2$ and $X_t^2$.

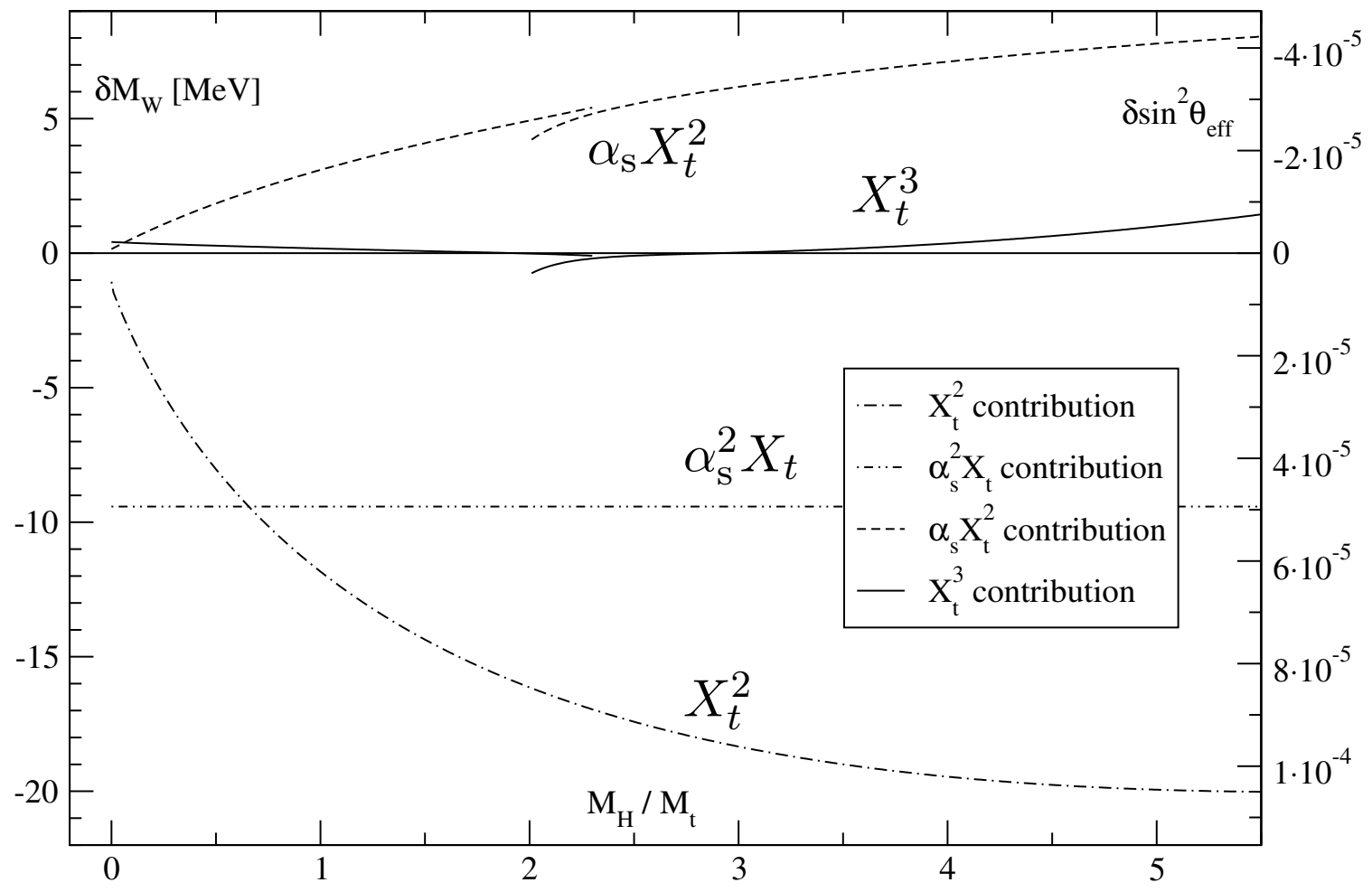

Fig. 3: Shifts in $M_W$ and in $\sin^2\Theta_W$ from various corrections to $\Delta\rho$ [23].

Let us emphasize that there are still various uncalculated terms of order 0.1 MeV – 0.5 MeV, for example the four-loop term of order $\alpha_s^2 X_t^2$ or higher order terms relating pole and $\overline{\rm MS}$-mass.

## Perspectives for $e + e^- \to$ hadrons above the $Z$-resonance

No detailed analysis of the cross section exists in the moment for the energy region above the $Z$-resonance. The situation is significantly more complicated than around the $Z$-boson or at low energies. As a consequence of Furry's theorem the interference between Born and one-loop correction vanishes, as long as only vector currents are involved. This is no longer true for a mixture of vector and axial currents, such that the interference (relative to the Born term) is proportional to order $\alpha$. In combination with the large radiative tail this enhances the radiative corrections drastically and no prediction for the $R$-ratio at a level required for a precise determination of the $Zf\bar{f}$ coupling is available at present.

## Perspectives for $e^+e^- \to Z + H(\to$ hadrons$)$

### a) $\mathbf{H \to b\bar{b}}$

Up to this point only QCD corrections for vector and axial currents have been considered, where the first terms are equal for both cases and given by $(1 + \alpha_s/\pi)$. In contrast, for the case of Higgs boson decay into a quark-antiquark pair, the leading QCD correction is of the form $(1+17/3\cdot\alpha_s/\pi)$. The full correction for the decay rate into $b$ quarks is given by

$$\Gamma(H \to b\bar{b}) = \frac{G_F M_H}{4\sqrt{2}\pi} m_b^2(M_H) R^S(s = M_H^2, \mu^2 = M_H^2) \tag{18}$$

$$R^S(M_H) = 1 + 1 + 5.667\, a_s + 29.147\, a_s^2 + 41.758\, a_s^3 - 825.7\, a_s^4 = 1.2298 \tag{19}$$

for $\alpha_s(M_H) = 0.108$, corresponding to $\alpha_s(M_Z) = 0.118$. The theory uncertainty of the correction factor has been reduced from 5 (four loop) to 1.5 (five loop) permille, where a variation of $\mu$ between $M_H/3$ and $3M_H$ has been assumed and $\alpha_s(M_H)$ has been fixed to 0.108. The parametric uncertainties can be traced to the value of $\alpha_s$, of $m_b$ at low energies, say 10 GeV, and to the running of $m_b$ from 10 GeV to $M_H$. For the values of $\alpha_s$ and $m_b$ we adopt $\alpha_s(M_Z) = 0.1189 \pm 0.002$ and $m_b(10\,\text{GeV}) = 3610 - \frac{\alpha_s - 0.1189}{0.002}$. Running $m_b$ to the scale of $M_H$ one obtains $m_b(M_H) = 2759 \pm 8 \,|_{m_b} \pm 27 \,|_{\alpha_s}$ MeV. While the quark mass anomalous dimension $\gamma_4$ has been obtained in five loop approximation already some time ago [26] the corresponding result for the QCD beta-function is still missing at present. This corresponds to an uncertainty

$$\frac{\delta m_b^2(M_H)}{m_b^2(M_H)} = -1.4 \times 10^{-4}\, (\frac{\beta_4}{\beta_0} = 0) \quad | \quad -4.3 \times 10^{-4}\, (\frac{\beta_4}{\beta_0} = 100) \quad | \quad -7.3 \times 10^{-4}\, (\frac{\beta_4}{\beta_0} = 200) \tag{20}$$

which should be compared to $\delta\Gamma(H \to b\bar{b})/\Gamma(H \to b\bar{b}) = 2.0 \times 10^{-4}$ expected for the FCC-ee. Assuming an improvement of our knowledge of the strong coupling by a factor 10, corresponding to $\delta\alpha_s = 2 \times 10^{-4}$ and for $m_b$ by a factor 4 ($\delta m_b(M_H)/m(M_H) = 10^{-3}$) on finds

$$\frac{\delta\Gamma_{H\to b\bar{b}}}{\Gamma_{H\to b\bar{b}}} = \pm 2 \times 10^{-3}|_{m_b} \pm 1.3 \times 10^{-3}|_{\alpha_s,\text{running}} \pm 1 \times 10^{-3}|_{\text{theory}}. \tag{21}$$

### b) $\mathbf{H \to gg}$

Another hadronic decay mode of the Higgs boson with a large branching ratio is the one into two gluons, which starts at order $\alpha_s^2$ and is presently known [27] to order $\alpha_s^5$:

$$
\begin{aligned}
\Gamma(H \to gg) &= K \cdot \Gamma_{\text{Born}}(H \to gg) \qquad (22)\\
K = 1 &+ 17.9167\, a_s' + (156.81 - 5.7083 \ln \frac{M_t^2}{M_H^2})(a_s')^2 \\
&+ (467.68 - 122.44 \ln \frac{M_t^2}{M_H^2} + 10.94 \ln^2 \frac{M_t^2}{M_H^2})\ (a_s')^3. \\
K &= 1 + 17.9167\, a_s' + 152.5(a_s')^2 + 381.5(a_s')^3 \\
&= 1 + 0.65038 + 0.20095 + 0.01825.
\end{aligned}
$$

where $M_t = 175$ GeV, $m_H = 120$ GeV and $a_s' = \alpha_s^{(5)}(M_H)/\pi$ have been assumed. Note that an experimental precision for $\sigma(HZ) \times Br(H \to gg)$ of 1.4% has been claimed in [28]. Mass corrections to the Higgs-boson decay rate result can be found in [29].

### Summary and Conclusions

Improved measurements of the total cross section for electron-positron annihilation just below the charm and the bottom threshold would lead to model independent tests of the running of $\alpha_s$ over a wide energy range. At high energies the precise control of QCD corrections is crucial for the detailed and precise comparison of theory predictions with experimental results at a future circular or linear electron-positron collider.

A significant improvement in the determination of the $Z$ decay rate would immediately lead to a correspondingly improved value of the strong coupling constant. In combination with an improved determination of the fine structure constant at the scale of the $Z$-boson this would lead to a significantly improved prediction for $M_W$ at the level of 1 MeV, corresponding to the anticipated precision of a future measurement.

A collider with an energy around 250 GeV would be ideally suited for the measurement of $e^+e^- \to ZH$, leading to the precise determination of the branching ratios of the Higgs boson into its various decay channels. Examples are presented which involve corrections up to order $\alpha_s^4$ for the decay into $b\bar{b}$ and order $\alpha_s^5$ for decays into $gg$. These demonstrate that predictions with a precision around one percent are well within reach for the dominant decay modes.

# $\alpha_s$ from top-pair cross sections at hadron colliders

Siegfried Bethke,[1] Günther Dissertori,[2] Thomas Klijnsma[2] and **Gavin P. Salam**[3,*]

[1]Max-Planck-Institut für Physik (Werner-Heisenberg-Institut), München, Germany
[2]Institute for Particle Physics, ETH Zurich, Switzerland
[3]CERN, PH-TH, CH-1211 Geneva 23, Switzerland

**Abstract:** A preliminary value of the strong coupling $\alpha_s$ is extracted from the latest $t\bar{t}$ cross sections measurements at the LHC and Tevatron, using updated parton distribution functions.

## Introduction

There is currently one published extraction of the strong coupling constant $\alpha_s$ from collider $t\bar{t}$ cross section measurements [1]. From a measured cross section of $161.9 \pm 6.7\,(\text{stat.} + \text{syst.} + \text{lumi.}) \pm 2.9\,(E_{\text{beam}})$ pb in the di-lepton channel at a centre-of-mass energy of $\sqrt{s} = 7$ TeV [2], it determines an extracted $\alpha_s(\text{m}_\text{Z}^2)$ of $0.1151^{+0.0028}_{-0.0027}$. This was obtained using a theoretical prediction that incorporated next-to-next-to-leading-order (NNLO) and next-to-next-to-leading-logarithmic threshold (NNLL) corrections [3] and the NNPDF2.3 parton distribution functions (PDFs) [4].

Since that work, a number of other measurements of the $t\bar{t}$ cross section have appeared at the LHC, notably in the di-lepton channel where systematic uncertainties tend to be smallest: CMS has updated its original measurement with new results for both 7 and 8 TeV collisions [5] and ATLAS has published its results for 7 and 8 TeV collisions [6]. Furthermore CDF and D0 have published a combined cross section from Tevatron $p\bar{p}$ collisions at $\sqrt{s} = 1.96$ TeV [7]. Lately CMS and ATLAS also published their first measurements of the $t\bar{t}$ cross section at $\sqrt{s}$=13 TeV [8][9]; these measurements are not yet incorporated in our analysis. The results are summarised in Table 1.

| Experiment | $\sqrt{s}$ [GeV] | $\sigma_{t\bar{t}}$ [pb] | Exp. err. [pb] | $E_{beam}$ err. [pb] |
|---|---|---|---|---|
| ATLAS | 7000 | 182.9 | ±6.3 | ±3.3 |
| CMS | 7000 | 174.5 | ±6.1 | ±3.1 |
| ATLAS | 8000 | 242.4 | ±9.5 | ±4.2 |
| CMS | 8000 | 245.6 | ±9.0 | ±4.2 |
| ATLAS | 13000 | 825 | ±114 | |
| CMS | 13000 | 769 | ±123 | |
| Tevatron | 1960 | 260.0 | ±0.41 | |

Table 1: $t\bar{t}$ cross section measurements at CMS, ATLAS and Tevatron. Experimental errors reported in this table are calculated by taking the quadratic sum of the statistical, systematic and luminosity errors.

While Ref. [5] contained an updated extraction of the top mass, there has so far been no updated extraction of $\alpha_s$ based on the more recent $t\bar{t}$ production cross section results. In this note we discuss some preliminary results for the extraction of $\alpha_s$ based on the most recent $t\bar{t}$ production cross section measurements.

*On leave from CNRS, UMR 7589, LPTHE, F-75005, Paris, France.

### Uncertainties on the prediction of the cross section

To determine the uncertainty on the prediction of $\sigma_{t\bar{t}}$ for a given value for $\alpha_s$, three sources of uncertainty are taken into account: the PDF uncertainties (within a given set), the choice of renormalisation and factorisation scale, and the choice of the top pole mass, $m_{\rm top}^{\rm pole}$. All three uncertainties are currently estimated using the program Top++ 2.0 [10]. Top++ 2.0 calculates the cross section to NNLO in the strong coupling and has the option of including soft-gluon resummation contributions up to NNLL. Both results with and without NNLL soft-gluon resummation are considered in our analysis. All reported uncertainties are calculated at the 68% confidence level.

We show below results obtained with the PDF sets NNPDF2.3 [4] and CT14 [11], however we plan to continue our studies with a wider range of PDF sets. Uncertainties from the CT14 PDF set are provided at the 90% confidence level by default; for the analysis at hand, the uncertainties from this set are divided by a factor of $\sqrt{2}\,\mathrm{erf}^{-1}(0.90)$ to calculate the corresponding 68% confidence level uncertainties. Scale errors are found by varying the scales $\mu_R$ and $\mu_F$ independently up and down by a factor of two around a central value $\mu = m_{\rm top}$, with the restriction $1/2 \le \mu_R/\mu_F \le 2$. The minimum and maximum values of $\sigma_{t\bar{t}}$ are used to calculate the uncertainties. The uncertainties found in this way are treated as corresponding to a 68% Gaussian confidence interval. This differs from the approach adopted in Ref. [1] of treating scale errors as a 100% confidence interval (with a flat probability distribution inside the range). Both approaches have been advocated in various contexts, however for the purpose of $\alpha_s$ determinations it seems to us that our choice, which is more conservative, is also more standard.

Uncertainties due to the top pole mass uncertainty are calculated by first constructing a fit function for the dependence of $\sigma_{t\bar{t}}$ on $m_{\rm top}$. Since both measured and predicted cross section depend on $m_{\rm top}^{\rm pole}$, simply shifting $\sigma_{t\bar{t}}(m_{\rm top})$ by the uncertainty in the top pole mass is not sufficient. Instead, the uncertainties on $\sigma_{t\bar{t}}$ are effectively calculated as follows:

$$\sigma_{t\bar{t}}^{+} = \sigma_{t\bar{t}}(m_{\rm top} + \Delta m_{\rm top}^{\rm pole})_{\rm theory} \cdot \frac{\sigma_{t\bar{t}}(m_{\rm top})_{\rm experimental}}{\sigma_{t\bar{t}}(m_{\rm top} + \Delta m_{\rm top}^{\rm pole})_{\rm experimental}}$$

$$\sigma_{t\bar{t}}^{-} = \sigma_{t\bar{t}}(m_{\rm top} - \Delta m_{\rm top}^{\rm pole})_{\rm theory} \cdot \frac{\sigma_{t\bar{t}}(m_{\rm top})_{\rm experimental}}{\sigma_{t\bar{t}}(m_{\rm top} - \Delta m_{\rm top}^{\rm pole})_{\rm experimental}}$$

As in CMS's original analysis [1], we use $m_{\rm top} = 173.2 \pm 1.4\,\mathrm{GeV}$, where the central value corresponds to the world average from the Particle Data Group [12]. The error includes the experimental measurement errors (0.9 GeV, rounded up conservatively to 1 GeV) and a further 1 GeV uncertainty associated with the conversion between experimental top-mass measurements and the pole-mass scheme that is used in the theoretical prediction.

### Fit procedure

As part of our study, we have examined two procedures for extracting the strong coupling and its uncertainties. In the first procedure (the default for the final results shown here), the central fit value for $\alpha_{\rm s}$ is chosen such that the theoretical prediction reproduces the measured cross section. Given an uncertainty source, uncertainties on $\alpha_{\rm s}$ are then determined by evaluating the fit of the dependence of $\sigma_{t\bar{t}}$ on $\alpha_{\rm s}$ at the upper and lower bounds due to the uncertainty source. For theory uncertainties, higher upper bounds on $\sigma_{t\bar{t}}$ should lead to lower extracted values of $\alpha_{\rm s}$ and vice versa. For this reason, theory uncertainties on $\sigma_{t\bar{t}}$ are 'flipped', i.e. reversed in sign, before reading off the corresponding values for $\alpha_{\rm s}$. For each experimental measurement of $\sigma_{t\bar{t}}$, this procedure returns a central value for $\alpha_s$, and a set of errors associated with each of the sources of uncertainty. A final estimate of $\alpha_s$ then requires a combination of the results from the different measurements. The combination can be performed using the *Best Linear Unbiased Estimate* method [13][14], which takes into account the correlations between uncertainty sources. Results for the combination are not shown in this note, as determining correlations between uncertainty sources is still work in progress.

Our second procedure for the extraction of $\alpha_s$ uses a more elaborate statistical approach, and closely follows the procedure adopted in Ref. [1]. First, a Bayesian prior for the cross section measurement is constructed, using the theory uncertainties. A typical approach is to convolute separate sources of uncertainty. Figure 1 illustrates an example of a prior at $\alpha_s(m_Z^2) = 0.1180$.

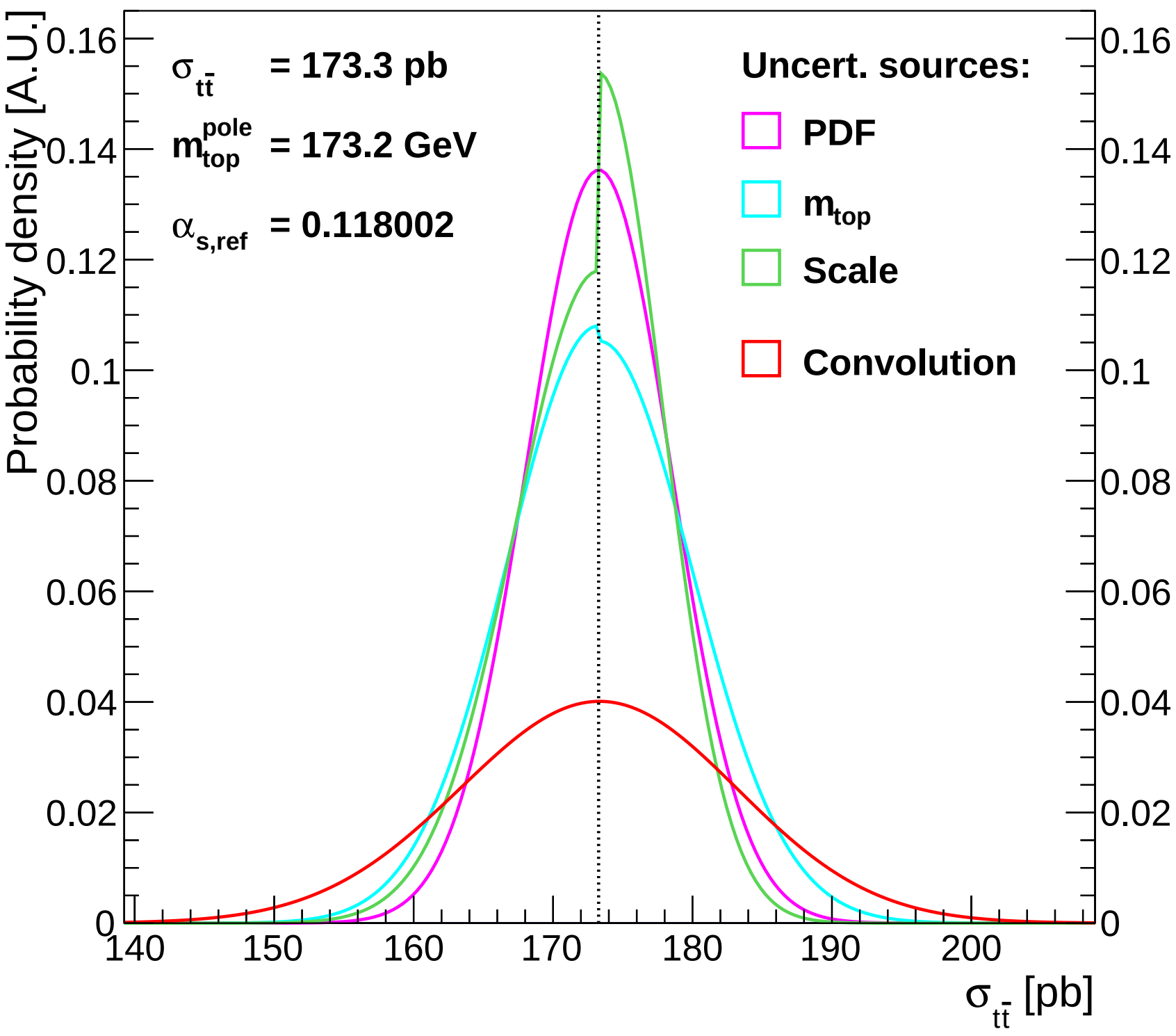

Figure 1: Example of the probability distribution functions for separate error sources and their numerical convolution, for the CMS result [5] at a centre-of-mass energy of 7 TeV. Discontinuities in the probability distribution functions account for the asymmetric errors on $\sigma_{t\bar{t}}$.

A similar probability distribution function is then constructed using the experimental cross section and its uncertainties. An overall likelihood function is then determined by multiplying the theory and experimental probability distribution functions, and integrating over $\sigma_{t\bar{t}}$:

$$P(\alpha_s) = \int f_{exp}(\sigma_{t\bar{t}}|\alpha_s) \cdot f_{theory}(\sigma_{t\bar{t}}|\alpha_s)\, d\sigma_{t\bar{t}} \,.$$

The dependence of $f_{exp}(\sigma_{t\bar{t}}|\alpha_s)$ on $\alpha_s$ appears to be very weak in Ref. [1] and here we ignore it.

Choosing the $\alpha_s$ value that maximises $P(\alpha_s)$ gives us our central fit result, while selecting the 68% confidence interval of the likelihood function $P(\alpha_s)$ provides the bounds on $\alpha_s$ from each experimental measurement. Combining these extractions requires a break down into separate error sources again, so that correlations can be taken into account.

Preliminary results (not shown in this note) indicate that final extracted values of $\alpha_s$ agree within a fraction of a percent between our two procedures.

## Preliminary results

Preliminary results are provided for individual experiments. Table 2 and 3 show preliminary results using the PDF set NNPDF2.3, with and without soft-gluon resummation (NNLO vs. NNLO+NNLL) respectively. Table 4 shows the results for the PDF set CT14 (NNLO).

| Experiment | $\sqrt{s}$ [GeV] | $\alpha_s(m_z^2)$ | Exp. | Scale | PDF | $m_{top}$ | $E_{beam}$ | Total |
|---|---|---|---|---|---|---|---|---|
| ATLAS | 7000 | 0.1205 | ±0.0016 | ±0.0014 | ±0.0013 | ±0.0017 | ±0.0008 | ±0.0032 |
| ATLAS | 8000 | 0.1168 | ±0.0018 | ±0.0014 | ±0.0013 | ±0.0018 | ±0.0008 | ±0.0033 |
| CMS | 7000 | 0.1184 | ±0.0016 | ±0.0014 | ±0.0013 | ±0.0017 | ±0.0007 | ±0.0031 |
| CMS | 8000 | 0.1174 | ±0.0017 | ±0.0014 | ±0.0013 | ±0.0018 | ±0.0008 | ±0.0032 |
| TEV | 1960 | 0.1200 | ±0.0031 | ±0.0012 | ±0.0009 | ±0.0013 | ±0.0000 | ±0.0038 |

Table 2: $\alpha_s$ extractions using CMS, ATLAS and Tevatron measurements for $\sigma_{t\bar{t}}$ with the NNPDF2.3 PDF set and NNLO+NNLL theory predictions. Asymmetric errors are symmetrised.

| Experiment | $\sqrt{s}$ [GeV] | $\alpha_s(m_z^2)$ | Exp. | Scale | PDF | $m_{top}$ | $E_{beam}$ | Total |
|---|---|---|---|---|---|---|---|---|
| ATLAS | 7000 | 0.1220 | ±0.0016 | ±0.0024 | ±0.0014 | ±0.0018 | ±0.0009 | ±0.0038 |
| ATLAS | 8000 | 0.1182 | ±0.0019 | ±0.0025 | ±0.0013 | ±0.0018 | ±0.0008 | ±0.0040 |
| CMS | 7000 | 0.1198 | ±0.0016 | ±0.0024 | ±0.0014 | ±0.0018 | ±0.0008 | ±0.0038 |
| CMS | 8000 | 0.1188 | ±0.0018 | ±0.0025 | ±0.0013 | ±0.0018 | ±0.0008 | ±0.0039 |
| TEV | 1960 | 0.1214 | ±0.0033 | ±0.0026 | ±0.0010 | ±0.0014 | ±0.0000 | ±0.0046 |

Table 3: $\alpha_s$ extractions using CMS, ATLAS and Tevatron measurements for $\sigma_{t\bar{t}}$ with the NNPDF2.3 PDF set and NNLO theory predictions. Asymmetric errors are symmetrised.

| Experiment | $\sqrt{s}$ [GeV] | $\alpha_s(m_z^2)$ | Exp. | Scale | PDF | $m_{top}$ | $E_{beam}$ | Total |
|---|---|---|---|---|---|---|---|---|
| ATLAS | 7000 | 0.1205 | ±0.0015 | ±0.0024 | ±0.0033 | ±0.0018 | ±0.0008 | ±0.0048 |
| ATLAS | 8000 | 0.1171 | ±0.0020 | ±0.0024 | ±0.0031 | ±0.0018 | ±0.0009 | ±0.0049 |
| CMS | 7000 | 0.1184 | ±0.0016 | ±0.0024 | ±0.0033 | ±0.0018 | ±0.0008 | ±0.0049 |
| CMS | 8000 | 0.1177 | ±0.0018 | ±0.0024 | ±0.0031 | ±0.0018 | ±0.0008 | ±0.0048 |
| TEV | 1960 | 0.1203 | ±0.0033 | ±0.0030 | ±0.0031 | ±0.0016 | ±0.0000 | ±0.0058 |

Table 4: $\alpha_s$ extractions using CMS, ATLAS and Tevatron measurements for $\sigma_{t\bar{t}}$ with the CT14 PDF set and NNLO theory predictions. Asymmetric errors are symmetrised.

Overall these preliminary results imply a somewhat larger $\alpha_s$ than the previous extraction from CMS, which reported $\alpha_s(m_z^2) = 0.1151^{+0.0028}_{-00027}$ [1]. This can be adequately explained by the higher value of $\sigma_{t\bar{t}}^{exp}$ in the new data.

The differences between NNLO+NNLL and NNLO theory are at the level of about 0.0015, and a change of PDF from NNPDF2.3 to CT14 also has an effect that is of a similar order of magnitude (with a significantly larger PDF uncertainty).

### Conclusions

In this text the extraction of $\alpha_\mathrm{s}$ from $\sigma_{t\bar{t}}$ was described using new experimental data and updated parton distribution functions. Currently our analysis involves extractions with and without soft-gluon resummation at next-to-next-to-logarithmic order (NNLL). We have shown results with the NNPDF2.3 PDF set and the CT14 set, however a final choice of PDF set(s) remains to be made.[†] The results obtained with the NNPDF2.3 PDF set with NNLO+NNLL theory can be compared directly to the results obtained in Ref. [1]. The new data included here leads to extracted $\alpha_\mathrm{s}$ values that are somewhat higher than that from Ref. [1].

We have refrained from quoting a unique final $\alpha_s$ extraction in this note. In order to do so, it will be necessary to conclude a study of the correlations between the different experimental inputs, to compare the two different statistical extraction procedures discussed above and to extend our study of the dependence of the results on the choice of PDF.

[†] Any other PDF set that is used should satisfy the requirement that experimental data from $\sigma_{t\bar{t}}$ measurements has not been included as part of the global fit.

# Future prospects of $\alpha_s$ from NNLO jets at the LHC and beyond

Joao Pires[1,2]

[1]Università degli Studi di Milano, Milano, Italy
[2]Max-Planck-Institut for Physics, München, Germany

**Abstract:** The extraction of the strong coupling $\alpha_s$ using inclusive jet production in proton-proton collisions is reviewed, including upcoming NNLO theoretical developments.

The measurement of inclusive jet production in proton-proton collisions is an important part of the physics program at the LHC, where these measurements can provide a precise determination of the strong coupling constant at the highest possible energy scales and also be used to obtain information about the structure of the proton. For this observable, the cross section has a simple and clean definition,

$$\frac{\mathrm{d}^2\sigma_{\mathrm{inc.jet}}}{\mathrm{d}p_{\mathrm{T}}\mathrm{d}y} = \frac{1}{\mathcal{L}}\frac{N_{\mathrm{jets}}}{\Delta p_T \Delta_y} \tag{1}$$

where we count the number of observed jets within a given transverse momentum $p_{\mathrm{T}}$ and rapidity $y$ interval per unity of integrated luminosity. The mechanism responsible for jet production at a hadron collider is the scattering at high-energy of the proton constituents, quarks and gluons, that produces a narrow spray (or jet) of colourless strongly interacting hadrons, clustered around the direction determined by the high-energy scattering event. The precise definition of a jet is then obtained with a jet algorithm, a procedure that performs a sequential recombination of the candidate jet constituents using the relative geometric distance $\Delta_R = \sqrt{\Delta\Phi^2 + \Delta y^2}$ (where $\Phi$ and $y$ are their azimuthal angle and rapidity), to collect all the particles belonging to a specific jet, determining at the end of the procedure its properties, such as the jet four-momentum, $p_{\mathrm{T}}$ and rapidity. At the LHC, the jet algorithm of choice is the anti-$k_T$ algorithm [1] with jet resolution parameters in the range $R = 0.4 \sim 0.7$. As an example we show in Fig. 1 the double differential inclusive jet cross section measurement by CMS [2] with a total integrated luminosity of 5 fb$^{-1}$ from $\sqrt{s} = 7$ TeV $pp$ collisions together with the experimental systematic uncertainties. We observe that the dominant systematic uncertainty comes from the jet energy scale uncertainty, estimated to be 2.0–2.5% [3]. Due to the steep slope of the $p_{\mathrm{T}}$ spectrum, this translates into a $< 10\%$ uncertainty on the cross section measurement, opening the possibility to do precision tests of QCD and ultimately $\alpha_s$ extractions from jet data using predictions from perturbative QCD.

| perturbative QCD accuracy | $\frac{\mathrm{d}^2\sigma_{incl.jet}}{dp_{\mathrm{T}}d_y} \propto \alpha_s^2$ | $R_{32} = \frac{\mathrm{d}\sigma_{3jet/2jet}}{d\bar{p}_{T_{1,2}}} \propto \alpha_s$ | $\frac{\mathrm{d}^2\sigma_{3jet}}{dm_{jjj}dy_{max}} \propto \alpha_s^3$ |
|---|---|---|---|
| NLO | ✓ [4,5,6] | ✓ [6] | ✓ [6] |
| NNLO | (partial) [7] | × | × |

Table 1: Three observables to extract $\alpha_s$ from LHC jet data: the double differential inclusive jet cross section, the three-jet to two-jet cross section ratio differential in the average $p_{\mathrm{T}}$ of the two leading jets and the three jet double-differential cross section in the three jet mass and maximum rapidity of the three jet system. The references correspond to the state of the art of the corresponding theory prediction.

In Table 1 we summarise the observables that have been proposed to extract $\alpha_s$ with jet data from the LHC together with the state of the art accuracy in the theory prediction from perturbative

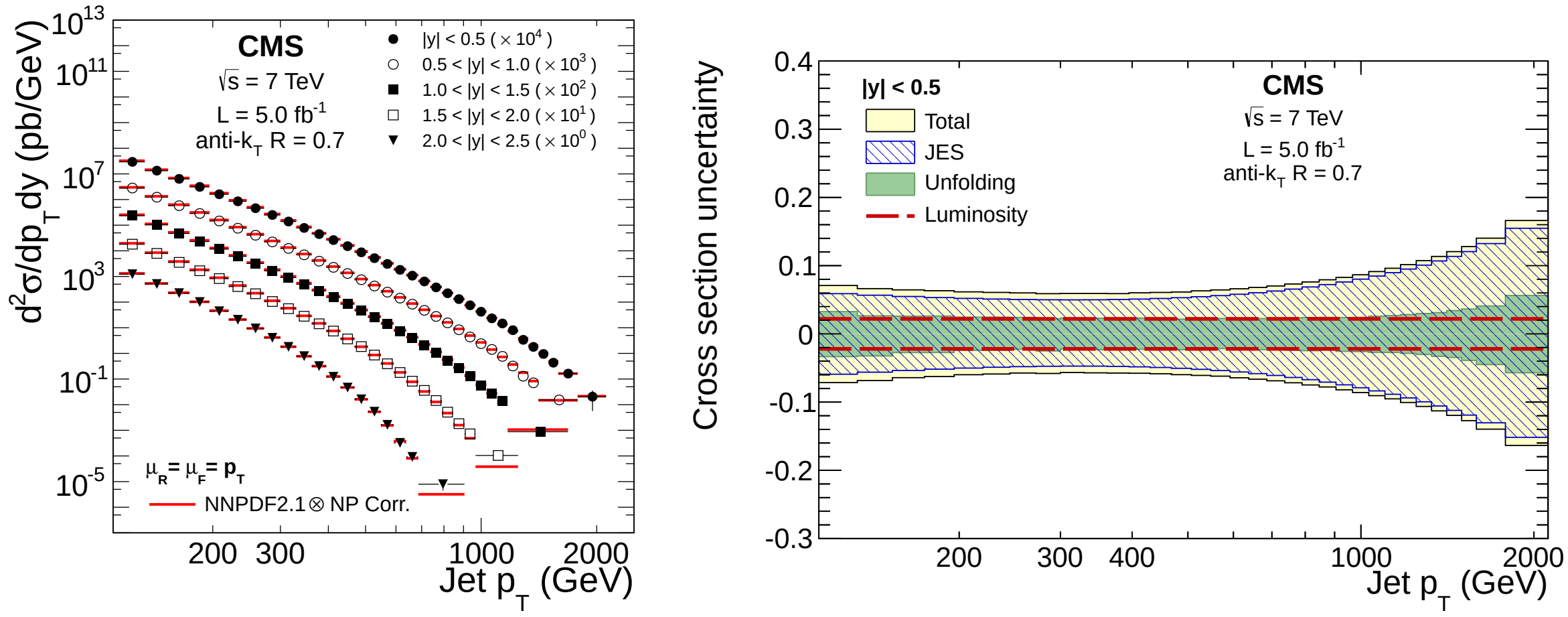

Figure 1: Inclusive jet cross sections (left) for five different rapidity bins, for data (markers) and theory (thick lines) using the NNPDF2.1 PDF set and relative experimental uncertainties (right) of the inclusive jet measurement [2].

QCD. For the remainder of this contribution we will concentrate on the double differential inclusive jet cross section.

In reference [8] a first $\alpha_s$ extraction from this observable was performed using 37 pb$^{-1}$ of data recorded by the ATLAS detector from $pp$ collisions at $\sqrt{s} = 7$ TeV [9]. In this study, 42 data points of the double differential inclusive jet cross section in the $p_T$ range 45 GeV to 600 GeV together with their statistical and systematic uncertainties and their correlation were used, and the respective $\alpha_s(M_Z)$ obtained values are shown in Fig. 2. Theoretical predictions for the inclusive jet cross sections are calculated in perturbative QCD at next-to-leading order (NLO) corrected by non-perturbative contributions to obtain the prediction at the particle level. The final result of the fit reads [8],

$$\begin{aligned}\alpha_s(M_Z) \quad = \quad & 0.1151 \pm 0.0001\ (\text{stat.}) \pm 0.0047\ (\text{exp. syst.}) \pm 0.0014\ (\text{p}_\text{T}\ \text{range}) \\ & \pm 0.0060\ (\text{jet size})^{+0.0044}_{-0.0011}\ (\text{scale})^{+0.0022}_{-0.0015}\ (\text{PDF choice}) \\ & \pm 0.0010\ (\text{PDF eig.})^{+0.0009}_{-0.0034}\ (\text{NP corrections}), \qquad (2)\end{aligned}$$

where the dominant systematic uncertainty arises in the difference between the results obtained with jet sizes $R = 0.4$ and $R = 0.6$ followed by experimental systematics (dominated by the jet energy scale) and the scale uncertainty of the perturbative prediction.

More recently, the CMS collaboration presented in [10] an $\alpha_s$ extraction based on the full 5 fb$^{-1}$ $\sqrt{s} = 7$ TeV dataset [2], using 133 data points in the $p_T$ range 114 GeV to 2116 GeV of the inclusive jet cross section with jet size $R = 0.7$. In Table 2 the obtained $\alpha_s$ values in bins of rapidity up to $|y| < 2.5$ are shown [10]. Using all rapidity bins the strong coupling constant has been determined to be [10],

$$\alpha_s(M_Z) = 0.1185 \pm 0.0019\,(\text{exp}) \pm 0.0028\,(\text{PDF}) \pm 0.0004\,(\text{NP})\,^{+0.0053}_{-0.0024}\,(\text{scale}),$$

where the scale uncertainty of the NLO QCD prediction dominates the total uncertainty in the $\alpha_s$ value followed by the PDF uncertainty.

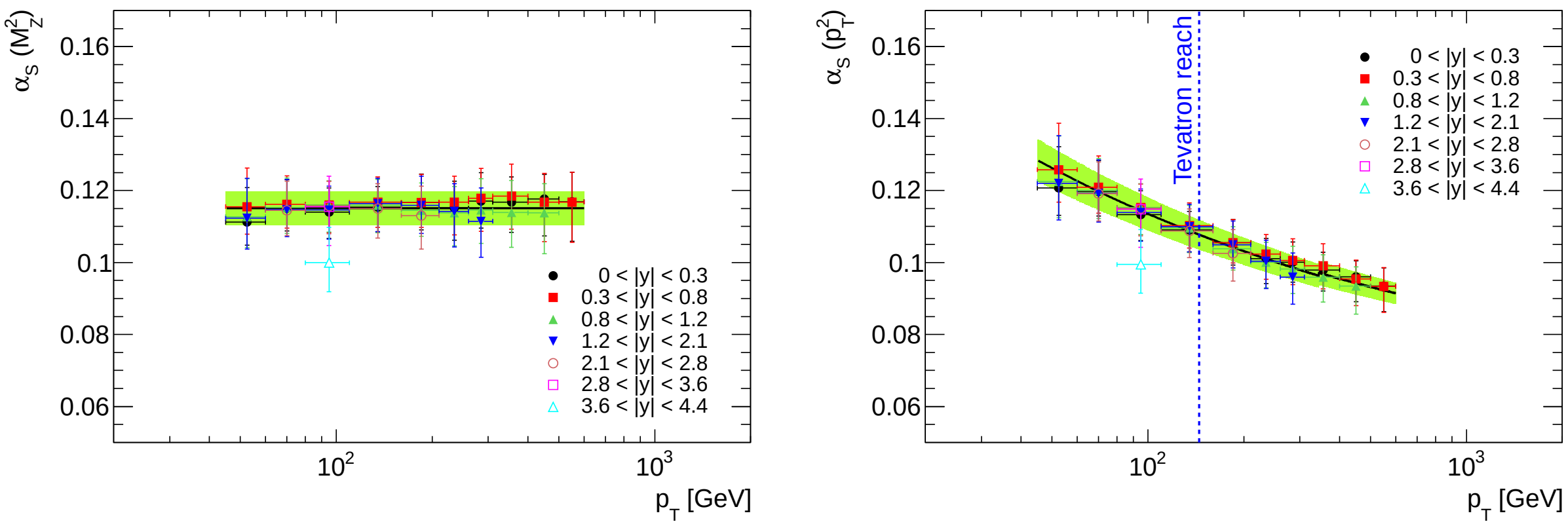

Figure 2: Global weighted $\alpha_s(M_Z^2)$ (left) and evolved to the corresponding $p_T$ scale (right) averages (green bands) together with the $\alpha_s$ values (with asymmetric uncertainties) obtained in all the ($p_T$; $|y|$) bins used in the combination. The blue vertical line indicates the highest $p_T$ value used in the Tevatron $\alpha_s(M_Z^2)$ determination, from the inclusive jet cross section [8].

While both determinations above are consistent within uncertainties with the world average value of $\alpha_s(M_Z) = 0.1185 \pm 0.0006$ [11], it is clear that the current theory uncertainties limit the achievable precision. The theoretical prediction may be improved by including the next-to-next-to-leading order (NNLO) perturbative predictions. This has the effect of (a) reducing the renormalisation and factorization scale dependence of the prediction and (b) improving the matching of the parton level theoretical jet algorithm with the hadron level experimental jet algorithm because the jet structure can be modeled by the presence of a third parton, directly addressing the currently dominant systematic uncertainties on the $\alpha_s$ determinations above.

| $\lvert y\rvert_{\rm range}$ | No. of data points | $\alpha_s(M_Z)$ | $\chi^2/n_{\rm dof}$ |
|---|---|---|---|
| $\lvert y\rvert < 0.5$ | 33 | $0.1189 \pm 0.0024\,(\text{exp}) \pm 0.0030\,(\text{PDF}) \pm 0.0008\,(\text{NP})^{+0.0045}_{-0.0027}\,(\text{scale})$ | 16.2/32 |
| $0.5 \leq \lvert y\rvert < 1.0$ | 30 | $0.1182 \pm 0.0024\,(\text{exp}) \pm 0.0029\,(\text{PDF}) \pm 0.0008\,(\text{NP})^{+0.0050}_{-0.0025}\,(\text{scale})$ | 25.4/29 |
| $1.0 \leq \lvert y\rvert < 1.5$ | 27 | $0.1165 \pm 0.0027\,(\text{exp}) \pm 0.0024\,(\text{PDF}) \pm 0.0008\,(\text{NP})^{+0.0043}_{-0.0020}\,(\text{scale})$ | 9.5/26 |
| $1.5 \leq \lvert y\rvert < 2.0$ | 24 | $0.1146 \pm 0.0035\,(\text{exp}) \pm 0.0031\,(\text{PDF}) \pm 0.0013\,(\text{NP})^{+0.0037}_{-0.0020}\,(\text{scale})$ | 20.2/23 |
| $2.0 \leq \lvert y\rvert < 2.5$ | 19 | $0.1161 \pm 0.0045\,(\text{exp}) \pm 0.0054\,(\text{PDF}) \pm 0.0015\,(\text{NP})^{+0.0034}_{-0.0032}\,(\text{scale})$ | 12.6/18 |
| $\lvert y\rvert < 2.5$ | 133 | $0.1185 \pm 0.0019\,(\text{exp}) \pm 0.0028\,(\text{PDF}) \pm 0.0004\,(\text{NP})^{+0.0053}_{-0.0024}\,(\text{scale})$ | 104.1/132 |

Table 2: Determination of $\alpha_s(M_Z)$ in bins of rapidity using the CT10-NLO PDF set. The last row presents the result of a simultaneous fit in all rapidity bins [10].

In order to perform this computation one needs to systematically combine all radiative corrections to the perturbative cross section at the second order in the strong coupling. This includes the six parton tree-level, five-parton one-loop and four-parton two-loop matrix elements for parton-parton scattering in QCD. For the partonic subprocess of pure gluon scattering this was achieved in [7], making use of the antenna subtraction scheme [12] to perform the analytic cancellation of IR singularities between real and virtual corrections at NNLO [13]. For hadron collider observables this includes contributions due to radiative corrections from partons in the initial state [14]. In this calculation, the QCD matrix elements are included in a parton-level generator NNLOJET, in development for the calculation of dijet production at NNLO in QCD, which integrates them over the exact full phase space to compute any infrared safe two-jet observable to NNLO accuracy.

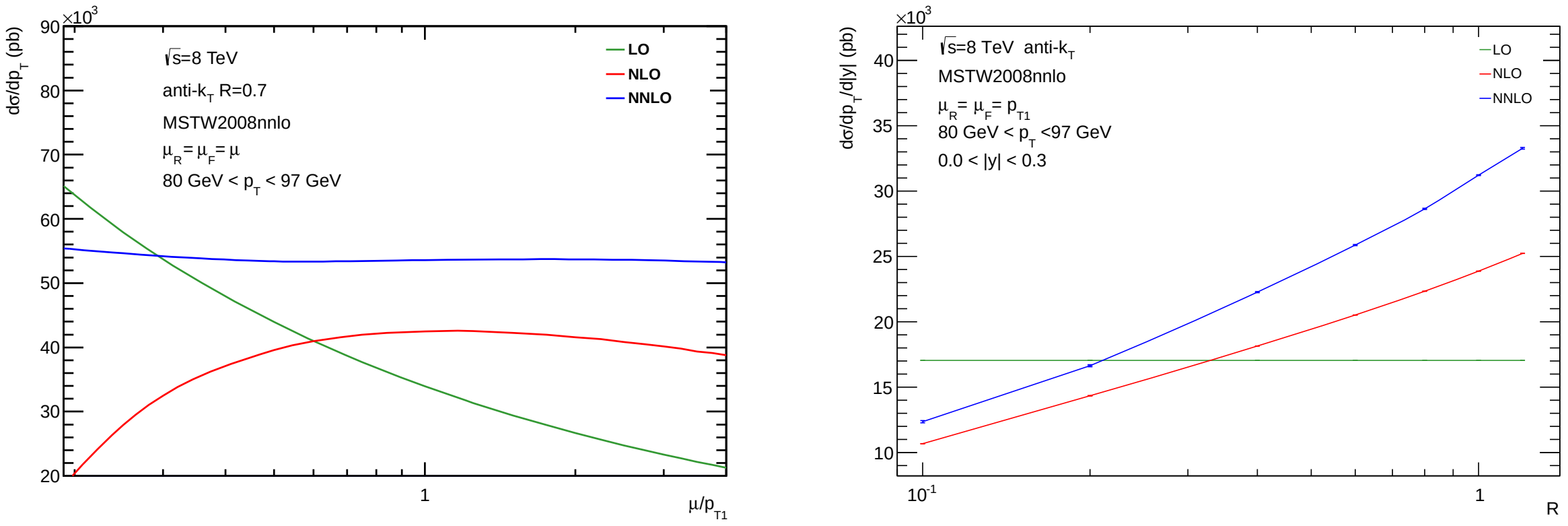

Figure 3: Scale dependence (left) and jet size parameter $R$ dependence (right) of the inclusive jet cross section for the pure gluonic scattering subprocess at each order in perturbation theory, for 80 GeV $< p_{\rm T} <$ 97 GeV jet production with the anti-$k_T$ algorithm and $\sqrt{s} = 8$ TeV [7].

In Fig. 3 we plot the results of this exercise where we show the perturbative predictions up to NNLO in QCD for jet production [7]. We stress that the predictions plotted for each order in the perturbative expansion contain only the pure gluonic scattering subprocess, and also in this example the central scale choice used to obtain the perturbative prediction was set to $\mu_R = \mu_F = p_{T1}$ (the $p_{\rm T}$ of the leading jet), with the MSTW2008NNLO PDF set fixed including for the evaluation of the LO and NLO contributions. We can observe that simultaneous variations of the factorisation and renormalisation scales at NNLO result in a rapidity integrated inclusive jet cross section with a flat scale dependence. Moreover we show the size of the NNLO corrections as a function of the jet size $R$, with the aim of addressing the currently dominant systematic uncertainties in $\alpha_{\rm s}$ extractions from jet cross section measurements at the LHC.

In order for these improvements to materialise in future extractions of $\alpha_{\rm s}$ from NNLO jets at the LHC, all the partonic subprocesses that contribute at NNLO are being added to the partial results in reference [7], to analyse data from Run II of the LHC. It is anticipated that these theoretical developments will contribute to reduce current theoretical uncertainties, namely the uncertainty from missing higher orders, the description of the perturbative cross section for different jet sizes and the perturbative prediction scale uncertainty.

# $\alpha_s$ determinations from ATLAS (status and plans)

Bogdan Malaescu, on behalf of the ATLAS Collaboration

Laboratoire de Physique Nucléaire et des Hautes Energies, IN2P3-CNRS et Universités
Pierre-et-Marie-Curie et Denis-Diderot, Paris, France

**Abstract:** The determinations of the strong coupling constant using ATLAS data at $\sqrt{s} = 7$ TeV, from the inclusive jet cross section, multi-jet cross section ratios and transverse energy-energy correlations in multi-jet events are presented.

The strong coupling constant $\alpha_s$ is the least well known coupling in the Standard Model (SM) of particle physics. The test of the energy dependence (running) of $\alpha_s$ provides an implicit test of QCD and probes potential effects from "New Physics". The ATLAS collaboration has extracted the strong coupling using data at $\sqrt{s} = 7$ TeV, from the inclusive jet cross section [1,2], multi-jet cross section ratios [3] and transverse energy-energy correlations in multi-jet events [4]. The first two studies use an integrated luminosity of 37 pb$^{-1}$, while the third benefits from 158 pb$^{-1}$. The measurements are performed for anti-$k_t$ jets [5] with a distance parameter R = 0.4 and/or R = 0.6. These studies for extracting $\alpha_s$ take into account the full set of experimental uncertainties, together with their information on the bin-to-bin correlations. They allow to test renormalization group equation (RGE) prediction for the evolution of $\alpha_s$ up to the TeV scale.

## The determination of $\alpha_s$ from inclusive jets

This study uses the unfolded double-differential ATLAS inclusive jet cross section data, in bins of the absolute rapidity ($|y|$) and as a function of the jet transverse momentum ($p_T$), with their uncertainties [1]. This measurement, performed for jets with a distance parameter R = 0.4 and R = 0.6 respectively, covers a large phase-space region going up to 4.4 in rapidity and from 20 GeV to more than 1 TeV in $p_T$. The dominant uncertainty is due to the jet energy scale (JES) calibration. Other smaller systematic uncertainties originate from the luminosity measurement, multiple proton-proton interactions, trigger efficiency, jet reconstruction and identification, jet energy resolution (JER) and deconvolution of detector effects. The amplitudes of these (asymmetric) uncertainties and their (anti-) correlations between the various phase-space regions are provided using a set of independent nuisance parameters. The statistical uncertainties and their (anti-) correlations are described using a covariance matrix for each rapidity bin.

It has been pointed out in Ref. [1] that, when performing the comparison between data and the theoretical prediction (using the world average $\alpha_s$ value), somewhat larger differences are observed for R = 0.4 than for R = 0.6 jets. In this study, the measurement of jets with R = 0.6 is used for determining the nominal result, while the comparison with the result using jets with R = 0.4 provides a systematic uncertainty.

Theoretical predictions for the inclusive jet cross sections are calculated in perturbative QCD (pQCD) at next-to-leading order (NLO), using the $\alpha_s$ scans performed in the PDF fits. A fast evaluation of the cross section allows for an efficient evaluation of the PDF and scale uncertainties. Non-perturbative (NP) corrections and their uncertainties are evaluated using leading-logarithmic parton shower generators with their various tunes. They have an important dependence on the jet size parameter, which was a motivation for performing the study for both R = 0.4 and R = 0.6 (see Ref. [2] and references therein).

The theoretical prediction provides, in each ($p_T$; $|y|$) bin, a one-to-one mapping between $\alpha_s$ and the inclusive jet cross section. An $\alpha_s$ value is then obtained by comparing the theoretical prediction and the experimental cross section. A series of pseudo-experiments are used for propagating the experimental uncertainties and correlations, from the cross section to the $\alpha_s$ values. Several methods for performing the combination of the $\alpha_s$ values from the various ($p_T$; $|y|$) have been tested. An unbiased averaging procedure is used and the uncertainties are propagated to the corresponding result, through pseudo-experiments [2].

The determination of the strong coupling constant at the Z scale, using inputs with transverse momenta between 45 and 600 GeV, yields [2] :

$$\alpha_s(m_Z^2) = 0.1151 \pm 0.0001 \text{ (stat.)} \pm 0.0047 \text{ (exp. syst.)} \pm 0.0014 \text{ (p}_T\text{ range)}^{+0.0044}_{-0.0011} \text{ (scale)} \\ \pm 0.0060 \text{ (jet size)}^{+0.0022}_{-0.0015} \text{ (PDF choice)} \pm 0.0010 \text{ (PDF eig.)}^{+0.0009}_{-0.0034} \text{ (NP corrections)},$$

where all the statistical and systematic uncertainties are included. The experimental uncertainty is about 30% smaller compared to the one obtained in the most precise individual $|y|$ bin due to the rather small correlations of the systematic uncertainties between the central and the forward region. While the nominal result is obtained for anti-$k_t$ jets with R = 0.6, the result for anti-$k_t$ jets with R = 0.4 yields a lower value. This is due to a systematic difference seen between the comparisons of the two (strongly correlated) measurements with the corresponding theoretical predictions, as discussed above. The difference between the $\alpha_s$ determinations with the two jet sizes is indeed the largest uncertainty here and an improvement in the understanding of these differences is desirable.

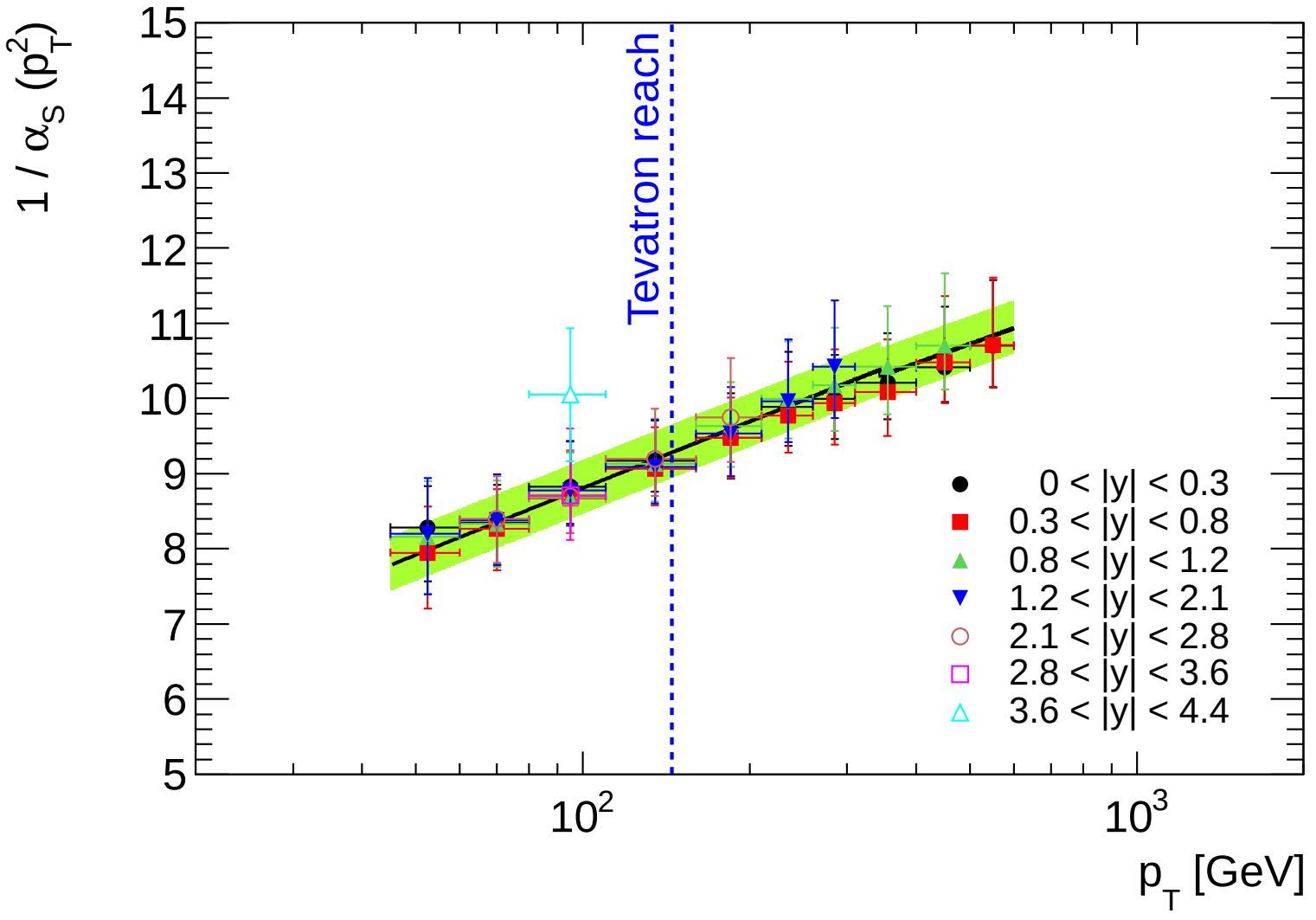

Figure 1: Weighted $\alpha_s$ average (green band) obtained from inclusive jets, evolved to the corresponding $p_T$ scale, together with the values obtained in all the ($p_T$; $|y|$) bins used in the combination. The blue vertical line indicates the highest $p_T$ value used in the Tevatron $\alpha_s(m_Z^2)$ determination, from the inclusive jet cross section. Taken from Ref. [2].

The nominal result for $\alpha_s$ evolved by RGE running, its uncertainty, and the comparison with the inputs used in the combination are shown in Fig. 1. The result of this combination corresponds to a $\chi^2$/dof = 0.54 (for 41 degrees of freedom), which represents an implicit test of the RGE in the $p_T$ range between 45 and 600 GeV. No evidence of a deviation from the QCD prediction has been observed. This result is in good agreement with the world average at the Z-boson scale [6] .

## $\alpha_s$ from multi-jet cross-section ratios

This study uses jets with a distance parameter R = 0.6, with $p_T > 40$ GeV and $|y| < 2.8$. Selected events must have at least two jets and the leading jet in the event must satisfy $p_T^{\text{lead}} > 60$ GeV. This provides an unbiased trigger selection and good stability of the NLO pQCD calculation. The Monte Carlo simulation is based on $(2 \to n,\ n \in \{2-6\})$ LO pQCD matrix element, interfaced with a showering and hadronization algorithm [3].

The inclusive three-to-two jet cross section ratios considered in this study are

$$R_{3/2}\left(p_T^{\text{lead}}\right) = \frac{d\sigma_{N_{\text{jet}}\geq 3}/dp_T^{\text{lead}}}{d\sigma_{N_{\text{jet}}\geq 2}/dp_T^{\text{lead}}};\ N_{3/2}\left(p_T^{(\text{alljets})}\right) = \frac{\sum_i^{N_{\text{jet}}}\left(d\sigma_{N_{\text{jet}}\geq 3}/dp_{T,i}\right)}{\sum_i^{N_{\text{jet}}}\left(d\sigma_{N_{\text{jet}}\geq 2}/dp_{T,i}\right)},$$

where $N_{\text{jet}}$ is the number of jets in each event satisfying the kinematic requirements defined above. The numerator and denominator used to define $R_{3/2}$ receive a single entry per event, while for $N_{3/2}$ they receive one entry per jet. The unfolded measurements of $R_{3/2}$ and $N_{3/2}$ are sensitive to $\alpha_s$.

The dominant systematic uncertainty for these measurements is the JES. Other sources of systematics are the JER, pile-up, trigger efficiency, jet quality and unfolding. The statistical uncertainty on the ratios is evaluated using the bootstrap method, which allows to take into account the correlations between the numerator and denominator, for $R_{3/2}$ and $N_{3/2}$ [3].

The scale dependence of the NLO pQCD predictions for the three-to-two ratios have been studied in detail [3]. The scale choice made for these predictions are $\mu_R = \mu_F = p_T^{\text{lead}}$ ($p_T^{(\text{alljets})}$), consistently for the numerator and denominator of $R_{3/2}$ ($N_{3/2}$). $N_{3/2}$ is found to be less sensitive to the scale choice compared to $R_{3/2}$, while also having a somewhat better sensitivity to $\alpha_s$. $N_{3/2}$ is used to extract $\alpha_s$, for $p_T^{(\text{alljets})} > 210$ GeV. Predictions obtained for jets with a distance parameter R = 0.4 are found to be much more sensitive to the scale choice and are not used in the rest of this study.

The $\alpha_s(m_Z^2)$ value is extracted by performing a $\chi^2$ fit in the range 210 GeV $< p_T^{(\text{alljets})} <$ 800 GeV. The $\chi^2$ evaluation takes into account the experimental uncertainties and their correlations. The resulting $\chi^2 = 7.1$, for 5 degrees of freedom, indicates a successful test of the RGE. The theoretical uncertainties are propagated to the $\alpha_s$ evaluation through shifts of the pQCD prediction by one standard deviation of each uncertainty component. The scale uncertainty dominates for this $\alpha_s$ evaluation. The differences between the $\alpha_s(m_Z^2)$ values obtained for the various PDF sets considered in the study are of the same order of magnitude as the experimental uncertainties. This study yields

$$\alpha_s(m_Z^2) = 0.111 \pm 0.006\ (\text{exp.})^{+0.016}_{-0.003}\ (\text{theory}),$$

where the uncertainties are experimental and respectively from the theory prediction.

## $\alpha_s$ from transverse energy-energy correlations in multi-jet events

The jets used in this study are reconstructed with a distance parameter R = 0.4 [4]. The event selection requires at least two jets with $p_T > 50$ GeV and $|\eta| < 2.5$, as well as $p_{T,1} + p_{T,2} > 500$ GeV.

The transverse energy-energy correlation (TEEC) observable is defined as an angular distribution in bins of $\cos\phi_{ij}$, where $\phi_{ij}$ is the azimuthal angular difference between the two jets, weighted by the product of the transverse energies of the two jets. All the pairs of two jets selected in the event are used in the TEEC evaluation. The ATEEC is defined as the asymmetry of the TEEC, between $\phi$ and $\pi - \phi$ [4].

The measurements are corrected to particle level using a bin-by-bin procedure, cross-checked with an iterative Bayesian method. The full set of uncertainties and their correlations are propagated to the corrected distributions. The dominant systematic uncertainties are due to the JES, the parton-shower modelling in the unfolding, and pile-up.

The theoretical predictions are provided by NLO pQCD with NP corrections. The scale choice made here is $\mu_R = \mu_F = (p_{\mathrm{T},1} + p_{\mathrm{T},2})/2$, ranging between 250 and 1300 GeV. The theoretical systematics originate from the scale uncertainty, PDFs and NP corrections.

The $\alpha_{\mathrm{s}}(\mathrm{m}_{\mathrm{Z}}^2)$ value is extracted by performing a $\chi^2$ fit, taking into account the experimental uncertainties and their correlations. Here also, each theoretical uncertainty is propagated through a shift of the pQCD prediction by one standard deviation. The nominal result is obtained from the TEEC measurement, which has a good experimental precision, employing the CT10 PDF, the eigenvectors of which cover the differences with the results obtained using other PDF sets. This yields [4]

$$\alpha_{\mathrm{s}}(\mathrm{m}_{\mathrm{Z}}^2) = 0.1173 \pm 0.0010\ (\mathrm{exp.})^{+0.0063}_{-0.0020}\ (\mathrm{scale}) \pm 0.0017\ (\mathrm{PDF}) \pm 0.0002\ (\mathrm{NP\ corrections}),$$

where the uncertainties are experimental, from the scale choice, PDFs and NP corrections respectively. The good fit quality ($\chi^2 = 28.4$ (21 dof)) represents a test of the RGE.

## Conclusions

Figure 2 shows the comparison between the $\alpha_{\mathrm{s}}(\mathrm{m}_{\mathrm{Z}}^2)$ evaluations based on ATLAS jet measurements with the ones from other experiments. The comparison of the uncertainties of the various results has to be done with care. First, the study of the dependence of the result on the choice of the distance parameter R, as performed in Ref. [2], indicates an important sensitivity to this choice, yielding a large uncertainty. Similar studies are desirable for the other jet-based evaluations of $\alpha_{\mathrm{s}}(\mathrm{m}_{\mathrm{Z}}^2)$. Second, the difference between the results obtained with various PDF sets is not always included in the total uncertainty of a given study. Testing the compatibility among these results cannot be done through a simple comparison between their difference and the uncertainties from a single PDF set. Indeed, the strong correlations between the uncertainties of the various PDF sets, due to their common input data, need to be carefully taken into account. Actually, one of the dominant sources of differences between the PDF sets originates in the parameterizations used in the various fits and is not covered by the uncertainties of a single set. Any comparison or combination of the existing results needs to account for these differences in the uncertainty evaluation.

It is also interesting to point out that the PDF uncertainties are not negligible for the $\alpha_{\mathrm{s}}$ evaluations based on multi-jet cross section ratios and on the (A)TEEC observables. Indeed, for these observables, the sensitivity to $\alpha_{\mathrm{s}}$ originates in the probability of having extra radiation in the final state, which does depend on the partons in the initial state. Actually, for the precision of the $\alpha_{\mathrm{s}}$ extraction from a given observable, the comparison of its sensitivity to PDFs, to experimental uncertainties and to $\alpha_{\mathrm{s}}$ respectively is relevant.

The ATLAS plans for the $\alpha_{\mathrm{s}}$ studies involve evaluations at higher scales and using more observables.

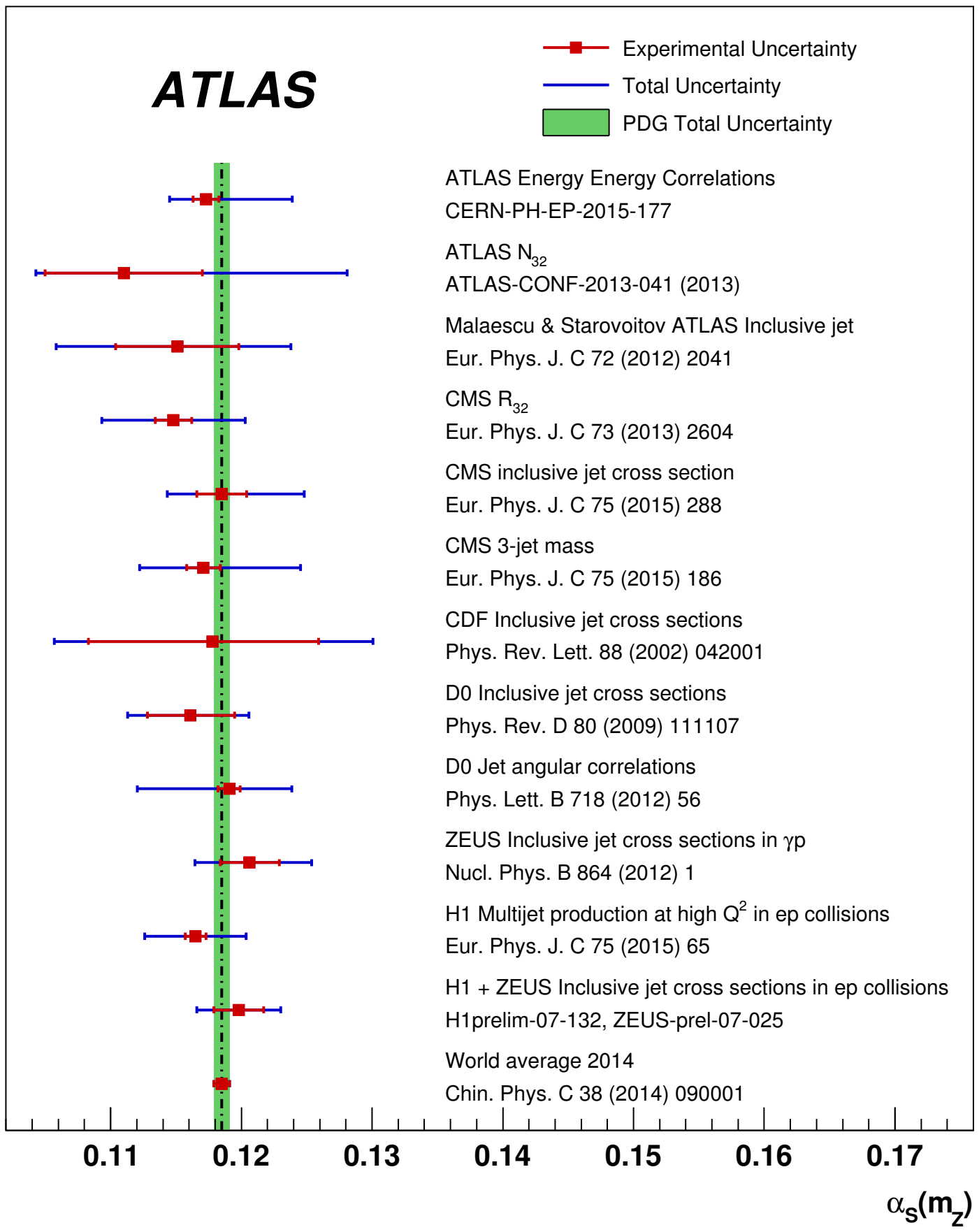

Figure 2: Comparison of the value of $\alpha_s(m_Z^2)$ obtained by ATLAS with values from jet observables measured by other experiments in hadron-hadron and ep colliders. The red error bar represents the experimental uncertainties, while the blue bar includes both experimental and theoretical errors. The vertical green band represents the PDG world average for 2014. Taken from Ref. [4].

# $\alpha_s$ determinations from CMS (status and plans)

Klaus Rabbertz (on behalf of CMS)

KIT, Karlsruhe, Germany

**Abstract:** The extractions of the strong coupling $\alpha_s$ from $p_T$-differential jet cross sections, the 3-jet mass $m_3$, 3- to 2-jet cross section ratio $R_{3/2}$, and top-pair ($t\bar{t}$) cross sections, carried out by the CMS experiment in p-p collisions at the LHC are reviewed.

Numerous extractions of the strong coupling constant $\alpha_s$ have been performed at hadron colliders, in particular from jet cross sections. The latest results achieved by the H1 and ZEUS experiments at the $ep$ collider HERA and by the CDF and D0 experiments at the $p\bar{p}$ collider Tevatron are reported in Refs. [1,2,3] and [4,5,6] respectively. The range in scale $Q$ for the running coupling constant $\alpha_s(Q)$ covered by these measurements runs from about 5 GeV up to 400 GeV. Evolving the fit results to the reference scale of the $Z$ boson mass, $M_Z = 91.1876$ GeV, $\alpha_s(M_Z)$ values between 0.1160 and 0.1206 are obtained. Within their rather large uncertainties up to 10%, they are compatible with each other. The dominant contributions to the uncertainty are of theoretical nature, namely scale uncertainties accounting for missing higher orders in the next-to-leading order predictions (NLO), and uncertainties caused by the limited knowledge of the proton structure that is encoded in parton distribution functions (PDFs). Lacking next-to-next-to-leading order (NNLO) theory, the quoted $\alpha_s(M_Z)$ determinations are not included in the derivation of the world average of $\alpha_s(M_Z) = 0.1185 \pm 0.0006$ [7], nevertheless they are compatible with it.

With the advent of the LHC collider and $pp$ collisions at center-of-mass energies of $\sqrt{s} = 2.76$, 7, and 8 TeV, much higher $Q$ scales become accessible. In addition, modern particle detectors like ATLAS [8] and CMS [9] in combination with more elaborate jet calibration and detector simulation techniques easily outperform the much older Tevatron experiments leading to jet calibration uncertainties below percent level in the best case. $\alpha_s$ extractions using ATLAS data have been reported in Refs. [10] and [11] and are further discussed in [12]. The CMS Collaboration determined $\alpha_s(Q)$ from scales $Q$ up to 1.4 TeV, more than threefold the previous upper limit, from jet cross sections as functions of the inclusive jet $p_T$ [13] and the 3-jet mass $m_3$ [14], and from the 3- to 2-jet cross section ratio $R_{3/2}$ [15]. Again, scale and PDF uncertainties, amounting to roughly 4% and 2%, are the limiting factors impeding a better accuracy. The potential gain through NNLO theory is demonstrated in another analysis by CMS, where for the first time the $t\bar{t}$ cross section is used to extract the strong coupling constant [16]. Theoretical calculations, even available at NNLO+NNLL (next-to-next-to-leading-log) [17], reduce the scale uncertainty in this case to 0.7%. Further prospects with respect to $t\bar{t}$ production are discussed in [18,19].

Table 1 summarizes the uncertainties for the main result on $\alpha_s(M_Z)$ reported in the references given above. Except for scale uncertainties where always the larger deviation is considered (scale), asymmetric uncertainties have been symmetrized for simpler comparison and are presented as $\pm$ percentual uncertainty on the respective central value. Separately detailed uncertainties of experimental origin, including statistical ones, are quadratically added together (exp). The same is done with multiple uncertainties caused by similar sources like a PDF uncertainty quoted for one PDF set and an additional uncertainty quoted for using different PDF sets (PDF). The entry of “scl” in the column for the nonperturbative (NP) effects from multiple-parton interactions and hadronization indicates that they are included in the scale uncertainty. The extra column “other” lists additional uncertainties that have been attributed to the choice of infrared-unsafe(!) jet algorithm (CDF), to the choice of the jet size $R = 0.4$ or 0.6 (ATLAS incl. jets), and to the top pole-mass $M_t^{\text{pole}}$ as input parameter (CMS $\sigma(t\bar{t})$).

| Process | LO | $\sqrt{s}$ | Q | $N_p$ | $\alpha_s(m_Z)$ | $\Delta\alpha_s(m_Z)/\alpha_s(m_Z)$ [%] | | | | |
|---|---|---|---|---|---|---|---|---|---|---|
| $ep$, $p\bar{p}$, $pp$ | $\alpha_s^n$ | [TeV] | [GeV] | | | exp | PDF | scale | NP | other |
| H1 jets low $Q^2$ | 1 | 0.32 | 5–57 | 62 | 0.1160 | 1.2 | 1.4 | 8.0 | scl | – |
| ZEUS $\gamma p$ jets | 1 | 0.32 | 21–71 | 18 | 0.1206 | 1.9 | 1.9 | 2.5 | 0.4 | – |
| H1 jets high $Q^2$ | 1 | 0.32 | 10–94 | 64 | 0.1165 | 0.7 | 0.8 | 3.1 | 0.7 | – |
| CDF incl. jets | 2 | 1.8 | 40–250 | 27 | 0.1178 | 7.5 | 5.0 | 5.0 | – | 2.5 |
| D0 incl. jets | 2 | 1.96 | 50–145 | 22 | 0.1161 | 2.9 | 1.0 | 2.5 | 1.1 | – |
| D0 ang. corr. | 1 | 1.96 | 50–450 | 102 | 0.1191 | 0.7 | 1.2 | 5.5 | 0.1 | – |
| ATLAS incl. jets | 2 | 7 | 45–600 | 42 | 0.1151 | 4.3 | 1.8 | 3.8 | 1.9 | 5.2 |
| ATLAS EEC | 1 | 7 | 250–1300 | 22 | 0.1173 | 0.9 | 1.4 | 5.4 | 0.2 | – |
| CMS $R_{3/2}$ | 1 | 7 | 420–1390 | 21 | 0.1148 | 1.2 | 1.6 | 4.4 | scl | – |
| CMS $\sigma(t\bar{t})$ | 2 | 7 | $M_t^{\text{pole}}$ | 1 | 0.1151 | 2.2 | 1.5 | 0.7 | – | 1.1 |
| CMS 3-jet mass | 3 | 7 | 332–1635 | 46 | 0.1171 | 1.1 | 2.0 | 5.9 | 0.7 | – |
| CMS incl. jets | 2 | 7 | 114–2116 | 133 | 0.1185 | 1.6 | 2.4 | 4.5 | 0.3 | – |

Table 1: Summary of latest $\alpha_s(M_Z)$ determinations at hadron colliders. For each process the power in $\alpha_s$ of the leading order (LO), the center-of-mass energy, the accessed range of scale $Q$, and the number of fitted data points is given. Theory is used at NLO accuracy except for the D0 inclusive jets, where additional threshold corrections are considered, and for CMS $\sigma(t\bar{t})$, where theory is known to NNLO+NNLL precision. The experimental, PDF, scale, NP, and additional uncertainies are presented as percentual uncertainties on the extracted $\alpha_s(M_Z)$ value. All numbers are derived from Refs. [1]-[6] and Refs. [10]-[16].

Within CMS the focus was on testing the running of $\alpha_s(Q)$ up to the highest scales possible. As shown in Fig. 1, no indication of any significant deviation was found up to 1.4 TeV. A preliminary result from inclusive jets at 8 TeV, published just after the workshop, increases the investigated range in scale $Q$ even to 1.5 TeV [20]. With 13 TeV data it can be expected to extend these tests into the multi-TeV range, where new colored matter potentially changes the running of $\alpha_s$, cf. also [21]. The CMS jet cross section ratio $R_{3/2}$ has been employed to set limits on such models of new physics [22]. The best observables and strategies to detect such deviations, however, need still to be established.

On the other hand there is still a significant optimization potential for a precision determination of $\alpha_s(M_Z)$ within CMS. Concentrating on the best understood detector parts at central rapidity and within a jet $p_T$ range restricted to the most precisely calibrated jets, experimental uncertainties can be reduced. At the same time the uncomfortable increase in uncertainty when evolving downwards to $Q = M_Z$ is smaller for medium jet $p_T$'s and the PDFs are more precisely known as well. It has to be assured though that the same data are not used twice, first for constraining the proton structure and then to extract the strong coupling constant with these PDFs.

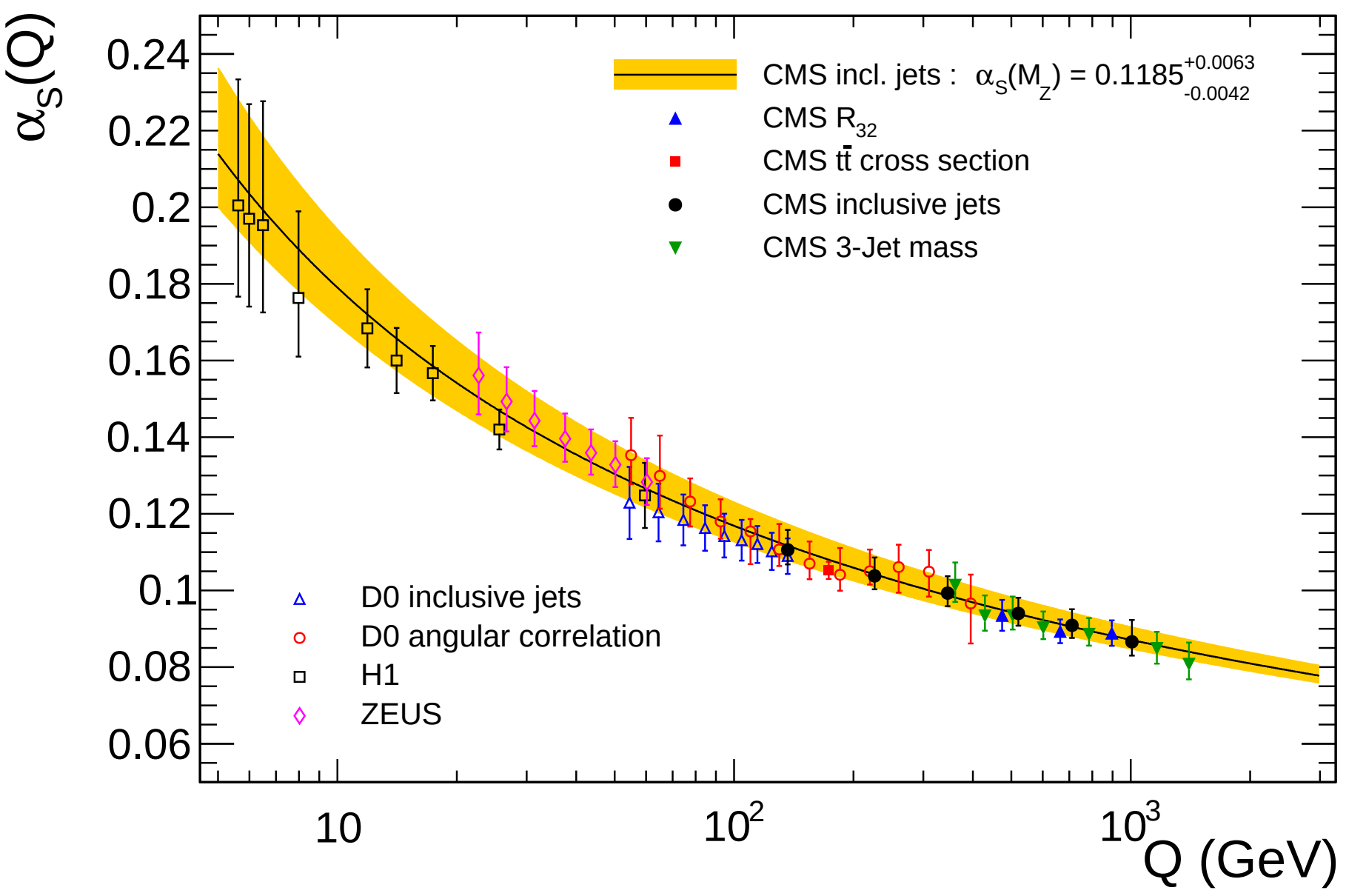

Figure 1: The strong coupling $\alpha_s(Q)$ (full line) and its total uncertainty (band) as a function of the respective choice for the scale $Q$. The four analyses of $\alpha_s(Q)$ from CMS [13]-[16] are shown together with results from the H1 [1,3], ZEUS [2], and D0 [5,6] experiments at the HERA and Tevatron colliders. Figure taken from auxiliary material of Ref. [14].

The most important improvement beyond CMS comes from NNLO predictions for inclusive and dijet cross sections, cf. also [23]. This will help to reduce significantly the dominating theoretical uncertainty of several percent on many jet-based $\alpha_s(M_Z)$ determinations at hadron colliders. Corresponding electroweak corrections have been calculated already [24]. The technique of cross section ratios to cancel at least partially some systematic effects might be helpful as well. Corresponding cross sections are known to NLO up to jet multiplicities of five [25,26]. For ratios like $R_{3/2}$ at NNLO the 3-jet cross sections are not yet available at this level of accuracy.

The extraction of the strong coupling constant $\alpha_s(M_Z)$ at hadron colliders, in particular from jet cross sections, has a long tradition, but was hampered by the complex and strongly interacting initial states and theoretical predictions that were limited in accuracy to NLO. This will be cured in the near future so that further advances in precision can only be obtained by addressing all heretofore subdominant uncertainties including nonperturbative effects. For the ultimate precision possible with hadron collisions the further refinements of combining multiple observables to disentangle e.g. correlations between the top mass $M_t$, the gluon PDF $g(x)$, and the strong coupling constant $\alpha_s(M_Z)$, and combining multiple experiments, i.e. ATLAS, CMS, and the HERA and Tevatron experiments, will be necessary.

To conclude and as a near term perspective the following challenge of "Triple Five" is presented: Within the next **FIVE** years, check the running of $\alpha_s(Q)$ up to **FIVE** TeV, and determine $\alpha_s(M_Z)$ to **FIVE** permille accuracy.

# Workshop summary, future prospects, and FCC-ee impact on $\alpha_s$

David d'Enterria

CERN, PH Department, Geneva, Switzerland

The workshop on "High-precision $\alpha_s$ measurements from LHC to FCC-ee" –organized in the context of the preparation of the Future Circular Collider (FCC) Conceptual Design Report, within the FCC-$e^+e^-$ "QCD and photon-photon" physics working group activities– discussed the latest advances in the measurement of the strong interaction coupling $\alpha_s$. The meeting brought together leading experts in the field to explore in depth recent theoretical and experimental developments on the determination of $\alpha_s$, new ways to measure this coupling in lepton-lepton, lepton-hadron and hadron-hadron collisions, and in particular the improvements expected at the proposed Future Circular Collider $e^+e^-$ (FCC-ee) facility.

## Workshop summary

The workshop was organized along four sessions. In the introductory session, Siggi Bethke [2] presented the preliminary 2015 update of the Particle-Data-Group (PDG) world-average $\alpha_s$ obtained from comparison of next-to-next-to-leading-order (NNLO) pQCD calculations to a set of 6 groups of experimental data. Enlarged uncertainties from lattice QCD and tau-lepton decays, as well as the first NNLO extraction from top-pair cross sections in p-p at the LHC, have resulted in a factor of two increase of the uncertainty on $\alpha_s$ which will move from the 2014 value of $\alpha_s(m_Z^2) = 0.1185 \pm 0.0006$ [1] to $\alpha_s(m_Z^2) = 0.1177 \pm 0.0013$. Luminita Mihaila [3] reviewed the impact on Higgs physics of $\alpha_s$ which is the second major contributor –after the bottom mass– to the parametric uncertainties of its dominant $H \to b\bar{b}$ partial decay, and the largest source of uncertainty for the $c\bar{c}$ and $g\,g$ decay modes. An accurate knowledge of the running of $\alpha_s$ at TeV energy scales is also crucial for searches of physics beyond the SM. Francesco Sannino [4] presented generic exclusion bounds on masses of new coloured particles based on LHC data.

The second session of the workshop was devoted to measurements that allow one to extract $\alpha_s$ at low energies. Those include results from lattice QCD (reviewed by Paul Mackenzie [5] and Xavier Garcia i Tormo [6]), pion decay factor (Jean-Loïc Kneur [7]), $\tau$ decay (Toni Pich [8]), $\Upsilon$ decays (Joan Soto [9]), and the evolution of the soft parton-to-hadron fragmentation functions FFs (Redamy Pérez-Ramos [10]). The comparison of pQCD predictions to computational lattice-QCD "data", yielding $\alpha_s(m_Z^2) = 0.1184 \pm 0.0012$, still provides the most precise $\alpha_s$ extraction with a $\delta\alpha_s(m_Z^2)/\alpha_s(m_Z^2) = 1\%$ uncertainty. Hadronic decays of the tau lepton yield $\alpha_s(m_Z^2) = 0.1187 \pm 0.0023$ (i.e. $\delta\alpha_s(m_Z^2)/\alpha_s(m_Z^2) = 1.9\%$), although the results of different theoretical approaches are still a matter of debate. The pion decay factor was proposed as a new observable to extract $\alpha_s(m_Z^2) = 0.1174 \pm 0.0017$, notwithstanding the low scales involved which challenge the pQCD applicability. Decays of the $b\bar{b}$ bound state ($\Upsilon$) used to constrain the QCD coupling until a few years ago ($\alpha_s(m_Z^2) = 0.1190 \pm 0.0070$), but their lower degree of theoretical (NLO only) accuracy should be improved in order to be included into future PDG updates. Similarly, the energy evolution of the distribution of hadrons in jets has proven to be a novel robust method to extract $\alpha_s(m_Z^2) = 0.1205 \pm 0.0022$, but the calculations need to go beyond their current approximate-NNLO accuracy.

The $\alpha_s$ determinations at higher energy scales –including global fits of parton distribution functions PDFs (reviewed by Johannes Blümlein [11]), hard parton-to-hadron FFs (Bernd Kniehl [12]), jets in deep-inelastic (DIS) and photoproduction in $e^{\pm}p$ collisions (Michael Klasen [13]), $e^+e^-$ event shapes (Stefan Kluth [14] and Andre Hoang [15]), jet cross sections in $e^+e^-$ (Andrea Banfi [16]), hadronic Z and W decays (Klaus Mönig [17] and Matej Srebre [18]), and the $e^+e^- \to$ hadrons cross section (Johannes Kühn [19])– were covered in the third section of the workshop. The NNLO analyses of PDFs have a good precision, $\delta\alpha_s(m_Z^2)/\alpha_s(m_Z^2) = 1.7\%$ given by the spread of different theoretical analyses, albeit yielding a central value lower than the rest of methods: $\alpha_s(m_Z^2) = 0.1154 \pm 0.0020$. Upcoming NNLO fits of the jet FFs will provide a value of the QCD coupling more accurate than the current one at NLO ($\alpha_s = 0.1176 \pm 0.0055$). Similarly, a full-NNLO analysis of jet production in $e^{\pm}p$ is needed to improve the current $\alpha_s(m_Z^2) = 0.121 \pm 0.003$ extraction from these observables. Electron-positron event shapes and jet rates yield $\alpha_s(m_Z^2) = 0.1174 \pm 0.0051$ with a $\delta\alpha_s(m_Z^2)/\alpha_s(m_Z^2) = 4.3\%$ uncertainty, but new $e^+e^-$ data at lower and higher energies than LEP is required for a better control of hadronization corrections. The hadronic decays of electroweak bosons are high-precision observables for the extraction of the strong coupling. The current Z data provide $\alpha_s(m_Z^2) = 0.1196 \pm 0.0030$, i.e. a $\delta\alpha_s(m_Z^2)/\alpha_s(m_Z^2) = 2.5\%$ uncertainty which can be reduced to about $\pm0.15\%$ with the huge statistical data sets expected at the FCC-ee. The W hadronic decay data are not (by far) as precise today ($\alpha_s(m_Z^2) = 0.117 \pm 0.043$, i.e. $\delta\alpha_s(m_Z^2)/\alpha_s(m_Z^2) = 37\%$), but promise the same $\alpha_s$ sensitivity with the expected FCC-ee high-statistics measurements.

The final session was dedicated to $\alpha_s$ extractions at hadron colliders. Important NNLO theoretical developments for top-quark pair and jets cross sections were reviewed by Alexander Mitov [20], Gavin Salam [21], and Joao Pires [22]. A lowish $\alpha_s(m_Z^2) = 0.1151 \pm 0.0028$ value with $\delta\alpha_s(m_Z^2)/\alpha_s(m_Z^2) = 2.5\%$ uncertainty is obtained using the only $t\bar{t}$ cross sections published so far by CMS, although inclusion of all existing data would increase it to a preliminary value of $\alpha_s(m_Z^2) = 0.1186 \pm 0.0033$. The imminent release of the NNLO calculation for jets will provide a huge lever-arm for PDFs, FFs, and cross sections studies in p-p, $e^{\pm}p$ and $\gamma$-p collisions. To date, the NLO combination of ATLAS, CMS and Tevatron jet results yields $\alpha_s(m_Z^2) = 0.1179 \pm 0.0023$. Existing and planned measurements of $\alpha_s$ at the LHC were also reviewed by Bogdan Malaescu (ATLAS) [23] and Klaus Rabbertz (CMS) [24], confirming asymptotic freedom at multi-TeV scales for the first time. Figure 1 shows a combined plot with all the $\alpha_s(m_Z^2)$ values discussed during the meeting.

## Future prospects and FCC-ee impact on $\alpha_s$

Beyond reviewing the current status of the different QCD coupling determinations available today, one of the key goals of the workshop was to assess the impact on the extraction of $\alpha_s$ of theoretical and experimental improvements expected in the coming years, and in particular at the FCC-ee. Table 1 summarizes the status and prospects for each one of the different observables considered:

1. Lattice QCD [5]: The expected improvements in computing power over the next 10 years would reduce the uncertainty of the $\alpha_s$ determination from its current 1% down to 0.3%. The calculation of a 4th order of perturbation theory may further improve the precision of this determination to $\sim$0.1%. Since this $\alpha_s$ extraction solely depends on lattice-versus-perturbative calculations, the impact of current or future experimental data is null.

2. Pion decay factor [7]: Since the precision of the experimental $F_\pi$ value propagates into a negligible $\alpha_s$ uncertainty, improvements need to be achieved in the theoretical side of the ex-traction. Incorporating 5-loop corrections and explicit quark mass $m_{u,d,s}$ dependencies within the renormalization-group optimized approach, would halve the ~1.5% theoretical error.

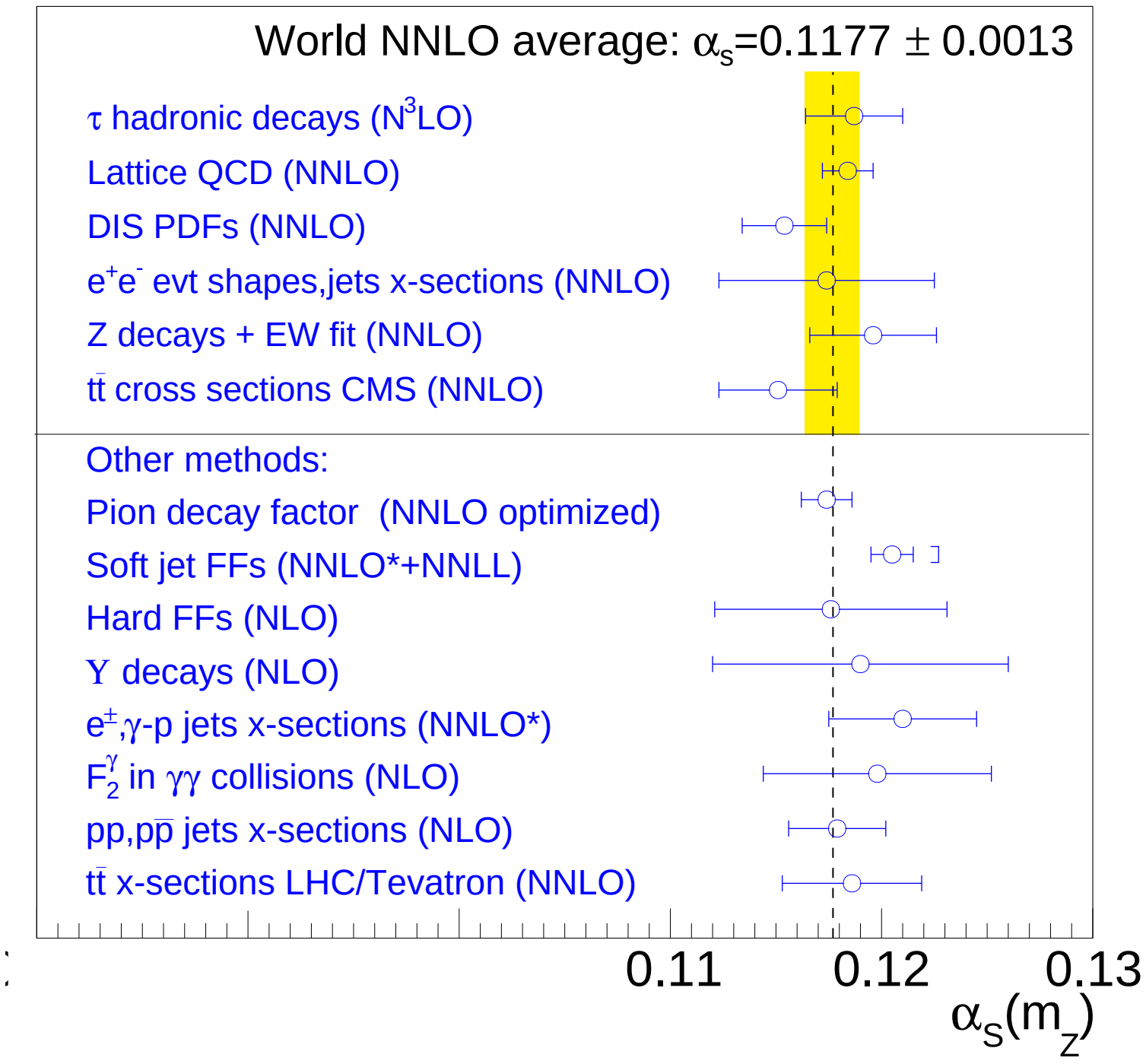

Figure 1: Summary of all the strong coupling extractions, $\alpha_s(m_Z^2)$ and associated uncertainties, discussed during the workshop. The (yellow) band indicates the uncertainty of the updated (2015) NNLO world-average quoted at the top (which does not include the results of the 'other methods' plotted in the bottom half).

3. $\tau$ decays [8]: The current $\alpha_s$ uncertainty of ~2% from $\tau$ data is shared roughly equally between experimental and theoretical systematics. The latter are driven by differences of the CIPT and FOPT implementations of N$^3$LO corrections: The updated PDG value is $\alpha_s(m_Z^2) = 0.1187 \pm 0.0023$, whereas the analysis presented here yields: $\alpha_s(m_Z^2) = 0.1120 \pm 0.0015$. High-statistics $\tau$ spectral functions (e.g. from B-factories now, or FCC-ee in the future) and solving CIPT-FOPT discrepancies (or calculations at N$^4$LO accuracy, within a ~10 years time scale) would reduce the relative $\alpha_s$ uncertainty to ~1% (and below that, at FCC-ee).

4. Quarkonia [9]: The current relative uncertainty on $\alpha_s$ from $\Upsilon$ decays is ~6%, shared in equal parts by experimental and theoretical (NLO) systematics. Calculations at NNLO accuracy with improved long-distance matrix elements (LDME), and more precise measurements of the photon spectrum of radiative $\Upsilon$ decays (and of the parton-to-photon FF) would allow an extraction with $\delta\alpha_s(m_Z^2)/\alpha_s(m_Z^2) \approx 2\%$ in a few years from now. FCC-ee would not provide much advantage here compared to lower-energy $e^+e^-$ colliders running at $\sqrt{s} = m_{Q\overline{Q}}$.

5. Soft parton-to-hadron FFs [10]: The study of the energy evolution of the soft (i.e. low-$z$) parton-to-hadron fragmentation functions, provides $\alpha_s$ at approximate-NNLO+NNLL accuracy with a ~2% uncertainty (of which the experimental systematics represents 0.7%). Extension of the calculations to NNLO+NNLL accuracy would reduce the scale uncertainty, yielding theoretical and experimental uncertainties of the same size, i.e. $\delta\alpha_s(m_Z^2)/\alpha_s(m_Z^2) \approx 1\%$. The lever-arm provided by FCC-ee in terms of the jet fragmentation evolution over $\sqrt{\mathrm{s}} = 90$–350 GeV would further improve the precision of the $\alpha_\mathrm{s}$ determination.

6. Hard parton-to-hadron FFs [12]: Extraction of $\alpha_\mathrm{s}$ from the scaling violations in the high-$z$ range of the parton-to-hadron FFs is done today at NLO with relative uncertainties of order 5%, mostly of experimental origin. Extension of the global FF fits at NNLO accuracy, and inclusion of new datasets (e.g. available already now from B-factories) would allow reaching $\delta\alpha_\mathrm{s}(\mathrm{m}_\mathrm{Z}^2)/\alpha_\mathrm{s}(\mathrm{m}_\mathrm{Z}^2) \approx 2\%$. At the FCC-ee, the availability of a huge statistics of high-precision measurements of jets in a wide $\sqrt{\mathrm{s}}$ range, finely-binned in the $z$ variable and with light-quark/gluon/heavy-quark discrimination, would result in $\alpha_\mathrm{s}$ extractions with <1% precision.

7. Global PDF fits [11]: The extraction of $\alpha_\mathrm{s}$ from nucleon structure functions is done today at NNLO (or N$^3$LO for some particular analyses) with a relative $\delta\alpha_\mathrm{s}(\mathrm{m}_\mathrm{Z}^2)/\alpha_\mathrm{s}(\mathrm{m}_\mathrm{Z}^2) \approx 1.7\%$ uncertainty, of which ∼1% is experimental and ∼1.5% theoretical driven by differences between various global fits and observables ($\mathrm{d}^2\sigma/\mathrm{d}x\mathrm{d}Q^2$, $F_\mathrm{L}$, $F_2^{c\bar{c}}$). Incorporation of N$^3$LO corrections, resolving the differences among fits, and/or full global fits of DIS+hadronic data (including consistent treatment of heavy-quark masses) would yield an $\alpha_\mathrm{s}$ extraction with ∼1% uncertainty. High-precision and large-statistics studies at a future DIS machine (such as LHeC or FCC-eh) would ultimately result in $\delta\alpha_\mathrm{s}(\mathrm{m}_\mathrm{Z}^2)/\alpha_\mathrm{s}(\mathrm{m}_\mathrm{Z}^2) \approx 0.15\%$ uncertainties.

8. Jets in DIS and photoproduction [13]: The NNLO$^*$ calculation of jet cross sections in DIS and photoproduction provides an $\alpha_\mathrm{s}$ extraction with $\delta\alpha_\mathrm{s}(\mathrm{m}_\mathrm{Z}^2)/\alpha_\mathrm{s}(\mathrm{m}_\mathrm{Z}^2) \approx 2.5\%$ precision today. Upcoming full-NNLO analyses (combined with new extra $\gamma$-p data, from e.g. "ultraperipheral" hadronic collisions at the LHC) could reduce the $\alpha_\mathrm{s}$ uncertainty to the ∼1.5% level. A future DIS machine (such as LHeC or FCC-eh) would further bring this uncertainty below 1%.

9. Photon structure function [13]: Extractions of $\alpha_\mathrm{s}$ from $\mathrm{F}_2^\gamma(x, Q^2)$ are today limited to NLO analyses of old LEP $\gamma$-$\gamma$ data, with $\delta\alpha_\mathrm{s}(\mathrm{m}_\mathrm{Z}^2)/\alpha_\mathrm{s}(\mathrm{m}_\mathrm{Z}^2) \approx 4.5\%$ roughly equally shared between theoretical and experimental systematics. Extension to NNLO accuracy (and inclusion of possible new B-factories data) would reduce this uncertainty to ∼2%. High-statistics/precision FCC-ee inclusive data on $F_2^\gamma$ can have a big impact on $\alpha_\mathrm{s}$, since the pointlike contribution dominates at high $Q^2$, making the fits independent of the photon PDFs, and resulting in final uncertainties (well) below 1%.

10. Event shapes in $e^+e^-$ [14,15]: The uncertainties in different extractions of $\alpha_\mathrm{s}$ from event shapes in $e^+e^-$ vary in the range $\delta\alpha_\mathrm{s}(\mathrm{m}_\mathrm{Z}^2)/\alpha_\mathrm{s}(\mathrm{m}_\mathrm{Z}^2) \approx (1.5$–$4)\%$ today. There are discrepancies at the same level in the NNLO event-shape (thrust, C-parameter) analyses due to different approaches to correct for (important) hadronization effects (Monte Carlo or combined fits to data). Full resolution of the methodology discrepancies and reduction of the non-pQCD uncertainties (e.g. incorporating off-resonance data from B-factories) should provide $\alpha_\mathrm{s}$ determinations with ∼1.5% precision. High-precision/statistics studies of flavour-tagged event shapes at FCC-ee, both below and above the Z peak, would provide important constraints on the hadronization corrections (whose impact decreases roughly as $1/\sqrt{\mathrm{s}}$), resulting in final $\alpha_\mathrm{s}$ uncertainties (well) below 1%.

11. Jet rates in $e^+e^-$ [16]: Measurements of different jet-rates observables in $e^+e^-$ provide $\alpha_\mathrm{s}$ extractions at NNLO+NLL accuracy with $\delta\alpha_\mathrm{s}(\mathrm{m}_\mathrm{Z}^2)/\alpha_\mathrm{s}(\mathrm{m}_\mathrm{Z}^2) \approx (2$–$5)\%$ uncertainties, mostly of theoretical origin. Improved resummations (NNLL or N$^3$LL) would reduce those to the experimental (∼1.5%) level. At the FCC-ee, increasing the energy from 91 GeV to 350 GeV would further reduce the hadronization corrections and improve the $\alpha_\mathrm{s}$ uncertainty below 1%.

12. W hadronic decays [18]: Although the current state-of-the-art $N^3LO$ calculations of W boson decays would allow one to extract $\alpha_s$ with a ∼0.7% theoretical uncertainty, the relevant (LEP $W^+W^-$) data are poor and provide an extraction with a huge ∼37% uncertainty. Dedicated W-boson studies at the LHC could reduce that to the ∼10% level, but a competitive determination of $\alpha_s$ can only be made with the huge W statistical samples expected at FCC-ee which, combined with improved $N^4LO$ corrections, can ultimately yield $\delta\alpha_s(m_Z^2)/\alpha_s(m_Z^2) \approx 0.15\%$.

13. Z hadronic decays [17,19]: The three closely-related Z boson observables measured at LEP ($R_\ell^0 = \Gamma_{had}/\Gamma_\ell$, $\sigma_0^{had} = 12\pi/m_Z \cdot \Gamma_e\Gamma_{had}/\Gamma_Z^2$, and $\Gamma_Z$) combined with $N^3LO$ calculations, yield an $\alpha_s$ extraction with $\delta\alpha_s(m_Z^2)/\alpha_s(m_Z^2) \approx 2.5\%$ precision, mostly dominated by the experimental systematics. At the FCC-ee, with high-precision measurements of $\Delta m_Z = 0.1$ MeV, $\Delta\Gamma_Z = 0.05$ MeV, $\Delta R_\ell^0 = 10^{-3}$ (achievable thanks to the possibility to perform a threshold scan including energy self-calibration with resonant depolarization) reduces the uncertainty on $\alpha_s$ to $\sim 0.15\%$.

14. Jet cross sections at hadron colliders [22,23,24]: The various jet-rates observables measured at hadron colliders, combined with the corresponding NLO calculations, provide $\alpha_s$ with relative uncertainties in the range $\delta\alpha_s(m_Z^2)/\alpha_s(m_Z^2) \approx (4–5)\%$ (of which the theoretical uncertainties account for ∼3.5%). The imminent incorporation of NNLO corrections and a consistent combination (including correlations) of the multiple datasets available at Tevatron and LHC, may reduce the $\alpha_s$ uncertainties to the 1.5% level in the upcoming years.

15. Top-pair cross sections at hadron colliders [20,21]: The recent availability of NNLO+NNLL calculations of $t\bar{t}$ cross sections in p-p, p-$\bar{p}$ collisions, allowed CMS to extract $\alpha_s$ with ∼2.5% precision (roughly divided between experimental and theoretical systematics). The consistent combination (including correlations) of the multiple $t\bar{t}$ measurements available at Tevatron and LHC, combined with improved determination of the top mass and gluon PDFs, may reduce the $\alpha_s$ uncertainties to the 1.5% level in the upcoming years*.

Assuming that all methods provide consistent $\alpha_s(m_Z^2)$ determinations, a simple weighted-average of the uncertainties expected within the next years for the 15 extraction methods listed in the last column of Table 1†, yields $\delta\alpha_s(m_Z^2)/\alpha_s(m_Z^2) \approx 0.35\%$, i.e. a reduction of a factor of ∼3 with respect to the current (2015) $\alpha_s$ uncertainty. Measurements at FCC-ee would provide $\alpha_s$ with permille-level precision, i.e. $\delta\alpha_s(m_Z^2)/\alpha_s(m_Z^2) \approx 0.1\%$, mostly thanks to the high-precision Z and W measurements expected, complemented with state-of-the-art pQCD theory with small and controlled uncertainties.

## Conclusions

The workshop ended with the summary of the presentations described here. Originally, Guido Altarelli had agreed to summarize the results of the workshop and was in touch with us until a few days before he passed away. His loss is an enormous loss for the QCD and high-energy particle physics community at large.

All in all, the meeting was well-attended and featured lively discussions among different experts on details of their work that have an important impact on $\alpha_s$ measurements. Novel ideas to extract $\alpha_s$,

*This is a statement of the author of this Summary. The authors of [21] believe that it is not possible right now to provide reliable near and medium-term prospects for $\alpha_s$ extractions from $t\bar{t}$ cross sections as it is unclear how the range of current systematics will evolve.

†Conservatively keeping the current ∼1% and 2.5% precision of lattice and Z decays extractions in the next years.

| Method | Current $\delta\alpha_s(m_Z^2)/\alpha_s(m_Z^2)$ uncertainty (theory & experiment state-of-the-art) | Future $\delta\alpha_s(m_Z^2)/\alpha_s(m_Z^2)$ uncertainty (theory & experiment progress) |
|---|---|---|
| lattice | $\approx$ 1% (latt. stats/spacing, N$^3$LO pQCD) | $\approx$ 0.1% ($\sim$10 yrs) (improved computing power, N$^4$LO pQCD) |
| $\pi$ decay factor | $1.5\%_{th} \oplus 0.05\%_{exp} \approx$ 1.5% (N$^3$LO RGOPT) | $1\%_{th} \oplus 0.05\%_{exp} \approx$ 1% (few yrs) (N$^4$LO RGOPT, explicit $m_{u,d,s}$) |
| $\tau$ decays | $1.4\%_{th} \oplus 1.4\%_{exp} \approx$ 2% (N$^3$LO CIPT vs. FOPT) | $0.7\%_{th} \oplus 0.7\%_{exp} \approx$ 1% (+B-factories), <1% (FCC-ee) (N$^4$LO, $\sim$10 yrs. Improved spectral function data) |
| $Q\overline{Q}$ decays | $4\%_{th} \oplus 4\%_{exp} \approx$ 6% (NLO only. $\Upsilon$ only) | $1.4\%_{th} \oplus 1.4\%_{exp} \approx$ 2% (few yrs) (NNLO. More precise LDME and $R_\gamma^{exp}$) |
| soft FFs | $1.8\%_{th} \oplus 0.7\%_{exp} \approx$ 2% (NNLO$^*$ only (+NNLL), npQCD small) | $0.7\%_{th} \oplus 0.7\%_{exp} \approx$ 1% ($\sim$2 yrs), <1% (FCC-ee) (NNLO+NNLL. More precise $e^+e^-$ data: 90–350 GeV) |
| hard FFs | $1\%_{th} \oplus 5\%_{exp} \approx$ 5% (NLO only. LEP data only) | $0.7\%_{th} \oplus 2\%_{exp} \approx$ 2% (+B-factories), <1% (FCC-ee) (NNLO. More precise $e^+e^-$ data) |
| global PDF fits | $1.5\%_{th} \oplus 1\%_{exp} \approx$ 1.7% (Diff. NNLO PDF fits. DIS+DY data) | $0.7\%_{th} \oplus 0.7\%_{exp} \approx$ 1% (few yrs), 0.15% (LHeC/FCC-eh) (N$^3$LO. Full DIS+hadronic data fit) |
| jets in e$^\pm$p, $\gamma$-p | $2\%_{th} \oplus 1.5\%_{exp} \approx$ 2.5% (NNLO$^*$ only) | $1\%_{th} \oplus 1\%_{exp} \approx$ 1.5% (few yrs), < 1% (FCC-eh) (NNLO. Combined DIS + (extra?) $\gamma$-p data) |
| $F_2^\gamma$ in $\gamma$-$\gamma$ | $3.5\%_{th} \oplus 3\%_{exp} \approx$ 4.5% (NLO only) | $1\%_{th} \oplus 2\%_{exp} \approx$ 2% ($\sim$2 yrs), <1% (FCC-ee) (NNLO. More precise new $F_2^\gamma$ data) |
| $e^+e^-$ evt shapes | $(1.5\text{–}4)\%_{th} \oplus 1\%_{exp} \approx$ (1.5–4)% (NNLO+N$^{(3)}$LL, npQCD significant) | $1\%_{th} \oplus 1\%_{exp} \approx$ 1.5% (+B-factories), < 1% (FCC-ee) (NNLO+N$^3$LL. Improved npQCD via $\sqrt{s}$-dep. New data) |
| jets in $e^+e^-$ | $(2\text{–}5)\%_{th} \oplus 1\%_{exp} \approx$ (2–5)% (NNLO+NLL, npQCD moderate) | $1\%_{th} \oplus 1\%_{exp} \approx$ 1.5% (few yrs), < 1% (FCC-ee) (NNLO+NNLL. Improved npQCD. New high-$\sqrt{s}$ data) |
| W decays | $0.7\%_{th} \oplus 37\%_{exp} \approx$ 37% (N$^3$LO, npQCD small. Low-stats data) | $(0.7\text{–}0.1)\%_{th} \oplus (10\text{–}0.1)\%_{exp} \approx$ (10–0.15)% (LHC,FCC-ee) (N$^4$LO, $\sim$10 yrs. High-stats/precise W data) |
| Z decays | $0.7\%_{th} \oplus 2.4\%_{exp} \approx$ 2.5% (N$^3$LO, npQCD small) | $0.1\%_{th} \oplus (0.5\text{–}0.1)\%_{exp} \approx$ (0.5–0.15)% (ILC,FCC-ee) (N$^4$LO, $\sim$10 yrs. High-stats/precise Z data) |
| jets in p-p, p-$\bar{p}$ | $3.5\%_{th} \oplus (2\text{–}3)\%_{exp} \approx$ (4–5)% (NLO only. Combined exp. observables) | $1\%_{th} \oplus 1\%_{exp} \approx$ 1.5% (Tevatron+LHC, $\sim$2 yrs) (NNLO. Multiple datasets+observables) |
| $t\bar{t}$ in p-p, p-$\bar{p}$ | $1.5\%_{th} \oplus 2\%_{exp} \approx$ 2.5% (NNLO+NNLL. CMS only) | $1\%_{th} \oplus 1\%_{exp} \approx$ 1.5% (Tevatron+LHC, $\sim$2 yrs) (Improved $m_{top}^{pole}$ & PDFs. Multiple datasets) |

Table 1: Summary of approximate current and future (theoretical, experimental, total) relative uncertainties in the 15 different $\alpha_s(m_Z^2)$ extraction methods considered in these proceedings. Acronyms and symbols used: $\oplus$=‘quadratic sum’, CPIT=‘contour-improved perturbation theory’, FOPT=‘fixed-order perturbation theory’, RGOPT=‘renormalization-group optimized perturbative theory’, LDME=‘long-distance matrix elements’, npQCD=‘non-perturbative QCD’, NNLO$^*$=‘approximate NNLO’.

estimation of expected reductions in the theoretical and experimental uncertainties of each method, as well as issues to be addressed in the experimental FCC-ee design were discussed. The main results of the workshop, published in these (online) proceedings, will be ultimately incorporated into the FCC-ee Conceptual Design Report under preparation. There was a common agreement of the usefulness of organizing similar dedicated $\alpha_s$ workshops every 1–2 years. Whereas the strong force decreases with energy, the scientific interest in the QCD interaction clearly proves constant, if not increasing, with time.